%% file: paper.tex
\pdfoutput=1
\newcommand*{\ATLASLATEXPATH}{latex/}
\documentclass[UKenglish,cernpreprint,texlive=2016]{\ATLASLATEXPATH atlasdoc}
\def \figurefolder {public}

\usepackage{\ATLASLATEXPATH atlaspackage}
\usepackage{float}
\usepackage{comment}
\usepackage{enumitem}
\usepackage{makecell}
\usepackage{multirow}
\usepackage{caption} 
\usepackage[utf8]{inputenc} 
\usepackage{\ATLASLATEXPATH atlasbiblatex}

\usepackage{\ATLASLATEXPATH atlascontribute}

\usepackage[BSM,process,xref]{\ATLASLATEXPATH atlasphysics}

\usepackage{pdflscape}
\usepackage{adjustbox}

\graphicspath{{logos/}{figures/}}

\usepackage{paper-defs}

\newcommand{\AtlasCoordFootnote}{
ATLAS uses a right-handed coordinate system with its origin at the nominal interaction point (IP)
in the centre of the detector and the $z$-axis along the beam pipe.
The $x$-axis points from the IP to the centre of the LHC ring,
and the $y$-axis points upwards.
Cylindrical coordinates $(r,\phi)$ are used in the transverse plane, 
$\phi$ being the azimuthal angle around the $z$-axis.
The pseudorapidity is defined in terms of the polar angle $\theta$ as $\eta = -\ln \tan(\theta/2)$.
Angular distance is measured in units of $\Delta R \equiv \sqrt{(\Delta\eta)^{2} + (\Delta\phi)^{2}}$.}

\input{paper-metadata}

\hypersetup{pdftitle={ATLAS document},pdfauthor={The ATLAS Collaboration}}

\begin{document}

\maketitle


\clearpage



\input{texfiles/introduction.tex}

\input{texfiles/strategy.tex}

\input{texfiles/samples.tex}
\input{texfiles/simulation.tex}

\input{texfiles/objects.tex}

\input{texfiles/variables.tex}

\input{texfiles/signalregions.tex}

\input{texfiles/backgrounds.tex}

\input{texfiles/systematics.tex}

\input{texfiles/results.tex}

\FloatBarrier
\input{texfiles/conclusion.tex}

\section*{Acknowledgements}
\input{acknowledgements/Acknowledgements}


%

\printbibliography

\clearpage
\input{atlas_authlist}

\end{document}

%% file: paper-metadata.tex
\AtlasTitle{Search for top-squark pair production in final states with one lepton, jets, and missing transverse momentum using 36~fb$^{-1}$ of \sqrtS $pp$ collision data with the ATLAS detector}

\date{\today}

\AtlasRefCode{SUSY-2016-16}

\AtlasNote{SUSY-2016-16}
\PreprintIdNumber{CERN-EP-2017-246}
\AtlasJournalRef{JHEP 06 (2018) 108}
\AtlasDOI{10.1007/JHEP06(2018)108}

\AtlasJournal{JHEP}
\AtlasAbstract{%
The results of a search for the direct pair production of top squarks, the supersymmetric partner of the top quark, in final states with one isolated electron or muon, several energetic jets, and missing transverse momentum are reported. 
The analysis also targets spin-0 mediator models, where the mediator decays into a pair of dark-matter particles and is produced in association with a pair of top quarks. 
The search uses data from proton--proton collisions delivered by the Large Hadron Collider in 2015 and 2016 at a centre-of-mass energy of \sqrtS and recorded by the ATLAS detector, corresponding to an integrated luminosity of 36 fb$^{-1}$. A wide range of signal scenarios with different mass-splittings between the top squark, the lightest neutralino and possible intermediate supersymmetric particles are considered, including cases where the $W$ bosons or the top quarks produced in the decay chain are off-shell.
No significant excess over the Standard Model prediction is observed. 
The null results are used to set exclusion limits at 95\% confidence level in several supersymmetry benchmark models. For pair-produced top-squarks decaying into top quarks, top-squark masses up to 940\,$\GeV$ are excluded.
Stringent exclusion limits are also derived for all other considered top-squark decay scenarios.
For the spin-0 mediator models, upper limits are set on the visible cross-section.
}

%% file: texfiles/introduction.tex
\section{Introduction}
\label{sec:introduction}

The hierarchy problem~\cite{Weinberg:1975gm,Gildener:1976ai,Weinberg:1979bn,Susskind:1978ms} has gained additional attention with the observation of a particle consistent with the Standard Model (SM) Higgs boson~\cite{HIGG-2012-27,CMS-HIG-12-028} at the Large Hadron Collider (LHC)~\cite{LHC:2008}. Supersymmetry (SUSY)~\cite{Miyazawa:1966,Ramond:1971gb,Golfand:1971iw,Neveu:1971rx,Neveu:1971iv,Gervais:1971ji,Volkov:1973ix,Wess:1973kz,Wess:1974tw}, which extends the SM by introducing supersymmetric partners for every SM particle, can provide an elegant solution to the hierarchy problem. The partner particles have identical quantum numbers except for a half-unit difference in spin. The superpartners of the left- and right-handed top quarks, \tleft\ and \tright, mix to form the two mass eigenstates \tone\ and \ttwo\ (top squark or stop), where $\tone$ is the lighter of the two.\footnote{Similarly the \bone\ and \btwo\ (bottom squark or sbottom) are formed by the superpartners of the bottom quarks, \bleft\ and \bright.} If the supersymmetric partners of the top quarks have masses $\lesssim$ 1\,\TeV, loop diagrams involving top quarks, which are the dominant divergent contribution to the Higgs-boson mass, can largely cancel out~\cite{Dimopoulos:1981zb,Witten:1981nf,Dine:1981za,Dimopoulos:1981au,Sakai:1981gr,Kaul:1981hi,Barbieri:1987fn,deCarlos:1993yy}.

Significant mass-splitting between the \tone\ and \ttwo\ is possible due to the large top-quark Yukawa coupling.
Furthermore, effects of the renormalisation group equations are strong for the third-generation squarks, usually driving their masses to values significantly lower than those of the other generations. These considerations suggest a light stop\footnote{The soft mass term of the superpartner of the left-handed bottom quark can be as light as that of the superpartner of the left-handed top quark in certain scenarios as they are both governed mostly by a single mass parameter in SUSY models at tree level. The mass of the superpartner of the right-handed bottom quark is governed by a separate mass parameter from the stop mass parameters, and it is assumed to be larger than 3 \TeV\, having no impact on the signal models considered in this paper.}~\cite{Inoue:1982pi,Ellis:1983ed} which, together with the stringent LHC limits excluding other coloured supersymmetric particles with masses below the \TeV\ level, motivates dedicated stop searches.

The conservation of baryon number and lepton number can be violated in SUSY models, resulting in a proton lifetime shorter than current experimental limits~\cite{Regis:2012sn}. This is commonly resolved by introducing a multiplicative quantum number called $R$-parity, which is $1$ and $-1$ for all SM and SUSY particles (sparticles), respectively. A generic $R$-parity-conserving minimal supersymmetric extension of the SM (MSSM)~\cite{Fayet:1976et,Fayet:1977yc,Farrar:1978xj,Fayet:1979sa,Dimopoulos:1981zb} predicts pair production of SUSY particles and the existence of a stable lightest supersymmetric particle (LSP). 

The charginos \chinoOneTwopm\ and neutralinos \ninoOneTwoThreeFour\ are the mass eigenstates formed from the linear superposition of the charged and neutral SUSY partners of the Higgs and electroweak gauge bosons (higgsinos, winos and binos). They are referred to in the following as electroweakinos. In a large variety of SUSY models, the lightest neutralino (\ninoone) is the LSP, which is also the assumption throughout this paper. The LSP provides a particle dark-matter (DM) candidate, as it is stable and interacts only weakly~\cite{Goldberg:1983nd,Ellis:1983ew}. 

This paper presents a search for direct \tone\ pair production in final states with exactly one isolated charged lepton (electron or muon,\footnote{Electrons and muons from $\tau$ decays are included.} henceforth referred to simply as `lepton') from the decay of either a real or a virtual $W$ boson. In addition the search requires several jets and a significant amount of missing transverse momentum $\Ptmiss$, the magnitude of which is referred to as \met, from the two weakly interacting LSPs that escape detection.
Results are also interpreted in an alternative model where a spin-0 mediator is produced in association with top quarks and subsequently decays into a pair of DM particles.

Searches for direct \tone\ pair production were previously reported by the ATLAS~\cite{Aaboud:2017wqg,Aaboud:2017nfd,SUSY-2016-20,SUSY-2015-02,SUSY-2014-07} and CMS~\cite{Sirunyan:2018iwl,Sirunyan:2018vjp,Sirunyan:2017leh,CMS-SUS-16-033,CMS-SUS-16-036,CMS-SUS-16-051,CMS-SUS-16-049,CMS-SUS-16-008,CMS-SUS-15-005,CMS-SUS-15-004,CMS-SUS-14-006,CMS-SUS-13-024,CMS-SUS-13-014,CMS-SUS-13-011,CMS-SUS-12-028,CMS-SUS-12-005} collaborations, as well as by the CDF and D\O\ collaborations (for example~\cite{PhysRevLett.104.251801, D0_stopSearch}) and the LEP collaborations~\cite{lepsusy_web_stop}.
The exclusion limits obtained by previous ATLAS searches for stop models with massless neutralinos reach $\sim 950$ \GeV\ for direct two-body decays \topLSP, $\sim 560$ \GeV\ for the three-body process \threeBody, and $\sim 400$ \GeV\ for four-body decays \fourBody, all at the 95\% confidence level.
Searches for spin-0 mediators decaying into a pair of DM particles and produced in association with heavy-flavour quarks have also been reported with zero or two leptons in the final state by the ATLAS collaboration~\cite{DMhfRun2}, and by the CMS collaboration~\cite{CMS-EXO-16-005,Sirunyan:2017leh}.

%% file: texfiles/strategy.tex
\section{Search strategy}
\label{sec:strategy}

\subsection{Signal models}
\label{subsec:signal_models}

\begin{figure}[htbp]
\begin{center}
\includegraphics[width=0.40\textwidth]{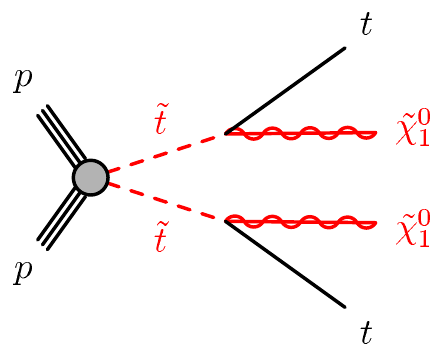}
\includegraphics[width=0.40\textwidth]{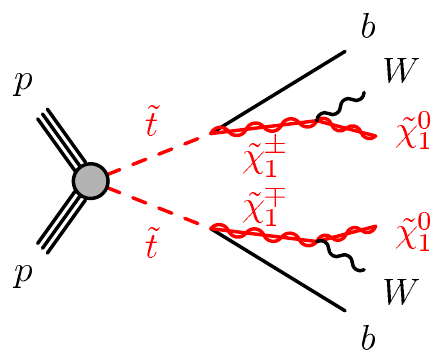}
\caption{
Diagrams illustrating the stop decay modes, which are referred to as (left) \topLSP\ and (right) \bChargino. 
Sparticles are shown as red lines. In these diagrams, the charge-conjugate symbols are omitted for simplicity. The direct stop production begins with a top squark--antisquark pair. 
}
\label{fig:diagram}
\end{center}
\end{figure}

The experimental signatures of stop pair production can vary dramatically, depending on the spectrum of low-mass SUSY particles. Figure~\ref{fig:diagram} illustrates two typical stop signatures: \topLSP\ and \bChargino. Other decay and production modes such as $\topNLSP$ and $\topNNLSP$, and sbottom direct pair production are also considered. 
The analysis attempts to probe a broad range of possible scenarios, taking the approach of defining dedicated search regions to target specific but representative SUSY models. The phenomenology of each model is largely driven by the composition of its lightest sparticles, which are considered to be some combination of the electroweakinos. In practice, this means that the most important parameters of the SUSY models considered are the masses of the electroweakinos and of the colour-charged third-generation sparticles.

In this search, the targeted signal scenarios are either simplified models~\cite{Alwall:2008ve,Alwall:2008ag,Alves:2011wf}, in which the masses of all sparticles are set to high values except for the few sparticles involved in the decay chain of interest, or models based on the phenomenological MSSM (pMSSM)~\cite{Djouadi:1998di,Berger:2008cq}, in which all of the 19 pMSSM parameters are set to fixed values, except for two which are scanned. The set of models used are chosen to give a broad coverage of the possible stop decay patterns and phenomenology that can be realised in the MSSM, in order to best demonstrate the sensitivity of the search for direct stop production. The simplified models used are designed with a goal of covering distinct phenomenologically different regions of pMSSM parameter space. 

The pMSSM parameters $m_{tR}$ and $m_{q3L}$ specify the \tright\ and \tleft\ soft mass terms, with the smaller of the two controlling the \tone\ mass. In models where the \tone\ is primarily composed of \tleft, the production of light sbottoms (\bone) with a similar mass is also considered. The mass spectrum of electroweakinos and the gluino is given by the running mass parameters $M_1$, $M_2$, $M_3$, and $\mu$, which set the masses of the bino, wino, gluino, and higgsino, respectively. If the mass parameters, $M_1$, $M_2$, and $\mu$, are comparably small, the physical LSP is a mixed state, composed of multiple electroweakinos. 
Other relevant pMSSM parameters include $\beta$, which gives the ratio of vacuum expectation values of the up- and down-type Higgs bosons influencing the preferred decays of the stop, the SUSY breaking scale ($M_S$) defined as $M_S = \sqrt{m_{\tone}m_{\ttwo}}$, and the top-quark trilinear coupling ($A_t$). In addition, a maximal $\tleft$--$\tright$ mixing condition, $X_t/M_S \sim \sqrt{6}$ (where $X_t = A_t - \mu/\tan\beta$), is assumed to obtain a low-mass stop (\tone) while the models remain consistent with the observed Higgs boson mass of 125\,\GeV~\cite{HIGG-2012-27,CMS-HIG-12-028}.

\begin{figure}[htbp]
\begin{center}
\includegraphics[width=0.99\textwidth]{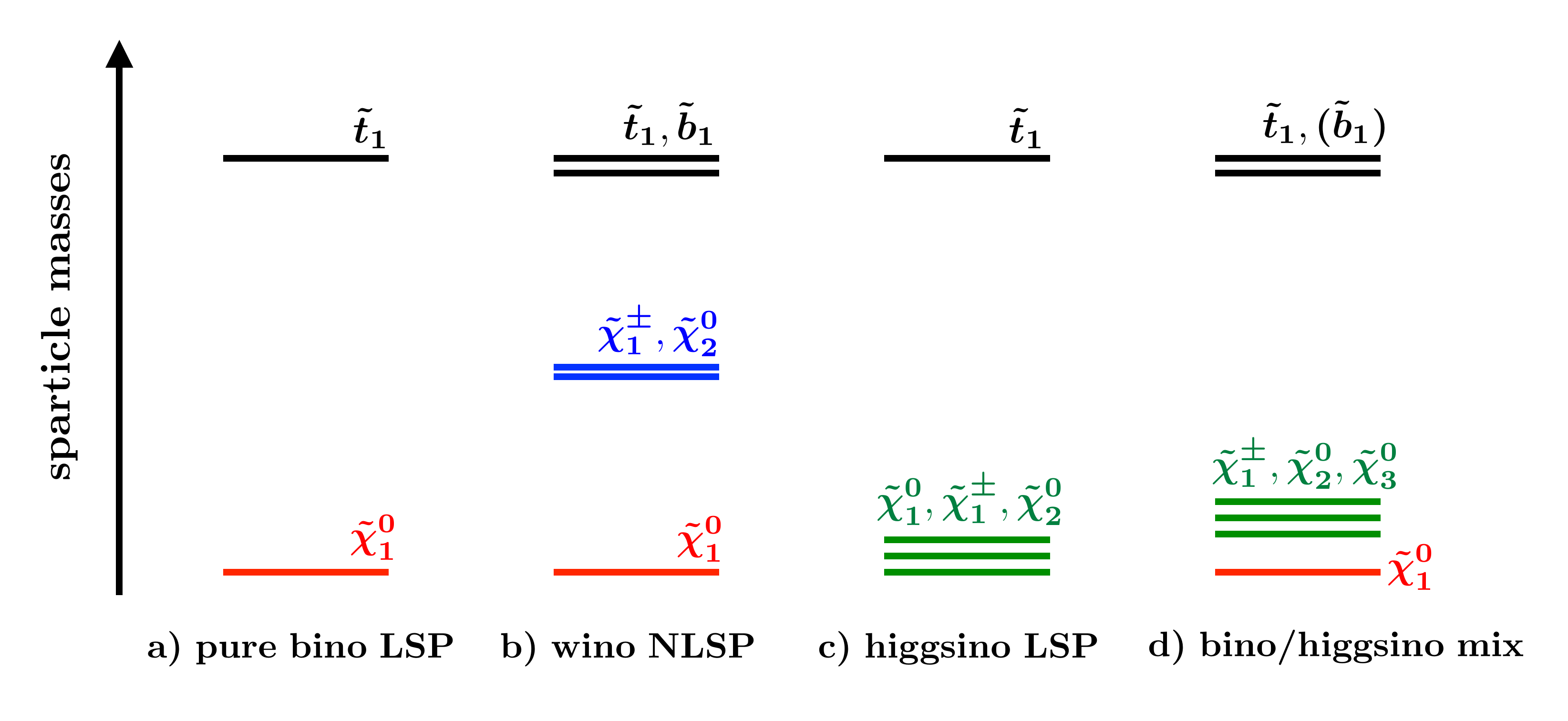}
\caption{Illustration of the sparticle mass spectrum for various LSP scenarios: a) pure bino LSP, b) wino NLSP, c) higgsino LSP, and d) bino/higgsino mixed LSP. The \tone\ and \bone, shown as black lines, decay into various electroweakino states: the bino state (red lines), wino state (blue lines), or higgsino state (green lines), possibly with the subsequent decay into the LSP. The light sbottom (\bone) is considered only for pMSSM models with $m_{q3L}<$ $m_{tR}$.}
\label{fig:sparticle_mass_spectrum}
\end{center}
\end{figure}

In this search, four scenarios\footnote{For the higgsino LSP scenarios, three sets of model assumptions are considered, each giving rise to different stop BRs for $\bChargino$, $\topLSP$, and \topNLSP.} are considered, where each signal scenario is defined by the nature of the LSP and the next-to-lightest supersymmetric particle (NLSP): (a) pure bino LSP, (b) bino LSP with a light wino NLSP, (c) higgsino LSP, and (d) mixed bino/higgsino LSP, which are detailed below with the corresponding sparticle mass spectra illustrated in Figure~\ref{fig:sparticle_mass_spectrum}. Complementary searches target scenarios where the LSP is a pure wino (yielding a disappearing track signature~\cite{SUSY-2013-01,CMS-EXO-12-034} common in anomaly-mediated models~\cite{Giudice:1998xp,Randall:1998uk} of SUSY breaking) as well as other LSP hypotheses (such as gauge-mediated models~\cite{Dine:1981gu,AlvarezGaume:1981wy,Nappi:1982hm}), which are not discussed further.

\begin{enumerate}[label=(\alph*)]

\item Pure bino LSP model: 

A simplified model is considered for the scenario where the only light sparticles are the stop (composed mainly of \tright) and the lightest neutralino. 
When the stop mass is greater than the sum of the top quark and LSP masses, the dominant decay channel is via \topLSP. If this decay is kinematically disallowed, the stop can undergo a three-body decay, $\threeBody$, when the stop mass is above the sum of masses of the bottom quark, $W$ boson, and $\ninoone$. Otherwise the decay proceeds via a four-body process, $\fourBody$, where $f$ and $f'$ are two distinct fermions, or via a flavour-changing neutral current (FCNC) process, such as the loop-suppressed \charmDecay. Given the very different final state, the FCNC decay is not considered further in this search, and therefore a 100\% branching ratio (BR) to $\fourBody$ is assumed. For very small splittings between the stop and neutralino masses the stop lifetime can become significant~\cite{Grober:2014aha}. In the simplified model considered in this paper the stop is always assumed to decay promptly, regardless of the mass splitting. The various \tone\ decay modes in this scenario are illustrated in Figure~\ref{fig:stopDecays}. The region of phase space along the line of $m_{\tone} = m_{\ninoone} + m_\mathrm{top}$ is especially challenging to target because of the similarity of the stop signature to the \ttbar\ process, and is referred to in the following as the `diagonal region'. 

\begin{figure}
\center
\includegraphics[width=0.99\textwidth]{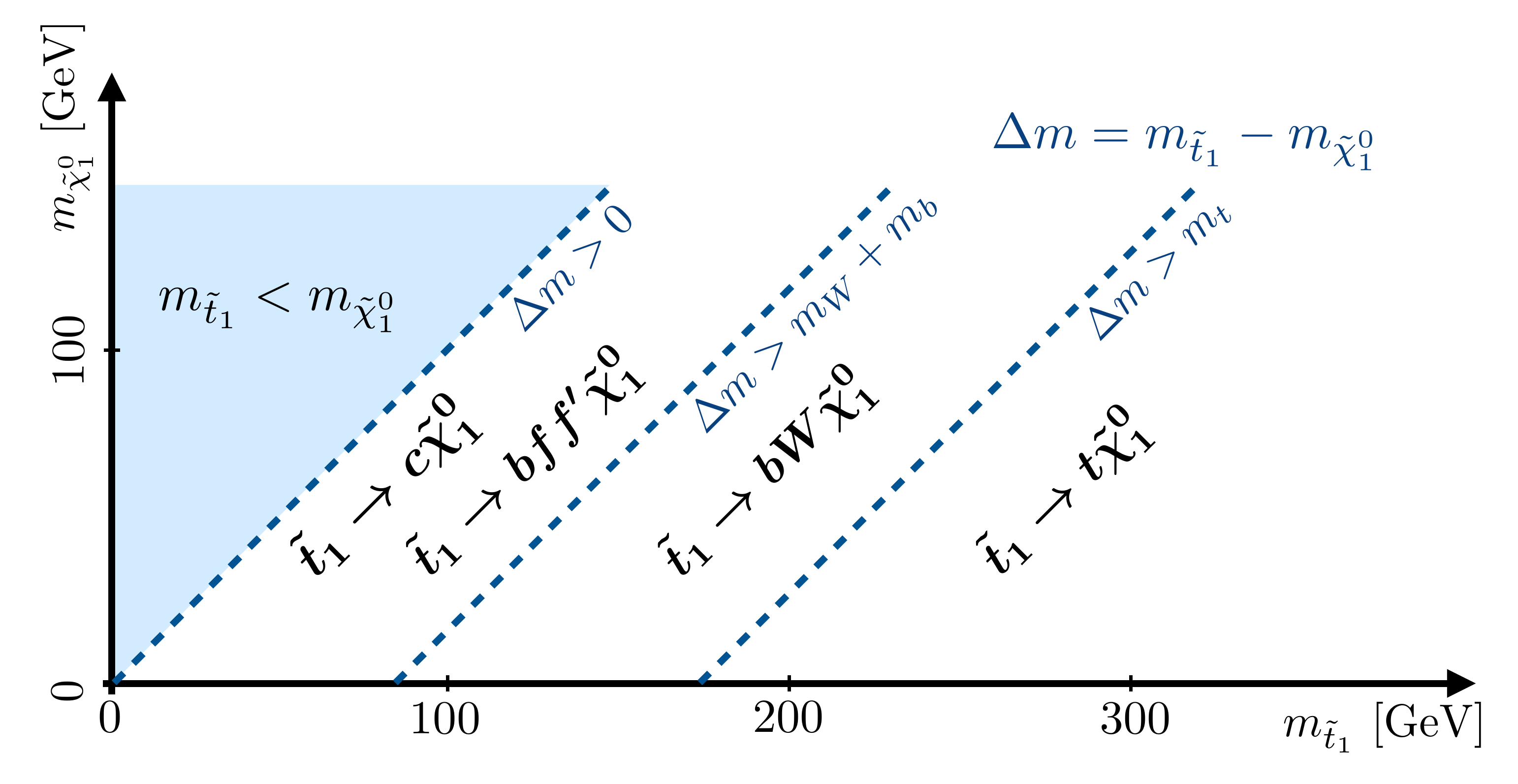}
\caption{
Illustration of the preferred stop decay modes in the plane spanned by the masses of the stop (\tone) and the lightest neutralino (\ninoone), where the latter is assumed to be the lightest supersymmetric particle. Stop decays into supersymmetric particles other than the lightest supersymmetric particle are not displayed.
\label{fig:stopDecays}}
\end{figure}

\item Wino NLSP model: 

A pMSSM model is designed such that a wino-like chargino (\chinoonepm) and neutralino (\ninotwo) are mass-degenerate, with the bino as the LSP. This scenario is motivated by models with gauge unification at the GUT scale such as the cMSSM or mSugra~\cite{Chamseddine:1982jx,Barbieri:1982eh,Kane:1993td}, where $M_2$ is assumed to be twice as large as $M_1$, leading to the \chinoonepm\ and \ninotwo\ having masses nearly twice as large as that of the bino-like LSP.

In this scenario, additional decay modes for the stop (composed mainly of \tleft) become relevant, such as the decay into a bottom quark and the lightest chargino (\bChargino) or the decay into a top quark and the second neutralino (\topNLSP). The $\chinoonepm$ and $\ninotwo$ subsequently decay into $\ninoone$ via emission of a (potentially off-shell) $W$ boson or $Z$/Higgs ($h$) boson, respectively. The \bChargino\ decay is considered for a chargino mass above about $100$\,\GeV\ since the LEP limit on the lightest chargino is $m_{\chinoonepm} > 103.5$\,\GeV~\cite{lepsusy_web_chargino}.

An additional \bChargino\ decay signal model (simplified model) is designed, motivated by a scenario with nearly equal masses of the \tone and \chinoonepm.  The model considered assumes the mass-splitting between the $\tone$ and $\chinoonepm$, $\Delta m(\tone,\chinoonepm)=10$ \GeV\ and that the top squark decays via the process $\bChargino$ with a BR of 100\%. In this scenario, the jets originating from the bottom quarks are too low in energy (soft) to be reconstructed and hence the signature is characterised by large $\met$ and no jets initiated by bottom quarks (referred to as $b$-jets).

\item Higgsino LSP model: 

`Natural' models of SUSY~\cite{naturalSUSY,Barbieri:1987fn,deCarlos:1993yy} suggest low-mass stops and a higgsino-like LSP. In such scenarios, a typical $\Delta m(\chinoonepm,\ninoone)$ varies between a few hundred \MeV\ to several tens of \GeV\, depending mainly on the mass relations amongst the electroweakinos. For this analysis, a simplified model is designed for various $\Delta m(\chinoonepm,\ninoone)$ of up to 30 \GeV\ satisfying the mass relation as follows:
\[
\Delta m(\chinoonepm,\ninoone) = 0.5 \times \Delta m(\ninotwo,\ninoone).
\]

The stop decays into either $b \chinoonepm$, $t \ninoone$, or $t \ninotwo$, followed by the \chinoonepm\ and \ninotwo\ decay through the emission of a highly off-shell $W/Z$ boson. Hence the signature is characterised by low-momentum leptons or jets from off-shell $W/Z$ bosons, and the analysis benefits from reconstructing low-momentum leptons (referred to as soft leptons). The stop decay BR strongly depends on the \tright\ and \tleft\ composition of the stop. Stops composed mainly of \tright\ have a large BR $\mathcal{B}(\bChargino)$, whereas stops composed mainly of \tleft\ have a large $\mathcal{B}(\topLSP)$ or $\mathcal{B}(\topNLSP)$. In this search, the three cases are considered separately: $\tone\sim\tright$, $\tone\sim\tleft$, and a case in which the stop decays democratically into the three decay modes.

\item Bino/higgsino mix model: 

The `well-tempered neutralino'~\cite{ArkaniHamed:2006mb} scenario seeks to provide a viable dark-matter candidate while simultaneously addressing the problem of naturalness by targeting an LSP that is an admixture of bino and higgsino.  The mass spectrum of the electroweakinos (higgsinos and bino) is expected to be slightly compressed, with a typical mass-splitting between the bino and higgsino states of $20$--$50$\,$\GeV$. A pMSSM signal model is designed such that only a low level of fine-tuning~\cite{Barbieri:1987fn,SUSY-2014-08} of the pMSSM parameters is needed and the annihilation rate of neutralinos is consistent with the observed dark-matter relic density\footnote{The quantities $\Omega$ and $H_0$ are the density parameter and Hubble constant, respectively.} ($0.10<\Omega H_0^2<0.12$)~\cite{relic_density}.

\end{enumerate}

\begin{figure}[htbp]
\begin{center}
\includegraphics[width=0.45\textwidth]{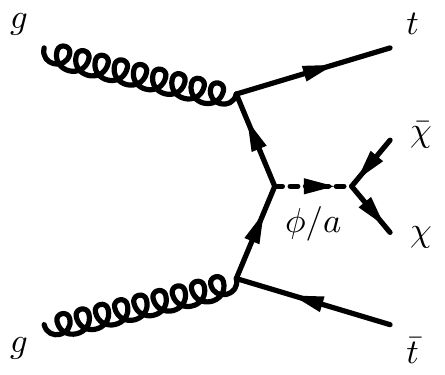}
\caption{A representative Feynman diagram for spin-0 mediator production. The $\phi$/$a$ is the scalar/pseudoscalar mediator, which decays into a pair of dark-matter ($\chi$) particles.}
\label{fig:diagram_DM}
\end{center}
\end{figure}

The final state produced by many of the models described above is consistent with a $\ttbar+\met$ final state. Exploiting the similarity, signal models with a spin-0 mediator decaying into dark-matter particles produced in association with \ttbar\ are also studied assuming either a scalar ($\phi$) or a pseudoscalar ($a$) mediator~\cite{Abercrombie:150700966,DMhfRun2}.
An example diagram for this process is shown in Figure~\ref{fig:diagram_DM}.

\subsection{Analysis strategy}
\label{subsec:analysis_strategy}

The search presented is based on 16 dedicated analyses that target the various scenarios mentioned above. Each of these analyses corresponds to a set of event selection criteria, referred to as a signal region (SR), and is optimised to target one or more signal scenarios. Two different analysis techniques are employed in the definition of the SRs, which are referred to as `cut-and-count' and `shape-fit'. 
The former is based on counting events in a single region of phase space, and is employed in the 16 analyses. The latter is used in some SRs in addition to the `cut-and-count` technique and employs SRs split into multiple bins in a specific discriminating kinematic variable, that can cover a range that is larger than the `cut-and-count' SR. 
By utilising different signal-to-background ratios in the various bins, the search sensitivity is enhanced in challenging scenarios where it is particularly difficult to separate signal from background.

The main background processes after the signal selections include \ttbar, single-top $Wt$, $\ttbar+Z(\rightarrow \nu\bar{\nu})$, and $W$+jets. Each of those SM processes are estimated by building dedicated control regions (CRs) enhanced in each of the processes, making the analysis more robust against potential mis-modelling effects in simulated events and reducing the uncertainties in the background estimates. The backgrounds are then simultaneously normalised in data using a likelihood fit for each SR with its associated CRs. The background modelling as predicted by the fits is tested in a series of validation regions (VRs).

%% file: texfiles/samples.tex
\section{ATLAS detector and data collection}
\label{sec:atlasdetector}

The ATLAS detector~\cite{PERF-2007-01} is a multipurpose particle physics detector with nearly $4\pi$ coverage in solid angle around the collision point.\footnote{\AtlasCoordFootnote  The transverse momentum, \pt, is defined with respect to the beam axis ($x$--$y$ plane).} It consists of an inner tracking detector (ID),  surrounded by a superconducting solenoid providing a \SI{2}{\tesla} axial magnetic field, a system of calorimeters, and a muon spectrometer (MS) incorporating three large superconducting toroid magnets.  

The ID provides charged-particle tracking in the range $|\eta| < 2.5$. During the LHC shutdown between Run 1 (2010--2012) and Run 2 (2015--2018), a new innermost layer of silicon pixels was added~\cite{ATLAS-TDR-19}, which improves the track impact parameter resolution, vertex position resolution and $b$-tagging performance~\cite{ATL-PHYS-PUB-2016-012}.  

High-granularity electromagnetic and hadronic calorimeters cover the region $|\eta| < 4.9$. The central hadronic calorimeter is a sampling calorimeter with scintillator tiles as the active medium and steel absorbers. All the electromagnetic calorimeters, as well as the endcap and forward hadronic calorimeters, are sampling calorimeters with liquid argon as the active medium and lead, copper, or tungsten absorbers.
The MS consists of three layers of high-precision tracking chambers with coverage up to $|\eta|=2.7$ and dedicated chambers for triggering in the region $|\eta|<2.4$. 
Events are selected by a two-level trigger system~\cite{Aaboud:2016leb}: the first level is a hardware-based system and the second is a software-based system.

This analysis is based on a dataset collected in 2015 and 2016 at a collision energy of $\sqrt{s} = 13$\,\TeV. The data contain an average number of simultaneous $pp$ interactions per bunch crossing, or ``pile-up'', of approximately 23.7 across the two years.
After the application of beam, detector and data-quality requirements, the total integrated luminosity is \ourLumi\ with an associated uncertainty of 3.2\%. The uncertainty is derived following a methodology similar to that detailed in Ref.~\cite{DAPR-2011-01} from a preliminary calibration of the luminosity scale using a pair of $x$--$y$ beam separation scans performed in August 2015 and June 2016.

The events were primarily recorded with a trigger logic that accepts events with \met\ above a given threshold. 
The trigger is fully efficient for events passing an offline-reconstructed $\met > 230$\,\GeV\ requirement, which is the minimum requirement deployed in the signal regions and control regions relying on the \met\ triggers.
To recover acceptance for signals with moderate \met, events having a well-identified lepton with a minimum \pt\ at trigger level are also accepted for several selections.
Events in which the offline reconstructed $\met $ is measured to be less than $ 230$\,\GeV\ are collected using single-lepton triggers, where the thresholds are set to obtain a constant efficiency as a function of the lepton \pT\ of $\approx$90\% ($\approx$80\%) for electrons (muons).

%% file: texfiles/simulation.tex
\section{Simulated event samples}
\label{sec:simulation}

Samples of Monte Carlo (MC) simulated events are used for the description of the SM background processes and to model the signals. 
Details of the simulation samples used, including the matrix element (ME) event generator and parton distribution function (PDF) set, the parton shower (PS) and hadronisation model, the set of tuned parameters (tune) for the underlying event (UE) and the order of the cross-section calculation, are summarised in Table~\ref{tab:mc_samples1}.

\begin{table}[htbp]
\footnotesize
\centering
\caption{Overview of the nominal simulated samples.
}
\vspace{3mm}
\makebox[\textwidth][c]{
\begin{tabular}{| l | lllll |}
\hline\hline
Process & ME event generator & ME PDF  & PS and & UE tune   & Cross-section\\
        &           & & hadronisation &  & calculation \\
\hline
$\ttbar$ & \powbox~v2~\cite{Alioli:2010xd} & CT10~\cite{Lai:2010vv}& \PYTHIA 6~\cite{Sjostrand:2006za} & P2012~\cite{Skands:2010ak} &NNLO+NNLL~\cite{Czakon:2013goa,Czakon:2012pz,Czakon:2012zr,Baernreuther:2012ws,Cacciari:2011hy,Czakon:2011xx} \\
Single-top  & & & & & \\
\quad $t$-channel  & \powbox~v1 & CT104f & \PYTHIA 6 & P2012 &NNLO+NNLL~\cite{Kidonakis:2011wy} \\
\quad $s$- and $Wt$-channel & \powbox~v2 & CT10 & \PYTHIA 6 & P2012 &NNLO+NNLL~\cite{Kidonakis:2010ux,Kidonakis:2010tc} \\
$V$+jets $(V=W/Z)$ & \SHERPA~2.2.0~\cite{Gleisberg:2008ta} & NNPDF3.0~\cite{Ball:2014uwa} & \SHERPA & Default & NNLO~\cite{Catani:2009sm}\\
Diboson & \SHERPA~2.1.1 -- 2.2.1 & CT10/NNPDF3.0 & \SHERPA & Default & NLO\\
$\ttbar+V $ & \MGaMC~2.2.2~\cite{Alwall:2014hca} & NNPDF3.0 & \PYTHIA 8~\cite{Sjostrand:2007gs} & A14~\cite{ATL-PHYS-PUB-2014-021} & NLO~\cite{Alwall:2014hca}\\
SUSY signal & \MGaMC~2.2 -- 2.4 & NNPDF2.3~\cite{Ball:2012cx} & \PYTHIA 8 & A14 & NLO+NLL~\cite{Borschensky:2014cia}\\
DM signal & \MGaMC~2.3.3 & NNPDF2.3 & \PYTHIA 8 & A14 & NLO\\
\hline\hline
\end{tabular}
}
\label{tab:mc_samples1}
\end{table}

The samples produced with \MGaMC~\cite{Alwall:2014hca} and {\textsc{Powheg-Box}}~\cite{Alioli:2010xd,Re:2010bp,Frixione:2007nw,Frederix:2012dh,Alioli:2009je} used {\textsc{EvtGen}} v1.2.0~\cite{EvtGen} for the modelling of $b$-hadron decays.
The signal samples were all processed with a fast simulation~\cite{SOFT-2010-01}, whereas all background samples were processed with the full simulation of the ATLAS detector~\cite{SOFT-2010-01} based on GEANT4~\cite{Agostinelli:2002hh}. All samples were produced with varying numbers of minimum-bias interactions overlaid on the hard-scattering event to simulate the effect of multiple $pp$ interactions in the same or nearby bunch crossings. 
The number of interactions per bunch crossing was reweighted to match the distribution in data.

\subsection{Background samples}
\label{sec:background_samples}

The nominal \ttbar~sample and single-top sample cross-sections
were calculated to next-to-next-to-leading order (NNLO) with the resummation of soft gluon emission at next-to-next-to-leading-logarithm (NNLL) accuracy and were generated with {\textsc{Powheg-Box}} (NLO) interfaced to {\textsc{Pythia}}6 for parton showering and hadronisation. Additional $t\bar{t}$ samples were generated with \MGaMC\ (NLO)+{\textsc{Pythia}}8, {\textsc{Sherpa}}, and {\textsc{Powheg-Box}}+{\textsc{Herwig++}}~\cite{Bahr:2008pv,Bellm:2015jjp} for modelling comparisons and evaluation of systematic uncertainties.

Additional samples for $WWbb$, $Wt+b$, and $\ttbar$ were generated with \MGaMC\ leading order (LO) interfaced to {\textsc{Pythia}}8, in order to assess the effect of interference between the singly and doubly resonant processes as a part of the $Wt$ theoretical modelling systematic uncertainty.

Samples for \Wjets, \Zjets\ and diboson production were generated with {\textsc{Sherpa}} 2.2.0~\cite{Gleisberg:2008ta} (and {\textsc{Sherpa}} 2.1.1 -- 2.2.1 for the latter) using Comix~\cite{Gleisberg:2008fv} and OpenLoops~\cite{Cascioli:2011va}, and merged with the {\textsc{Sherpa}} parton shower~\cite{Schumann:2007mg} using the {\textsc{ME+PS@NLO}} prescription~\cite{Hoeche:2012yf}. The NNPDF30 PDF set~\cite{Ball:2014uwa} was used in conjunction with a dedicated parton shower tuning developed by the {\textsc{Sherpa}} authors. The $W/Z$ + jets events were further normalised with the NNLO cross-sections.

The $\ttbar+V $ samples were generated with \MGaMC\ (NLO) interfaced to {\textsc{Pythia}}8 for parton showering and hadronisation. {\textsc{Sherpa}} (NLO) samples were used to evaluate the systematic uncertainties related to the modelling of $\ttbar+V $ production.

More details of the \ttbar, \Wjets, \Zjets, diboson and $\ttbar+V$ samples can be found in Refs.~\cite{ATL-PHYS-PUB-2016-004,ATL-PHYS-PUB-2016-003,ATL-PHYS-PUB-2016-002,ATL-PHYS-PUB-2016-005}.

\subsection{Signal samples}
\label{sec:signal_samples}

Signal SUSY samples were generated at leading order (LO) with \MGaMC\, including up to two extra partons, and interfaced to {\textsc{Pythia}}8 for parton showering and hadronisation.
For the pMSSM models, the sparticle mass spectra were calculated using Softsusy 3.7.3~\cite{Allanach:2001kg,Allanach:2013kza}. The output mass spectrum was then interfaced to {\textsc{HDECAY}} 3.4~\cite{hdecay} and {\textsc{SDECAY}} 1.5/1.5a~\cite{sdecay} to generate decay tables for each of the sparticles. The decays of the \ninotwo\ and \chinoonepm\ via highly off-shell $W/Z$ bosons were computed by taking into account the mass of $\tau$ leptons and charm quarks in the low $\Delta m(\chinoonepm/\ninotwo,\ninoone)$ regime. 
For all models considered the decays of SUSY particles are prompt. 
The details of the various simulated samples in the four LSP scenarios targeted are given below. The input parameters for the pMSSM models are summarised in Table~\ref{tab:pMSSMparm}.

\begin{enumerate}[label=(\alph*)]

\item Pure bino LSP: 

For the \topLSP\ samples, the stop was decayed in {\textsc{Pythia}}8 using only phase space considerations and not the full matrix element. Since the decay products of the samples generated did not preserve spin information, a polarisation reweighting was applied\footnote{A value of cos$\theta_{t}=0.553$ is assumed, corresponding to a \tone\ composed mainly of \tright ($\sim$70\%)} following Refs.~\cite{stopPol1,stopPol2}. For the \threeBody\ and \fourBody\ samples, the stop was decayed with MadSpin~\cite{Artoisenet:2012st}, interfaced to {\textsc{Pythia}}8. MadSpin emulates kinematic distributions such as the mass of the $bW$ system to a good approximation without calculating the full ME. For the MadSpin samples, the stop was assumed to be composed mainly of \tright ($\sim$70\%), consistent with the \topLSP\ samples.

\item Wino NLSP: 

In the wino NLSP model, the \tone\ was assumed to be composed mainly of \tleft\ (i.e. $m_{q3L}<$ $m_{tR}$). The stop was decayed according to $\mathcal{B}(\bChargino)\sim 66$\%, or $\mathcal{B}(\topNLSP)\sim 33$\%, followed by $\chinoonepm$ and $\ninotwo$ decays into the LSP, in a large fraction of the phase space. Since the coupling of \tleft\ to the wino states is larger than the one to the bino state, the stop decay into the bino state (\topLSP) is suppressed. The branching ratio (BR) can be significantly different in the regions of phase space where one of the decays is kinematically inaccessible. In the case that a mass-splitting between the \tone\ and \ninotwo\ is smaller than the top-quark mass ($\Delta m(\tone,\ninotwo)< m_\mathrm{top}$), for instance, the \topNLSP\ decay is suppressed, while the \bChargino\ decay is enhanced. Similarly, the \bChargino\ decay is suppressed near the boundary of $m_{\tone} = m_b + m_{\chinoonepm}$ while the \topLSP\ decay is enhanced.

The signal model was constructed by performing a two-dimensional scan of the pMSSM parameters $M_1$ and $m_{q3L}$.  For the models considered, $M_3$ $=$ 2.2\,\TeV\ and $M_S$ $=$ 1.2\,\TeV\ were assumed in order for the produced models to evade the current gluino and stop mass limits~\cite{Aaboud:2017ayj,Aaboud:2017hdf,SUSY-2015-10,SUSY-2015-06,Sirunyan:2018iwl,Sirunyan:2018vjp,Sirunyan:2017leh,CMS-SUS-16-033,CMS-SUS-16-036,CMS-SUS-16-051,CMS-SUS-16-049,CMS-SUS-16-008,CMS-SUS-16-042,CMS-SUS-16-037}. 

The \ninotwo\ decay modes are very sensitive to the sign of $\mu$. The \ninotwo\ decays into the lightest Higgs boson and the LSP (with $\mathcal{B}(\ninotwo \to h \ninoone) \sim 95$\%) if $\mu > 0$ and decays into a $Z$ boson and the LSP (with $\mathcal{B}(\ninotwo \to Z \ninoone)\sim 75$\%) if $\mu < 0$. Hence, the two $\mu$ scenarios were considered separately.\footnote{When the \ninotwo\ decay into the LSP via $Z$/Higgs boson is kinematically suppressed, the decay is instead determined by the LSP coupling to squarks. In the low-$m_{q3L}$ scenario considered, the decay via a virtual sbottom becomes dominant due to the large sbottom--bottom--LSP coupling, resulting in a $\ninotwo \rightarrow b\bar{b} \ninoone$ decay with a branching ratio up to 95\%.}

Both the stop and sbottom pair production modes were included. The stop and sbottom masses are roughly the same since they are both closely related to $m_{q3L}$. The sbottom decays largely via \tChargino\ and \bottomNLSP\ with a similar BR as for \bChargino\ and $\topNLSP$, respectively.

\item Higgsino LSP: 

For the higgsino LSP case, a simplified model was built. Similar input parameters to those of the wino NLSP pMSSM model were assumed when evaluating the stop decay branching ratios, except for the electroweakino mass parameters, $M_1$, $M_2$, and $\mu$. These mass parameters were changed to satisfy $\mu \ll M_1, M_2$.

The stop decay BR in scenarios with $m_{tR} < m_{q3L}$ were found to be $\sim 50$\% for $\mathcal{B}(\bChargino)$ and $\sim 25$\% for both $\mathcal{B}(\topLSP)$ and $\mathcal{B}(\topNLSP)$, independent of $\tan\beta$. On the other hand, in scenarios with $m_{q3L} <$ $m_{tR}$ and $\tan\beta = 20$, the $\mathcal{B}(\bChargino)$ was suppressed to $\sim 10$\% while $\mathcal{B}(\topLSP)$ and $\mathcal{B}(\topNLSP)$ were each increased to $\sim 45$\%. A third scenario with $\tan{\beta}=60$ and $m_{q3L}<$$m_{tR}$ was also studied. In this scenario, the stop BR was found to be $\sim 33$\% for each of the three decay modes. The \chinoonepm\ and \ninotwo\ subsequently decayed into the \ninoone\ via a highly off-shell $W/Z$ boson. The exact decay BR of \chinoonepm\ and \ninotwo\ depend on the size of the mass-splitting amongst the triplet of higgsino states. For the baseline model, $\Delta m(\chinoonepm,\ninoone)=5$\,\GeV\ and $\Delta m (\ninotwo,\ninoone)=10$\,\GeV\ were assumed, which roughly corresponds to $M_1 = M_2 \sim 1.2$--$1.5$\,$\TeV$. An additional signal model with $\Delta m(\chinoonepm,\ninoone)$ varying between 0 and 30\,\GeV\ was also considered. 

In the signal generation, the stop decay BR was set to 33\% for each of the three decay modes (\bChargino, \topNLSP, \topLSP). The polarisation and stop BR were reweighted to match the BR described above for each scenario. Samples were simulated down to $\Delta m(\chinoonepm,\ninoone)=2$\,$\GeV$ for the $\Delta m$ scan. The \topLSP\ samples generated for the pure bino scenario were used in the region below 2\,\GeV, scaling the cross section by $\bigl[\mathcal{B}(\topLSP)+\mathcal{B}(\topNLSP)\bigr]^2$, under the assumption that the decay products from \chinoonepm\ and \ninotwo\ are too soft to be reconstructed.

\item Bino/higgsino mix: 

For the well-tempered neutralino, the signal model was built in a similar manner to the wino NLSP model. Signals were generated by scanning in $M_1$ and $m_{q3L}$ parameter space, with $\tan\beta = 20$, $M_2 = 2.0$\,\TeV\ and $M_3 = 1.8$\,\TeV\ (corresponding to a gluino mass of $\sim 2.0$\,\TeV).\footnote{The light sbottom and/or stop become tachyonic when their radiative corrections are large in the low-$m_{q3L}$ regime, as the correction to squark masses is proportional to ($M_3$/$m_{q3L}$)$^2$, which can change the sign of the physical mass. This was an important consideration when choosing the value of $M_3$.} The value of $M_S$ was varied in the range of 700--1300\,\GeV\ in the large $\tleft$--$\tright$ mixing regime in order for the lightest Higgs boson to have a mass consistent with the observed mass. Since the dark-matter relic density is very sensitive to the mass-splitting $\Delta m(\mu,M_1)$, $\mu$ was chosen to satisfy $0.10<\Omega H_0^2<0.12$ given the value of $M_1$ considered ($-\mu \sim M_1$), which resulted in $\Delta m(\mu,M_1)=20$--$50$ \GeV.

The dark-matter relic density was computed using {\textsc{MicrOMEGAs}} 4.3.1f~\cite{micromegas1,micromegas2}. Softsusy-3.3.3 was used to evaluate the level of fine-tuning ($\Delta$)~\cite{finetune} of the pMSSM parameters. The signal models were required to have a low level of fine-tuning corresponding to $\Delta<100$ (at most 1\% fine-tuning).

For scenarios with $m_{tR}<$$m_{q3L}$, only stop pair production was considered while both stop and sbottom pair production were considered in scenarios with $m_{tR}>$$m_{q3L}$. The sbottom mass was found to be close to the stop mass as they were both determined mainly by $m_{q3L}$. The stop and sbottom decay largely into a higgsino state, $\chinoonepm$, $\ninotwo$, and $\ninothree$ with BR similar to those of the higgsino models. The stop and sbottom decay BR to the bino state were found to be small. 

\end{enumerate}

Signal cross-sections for stop/sbottom pair production were calculated to next-to-leading order in the strong coupling constant, adding the resummation of soft gluon emission at next-to-leading-logarithm accuracy (NLO+NLL)~\cite{Beenakker:1997ut,Beenakker:2010nq,Beenakker:2011fu}. The nominal cross-section and the uncertainty were taken from an envelope of cross-section predictions using different PDF sets and factorisation and renormalisation scales, as described in Ref.~\cite{Borschensky:2014cia}.

Signal events for the spin-0 mediator model were generated with \MGaMC\ (LO) with up to one additional parton, interfaced to {\textsc{Pythia}}8. The couplings of the mediator to the DM and SM particles ($g_{\chi}$ and $g_v$) were assumed to be equal and a common coupling with value $g=g_{\chi}=g_v=1$ is used. The kinematics of the decay was found not to depend strongly on the values of these couplings. The cross-section was computed at NLO~\cite{dMtt_xsec1,dMtt_xsec2} and decreased significantly when the mediator was produced off-shell.
\begin{table}[t]
  \centering
  \caption{Overview of the input parameters and typical stop decay branching ratios (BR) for the signal models. 
  Lists of mass parameters scanned are provided in between parentheses.
  The pMSSM mass parameters that are not shown below were set to values above 3\,\TeV.
  The table represents seven different models that are used in the interpretation of the results 
  (two for the wino NLSP, three for the higgsino LSP, and two for the bino/higgsino admixture).
  For the higgsino LSP scenarios, a simplified model is used instead of a pMSSM model, 
  although the stop decay BR are based on pMSSM scans with the parameters shown in the table.
  For the higgsino and bino/higgsino mix scenarios, the stop decay BR change 
  depending on the $\tleft$--$\tright$ composition of the \tone, 
  hence the BR for various scenarios corresponding to (a) $\tone\sim\tright$ and (b) $\tone\sim\tleft$ 
  (and (c) $\tone\sim\tleft$ with $\tan\beta=60$ in the higgsino model) are shown separately.
  For the wino NLSP model, only the $\tone\sim\tleft$ scenario is considered. 
  Sbottom pair production is also considered where $\bone\sim\bleft$ for the wino NLSP and bino/higgsino mix scenarios.
  }
  \vspace{3mm}
{\renewcommand{\arraystretch}{1.1}
  \begin{tabular*}{\textwidth}{@{\extracolsep{\fill}}| l | ccc |}
                \hline
    \hline
    Scenario                      & Wino NLSP              & Higgsino LSP                  & Bino/higgsino mix             \\ \hline
                Models                                   & pMSSM                  & simplified                    & pMSSM                         \\ \hline
    Mixing parameters                 & \multicolumn{3}{ c |}{$X_t/M_S \sim \sqrt{6}$}                                         \\
    $\tan\beta$                 & 20                     & 20 or 60                      &  20                           \\
    $M_S$ \,[\TeV]                   & 0.9--1.2                & 1.2                           & 0.7--1.3                       \\
    $M_3$ \,[\TeV]                   & 2.2                    & 2.2                           & 1.8                           \\
    Scanned mass parameters     & ($M_1$, $m_{q3L}$)     & ($\mu$, $m_{q3L}$/$m_{tR}$)   & ($M_1$, $ m_{q3L}$/$m_{tR}$)   \\
    Electroweakino masses \,[\TeV]     & $\mu = \pm 3.0$        & $M_2=M_1=1.5$                 & $M_2=2.0$                     \\
              & $M_2 = 2M_1 \ll |\mu|$ & $\mu\ll M_1 = M_2$            & $M_1 \sim - \mu$, $M_1 < M_2$ \\ \hline
    Additional requirements               & --                     & --                            & $0.10<\Omega H_0^2<0.12$        \\
              & --                     & --                            & $\Delta<100$                  \\
    Sbottom pair production               & considered        & --                            & considered                    \\
    \hline \hline
    \tone\ decay modes and their BR [\%]    & $\tone\sim\tleft$      & (a) / (b) / (c)            & (a) / (b)                     \\ \hline
             \topLSP       & $<5$                   & $\sim25$/$\sim45$/$\sim33$    & $<10$/$<10$                   \\
             \bChargino       & $\sim65$               & $\sim50$/$\sim10$/$\sim33$    & $\sim50$/$\sim10$             \\
             \topNLSP       & $\sim30$               & $\sim25$/$\sim45$/$\sim33$    & $\sim20$/$\sim40$             \\
             \topNNLSP       & --                     & --          & $\sim20$/$\sim40$             \\
    \hline \hline
    \bone decay modes and their BR [\%]     & $\bone\sim\tleft$      & --                 & $\bone\sim\bleft$             \\ \hline
             \bottomLSP      & $<5$                   & --           & $<5$                          \\
             \tChargino      & $\sim65$               & --           & $\sim85$                      \\
             \bottomNLSP      & $\sim30$               & --           & $<5$                          \\
             \bottomNNLSP      & --                     & --          & $<5$                          \\
    \hline \hline
  \end{tabular*}
}
  \label{tab:pMSSMparm}
\end{table}

%% file: texfiles/objects.tex
\section{Event reconstruction}
\label{sec:objects}

Events used in the analysis must satisfy a series of beam, detector and data-quality criteria. 
The primary vertex, defined as the reconstructed vertex with the highest $\sum_\text{tracks} \pt^2$, must have at least two associated tracks with \pt $>$ 400\,\MeV. 

Depending on the quality and kinematic requirements imposed, reconstructed physics objects are labelled either as {\textit{baseline}} or {\textit{signal}}, where the latter describes a subset of the former. Baseline objects are used when classifying overlapping physics objects and to compute the missing transverse momentum. Baseline leptons (electrons and muons) are also used to impose a veto on events with more than one lepton, which suppresses background contibutions from \ttbar\ and $Wt$ production where both $W$-bosons decay leptonically, referred to as dileptonic \ttbar\ or $Wt$ events. Signal objects are used to construct kinematic and multiplicity discriminating variables needed for the event selection.

Electron candidates are reconstructed from electromagnetic calorimeter cell clusters that are matched to ID tracks. Baseline electrons are required to have $\pT > 5$\,\GeV, $|\eta| < 2.47$, and to satisfy `VeryLoose' likelihood identification criteria that are defined following the methodology described in Ref.~\cite{ATL-PHYS-PUB-2015-041}. Signal electrons must pass all baseline requirements and in addition satisfy the `LooseAndBLayer' or `Tight' likelihood identification criteria depending on the signal region selection, and are classified as `loose' or `tight' signal electrons, respectively. 
They must also have a transverse impact parameter evaluated at the point of closest approach between the track
and the beam axis in the transverse plane ($d_0$) that satisfies $|d_0|/\sigma_{d_0} < 5$, where $\sigma_{d_0}$ is the uncertainty in $d_0$, and the distance from this point to the primary vertex along the beam direction ($z_0$) must satisfy $|z_0 \sin \theta| < 0.5$\,mm.
Furthermore, lepton isolation, defined as the sum of the transverse energy deposited in a cone with a certain size $\Delta R$ excluding the energy of the lepton itself, is required. The isolation criteria for `loose' electrons use only track-based information, while the `tight' electron isolation criteria rely on both track- and calorimeter-based information with a fixed requirement on the isolation energy divided by the electron's $\pt$. 

Muon candidates are reconstructed from combined tracks that are formed from ID and MS tracks, ID tracks matched to MS track segments, stand-alone MS tracks, or ID tracks matched to an energy deposit in the calorimeter compatible with a minimum-ionising particle (referred to as calo-tagged muon)~\cite{PERF-2015-10}. Baseline muons up to $|\eta|=2.7$ are used and they are required to have $\pt>4$\,\GeV\ and to satisfy the `Loose' identification criteria. Signal muons must pass all baseline requirements and in addition have impact parameters $| z_0 \sin \theta| < 0.5$\,mm and $|d_0|/\sigma_{d_0} < 3$, and satisfy the `Medium' identification criteria. Furthermore, signal muons must be isolated according to criteria similar to those used for signal electrons, but with a fixed requirement on track-based isolation energy divided by the muon's $\pt$. No separation into `loose' and `tight' classes is performed for signal muons.

Dedicated scale factors for the requirements of identification, impact parameters, and isolation are derived from $Z\rightarrow \ell\ell$ and $J/\Psi\rightarrow \ell\ell$ data samples for electrons and muons to correct for minor mis-modelling in the MC samples~\cite{ATLAS-CONF-2016-024,PERF-2015-10}. The \pt\ thresholds of signal leptons are raised to 25\,\GeV\ for electrons and muons in all signal regions except those that target higgsino LSP scenarios.

Jet candidates are built from topological clusters~\cite{PERF-2014-07,PERF-2011-03} in the calorimeters using the anti-$k_t$ algorithm~\cite{Cacciari:2008gp} with a jet radius parameter $R=0.4$ implemented in the FastJet package~\cite{Cacciari:2011ma}. Jets are corrected for contamination from pile-up using the jet area method~\cite{Cacciari:2007fd,Cacciari:2008gn,PERF-2014-03} and are then calibrated to account for the detector response~\cite{PERF-2012-01,PERF-2016-04}. Jets in data are further calibrated according to \textit{in situ} measurements of the jet energy scale~\cite{PERF-2016-04}. Baseline jets are required to have $\pt > 20$\,\GeV. Signal jets must have $\pt > 25$\,\GeV\ and $|\eta|<2.5$. Furthermore, signal jets with $\pT < 60$\,\GeV\ and $|\eta| < 2.4$ are required to satisfy track-based criteria designed to reject jets originating from pile-up~\cite{PERF-2014-03}. Events containing a jet that does not pass specific jet quality requirements (``jet cleaning'') are vetoed from the analysis in order to suppress detector noise and non-collision backgrounds~\cite{DAPR-2012-01,ATLAS-CONF-2015-029}.

Jets containing $b$-hadrons are identified using the MV2c10 $b$-tagging algorithm (and those identified are referred to as $b$-tagged jets), which incorporates quantities such as the impact parameters of associated tracks and reconstructed secondary vertices~\cite{ATL-PHYS-PUB-2016-012,Aad:2015ydr}. The algorithm is used at a working point that provides a 77\% $b$-tagging efficiency in simulated $\ttbar$ events, and corresponds to a rejection factor of about 130 for jets originating from gluons and light-flavour quarks (light jets) and about 6 for jets induced by charm quarks. Corrections derived from data control samples are applied to account for differences between data and simulation for the efficiency and mis-tag rate of the $b$-tagging algorithm~\cite{Aad:2015ydr}.

Jets and associated tracks are also used to identify hadronically decaying $\tau$ leptons using the `Loose' identification criteria described in Refs.~\cite{ATL-PHYS-PUB-2015-045, PERF-2016-04}, which have a 60\% (50\%) efficiency for reconstructing $\tau$ leptons decaying into one (three) charged pions. These $\tau$ candidates are required to have one or three associated tracks, with total electric charge opposite to that of the selected electron or muon, $\pt>20$ \GeV, and $|\eta|<2.5$. The $\tau$ candidate \pT\ requirement is applied after a dedicated energy calibration~\cite{ATL-PHYS-PUB-2015-045,PERF-2016-04}.

To avoid labelling the same detector signature as more than one object, an overlap removal procedure is applied. Table~\ref{tab:OR} summarises the procedure. Given a set of baseline objects, the procedure checks for overlap based on either a shared track, ghost-matching~\cite{Cacciari:2008gn}, or a minimum distance\footnote{Rapidity ($y \equiv 1/2\ln \left(E+p_{z}/E-p_{z}\right)$) is used instead of pseudorapidity ($\eta$) when computing $\DeltaR$ in the overlap removal procedure.} $\DeltaR$ between pairs of objects. For example, if a baseline electron and a baseline jet are separated by $\Delta R < 0.2$, then the electron is retained (as stated in the `Precedence' row) and the jet is discarded, unless the jet is $b$-tagged (as stated in the `Condition' row) in which case the electron is assumed to originate from a heavy-flavour decay and is hence discarded while the jet is retained. If the matching requirement in Table~\ref{tab:OR} is not met, then both objects under consideration are kept. The order of the steps in the procedure is given by the columns in Table~\ref{tab:OR}, which are executed from left to right. The second ($e j$) and the third ($\mu j$) steps of the procedure ensure that leptons and jets have a minimum $\Delta R$ separation of $0.2$. Jets overlapping with muons that satisfy one or more of the following conditions are not considered in the third step: the jet is $b$-tagged, the jet contains more than three tracks ($n_\text{track}^j>3$), or the ratio of muon \pt\ to jet \pt\ satisfies $\pt^\mu/\pt^j < 0.7$. Therefore, the fourth step ($\ell j$) is applied only to the jets that satisfy the above criteria or that are well separated from leptons with $\Delta R > 0.2$. 
For the remainder of the paper, all baseline and signal objects are those that have passed the overlap removal procedure. 

The missing transverse momentum is reconstructed from the negative vector sum of the transverse momenta of baseline electrons, muons, jets, and a soft term built from high-quality tracks that are associated with the primary vertex but not with the baseline physics objects~\cite{ATL-PHYS-PUB-2015-027,ATL-PHYS-PUB-2015-023}. Photons and hadronically decaying $\tau$ leptons are not explicitly included but enter either as jets, electrons, or via the soft term.

\begin{table}
{\small
  \centering
  \caption{Overlap removal procedure for physics objects.  The first two rows list the types of overlapping objects: electron ($e$), muon ($\mu$), electron or muon ($\ell$), jet ($j$), and hadronically decaying $\tau$ lepton ($\tau$). All objects refer to the baseline definitions, except for $\tau$ where no distinction between baseline and signal definition is made.  The third row specifies when an object pair is considered to be overlapping.  The fourth row describes an optional condition which must also be met for the pair of objects to be considered overlapping. The last row lists the object given precedence. Object 1 is retained and Object 2 is discarded if the condition is not met, and vice versa. More information is given in the text.}
  \vspace{3mm}
{\renewcommand{\arraystretch}{1.3}
\noindent\adjustbox{max width=\textwidth}{
  \begin{tabular}{| c | ccccc |}
  \hline
  \hline
    Object 1 & $e$    & $e$  & $\mu$  & $j$     & $e$        \\
    Object 2 & $\mu$  & $j$  & $j$    & $\ell$  & $\tau$     \\
  \hline
    \makecell{Matching \\ criteria} & shared track & $\Delta R<0.2$ & ghost-matched &$\Delta R< \min\left(0.4,0.04+\frac{10}{\pt^\ell/\text{GeV}}\right)$ & $\Delta R<0.1$ \\
      \hline
    Condition & calo-tagged $\mu$ & $j$ not $b$-tagged & \makecell{$j$ not $b$-tagged and \\ $\left(n_\text{track}^j<3 \ \text{or} \ \frac{\pt^\mu}{\pt^j }> 0.7\right)$} & -- & --\\
  \hline
    Precedence & $e$ & $e$ & $\mu$ & $j$ & $e$ \\
   \hline
  \hline
  \end{tabular}}
}
  \label{tab:OR}
  }
\end{table}

%% file: texfiles/variables.tex
\section{Discriminating variables}
\label{sec:variables}

The background processes contributing to a final state with one isolated lepton, jets and \met\ are primarily semileptonic \ttbar\ events with one of the $W$ bosons from two top quarks decaying leptonically, and $W$+jets events with a leptonic decay of the W boson. Both backgrounds can be effectively reduced by requiring the transverse mass of the event, $\mt$,\footnote{The transverse mass $\mt$ is defined as $\mt^2=2p_{\text{T}}^{\ell}\met[1-\cos(\Delta\phi)]$, where $\Delta\phi$ is the azimuthal angle between the lepton and missing transverse momentum directions. The quantity $\pt^{\ell}$ is the transverse momentum of the charged lepton.} to be larger than the $W$-boson mass. In most signal regions, the dominant background after this requirement arises from dileptonic \ttbar events, in which one lepton is not identified, is outside the detector acceptance, or is a hadronically decaying $\tau$ lepton. On the other hand, the \mt\ selection is not applied in the signal regions targeting the higgsino LSP scenarios, hence the background is dominated by semileptonic \ttbar\ events. A series of additional variables described below are used to discriminate between the \ttbar\ background and the signal processes.

\subsection{Common discriminating variables}
\label{subsec:common_variables}

\begin{figure}[htbp]
  \centering
  \includegraphics[width=.45\textwidth]{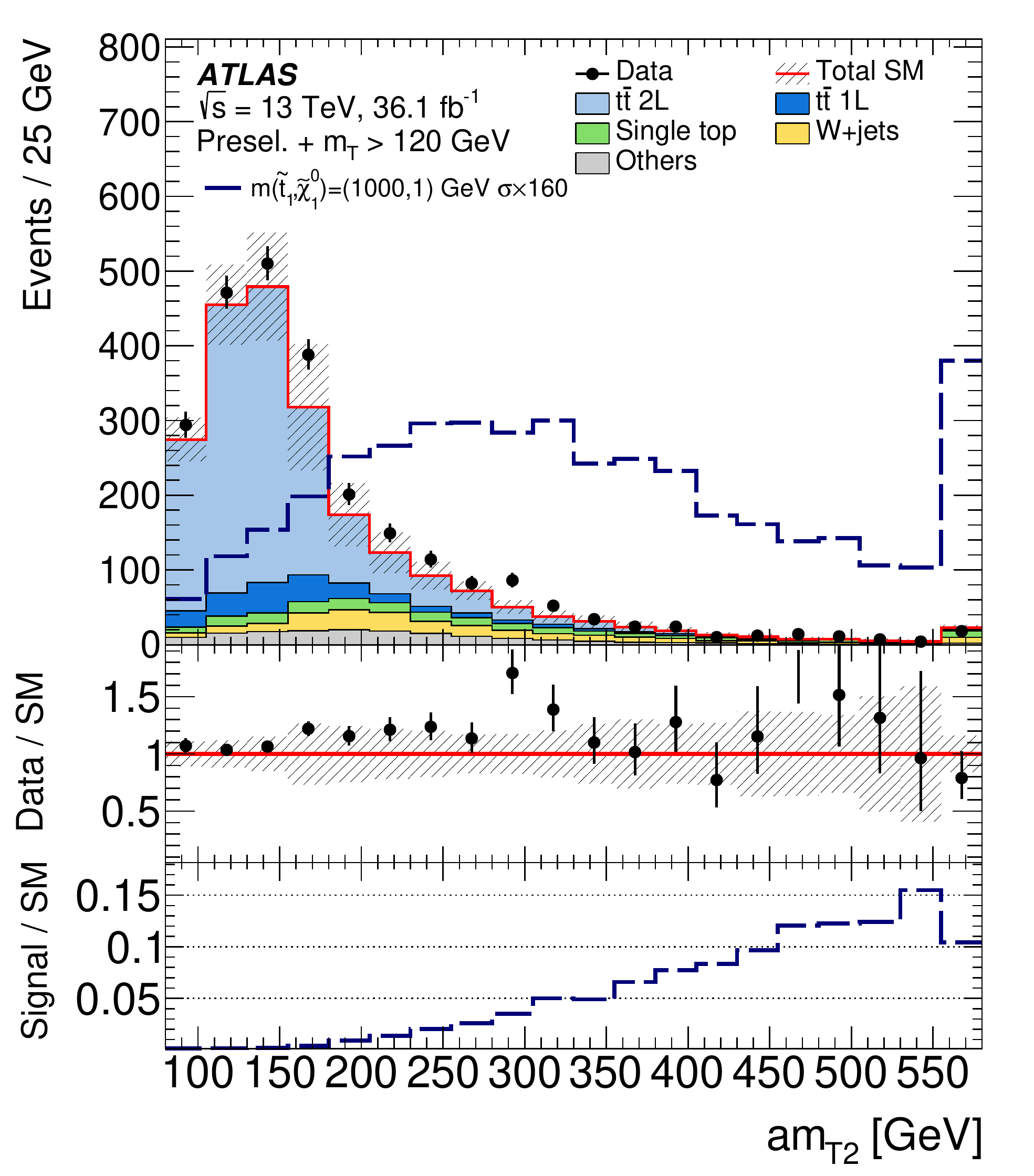}
  \includegraphics[width=.45\textwidth]{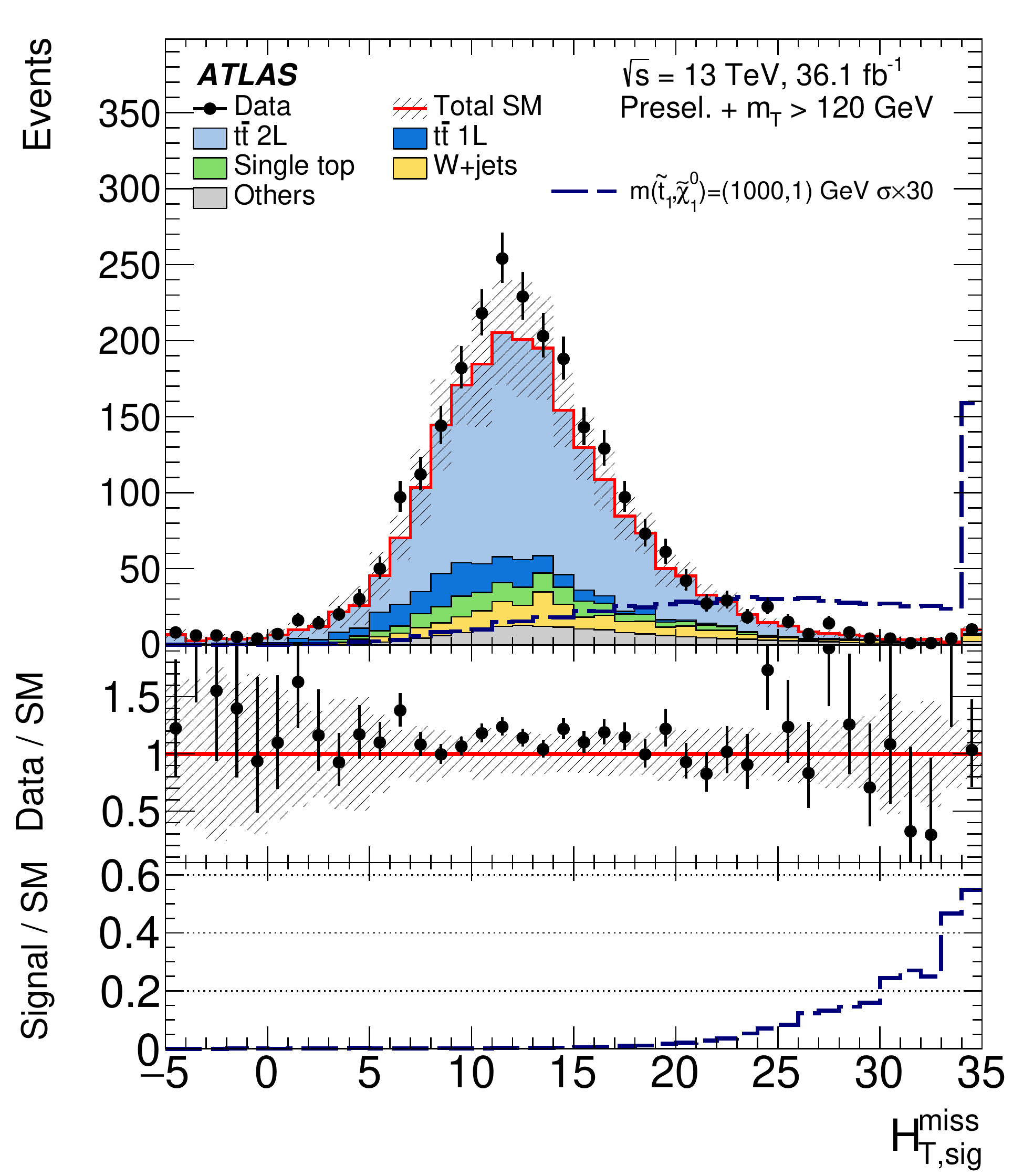}
  \caption{Distributions of discriminating variables: (left) \amtTwo\ and (right) $\HTmissSig$ 
  after the high-\met\ preselection shown in Table~\ref{tab:preselection} and $\mt>120$\,\GeV.
  In addition to the SM background prediction, a bino LSP signal model is shown for a stop mass of 1\,$\TeV$, 
  with a neutralino mass of 1\,$\GeV$, in the upper panel this component is scaled up by a factor of 160 (left) or 30 (right) for visibility.
  The $\ttbar$ 2L and $\ttbar$ 1L in the legend refer to dileptonic and semileptonic $\ttbar$, respectively.
  The lower panels show the ratio of data to total SM background and the ratio of expected signal to total SM background. 
  The category labelled `Others' stands for minor SM backgrounds that contribute less than 5\% of the total SM background.
  The hatched area around the total SM prediction and the hatched band in the Data/SM ratio include statistical and 
  experimental uncertainties. The last bin contains overflows. 
  }
  \label{fig:common_var}
\end{figure}

The asymmetric \mtTwo (\amtTwo)~\cite{Barr:2009jv,Konar:2009qr,Bai:2012gs,Lester:2014yga} and \mtTwoTau are both variants of the variable \mtTwo~\cite{Lester:1999tx}, a generalisation of the transverse mass applied to signatures where two particles are not directly detected. The \amtTwo\ variable targets dileptonic \ttbar\ events where one lepton is not reconstructed, while the \mtTwoTau variable targets \ttbar\ events where one of the two $W$ bosons decays via a hadronically decaying $\tau$ lepton.
In addition, the \HTmissSig\ variable is used in some signal regions to reject background processes without invisible particles in the final state. It is defined as:
\[
\HTmissSig = \frac{|\vec{H}_\text{T}^\text{miss}|-M}{\sigma_{|\vec{H}_\text{T}^\text{miss}|}},
\]
where $\vec{H}_\text{T}^\text{miss}$ is the negative vectorial sum of the momenta of the signal jets and signal lepton. The denominator is computed from the per-event jet energy uncertainties, while the lepton is assumed to be well measured. The offset parameter $M$, which is a characteristic scale of the background processes, is fixed at $100$\,\GeV\ in this analysis. These variables are detailed in Ref.~\cite{SUSY-2013-15}. Figure~\ref{fig:common_var} shows distributions of the $\amtTwo$ and $\HTmissSig$ variables.

Reconstructing the hadronic top-quark decay (top-tagging) can provide additional discrimination against dileptonic \ttbar\ events, which do not contain a hadronically decaying top quark. 
In events where the top quark is produced with moderate $\pt$, a $\chi^2$ technique is used to reconstruct candidate hadronic top-quark decays.
For every selected event with four jets of which at least one is $b$-tagged, the \mTopChi\ variable is defined as the invariant mass of the three jets in the event most compatible with the hadronic decay products of a top quark. The three jets are selected by a $\chi^2$ minimisation using the jet momenta and energy resolutions, and they have to contain exactly one $b$-tagged jet. 

After reconstructing the hadronic top-quark decay through the $\chi^2$ minimisation, the remaining $b$-tagged jet\footnote{If the event has exactly one $b$-tagged jet, the highest-\pt\ jet is used instead of the second highest-\pt\ $b$-tagged jet.} is paired with the lepton to reconstruct the semileptonically decaying top quark candidate (leptonic top quark).
Based on these objects, the azimuthal separation between the \pt\ of hadronic and of leptonic top-quark candidates, $\Delta \phi (t_{\textrm{had}}^{\chi},t_{\textrm{lep}}^{\chi})$ and between the missing transverse momentum vector and the \pt\ of hadronic top-quark candidate, $\Delta \phi (\Ptmiss,t_{\textrm{had}}^{\chi})$, are defined.

An alternative top-tagging method is used to target events where the top quark is produced with a significant boost.
The top-quark candidates are reconstructed by considering all small-radius jets in the event and clustering them into large-radius jets using the anti-$k_t$ algorithm with a radius parameter $R_0 = 3.0$. The radius of each jet is then iteratively reduced to an optimal radius, $R (\pt) = 2 \times m_\mathrm{top}/\pt$, that matches their \pt.
If a candidate loses a large fraction of its \pt\ in the shrinking process, it is discarded.\footnote{The algorithm procedure is as follows: (1) if $R_i> R_{i-1} + 0.3$, then discard the candidate (2) if $R_i < R_{i-1} - 0.5$, then continue iterating (3) else stop iterating and keep the candidate, where $R_i$ is the radius of the candidate in step $i$, and $R_0=3.0$.} 
In events where two or more top-quark candidates are found, the one with the mass closest to the top-quark mass is taken.
The same algorithm is also used to define boosted hadronic $W$-boson candidates, where only non-$b$-tagged jets are considered, and the mass of the $W$ boson is used to define the optimal radius. The masses of the reclustered top-quark and $W$-boson candidates are referred to as \mTopRecluster\ and \mWRecluster, respectively.

The \Ptmiss\ in semileptonic \ttbar\ events is expected to be closely aligned with the direction of the leptonic top quark. After boosting the leptonic top-quark candidate and the \Ptmiss\ into the \ttbar\ rest frame, computed from $t_{\textrm{had}}^{\chi}$ and $t_{\textrm{lep}}^{\chi}$, the magnitude of the perpendicular component of the \Ptmiss\ with respect to the leptonic top quark is computed.
This \perpmet\ is expected to be small for the background, as the dominant contribution to the total \met\ is due to the neutrino emitted in the leptonic top-quark decay.

\subsection{Discriminating variables for boosted decision trees}
\label{subsec:BDT_variables}

In the diagonal region where $m_{\tone} \approx m_\mathrm{top} + m_{\ninoone}$, the momentum transfer from the $\tone$ to the $\ninoone$ is small, and the stop signal is kinematically very similar to the \ttbar\ process. In order to achieve good separation between \ttbar\ and signal, a boosted decision tree (BDT) implemented in the TMVA framework~\cite{Hocker:2007ht} is used. Additional discriminating variables are developed to use as inputs to the BDT, or as a part of the preselection in the BDT analyses.

Some of the selections targeting the diagonal region 
in the pure bino LSP scenarios rely on the presence of high-\pt\ initial-state radiation (ISR) jets, which serves to  boost the di-stop system. 
A powerful technique to discriminate these signal models from the \ttbar background is to attempt to reconstruct the ratio of the transverse momenta of the di-neutralino and di-stop systems. This ratio $\alpha$ can be directly related to the ratio of the masses of the $\tone$ and the $\ninoone$ \cite{An:2015uwa,Macaluso:2015wja}:
\[
    \alpha \equiv \frac{m_{\ninoone}}{m_{\tone}} \sim \frac{\pt(\ninoone \ninoone)}{\pt(\tone \tone)}.
\]
The observed \met\ would also include a contribution from the neutrino produced in the leptonic $W$-boson decay, in addition to that due to the LSPs. 
A light \ninoone\ and a $\tone$ mass close to the mass of the top quark would result in the neutralinos having low momenta, 
making the reconstruction of the neutrino momentum and its subtraction from the \Ptmiss\ vital.
In the signal region targeting this scenario, a modified $\chi^2$ minimisation using jet momenta only is applied to define the hadronic top-quark candidate $t_{\mathrm{had}}^{\mathrm{ISR}}$. One or two light jets and one $b$-tagged jet are selected in such a way that they are most compatible with originating from hadronic $W$-boson and top-quark decays. The leading-\pt\ light jet is excluded, as it is assumed to originate from ISR.

Out of the two jets with the highest probabilities of being a $b$-jet according to the $b$-tagging algorithm, the one not assigned to $t_{\mathrm{had}}^{\mathrm{ISR}}$ is assigned to the leptonic top-quark candidate, together with the lepton.
For the determination of the neutrino momentum, two hypotheses are considered: that of a $\ttbar$ event and that of a signal event. 
For the $\ttbar$ hypothesis, the entire \Ptmiss\ is attributed to the neutrino. 
Under the signal hypothesis,
collinearity of each $\tone$ with both of its decay products is assumed. This results in the transverse-momentum vector of the neutrino from the leptonic $W$-boson decay being calculable by subtracting the momenta of the LSPs from \Ptmiss, when assuming a specific mass ratio $\alpha$:
\[
\vec{\pt}(\nu^{\alpha}) = (1-\alpha)\Ptmiss - \alpha \vec{\pt}(t_{\mathrm{had}}^{\mathrm{ISR}} + b_{\mathrm{lep}}  + \ell),
\]
where $\nu^{\alpha}$ represents the neutrino four-vector for a given value of $\alpha$, $b_{\mathrm{lep}}$ is the $b$-jet candidate assigned to the semileptonic top-quark candidate and $\ell$ is the charged lepton. The resulting momentum of $\nu^{\alpha}$ is then used to calculate further variables under the signal hypothesis, such as the leptonically decaying $W$ boson's transverse mass $\mt^{\alpha}$ or the mass of the top-quark candidate including the leptonic $W$-boson decay, $m(t_{\textrm{lep}}^{\alpha})$. The lepton pseudorapidity is used as a proxy for the neutrino pseudorapidity in the calculation. Further variables are the difference in $\mt$ between the calculation under the hypothesis of a $\ttbar$ event and under the signal hypothesis, $\Delta \mt^{\alpha} = \mt - \mt^{\alpha}$, where $\mt^{\alpha}$ is calculated using the lepton and $\nu^{\alpha}$, and the $\pt$ of the reconstructed $\ttbar$ system under the SM hypothesis, $\pt(\ttbar)$. The mass ratio $\alpha =$ 0.135 is used throughout the paper, as is calculated from $m_{\tone} = 200$\,\GeV\ and $m_{\ninoone} = 27$\,\GeV. This signal point was chosen since it is close to the exclusion limit from previous analyses.

Larger stop-mass values in compressed bino LSP scenarios 
boost the \ninoone\ such that neglecting the neutrino momentum in the determination of
$\alpha$ is a good approximation.
A recursive jigsaw reconstruction (RJR) technique~\cite{Jackson:2016mfb} is used to divide each event into an ISR hemisphere and a sparticle (S) hemisphere, where the latter contains both the invisible (I) and visible (V) decay products of the stops. Objects are grouped together according to their proximity in the lab frame’s transverse plane by maximising the $\pt$ of the S and ISR hemispheres over all choices of object assignment. In events with high-$\pt$ ISR jets, the axis of
maximum back-to-back $\pt$, also known as the thrust axis, should approximate the direction of the ISR and the di-stop system's back-to-back recoil.
 
The RJR variables used in the corresponding signal regions are the transverse mass of the S system, $M_{\textrm{T}}^{\textrm{S}}$, the ratio of the momenta of the I and ISR systems, $R_{\textrm{ISR}}$ (an approximation of $\alpha$), the azimuthal separation between the momenta of the ISR and I systems, $\Delta \phi ({\textrm{ISR}, I})$, and the number of jets assigned to the V system, $N_j^{\textrm{V}}$. 

\begin{figure}[htbp]
  \centering
  \includegraphics[width=.40\textwidth]{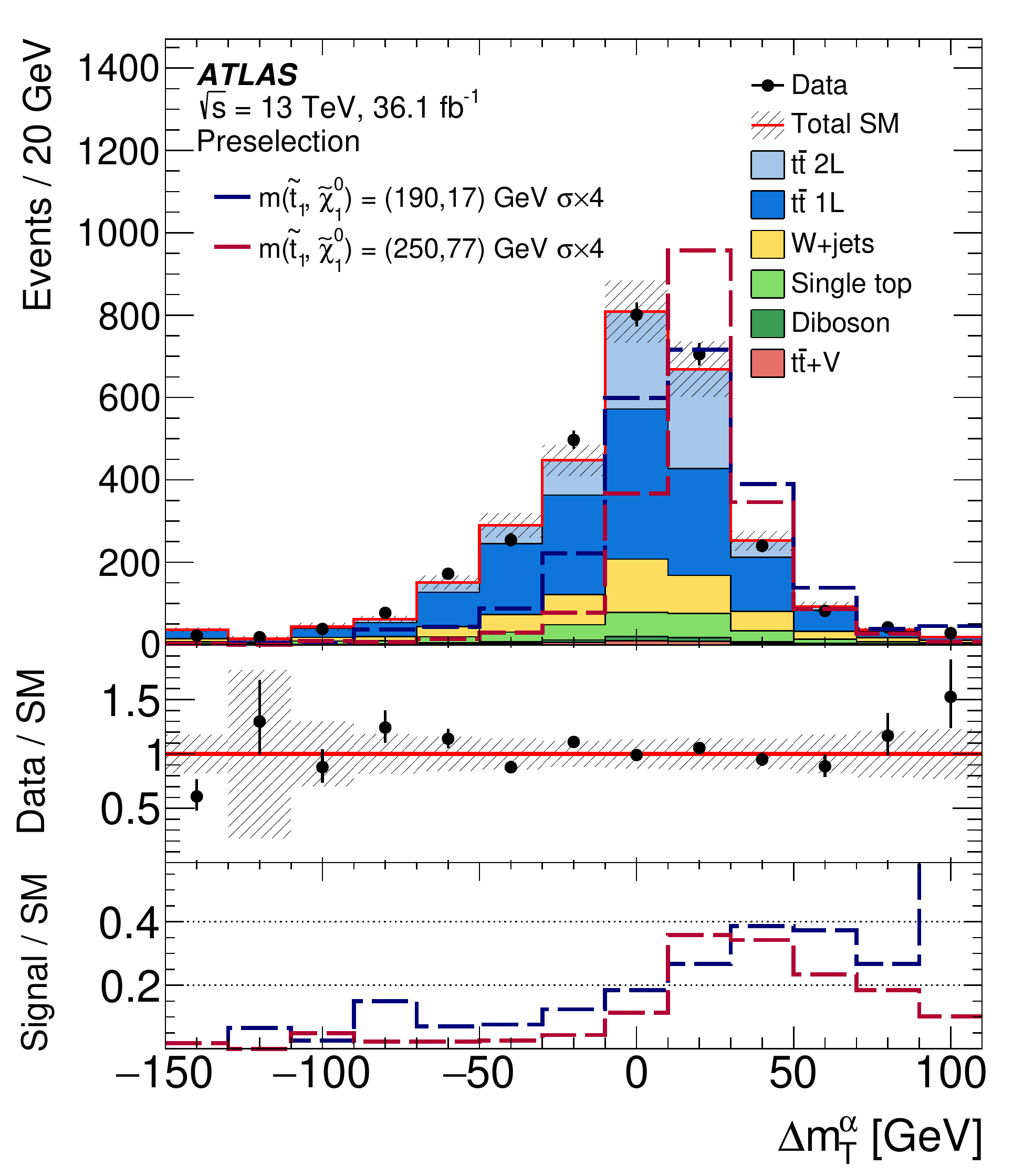}
  \includegraphics[width=.40\textwidth]{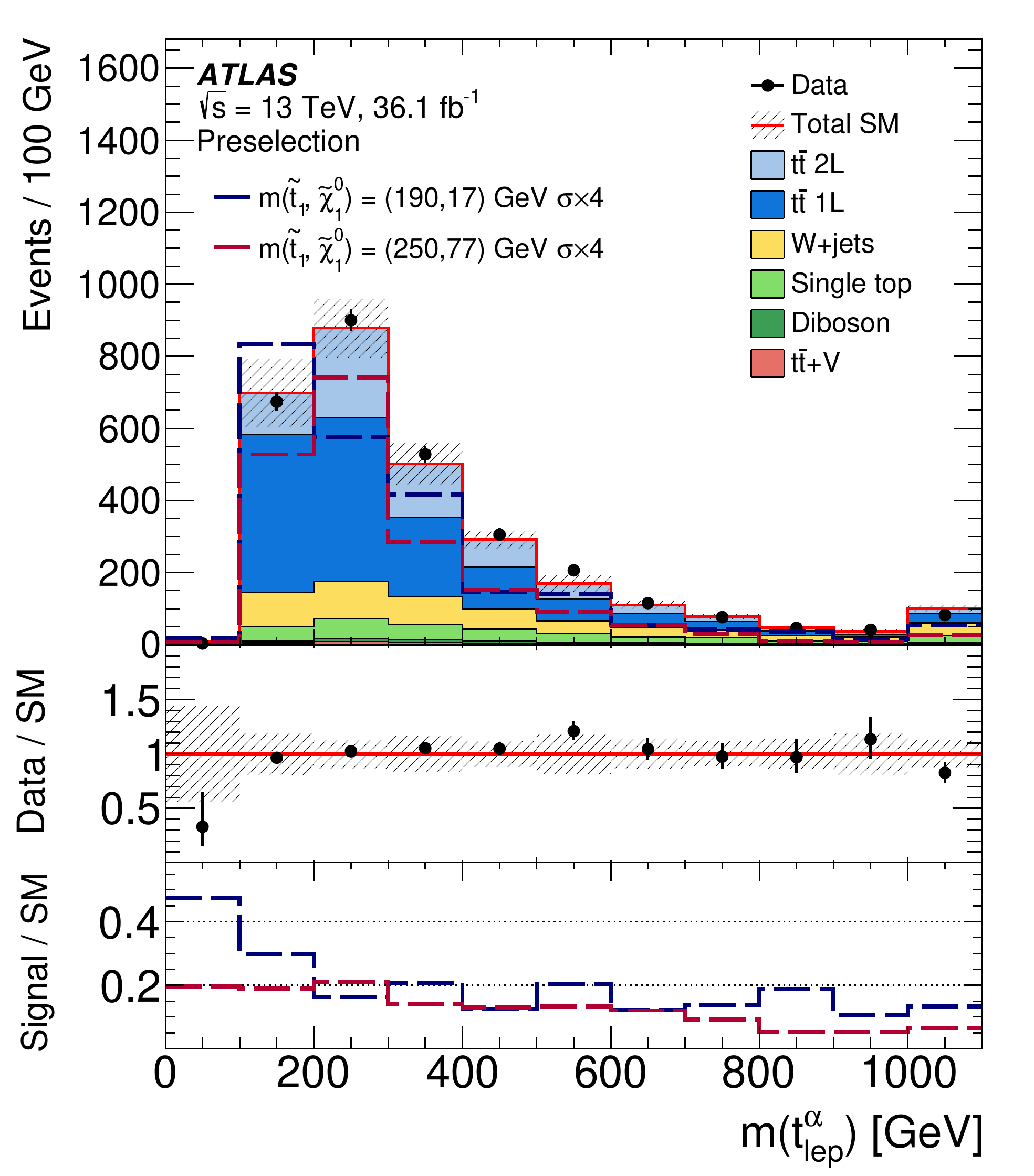}
  \caption{Distributions of discriminating variables: (left) $\Delta \mt^{\alpha}$ and (right) $m(t_{\textrm{lep}}^{\alpha})$. 
  They are used in the \tNdiaglow\ signal region, which is defined in Section~\ref{subsec:tN_diag}. 
  Preselection refers to the signal region selection but without any requirements on the BDT output score.
  In addition to the SM background prediction, signal models are shown, denoted by $m(\tone,\ninoone)$, 
  and scaled by a factor of four for visibility. 
  The lower panels show the ratio of data to total SM background and the ratio of expected signal to total SM background. 
  The hatched area around the total SM prediction and the hatched band in the Data/SM ratio include statistical and 
  experimental uncertainties. The last bin contains overflows. 
  }
  \label{fig:pre_tN_diag}
\end{figure}

\begin{figure}[htbp]
  \centering
  \includegraphics[width=.40\textwidth]{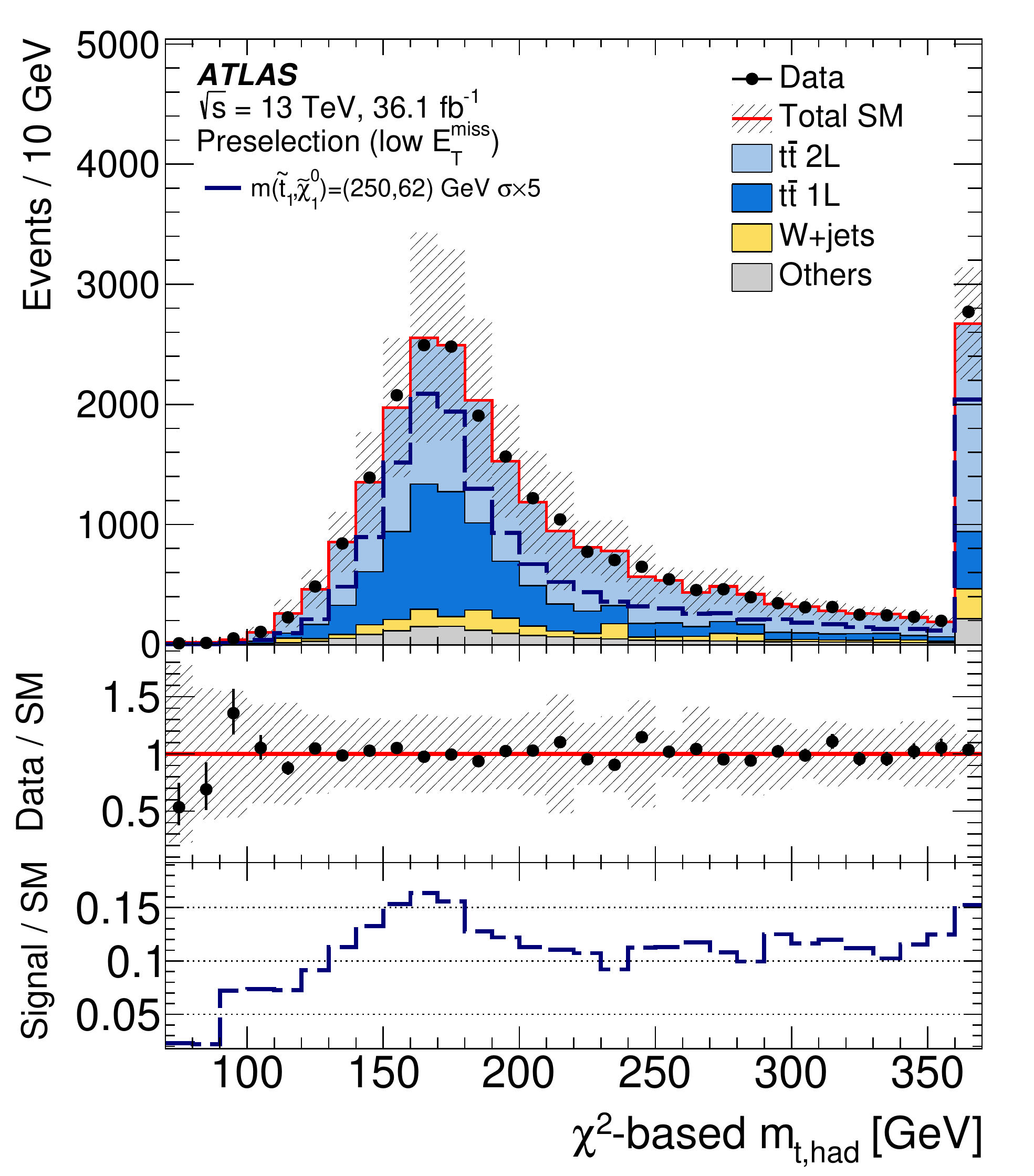}
  \includegraphics[width=.40\textwidth]{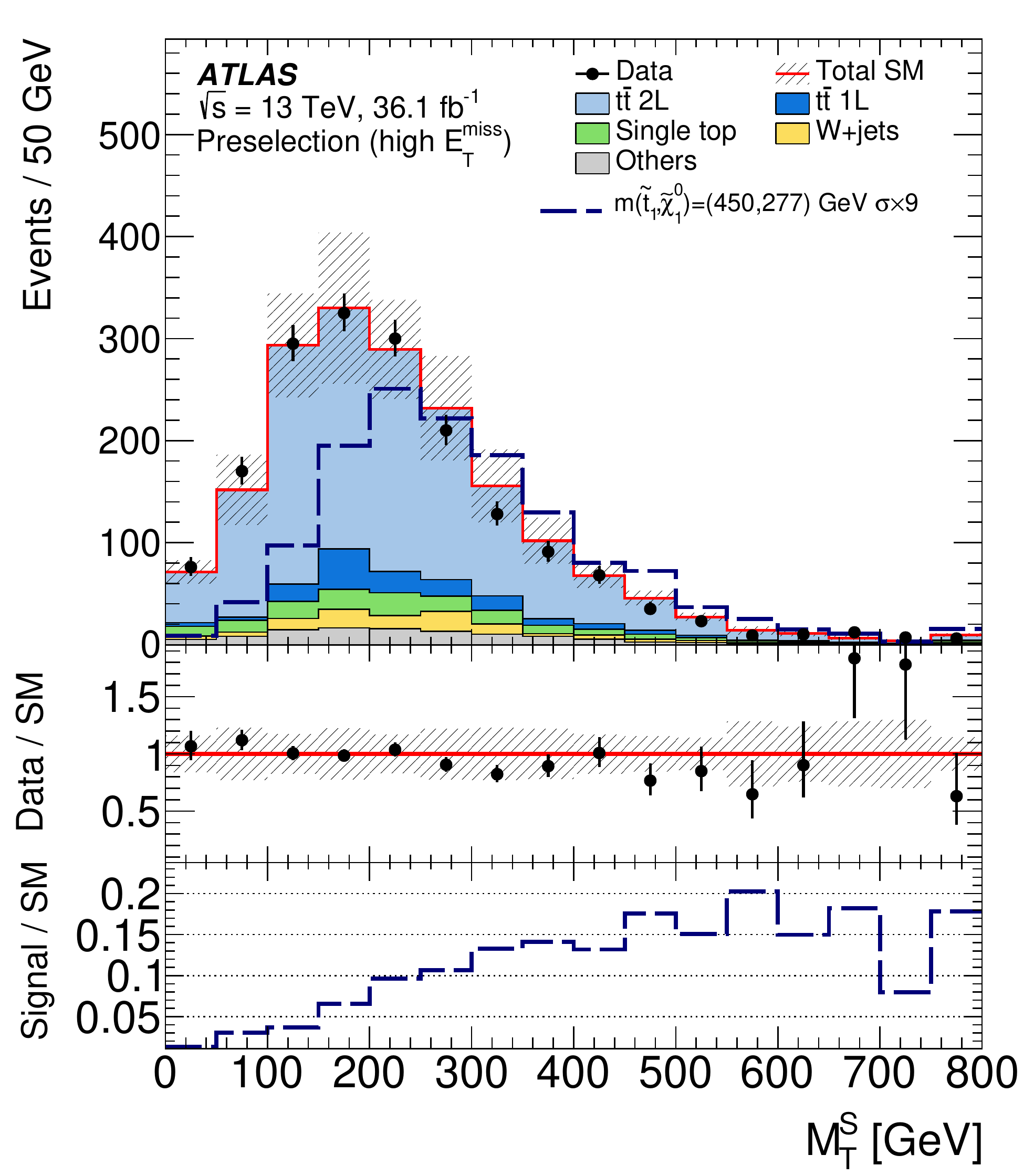}
  \includegraphics[width=.40\textwidth]{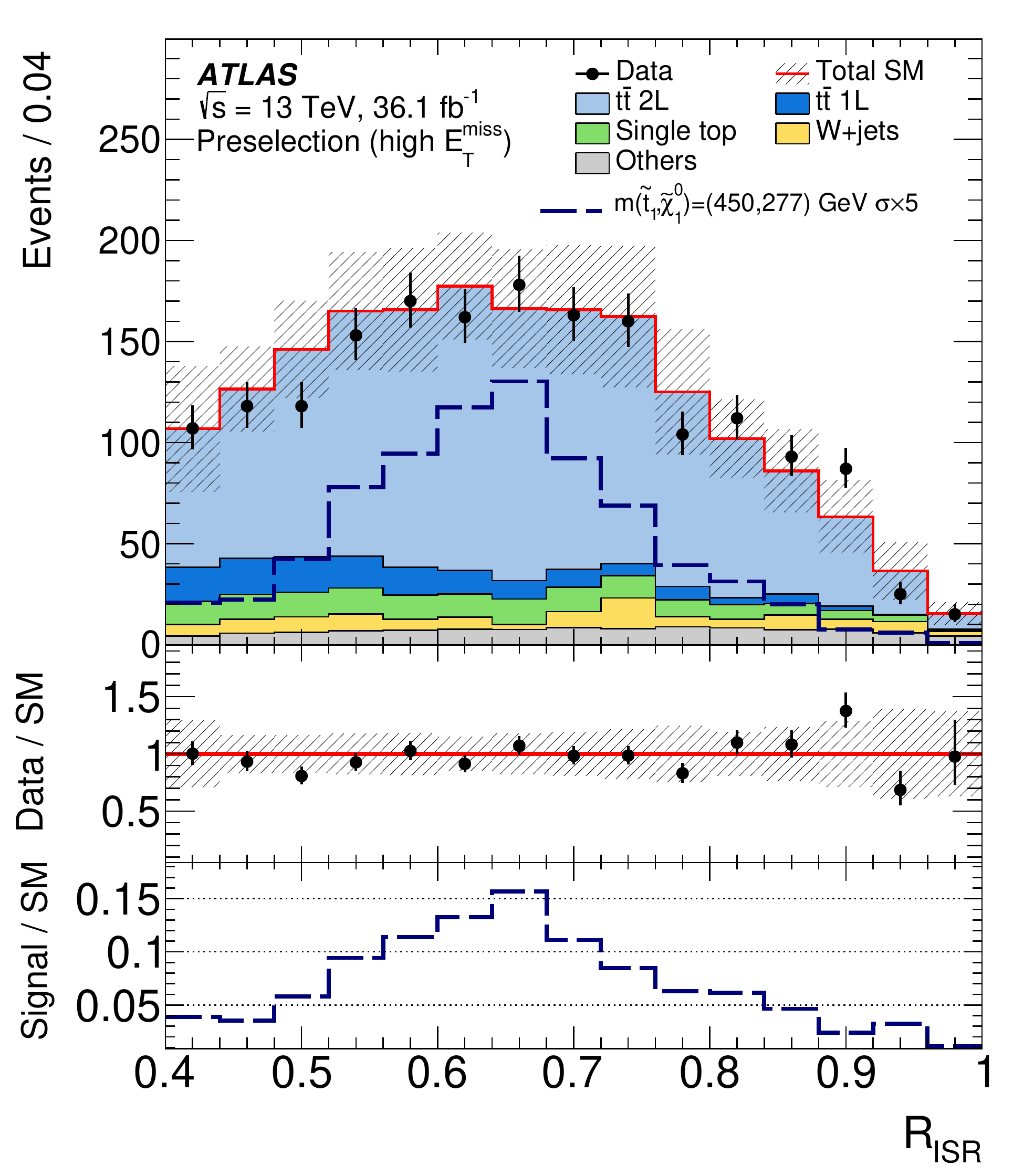}
  \includegraphics[width=.40\textwidth]{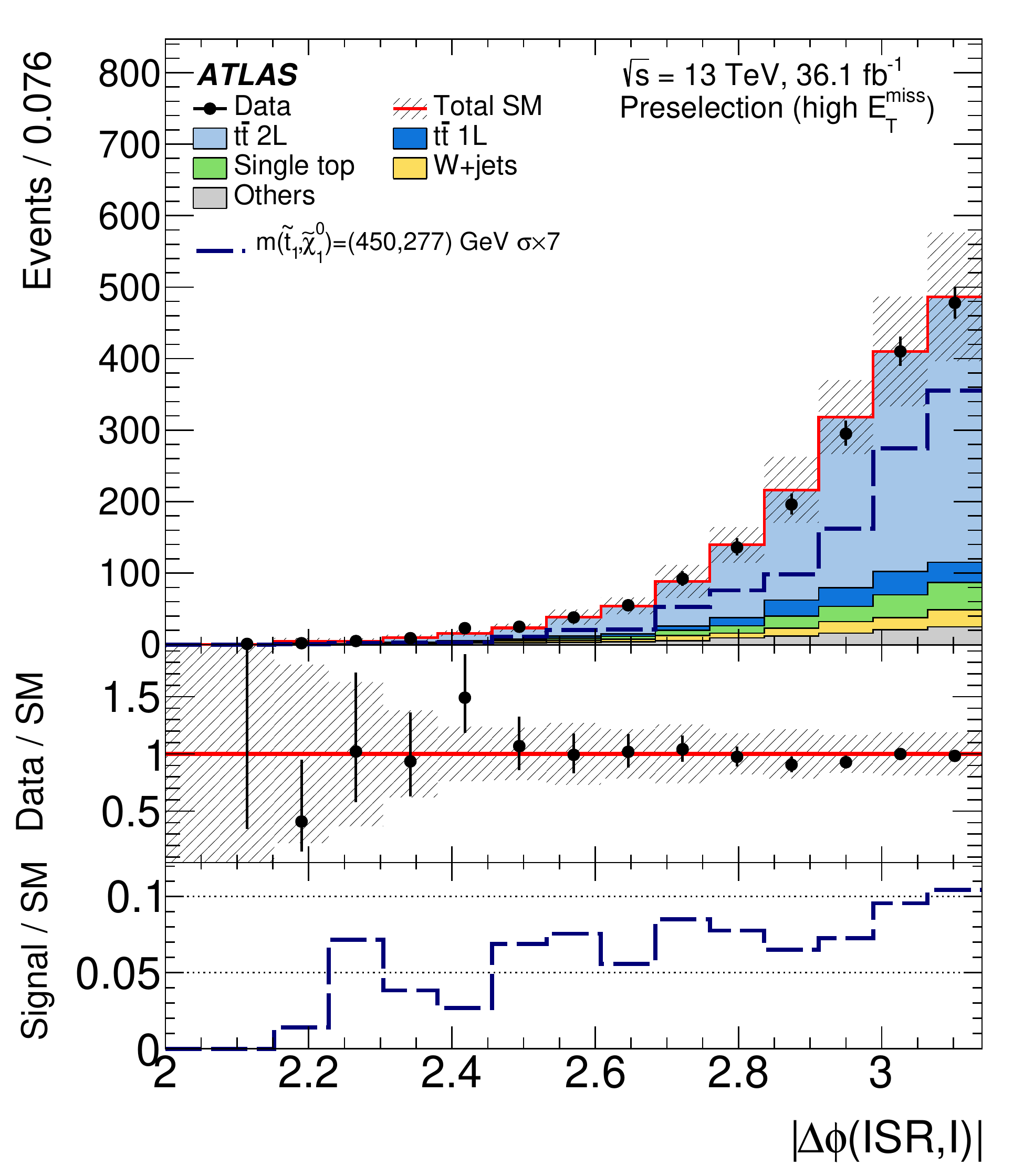}
  \caption{Distributions of discriminating variables: 
  (top left) reconstructed mass of the hadronic top-quark candidates with $\chi^2$-based minimisation method (\mTopChi), 
  (top right) $M_{\textrm{T}}^{\textrm{S}}$, (bottom left) $R_{\textrm{ISR}}$, and (bottom right) $|\Delta \phi ({\textrm{ISR}, I})|$.
  The $\mTopChi$ variable is used in the \tNdiagmed\ signal region and the others are used in the \tNdiaghigh\ signal region, which are defined in Section~\ref{subsec:tN_diag}.
  In addition to the SM background prediction, signal models are shown, denoted by $m(\tone,\ninoone)$, and scaled by a certain factor for visibility. 
  The lower panels show the ratio of data to total SM background and the ratio of expected signal to total SM background. 
  The category labelled `Others' stands for minor SM backgrounds that contribute less than 5\% of the total SM background.
  The hatched area around the total SM prediction and the hatched band in the Data/SM ratio include statistical and 
  experimental uncertainties. The last bin contains overflows. 
  }
  \label{fig:pre_tN_diag_2}
\end{figure}

Figures~\ref{fig:pre_tN_diag} and \ref{fig:pre_tN_diag_2} show example kinematic distributions of the variables used for the BDT trainings. 

%% file: texfiles/signalregions.tex
\section{Signal selections}
\label{sec:signalregions}

SR selections are optimised using simulated MC event samples. The metric of the optimisation is the discovery sensitivity for the various decay modes and for different regions of SUSY parameter space and masses in the spin-0 mediator models. A set of benchmark signal models, selected to cover the various stop and spin-0 mediator scenarios, is used for the optimisation. The optimisations of signal-region selections are performed using an iterative algorithm and considering all studied discriminating variables, accounting for statistical and systematic uncertainties.

\input{texfiles/common_sel.tex}

Table~\ref{tab:signalOverview} summarises all SRs with a brief description of the targeted signal scenarios. 
For the pure bino LSP scenario, seven SRs are considered in total. Five SRs target the $\topLSP$ decay. The corresponding SR labels begin with \texttt{tN}, which is an acronym for `top neutralino'. Additional text in the label describes the stop mass region. For example, \texttt{tN\_diag} targets the diagonal region where $m_{\tone}$ $\sim$ $m_{\ninoone}$ + $m_\mathrm{top}$. The third part of the labels \texttt{low}, \texttt{med}, and \texttt{high} denote the targeted stop mass range, relative to other regions of the same type (for example, \tNdiaglow\ targets a stop mass of 190 GeV, while \tNdiaghigh\ is optimised for $m_{\tone} = $ 450 GeV). Furthermore, two additional SRs labelled \texttt{bWN} and \texttt{bffN} are dedicated to the three-body (\threeBody) and four-body (\fourBody) decay searches, respectively. 

Six SRs target various \bChargino\ scenarios, and the SR labels follow the same logic: the first two characters \texttt{bC} stand for `bottom chargino'. The consecutive labels, \texttt{2x}, \texttt{bv}, or \texttt{soft}, denote the targeted electroweakino spectrum. For the wino NLSP scenario, three SRs are designed with the label \texttt{bC2x} denoting the mass relation $m_{\chinoonepm}$ $\sim$ $2\times$$m_{\ninoone}$ in the signal model. The label \texttt{bCbv} is used for the no $b$-tagged jets ($b$-veto) SR. For the higgsino LSP scenario, three SRs are labelled as \texttt{bCsoft} because their selections explicitly target soft-lepton signatures. 

Finally, three SRs labelled as \texttt{DM} target the spin-0 mediator scenario, with the consecutive labels, \texttt{low} and \texttt{low\_loose} for low mediator masses and \texttt{high} for high mediator masses.

With the exception of the \texttt{tN} and \texttt{bCsoft} regions, the above SRs are not designed to be mutually exclusive. A dedicated combined fit is performed using \tNmed\ and $\bCsoftmed$ (or $\bCsofthigh$) in the higgsino LSP and well-tempered neutralino scenarios in order to improve exclusion sensitivity. The SRs with the requirement of lepton $\pt>25$\,\GeV\ ($\pt>4$\,\GeV) are referred to as hard-lepton SRs (soft-lepton SRs) in the following sections.

\begin{table}
\begin{center}
\caption{Overview of all signal regions together with the targeted signal scenario, benchmarks used for the optimisation (with particle masses given in units of GeV), the analysis technique used for model-dependent exclusions, and a reference to the table with the event selection details. For the wino NLSP scenario, sbottom pair production (not shown) is also considered. 
}
\vspace{3mm}
\renewcommand{\arraystretch}{1.5}
{\scriptsize
\begin{tabular}{| l | c | c | c | c |}
\hline\hline
SR & Signal scenario & Benchmark & Exclusion technique & Table \\
\hline
\tNmed       & Pure bino LSP (\topLSP)                      & m($\tone,~\ninoone$)$=$(600,300)   & shape-fit (\met)          & \ref{tab:SRs_tN} \\
\tNhigh      & Pure bino LSP (\topLSP)                      & m($\tone,~\ninoone$)$=$(1000,1)    & cut-and-count             & \ref{tab:SRs_tN} \\
\tNdiaglow   & Pure bino LSP (\topLSP)                      & m($\tone,~\ninoone$)$=$(190,17)    & BDT cut-and-count         & \ref{tab:SRs_tN_diag_low_high} \\
\tNdiagmed   & Pure bino LSP (\topLSP)                      & m($\tone,~\ninoone$)$=$(250,62)    & BDT shape-fit             & \ref{tab:SRs_tN_diag_low_high} \\
\tNdiaghigh  & Pure bino LSP (\topLSP)                      & m($\tone,~\ninoone$)$=$(450,277)   & BDT shape-fit             & \ref{tab:SRs_tN_diag_low_high} \\
\bWN         & Pure bino LSP (\threeBody)                   & m($\tone,~\ninoone$)$=$(350,230)   & shape-fit (\amtTwo)       & \ref{tab:SRs_other} \\
\bffN        & Pure bino LSP (\fourBody)                    & m($\tone,~\ninoone$)$=$(400,350)   & shape-fit (\lepPtoverMET) & \ref{tab:SRs_other} \\
\hline
\bCmed       & Wino NLSP (\bChargino, \topNLSP)             & m($\tone,~\chinoonepm,~\ninoone$)$=$(750,300,150) & cut-and-count & \ref{tab:SRs_bC2x} \\ 
\bCdiag      & Wino NLSP (\bChargino, \topNLSP)             & m($\tone,~\chinoonepm,~\ninoone$)$=$(650,500,250) & cut-and-count & \ref{tab:SRs_bC2x} \\ 
\bCbv        & Wino NLSP (\bChargino, \topNLSP)             & m($\tone,~\chinoonepm,~\ninoone$)$=$(700,690,1)   & cut-and-count & \ref{tab:SRs_bC2x} \\ 
\hline
\bCsoftdiag  & Higgsino LSP (\topLSP, \topNLSP, \bChargino) & m($\tone,~\chinoonepm,~\ninoone$)$=$(400,355,350) & shape-fit (\lepPtoverMET) & \ref{tab:SRs_bCsoft} \\
\bCsoftmed   & Higgsino LSP (\topLSP, \topNLSP, \bChargino) & m($\tone,~\chinoonepm,~\ninoone$)$=$(600,205,200) & shape-fit (\lepPtoverMET) & \ref{tab:SRs_bCsoft} \\
\bCsofthigh  & Higgsino LSP (\topLSP, \topNLSP, \bChargino) & m($\tone,~\chinoonepm,~\ninoone$)$=$(800,155,150) & shape-fit (\lepPtoverMET) & \ref{tab:SRs_bCsoft} \\
\hline
\DMlowloose  & spin-0 mediator                              & m($\Phi/a,~\chi$)$=$(20,1)  & cut-and-count & \ref{tab:SRs_DM} \\ 
\DMlow       & spin-0 mediator                              & m($\Phi/a,~\chi$)$=$(20,1)  & cut-and-count & \ref{tab:SRs_DM} \\ 
\DMhigh      & spin-0 mediator                              & m($\Phi/a,~\chi$)$=$(300,1) & cut-and-count & \ref{tab:SRs_DM} \\ 
\hline\hline
\end{tabular}
}
\label{tab:signalOverview}
\end{center}
\end{table}

\subsection{Pure bino LSP scenario}\label{sub:tn-regions}
\input{texfiles/sr_tN.tex}

\input{texfiles/sr_tN_diag.tex}
\input{texfiles/sr_bWN.tex}

\subsection{Wino NLSP scenario}\label{sub:winoNLSP_intro}
\input{texfiles/sr_bC2x.tex}

\subsection{Higgsino LSP scenario}\label{sub:higgsino_intro}

\input{texfiles/sr_bCsoft.tex}

\subsection{Bino/higgsino mix scenario}\label{sub:well_tempered_intro}

\input{texfiles/sr_welltemp.tex}

\subsection{Spin-0 mediator scenario}\label{sub:dm-regions}
\input{texfiles/sr_DM.tex}

%% file: texfiles/common_sel.tex
All regions are required to have exactly one signal lepton (except for the $\ttbar Z(\to\ell\ell)$ control regions, where three signal leptons are required), no additional baseline leptons, and at least four (or in some regions two or three) signal jets. In most cases, at least one $b$-tagged jet is also required. A set of preselection criteria (high-\met, low-\met, and soft-lepton) is defined for monitoring the MC modelling of the kinematic variables. The preselection criteria are also used as the starting point for the SR optimisation.

In the SRs relying only on the \met\ trigger, all events are required to have $\met > 230$\,\GeV\ to ensure that the trigger is fully efficient. In SRs that use a combination of \met\ and lepton triggers, this requirement is relaxed to $\met > 100$\,\GeV. In order to reject multijet events, requirements are imposed on the transverse mass (\mt) and the azimuthal angles between the leading and sub-leading jets (in \pt) and \met\ ($\minDeltaPhi$) in most of SRs.
For events with hadronic $\tau$ candidates, the requirement $\mtTwoTau > 80$\,\GeV\ is applied in most SRs. 

\begin{figure}[!h]
  \centering
  \includegraphics[width=.40\textwidth]{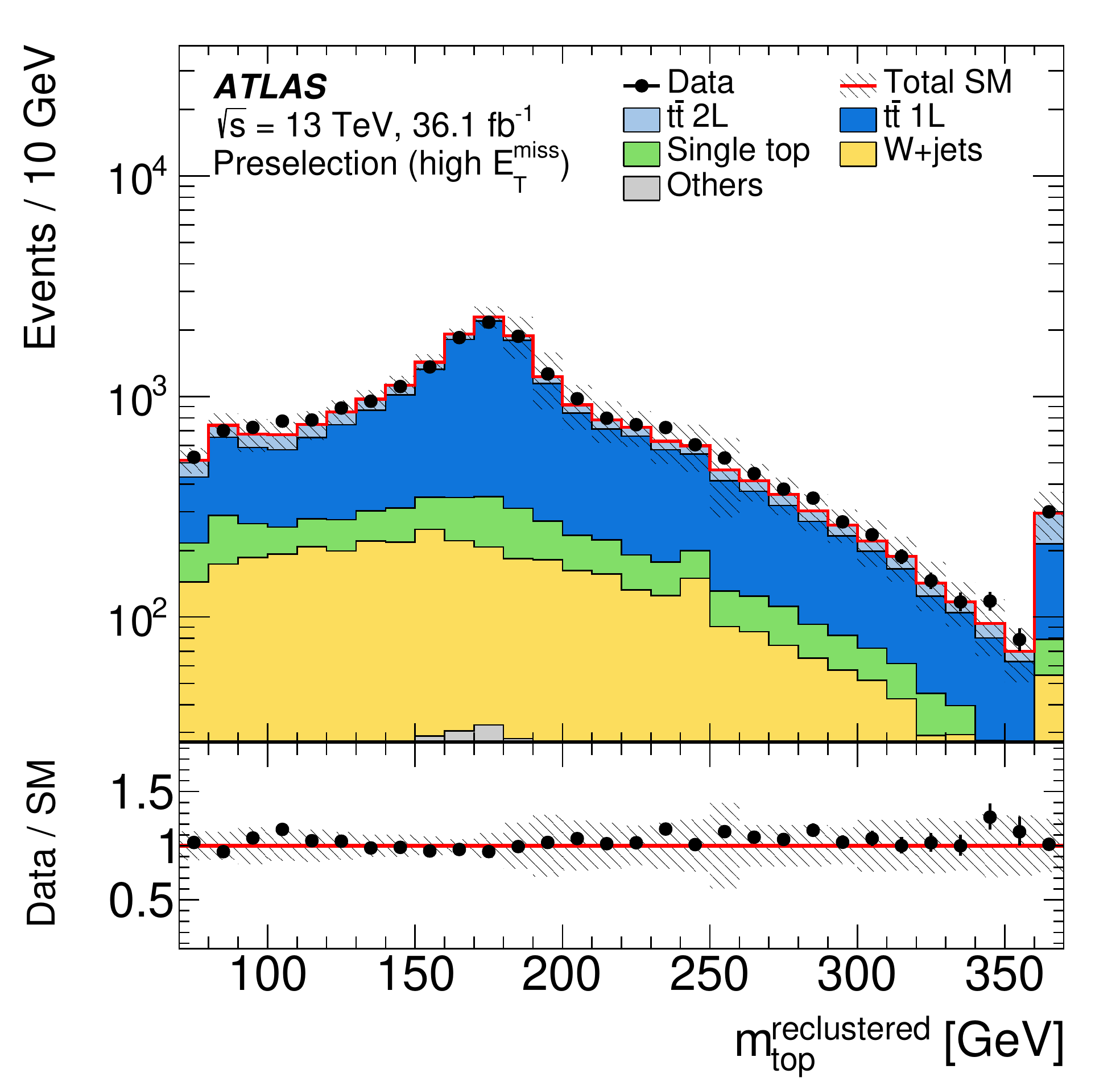}
  \includegraphics[width=.40\textwidth]{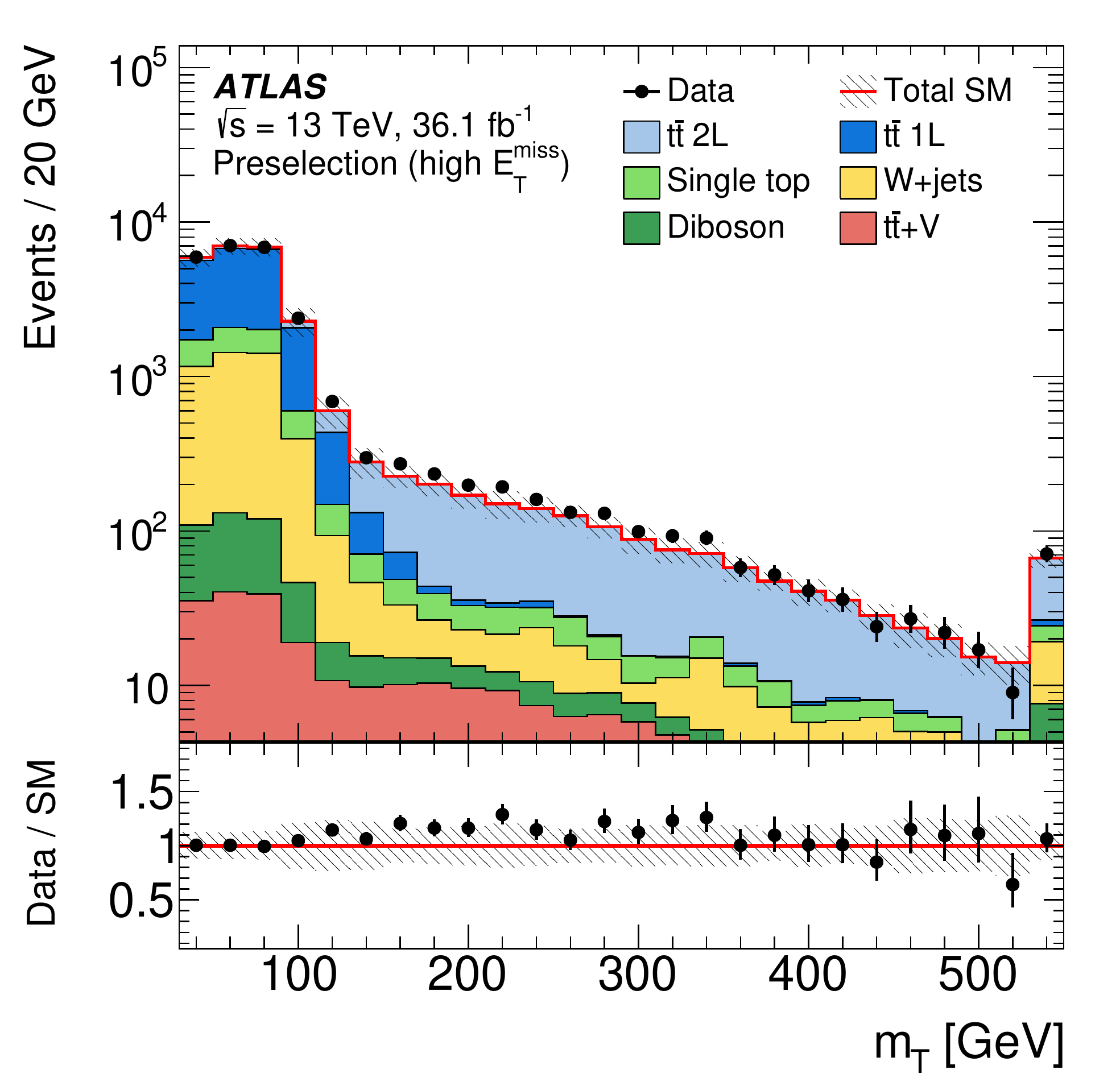}
  \includegraphics[width=.40\textwidth]{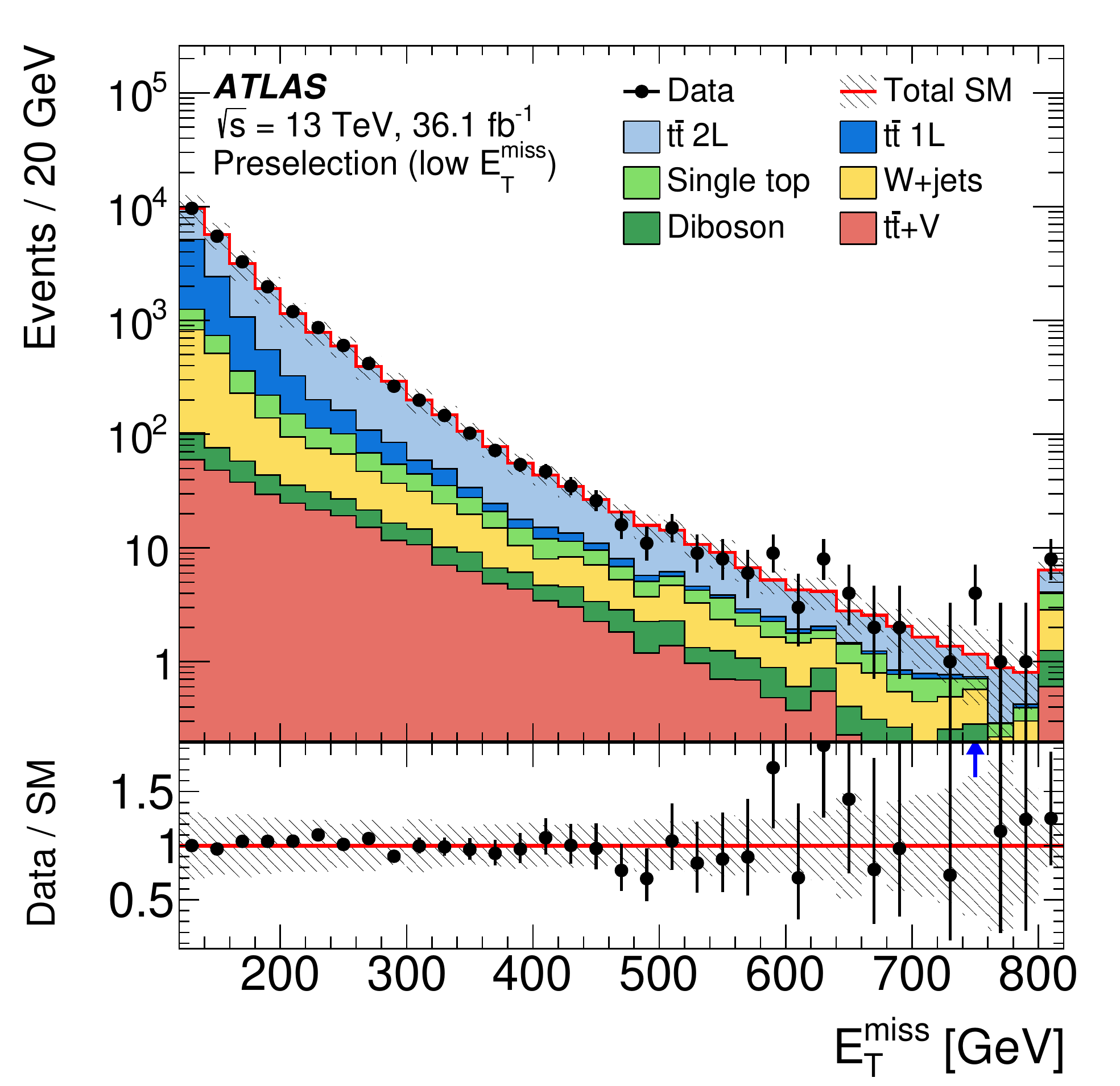}
  \includegraphics[width=.40\textwidth]{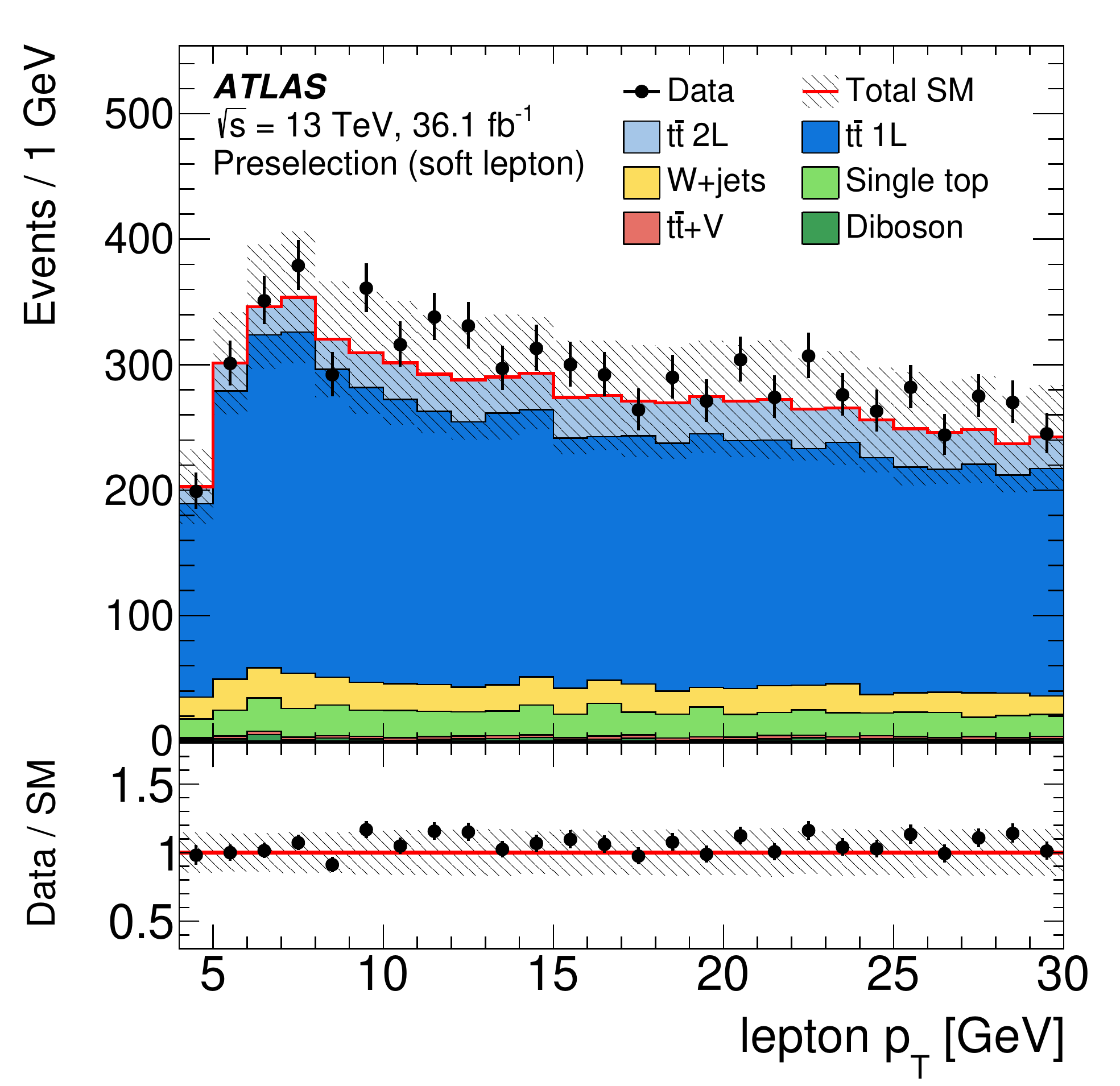}
  \caption{Kinematic distributions after the preselection: 
 (top left)  mass of the hadronic top-quark candidate with the recursive reclustering method (\mTopRecluster) after the high-\met\ preselection, 
 (top right)  \mt\ after the high-\met\ preselection,
 (bottom left)  \met after the low-\met\ preselection, and 
(bottom right)
  lepton $\pt$ after the soft-lepton preselection with an additional requirement of at least two $b$-tagged jets. 
  The SM background predictions are normalised with the theoretical cross-sections (pre-fit), except for in the \met distribution, 
  where the \ttbar\ events are scaled by the normalisation factors obtained from a simultaneous likelihood fit of the CRs.
  The category labelled `Others' in the top left panel stands for the sum of minor SM backgrounds that contribute less than 5\% of the total SM background.
  The hatched area around the total SM prediction and the hatched band in the Data/SM ratio include statistical and experimental uncertainties.
  The last bin contains overflows, except for the lepton \pt\ distribution.
  }
  \label{fig:Presel}
\end{figure}

The exact preselection criteria can be found in Table~\ref{tab:preselection}. The preselections do not include requirements on the $\minDeltaPhi$ and $\mtTwoTau$ variables, but these are often used to define SRs. Figure~\ref{fig:Presel} shows various relevant kinematic distributions at preselection level. The backgrounds are normalised with the theoretical cross-sections, except for the \met\ distribution where the \ttbar\ events are scaled with normalisation factors obtained from a simultaneous likelihood fit of the CRs, described in Section~\ref{sec:results}.

\begin{table}[t]
\begin{center}
\caption{Preselection criteria used for the high-\met\ signal regions (left), the low-\met\ signal regions (middle) and the soft-lepton signal regions (right). For the soft-lepton selection, \pt\ $\geq 5$\,\GeV\ is required for electrons. List
values are provided in between parentheses.}
\vspace{3mm}
\renewcommand{\arraystretch}{1.1}
\begin{tabular}{| l | ccc |}
\hline\hline
    Selection                             & \textbf{high-\met}               & \textbf{low-\met}                 & \textbf{soft-lepton} \\ \hline
    Trigger                               & \met\ triggers only              & \met\ and lepton triggers         & \met\ triggers only  \\
    Data quality                          & \multicolumn{3}{c|}{jet cleaning, primary vertex}                                           \\ 
    Second-lepton veto                    & \multicolumn{3}{c|}{no additional baseline leptons}                                         \\ \hline 
    Number of leptons, tightness          & $= 1$ `loose' lepton             & $= 1$ `tight' lepton              & $= 1$ `tight' lepton \\
    Lepton $\pt$ \,[\GeV]                 & $>25$                            & $>27$                             & $>4$ for $\mu$    \\ 
                &                                  &                                   & $>5$ for $e$      \\ 
    Number of (jets,~$b$-tags)       & ($\ge2$,~$\ge0$)                 & ($\ge4$,~$\ge1$)                  & ($\ge2$,~$\ge1$)     \\
    Jet $\pt$ \,[\GeV]         & $>(25,~25)$                      & $>(50,~25,~25,~25)$               & $>(25,~25)$          \\ \hline
    \met\,[\GeV]                          & $>230$                           & $>100$                            & $>230$               \\
    \mt\,[\GeV]                           & $>30$                            & $>90$                             & --                   \\ 
\hline\hline
\end{tabular}
\label{tab:preselection}
\end{center}
\end{table}

%% file: texfiles/sr_tN.tex
The signature of stop pair production with subsequent \tone\ decays is determined by the masses of the two sparticles, \tone\ and \ninoone. It often leads to a final state similar to that of \ttbar\ production, except for the additional \met\ due to the two additional neutralinos in the event. A set of event selections is defined targeting various signals.

Two signal regions are designed to target the majority of signal models with $\Delta m (\tone,\ninoone) > m_\mathrm{top}$, \tNmed\ and $\tNhigh$, which are optimised for medium and high \tone\ mass, respectively. For the compressed region with $m_{\tone} \approx m_\mathrm{top} + m_{\ninoone}$, three BDT selections ($\tNdiaglow$, $\tNdiagmed$, and $\tNdiaghigh$) target different $\tone$ masses. For the \threeBody\ region, a signal selection (\bWN) is defined by utilising the distinctive shape of the invariant mass of the $bW$ system. For the \fourBody\ region, the signal region (\bffN) is defined by making use of the soft-lepton selection designed for the higgsino LSP scenarios. The event selection for each signal region is detailed in the following subsections.

\subsubsection{$\topLSP$ decay}
\label{subsub:SRs_tN}

Table~\ref{tab:SRs_tN} details the event selections for the \tNmed\ and \tNhigh\ SRs. 
In addition to the high-\met\ preselection described in \Tab{\ref{tab:preselection}}, at least one reconstructed hadronic top-quark candidate based on the recursive reclustered jet algorithm is required in both SRs. Stringent requirements are also imposed on $\met$, $\mt$ and $\HTmissSig$. Furthermore, a requirement is placed on \amtTwo\ to reduce the dileptonic \ttbar\ background. 
The main background processes after all selection requirements are $\ttbar Z (\nu\nu)$, dileptonic \ttbar\ and $W$+heavy-flavour processes.

For the \tNmed\ SR, a shape-fit technique is employed, with the SR subdivided in bins of \met, which allows the model-dependent exclusion fits to be more sensitive than the cut-and-count analysis.

\begin{table}[t]
  \centering
  \caption{Overview of the event selections for the \tNmed\ and \tNhigh\ SRs. 
  List values are provided in between parentheses and square brackets denote intervals. 
  }
  \vspace{3mm}
{\renewcommand{\arraystretch}{1.1}
  \begin{tabular*}{\textwidth}{@{\extracolsep{\fill}}| l | cc |}
                \hline
    \hline
    Signal region                      & \tNmed        & \tNhigh                    \\ \hline
                Preselection                            & \multicolumn{2}{ c |}{high-\met\ preselection} \\ \hline
    Number of (jets,~$b$-tags)     & ($\ge4$,~$\ge1$)  & ($\ge4$,~$\ge1$)           \\
    Jet $\pt$ \,[\GeV]       & $>(60,~50,~40,~40)$  & $>(100,~80,~50,~30)$    \\
    \hline
    \met \,[\GeV]                     & $>250$       & $>550$                    \\
    \perpmet \,[\GeV]                   & $>230$       & --                    \\
    \HTmissSig                     & $>14$          & $>27$                    \\   
    \mt \,[\GeV]                 & \multicolumn{2}{ c |}{$>160$}                   \\
    \amtTwo  \,[\GeV]              & \multicolumn{2}{ c |}{$>175$}                  \\
    \mTopRecluster \,[\GeV]                 & $>150$           & $>130$                     \\
    $\DeltaR(b,\ell)$                & \multicolumn{2}{ c |}{ $<2.0$ }            \\
        $|\Delta\phi(j_{1,2},\Ptmiss)|$         & \multicolumn{2}{ c |}{$>0.4$}                  \\ 
        $\mtTwoTau$ based $\tau$-veto\,[\GeV]   & \multicolumn{2}{ c |}{$>80$}                   \\ 
    \hline \hline
      Exclusion technique      & shape-fit in \met & cut-and-count              \\ 
    Bin boundaries in \met \,[\GeV]        & $[250,350,450,600,\inf]$ &                   \\
    \hline \hline
  \end{tabular*}
}
  \label{tab:SRs_tN}
\end{table}

%% file: texfiles/sr_tN_diag.tex
\subsubsection{Compressed $\topLSP$ decay}
\label{subsec:tN_diag}

The three BDT selections ($\tNdiaglow$, $\tNdiagmed$, and $\tNdiaghigh$) are summarised in Table~\ref{tab:SRs_tN_diag_low_high} and detailed in the following.

\begin{table}[t]
        \centering
\small
        \caption{Overview of the signal selections using  BDTs to target compressed \topLSP\ scenarios. 
  List values are provided in between parentheses and square brackets denote intervals. 
  }
        \vspace{3mm}
        {\renewcommand{\arraystretch}{1.1}
        \begin{tabular*}{\textwidth}{@{\extracolsep{\fill}}| l | ccc |}
                \hline
                \hline
                Variable& \tNdiaglow & \tNdiagmed & \tNdiaghigh \\ 
                \hline
                Preselection & low-\met & low-\met & high-\met\\
                \hline
                Number of (jets,~$b$-tags)& ($\ge4$,~$\ge1$) & ($\ge4$,~$\ge1$) & ($\ge5$,~$\ge1$) \\
                Jet $\pt$ \,[\GeV] & $>\!(120,~25,~25,~25)$ & $>\!(100,~50,~25,~25)$ & $>\!(25,~25,~25,~25,~25)$ \\
                \hline
                $\met$ \,[\GeV]&  $> 100$ &  $> 120$ & $> 230$\\
                $\mt$ \,[\GeV]&  $> 90$ & $> 120$ & $> 120$ \\
                $R_{\textrm{ISR}} $ & -- & -- & > 0.4 \\
                $\pT(\ttbar)$ \,[\GeV]&  $> 400$ & -- & -- \\
                $|\Delta \phi(\ell, \ttbar)|$ &  $> 1.0$ & -- & -- \\
        $|\Delta\phi(j_{1,2},\Ptmiss)|$          & $>0.4$ & $>0.4$ & --    \\ 
        $\mtTwoTau$ based $\tau$-veto\,[\GeV]   & --     & $>80$  & --     \\ 
                BDT score &  BDT\_low $> 0.55$ &  BDT\_med $> 0.75$& BDT\_high$> 0.8$\\
                \hline \hline
                Exclusion technique & cut-and-count &shape-fit\,in\,BDT\,score&shape-fit\,in\,BDT\,score\\
                BDT\,score\,bin\,boundaries&  --  & $[0.4,0.5,0.6,0.7,0.8,1.0]$&$[0.6,0.7,0.8,1.0]$ \\
                \hline \hline
        \end{tabular*}
        }
        \label{tab:SRs_tN_diag_low_high}
\end{table}

\paragraph{Low $\tone$ mass}

For $\tone$ masses close to the top-quark mass a BDT is trained for the \tNdiaglow\ signal region. The preselection is based on the low-$\met$ selection in Table~\ref{tab:preselection}.

The variables input to the BDT are, in decreasing order of their importance for BDT performance: the difference $\Delta \mt^{\alpha}$ in $\mt$ between the SM and signal hypothesis, $\met$, the top-quark mass $m(t_{\mathrm{lep}}^{\alpha})$ of the leptonic top candidate under the signal hypothesis, $mt$, the azimuthal angles between the lepton and the $\ttbar$ system, as well as between the lepton and $\vec{p}_\mathrm{T}(\nu^{\alpha})$ and the mass $m(t_{\mathrm{had}}^{\mathrm{ISR}})$ of the hadronic top candidate.

The BDT output, from here on referred to as BDT\_low,
is used to define a single-bin cut-and-count signal region, using the optimal point of \mbox{BDT\_low $> 0.55$}, determined by maximising the expected discovery significance. 
To avoid a significant extrapolation between control and signal regions
an additional selection of $\pt(\ttbar) \geq 400$~GeV and $|\Delta \phi(\ell, \ttbar)| \geq 1.0$ is applied for all selected regions in the \tNdiaglow\ .

\paragraph{Medium $\tone$ mass}

Stop masses from about 200 to 400\,\GeV\ in the compressed scenario are targeted by a BDT using the 
low-$\met$ preselection given in Table~\ref{tab:preselection}. The input variables of the BDT, listed by decreasing order of importance are: $\Delta\phi(\Ptmiss,t_{\textrm{had}}^{\chi})$,
\mTopChi, $\met$, $\mt$, the number of jets, the angular variables $\Delta R (b,\ell)$, $\Delta\phi(t_{\textrm{had}}^{\chi}, t_{\textrm{lep}}^{\chi})$, 
as well as the fourth and third jet $\pt$, and \HTmissSig.

The BDT output score, referred to in the following as BDT\_med, is used to define a signal region called $\tNdiagmed$, based on the expected significance for a $\tone$ mass of 250\,$\GeV$.
The known signal shape is exploited for the exclusion of signal models, using five bins in the BDT score, including also BDT bins lower than the SR.

\paragraph{High $\tone$ mass}

For compressed bino LSP scenarios with high $\tone$ mass, a BDT is trained using the following variables, listed by decreasing order of importance: $R_{\textrm{ISR}}$, the angular variables $\Delta\phi(t_{\textrm{had}}^{\chi}, t_{\textrm{lep}}^{\chi})$, $\Delta R (b,\ell)$, and $\Delta \phi ({\textrm{ISR, I}})$, masses $\mt$, $M_{\textrm{T}}^{\textrm{S}}$, as well as the fourth jet $\pt$, \mTopChi\, third jet $\pt$, and the number of jets in the di-stop decay system, derived using the RJR techniques as described in Section \ref{sec:objects}. In addition to the high-\met\ preselection, a tightened selection of $\mt > 120$\,\GeV\ is imposed to control the multijet background. An additional selection of $R_{\textrm{ISR}} > 0.4$ is applied to further reduce the background while retaining high efficiency for the considered signal events.

The resulting BDT output score, hereafter called BDT\_high, is used to define the \tNdiaghigh\ signal region. In addition, three BDT bins are employed in a shape-fit to improve the exclusion sensitivity.

%% file: texfiles/sr_bWN.tex
\subsubsection{\threeBody\ and \fourBody\ decays}
\label{sec:SRs_other}

When the mass difference between the \tone\ and the \ninoone\ is smaller than the top-quark mass but greater than the sum of the $W$-boson and bottom-quark masses, the \stop\ decays dominantly through the three-body channel into a bottom quark, a $W$ boson, and a neutralino. 
The \bWN\ SR is optimised to search for these events.
Compared to the scenario with on-shell top quarks, the three-body decay yields the same final-state leptons and jets but with significantly lower momenta, although typically still above the reconstruction thresholds.

The \amtTwo\ variable is a powerful discriminant for separating dileptonic \ttbar\ background from signal models in this region of phase space.
Because $m_{\tone} - m_{\ninoone}$ is below the top-quark mass for signal, \amtTwo\ peaks at low values, while dileptonic \ttbar\ events typically saturate at values nearer to the top-quark mass.
A shape-fit technique is employed, using five bins of \amtTwo, similar to the shape-fit employed in the \tNmed\ SR.

When the \tone\ mass is much closer to the \ninoone\ mass, the stop undergoes a four-body decay with an off-shell $W$ boson, characterised by events having even lower momentum leptons and jets than in the three-body decay. A soft-lepton SR, $\bCsoftdiag$, designed for the higgsino LSP scenario with a relaxed \mt\ requirement, provides good sensitivity to this scenario.
A shape-fit is performed in the $\pt^{\ell}$/$\met$ variable, using three bins for the model-dependent exclusion fit.

The event selections for \bWN\ and \bffN\ are summarised in Table~\ref{tab:SRs_other}.

\begin{table}[t]
  \centering
  \caption{Overview of the event selections for the \bWN\ and \bffN\ SRs. 
  List values are provided in between parentheses and square brackets denote intervals. 
  The veto on the reclustered hadronic top-quark candidate is satisfied for events where no reclustered jet candidate is found, or where the mass of the hadronic top-quark candidate (\mTopRecluster) is below 150\,$\GeV$.  For the \bffN\ SR, the leading jet is required to not be $b$-tagged.
  }
  \vspace{3mm}
{\renewcommand{\arraystretch}{1.1}
  \begin{tabular*}{\textwidth}{@{\extracolsep{\fill}}| l | cc |}
                \hline
    \hline
                Signal region                   & \bWN                 & \bffN                             \\ \hline
                Preselection                    & high-\met\           & soft-lepton                       \\ \hline
                Number of (jets,~$b$-tags)      & ($\ge4$,~$\ge1$)     & ($\ge2$,~$\ge1$)                  \\
                Jet $\pt$ \,[\GeV]              & $>(50,~25,~25,~25)$  & $>(400,~25)$                      \\
                $b$-tagged jet $\pt$ \,[\GeV]   & $>25$                & $>25$                             \\
                \hline
                \met \,[\GeV]                   & $>300$               & $>300$                            \\
                \mt \,[\GeV]                    & $>130$               & $<160$                            \\
                \amtTwo  \,[\GeV]               & $<110$               & --                                \\
    \mTopRecluster \,[\GeV]         & --              & top veto                          \\
                $\pt^{\ell}$/\met               & --                   & $<0.02$                           \\
                \dPMETlep                       & $<2.5$               & --                                \\
                \dphiBPtmissMin                 & --                   & $<1.5$                            \\
        $|\Delta\phi(j_{1,2},\Ptmiss)|$  & \multicolumn{2}{ c |}{$>0.4$}                           \\ 
        $\mtTwoTau$ based $\tau$-veto\,[\GeV] & $>80$          & --                                \\ 
    \hline \hline
     Exclusion technique    & shape-fit in \amtTwo\ & shape-fit in $\pt^{\ell}$/$\met$ \\ 
    Bin boundaries in \amtTwo\ [\GeV] or $\pt^{\ell}$/$\met$ & $[0,91,97,106,118,130]$ & $[0,0.01,0.015,0.02]$          \\
    \hline \hline
  \end{tabular*}
}
  \label{tab:SRs_other}
\end{table}

%% file: texfiles/sr_bC2x.tex
If the wino mass parameter $M_2$ is small enough, the stop may decay directly into \chinoonepm\ and \ninotwo\ (in addition to the \ninoone, as the bino is still assumed to be the LSP). In this case, the decays \bChargino\ and \topNLSP\ become relevant, leading to a more complex phenomenology than that probed in the pure bino LSP scenario. 
The SRs targeting this scenario are referred to as \texttt{bC2x}.

Two SRs target the \bChargino\ decay: the \bCmed\ and \bCdiag\ SRs. The kinematics of the decay products are governed by the different mass-splittings, with high-$\pt$ $b$-jets produced from large $\Delta m$($\tone$,$\chinoonepm$) and high-$\pt$ $W$ bosons from large $\Delta m$($\chinoonepm$,$\ninoone$).
In addition to the high-\met\ preselection, two $b$-tagged jets and a hadronic $W$-boson candidate with a mass satisfying $\mWRecluster > 50$\,\GeV\ are required. Tight requirements on \mt\ and \amtTwo\ are placed to reduce the \ttbar\ background. The main backgrounds after the full signal selection are the $\ttbar Z (\nu\nu)$, dileptonic \ttbar, and single-top $Wt$ processes.

An additional SR, $\bCbv$, is designed for the simplified model $\bChargino$ scenario with $\Delta m$($\tone$,$\chinoonepm) = 10$\,$\GeV$, leading to a signature where the \bjets are too soft to be reconstructed. 

The event selections for \bCdiag, \bCmed\ and \bCbv\ are summarised in Table~\ref{tab:SRs_bC2x}.

\begin{table}[t]
  \centering
  \caption{Overview of the event selections for the $\bCmed$, $\bCdiag$, and \bCbv\ SRs. 
  List values are provided in between parentheses and square brackets denote intervals. 
  }
  \vspace{3mm}
{\renewcommand{\arraystretch}{1.1}
  \begin{tabular*}{\textwidth}{@{\extracolsep{\fill}}| l | ccc |}
                \hline
    \hline
                Signal region                   & \bCdiag           & \bCmed            & \bCbv         \\ \hline
                Preselection                    & \multicolumn{3}{ c |}{high-\met\ preselection} \\ \hline
                Number of (jets,~$b$-tags)      & ($\ge4$,~$\ge2$)  & ($\ge4$,~$\ge2$)  & ($\ge2$,~$=0$)      \\
                Jet $\pt$ \,[\GeV]              & $>(75,~75,~75,~30)$  & $>(200,~140,~25,~25)$& $>(120,~80)$\\
                $b$-tagged jet $\pt$ \,[\GeV]   & $>(30,~30)$       & $>(140,~140)$     & -- \\
                \hline
                \met \,[\GeV]                   & $>230$            & {$>230$}          & {$>360$} \\
                \HTmissSig                      & $>13$             & {$>10$}           & {$>16$} \\
                \mt \,[\GeV]                    & $>180$            & $>120$            & $>200$ \\
                \amtTwo  \,[\GeV]               & $>175$            & {$>300$}          & -- \\
                $|\Delta\phi(j_{1},\Ptmiss)|$             & $>0.7$            & $>0.9$            & $>2.0$ \\
                $|\Delta\phi(j_{2},\Ptmiss)|$             & $>0.7$            & $>0.9$            & $>0.8$ \\
                \mWRecluster \,[\GeV]           & $>50$             & $>50$             & $[70,100]$     \\
                \dPMETlep                       & --                & --                & $>1.2$ \\
                $|\Delta\phi(j_{1,2},\Ptmiss)|$          & \multicolumn{3}{ c |}{$>0.4$}    \\ 
                $\mtTwoTau$ based $\tau$-veto\,[\GeV]   & $>80$     & $>80$             & -- \\ 
                Lepton $\pt$ \,[\GeV]           & -- & -- & $>60$ \\
    \hline \hline
     Exclusion technique    & cut-and-count     & cut-and-count     & cut-and-count \\ 
    \hline \hline
  \end{tabular*}
}
  \label{tab:SRs_bC2x}
\end{table}

%% file: texfiles/sr_bCsoft.tex
The SRs optimised for the pure bino LSP scenarios such as $\tNmed$ have sensitivity to the higgsino model in events where a lepton is produced by a top quark from the stop decay. However, three additional SRs, $\bCsoftdiag$, $\bCsoftmed$, and $\bCsofthigh$, are designed to target the case when the lepton is soft, originating instead from a $\chinoonepm$ decay via a highly off-shell $W$ boson ($\chinoonepm \rightarrow \ninoone + W^{*}(\ell\nu)$). This is particularly important in scenarios with $m_{tR} < $ $m_{q3L}$ where the $\mathcal{B}(\bChargino)$ is large. These soft-lepton SRs are defined to be orthogonal to the \tNmed\ SR so that they can be statistically combined to profit from covering both decay chains.

The \bCsoftdiag\ SR targets a region where the mass difference between the stop and higgsinos is less than the mass of the top quark, so the stop must decay via the \bChargino\ mode.
Since none of the decay products receive a large momentum transfer, a high-$\pt$ ISR jet is required, resulting in a boost of the $\tone\tone$ system in order to achieve better separation between the signal and background. As a result, the signature is characterised by a high-$\pt$ jet, large \met, and a soft lepton. 
The main background after all selection requirements is semileptonic \ttbar\ and $W$+jets processes.
The \bCsoftdiag\ SR with relaxed \mt\ requirement is found to be sensitive to the \fourBody\ signature and is described further in Section~\ref{sec:SRs_other}.

The second SR, $\bCsoftmed$, targets generic higgsino models where each of the decays \bChargino, \topLSP, and \topNLSP\ are allowed.
In particular, it is designed to select the large fraction of events that produce ``mixed'' decays, where one \tone decays via a chargino and the other via a neutralino.  In such cases, the \bChargino\ decay produces a high-\pt\ $b$-jet, while the $b$-jet from the other branch, \topLSP\ or \topNLSP, can be much softer. The third SR, $\bCsofthigh$, targets the higher stop masses, 
focusing on the \bChargino\ signature. The $b$-jet is boosted due to the large mass difference between the stop and higgsino states, and therefore the signature is characterised by two high-$\pt$ $b$-jets, large \met, and a soft lepton. 
The remaining background after all signal selection requirements is dominated by semileptonic $\ttbar$, single-top $Wt$, and $W$+heavy-flavour jets events.

In all three SRs, $\pt^{\ell}$/$\met$ is a powerful discriminant as the higgsino signature is characterised by low-$\pt$ leptons and large $\met$, while the SM backgrounds are dominated by events where the \met arises from a leptonic $W$-boson decay, producing lepton \pt and \met of a similar magnitude. A shape-fit in $\pt^{\ell}$/\met\ is performed, similar to the shape-fits implemented for the \tNmed\ and \bWN\ SRs.

The event selections for $\bCsoftdiag$, $\bCsoftmed$, and $\bCsofthigh$ are detailed in Table~\ref{tab:SRs_bCsoft}.

\begin{table}[t]
  \centering
  \caption{Overview of the event selections for the $\bCsoftdiag$, $\bCsoftmed$, and \bCsofthigh\ SRs. 
  List values are provided in between parentheses and square brackets denote intervals. 
  The veto on the reclustered hadronic top-quark candidate is satisfied for events where no reclustered jet candidate is found, or where the mass of the hadronic top-quark candidate (\mTopRecluster) is below 150\,$\GeV$. 
  For the \bCsoftdiag\ SR, the leading jet is required not to be $b$-tagged.
  }
  \vspace{3mm}
{\renewcommand{\arraystretch}{1.1}
  \begin{tabular*}{\textwidth}{@{\extracolsep{\fill}}| l | ccc |}
                \hline
    \hline
                Signal region                   & \bCsoftdiag          & \bCsoftmed           & \bCsofthigh         \\ \hline
                Preselection                    & \multicolumn{3}{ c |}{soft-lepton preselection}                   \\ \hline
                Number of (jets,~$b$-tags)      & ($\ge2$,~$\ge1$)     & ($\ge3$,~$\ge2$)     & ($\ge2$,~$\ge2$)    \\
                Jet $\pt$ \,[\GeV]              & $>(400,~25)$         & $>(120,~60,~40)$     & $>(100,~100)$       \\
                $b$-tagged jet $\pt$ \,[\GeV]   & $>25$                & $>(120,~60)$         & $>(100,~100)$       \\
                \hline
                \met \,[\GeV]                   & $>300$               & $>230$               & $>230$              \\
                \mt \,[\GeV]                    & $<50$                & $<160$               & $<160$              \\
                $\pt^W$ \,[\GeV]                & --                   & $>400$               & $>500$              \\
                $\pt^{\ell}$/\met               & $<0.02$              & $<0.03$              & $<0.03$             \\
                \amtTwo  \,[\GeV]               & --                   & $>200$               & $>300$              \\
                \mTopRecluster \,[\GeV]         & top veto             & --                    & --                  \\
                \dphiBPtmissMin                 & $<1.5$               & $>0.8$               & $>0.4$              \\
                $\Delta R(b_1,b_2)$             & --                   & --                   & $>0.8$              \\
        $|\Delta\phi(j_{1,2},\Ptmiss)|$ & \multicolumn{3}{ c |}{$>0.4$}    \\ 
    \hline \hline
     Exclusion technique & shape-fit in $\pt^{\ell}$/\met & shape-fit in $\pt^{\ell}$/\met  & shape-fit in $\pt^{\ell}$/\met \\ 
    Bin boundaries in $\pt^{\ell}/\met $& $[0,0.01,0.015,0.02]$      & $[0,0.015,0.03,0.1]$       & $[0,0.015,0.03,0.1]$      \\ 
    \hline \hline
  \end{tabular*}
}
  \label{tab:SRs_bCsoft}
\end{table}

%% file: texfiles/sr_welltemp.tex
For the bino/higgsino mix scenario, 
the SRs designed for other scenarios are found to have good sensitivity for this scenario, and are therefore used.

%% file: texfiles/sr_DM.tex
Two SRs, \DMlow\ and $\DMhigh$, are designed to search for dark matter particles that are pair-produced via a spin-0 mediator (either scalar or pseudoscalar) produced in association with \ttbar. The \DMlow\ SR is optimised for mediator masses around $m_{\phi} = 20$\,\GeV, while the \DMhigh\ SR targets mediator masses around $m_{\phi} = 300$\,\GeV.

In addition, a predecessor to the \DMlow\ signal region, originally designed for a search using a smaller data set (13.2 \ifb), has been retained, as in that search the number of observed events exceeded the background prediction.
This signal region, which was previously called $\DMlow$, is referred to here as $\DMlowloose$.

Table~\ref{tab:SRs_DM} details the event selections for each of the three SRs. 
At least one reconstructed hadronic top-quark candidate is required with $\mTopRecluster>$130\,\GeV\ in the newly defined SRs. A high \amtTwo\ requirement and an angular selection requirement of $\minDeltaPhi$ are further imposed to reduce the \ttbar\ background. The main backgrounds after all signal selection requirements are the $\ttbar Z (\nu\nu)$, dileptonic \ttbar, and $W$+heavy-flavour processes.

The event selections for $\DMlowloose$, $\DMlow$, and $\DMhigh$ are summarised in Table~\ref{tab:SRs_DM}.

\begin{table}[t]
  \centering
  \caption{Overview of the event selections for the $\DMlowloose$, $\DMlow$, and $\DMhigh$ SRs. 
List values are provided in between parentheses.
  }
  \vspace{3mm}
{\renewcommand{\arraystretch}{1.1}
  \begin{tabular*}{\textwidth}{@{\extracolsep{\fill}}| l | ccc |}
                \hline
    \hline
    Signal region                      & \DMlowloose         & \DMlow               & \DMhigh              \\ \hline
                Preselection                            & \multicolumn{3}{ c |}{high-\met\ preselection}                    \\ \hline
    Number of (jets,~$b$-tags)     & ($\ge4$,~$\ge1$)    & ($\ge4$,~$\ge1$)     & ($\ge4$,~$\ge1$)     \\
    Jet $\pt $ \,[\GeV]       & $>(60,~60,~40,~25)$ & $>(120,~85,~65,~25)$ & $>(125,~75,~65,~25)$ \\
                $b$-tagged jet $\pt $ \,[\GeV]     & $>25$                & $>60$                & $>25$                \\
    \hline
    \met \,[\GeV]                     & $>300$         & $>320$                & $>380$               \\
    \mt \,[\GeV]                 & $>120$         & $>170$                & $>225$               \\
    \HTmissSig                     & $>14$            & $>14$                & --                   \\   
    \amtTwo  \,[\GeV]              & $>140$         & $>160$                & $>190$               \\
    \mTopRecluster \,[\GeV]                 & --             & $>130$               & $>130$               \\
    \dPMETlep                        & $>0.8$          & $>1.2$                & $>1.2$               \\
    \minDeltaPhi                & $>1.4$          & $>1.0$                & $>1.0$               \\
        $\mtTwoTau$ based $\tau$-veto\,[\GeV]   & \multicolumn{3}{ c |}{$>80$}                                      \\ 
    \hline \hline
     Exclusion technique      & cut-and-count   & cut-and-count     & cut-and-count               \\ 
    \hline \hline
  \end{tabular*}
}
  \label{tab:SRs_DM}
\end{table}

%% file: texfiles/backgrounds.tex
\section{Background estimates}
\label{sec:backgrounds}

The dominant background processes in this analysis originate from \ttbar, single-top $Wt$, $\ttbar+Z(\rightarrow \nu\bar{\nu})$, and $W$+jets production.
Most of the \ttbar\ and $Wt$ events in the hard-lepton signal regions have both $W$ bosons decaying leptonically, where one lepton is `lost' (meaning it is either not reconstructed, not identified, or removed by the overlap removal procedure) or one $W$ boson decaying leptonically and the other via a hadronically decaying $\tau$ lepton. This is in contrast to the soft-lepton signal regions, where most of the \ttbar\ and $Wt$ contribution arises from semileptonic decays. 

These \ttbar\ background decay components are treated separately, referred to as 1L and 2L, which also includes the dileptonic \ttbar\ process where a $W$ boson decays into a $\tau$ lepton that subsequently decays hadronically. The $\ttbar+Z$ background combined with the subdominant $\ttbar+W$ contribution is referred to as $\ttbar+V$. Other background contributions arise from dibosons, $Z$+jets, and multijet production. The multijet background is estimated from data using a fake-factor method~\cite{HIGG-2013-13}, and it is found to be negligible in all regions.

The main background processes are estimated via a dedicated CR, used to normalise the simulation to the data with a simultaneous fit, discussed in Section~\ref{sec:results}.
The CRs are defined with event selections that are kinematically close to the SRs but with a few key variable requirements inverted to significantly reduce the potential signal contribution and enhance the yield and purity of a particular background. 
Each SR has dedicated CRs for the background processes that have the largest contributions. The following background processes are normalised in dedicated CRs: semileptonic \ttbar\ (T1LCR), dileptonic \ttbar\ (T2LCR), $W$+jets (WCR), single-top (STCR), and $\ttbar+V$ (TZCR) processes. 
All other backgrounds are normalised with the most accurate theoretical cross-sections available. 

Several signal regions ($\bWN$, $\tNdiaglow$, and $\tNdiaghigh$) that are dominated exclusively by either semileptonic or dileptonic \ttbar\ events have only one associated CR, denoted generically TCR. Signal regions can have fewer associated CRs when the fractional contribution of the corresponding background is small. For the shape-fit analyses, the CRs of each background are not binned and only one normalisation factor is extracted for each background process, which is applied in all SR bins.\footnote{The binned CR approach has been tested by comparing the results to a one-bin CR. The normalisation factors were found to be consistent with each other within the statistical uncertainties.}

The background estimates are tested using VRs, which are disjoint from both the CRs and SRs. Background normalisations determined in the CRs are extrapolated to the VRs and compared with the observed data. Each SR has associated VRs for the \ttbar\ (T1LVR and T2LVR) and $W$+jets (WVR) processes, which are constructed by inverting or relaxing the selection requirements to be orthogonal to the corresponding SR and CRs. A single-top $Wt$ VR (STVR) is defined for the \bCsoftmed\ and \bCsofthigh\ SRs, where $Wt$ is one of the dominant background processes. 

The VRs are not used to constrain parameters in the fit, but provide a statistically independent test of the background estimates made using the CRs. The potential signal contamination in the VRs is studied for all considered signal models and mass ranges, and is found to be less than a few percent in most of the VRs, and less than 15\% in VRs for the \texttt{tN\_diag} SRs.

The background estimation techniques are categorised using several different approaches. The requirement of the presence of hadronic top-quark candidates (top-tagging) is used for the background estimate in the SRs targeting signals with high-\pt\ top quarks. Compared to previous analyses this background estimation technique has the advantage that the \ttbar\ background composition does not change in the extrapolation from CR to SR. Similarly hadronic $W$-boson reconstruction ($W$-tagging) is employed for the background estimate in the SRs targeting signals with high-\pt\ $W$ bosons decaying hadronically. In the following subsections the two approaches are described in detail together with the background estimates for the remaining SRs. Table~\ref{tab:bkgOverview} summarises the approaches for each SR with a brief description of the targeted signal scenarios, and each of those approaches are detailed in Sections~\ref{subsec:toptagCR}--\ref{subsec:softleptonCR}.

\begin{table}[t]
\begin{center}
\caption{Overview of various approaches for the background estimates in all signal regions together with the targeted signal scenario. The $\ttbar+Z(\ell\ell)$ control region (CR) described in Section~\ref{subsec:ttZ} is also defined in the top-tagging and $W$-tagging approaches, except for the \bCbv\ SR where the contribution of the $\ttbar+V$ background is negligible.
}
\vspace{3mm}
\renewcommand{\arraystretch}{1.5}
{\scriptsize
\begin{tabular}{| l | c | c | c | c |}
\hline\hline
SR             & Signal scenario & Background strategy                & Sections \\
\hline
\tNmed         & Pure bino LSP   & top-tagging + $\ttbar Z$ CR        & \ref{subsec:toptagCR} \\
\tNhigh        & Pure bino LSP   & top-tagging + $\ttbar Z$ CR        & \ref{subsec:toptagCR} \\
\tNdiaglow     & Pure bino LSP   & BDT                                & \ref{subsec:BDTCR} \\
\tNdiagmed     & Pure bino LSP   & BDT                                & \ref{subsec:BDTCR} \\
\tNdiaghigh    & Pure bino LSP   & BDT                                & \ref{subsec:BDTCR} \\
\bWN           & Pure bino LSP   & three-body                         & \ref{subsec:bWNCR} \\
\bffN          & Pure bino LSP   & soft-lepton                        & \ref{subsec:softleptonCR} \\
\hline
\bCmed         & Wino NLSP       & $W$-tagging + $\ttbar Z$ CR        & \ref{subsec:WtagCR} \\ 
\bCdiag        & Wino NLSP       & $W$-tagging + $\ttbar Z$ CR        & \ref{subsec:WtagCR} \\ 
\bCbv          & Wino NLSP       & $W$-tagging                        & \ref{subsec:WtagCR} \\ 
\hline
\bCsoftdiag    & Higgsino LSP    & soft-lepton                        & \ref{subsec:softleptonCR} \\
\bCsoftmed     & Higgsino LSP    & soft-lepton                        & \ref{subsec:softleptonCR} \\
\bCsofthigh    & Higgsino LSP    & soft-lepton                        & \ref{subsec:softleptonCR} \\
\hline
\DMlowloose    & DM+\ttbar       & \mt\ extrapolation + $\ttbar Z$ CR & \ref{subsec:toptagCR} \\ 
\DMlow         & DM+\ttbar       & top-tagging + $\ttbar Z$ CR        & \ref{subsec:toptagCR} \\ 
\DMhigh        & DM+\ttbar       & top-tagging + $\ttbar Z$ CR        & \ref{subsec:toptagCR} \\ 
\hline\hline
\end{tabular}
}
\label{tab:bkgOverview}
\end{center}
\end{table}

\input{texfiles/bkg_toptag}
\input{texfiles/bkg_tN_diag}
\input{texfiles/bkg_bWN}
\input{texfiles/bkg_bC2x}
\input{texfiles/bkg_bCsoft}
\FloatBarrier
\input{texfiles/bkg_ttZ}

%% file: texfiles/bkg_toptag.tex
\subsection{Hadronic top-tagging approach}
\label{subsec:toptagCR}

In SRs targeting signals with high-\pt\ top quarks ($\tNmed$, $\tNhigh$, $\DMlow$, and $\DMhigh$), a requirement is made that events contain a recursively reclustered jet with a mass consistent with the top-quark mass.
While the requirement on \mTopRecluster\ is powerful for identifying signals, it is also useful in defining CRs that are enriched in background processes with hadronically decaying top quarks (``top-tagged'') or depleted in such backgrounds (``top-vetoed'').

The CR for dileptonic \ttbar\ (T2LCR) requires \mt\ above the $W$-boson endpoint. The SR requirement on \amtTwo\ is inverted (to select events with values below the top-quark mass) and a hadronic top-quark veto is required to reduce the potential signal contamination and improve the purity. The semileptonic \ttbar\ CR (T1LCR) requires a tagged hadronic top-quark candidate and that the \mt\ be within a window around the $W$-boson mass. The background from semileptonic \ttbar\ events is negligible in the SR but can be sizeable in the other CRs.

The CRs for $W$+jets (WCR) and single-top (STCR) require \mt\ to be below the $W$-boson mass. Both CRs also require large \amtTwo\ and a hadronic top-quark veto, which is necessary to suppress the large semileptonic \ttbar\ background. The STCR also requires two $b$-tagged jets to reduce the $W$+jets contribution, and a minimum separation between the $b$-tagged jets, $\DeltaR(b_1,b_2)>1.2$. This latter requirement is useful to suppress the semileptonic \ttbar\ contribution, which can evade the \amtTwo\ endpoint when a charm quark from the hadronic $W$-boson decay is misidentified as a $b$-tagged jet, often leading to a small separation between the two identified $b$-tagged jets. Events with exactly one $b$-tagged jet or $\Delta R(b_1,b_2)<1.2$ are assigned to the WCR. In order to increase the $W$+jets purity, only events with a positively charged lepton are selected. This requirement exploits the asymmetry in the production of $W^+$ over $W^-$ events in LHC proton--proton collisions. The asymmetry is further enhanced by the requirement of large \met, as neutrinos from decays of the mostly left-handed $W^+$ boson are preferentially emitted in the momentum direction of the $W$ boson.

In addition, the background contribution from $\ttbar+V$ is large and a dedicated control region is designed, and is described in Section~\ref{subsec:ttZ}. 

\begin{figure}[htbp]
  \centering
  \includegraphics[width=.40\textwidth]{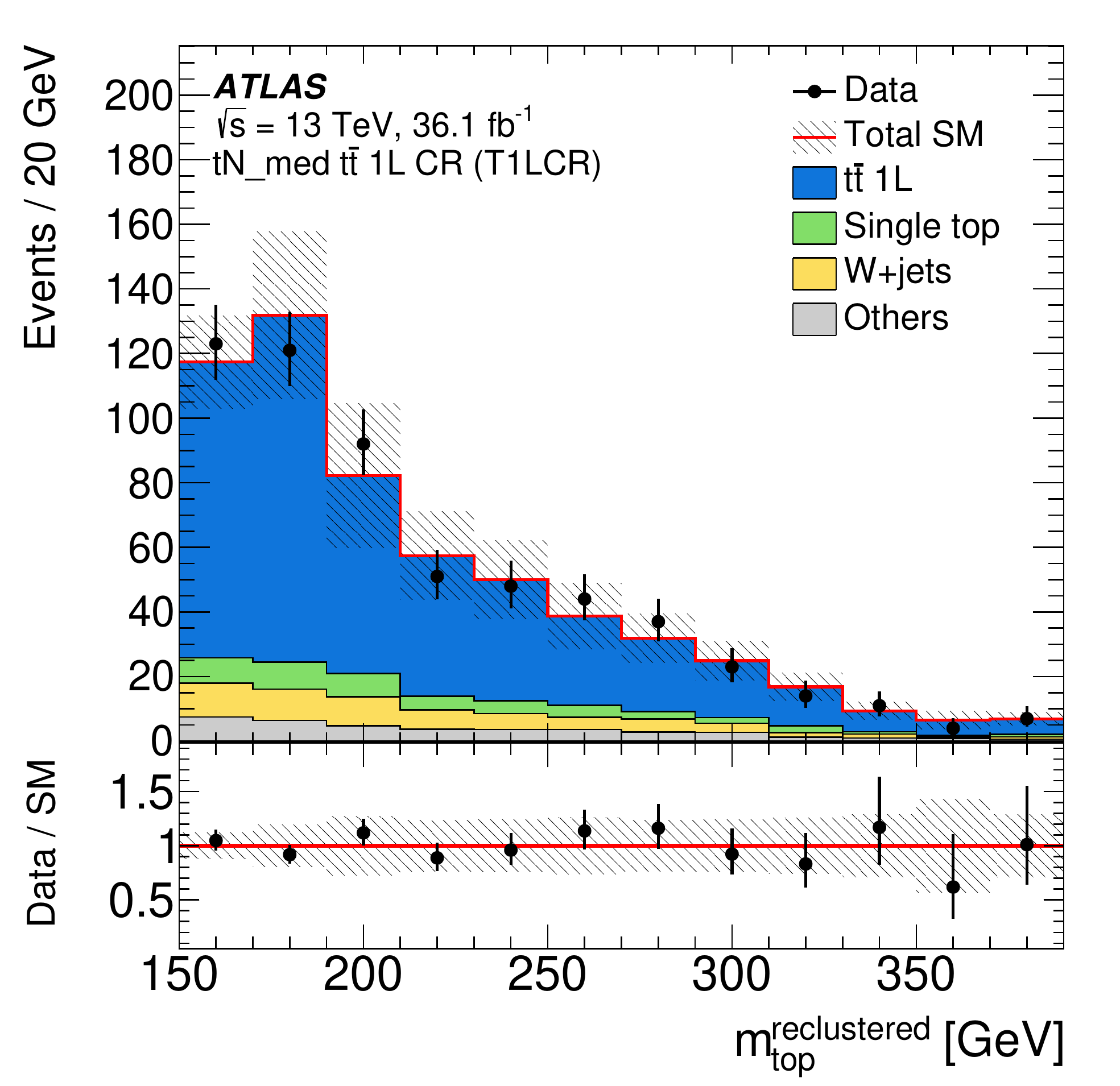}
  \includegraphics[width=.40\textwidth]{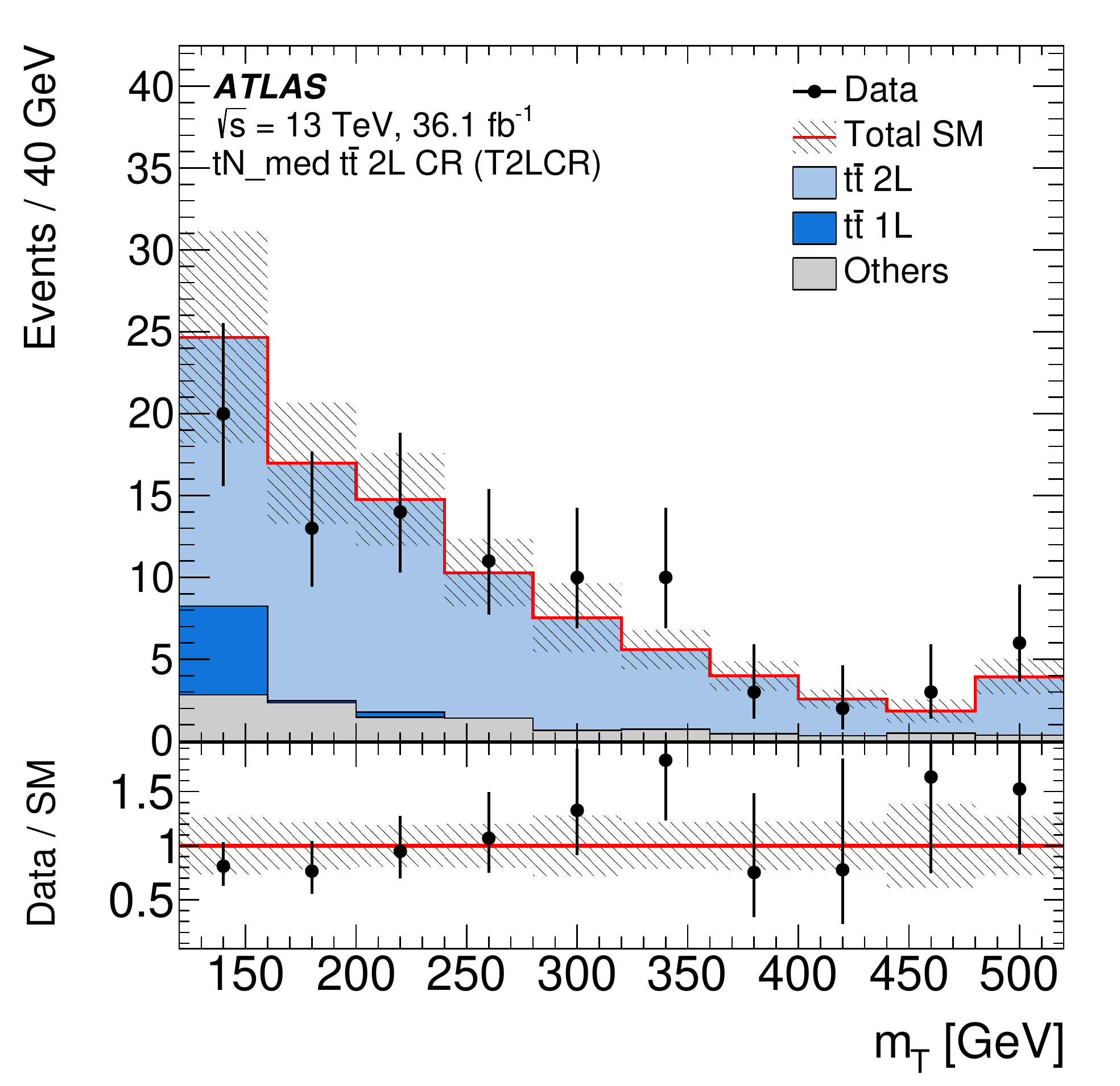}
  \includegraphics[width=.40\textwidth]{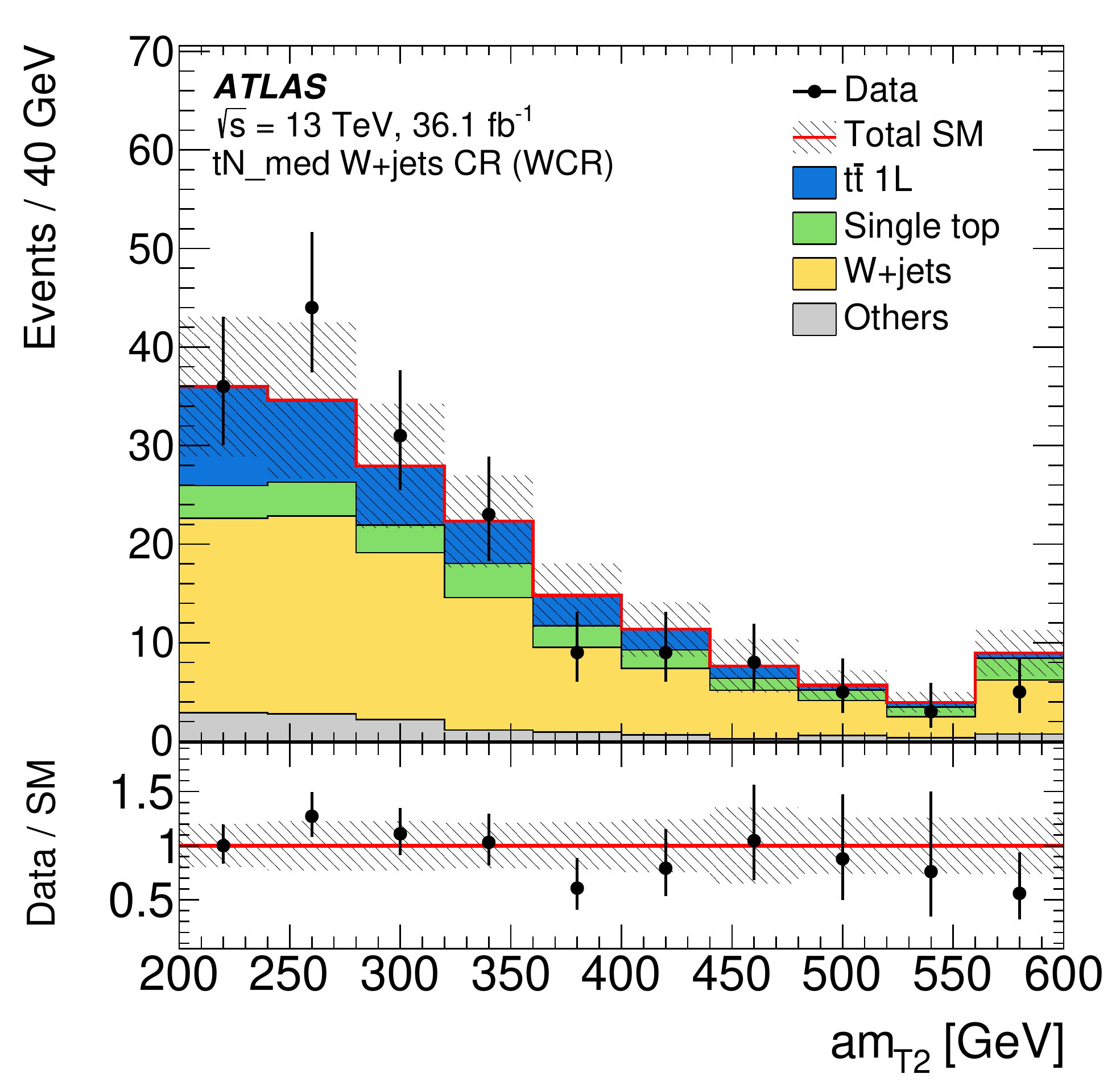}
  \includegraphics[width=.40\textwidth]{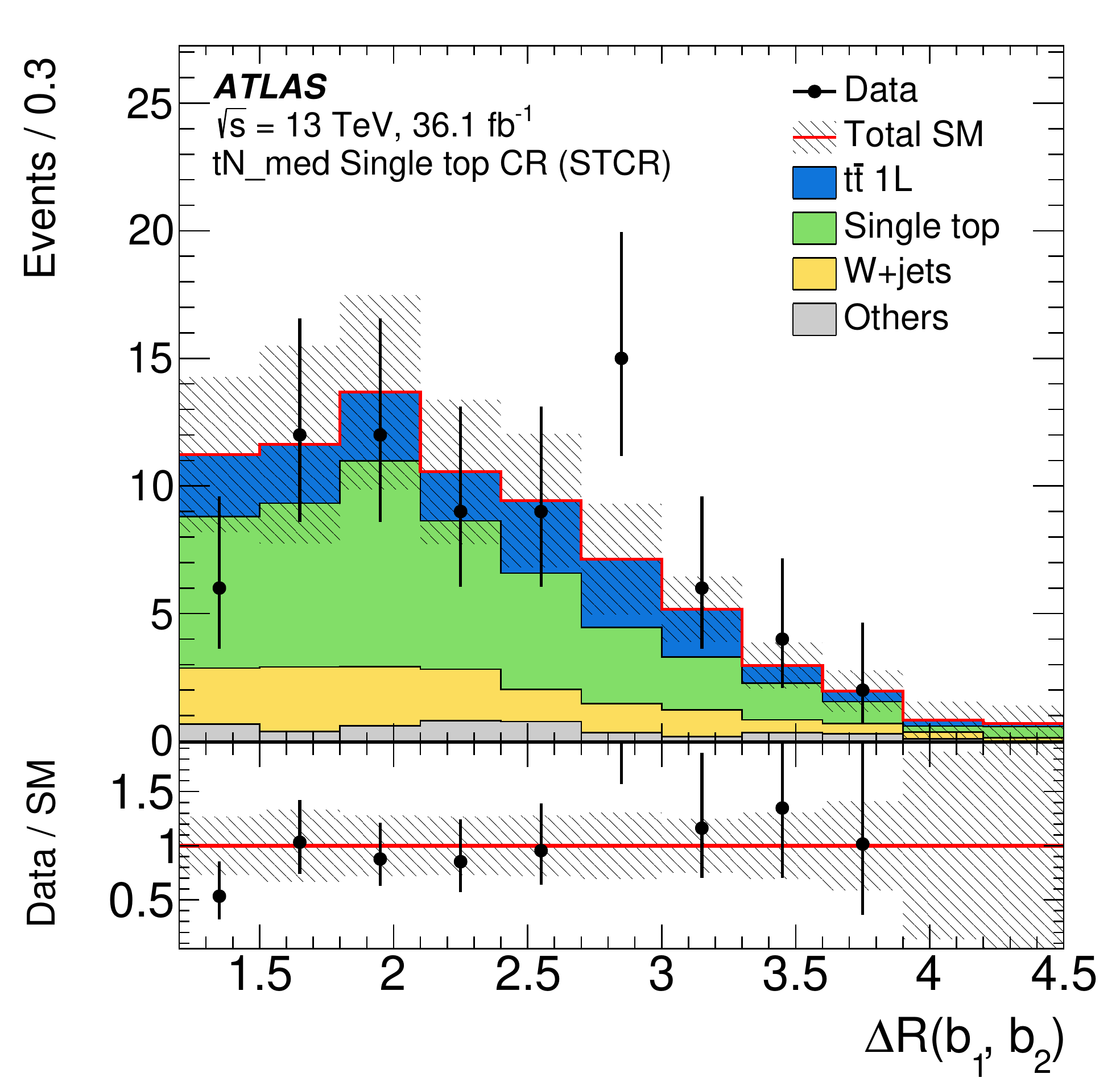}
  \includegraphics[width=.40\textwidth]{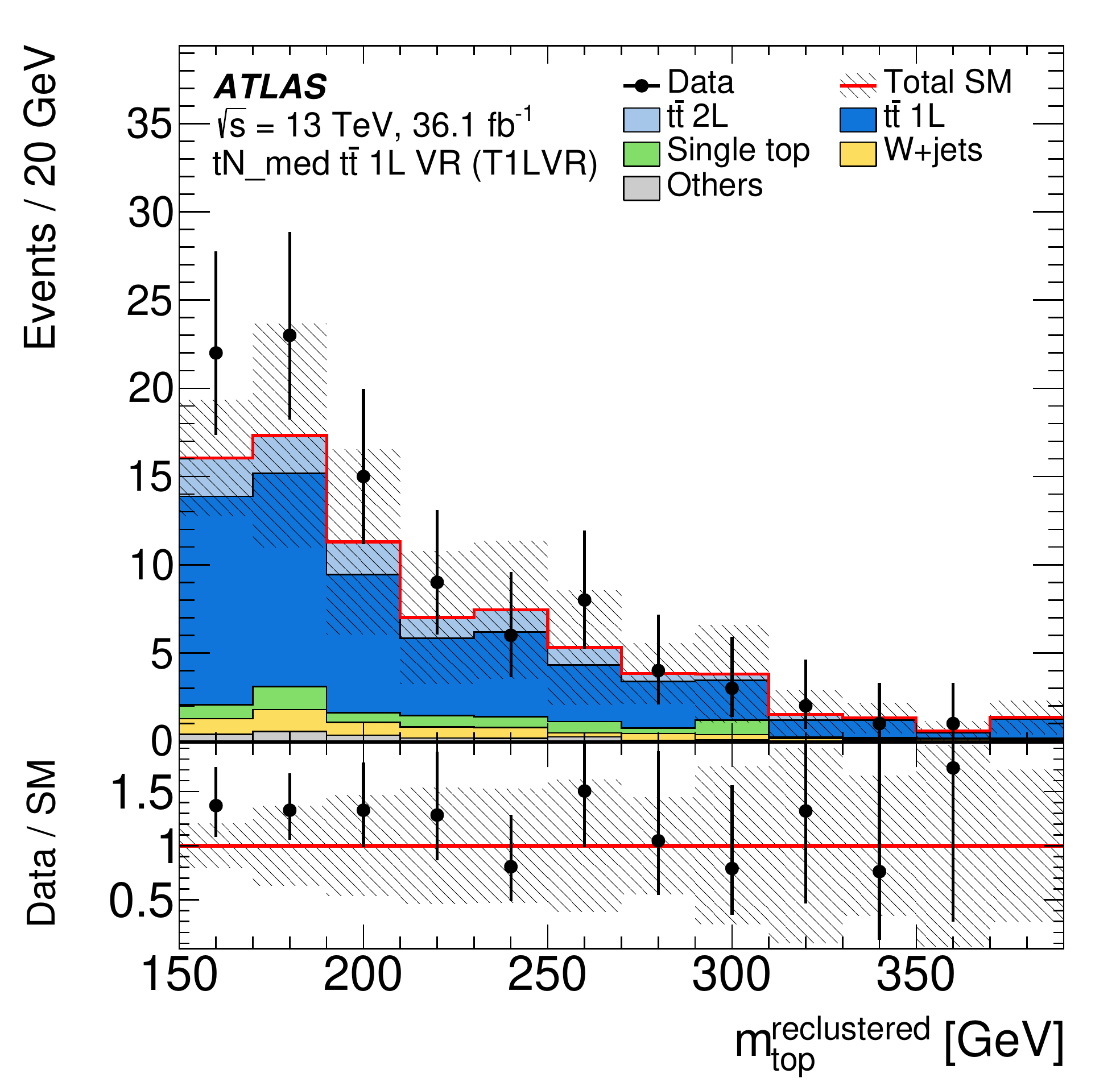}
  \includegraphics[width=.40\textwidth]{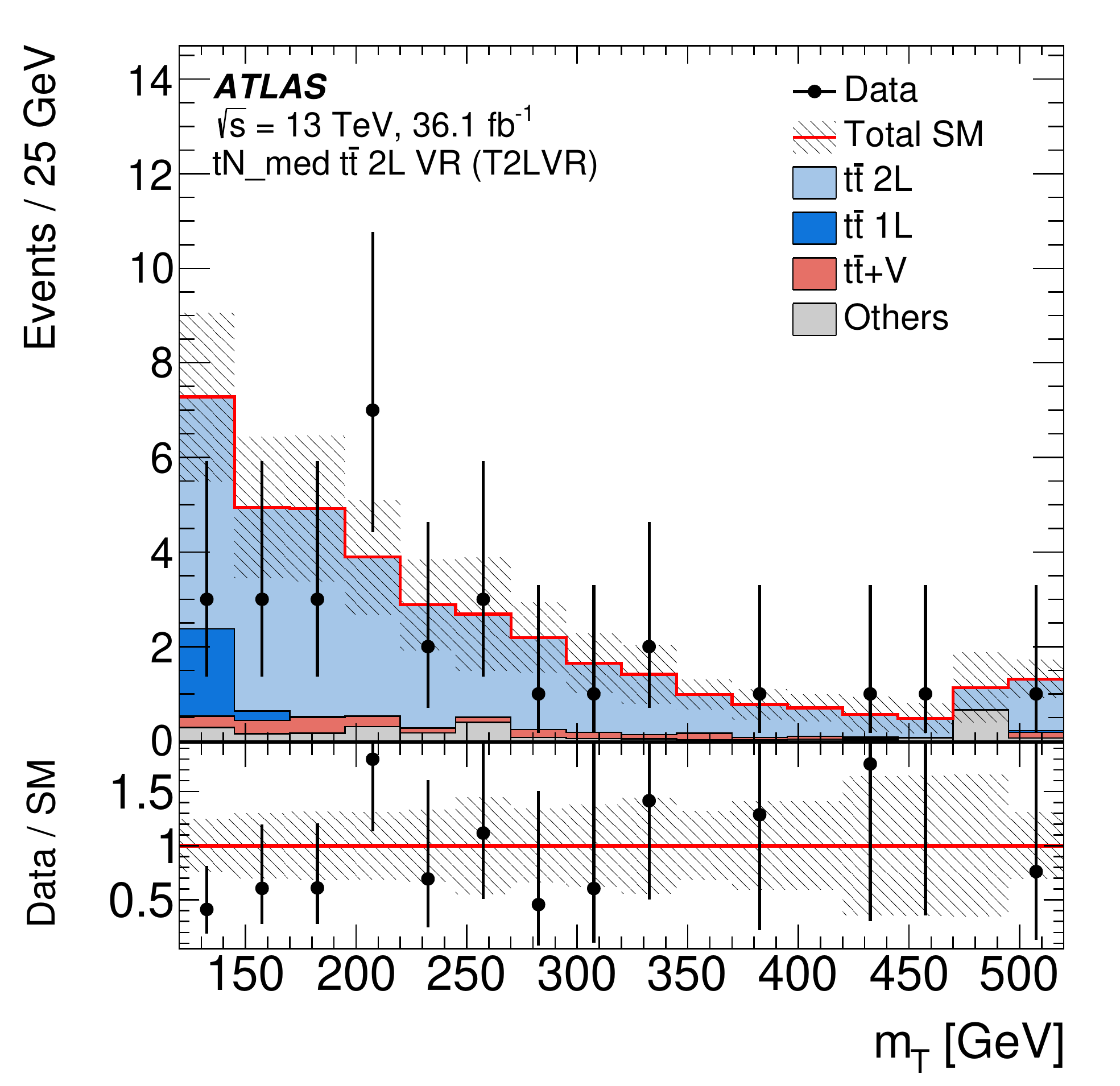}
  \caption{Various kinematic distributions in the \tNmed\ control and validation regions: (top left) reclustered jet mass (\mTopRecluster) in the semileptonic \ttbar\ control region, (top right) \mt\ in the dileptonic \ttbar\ control region, (middle left) \amtTwo\ in the $W$+jets control region, (middle right) $\DeltaR$($b_1$,$b_2$) in the single-top control region, (bottom left) reclustered jet mass (\mTopRecluster) in the semileptonic \ttbar\ validation region, and (bottom right) \mt\ in the dileptonic \ttbar\ validation region. Each of the backgrounds is scaled by a normalisation factor obtained from a simultaneous likelihood fit of the CRs. The category labelled `Others' stands for minor SM backgrounds that contribute less than 5\% of the total SM background. The hatched area around the total SM prediction and the hatched band in the Data/SM ratio include statistical and experimental uncertainties. The last bin contains overflows.
  }
  \label{fig:CRVRs_tN}
\end{figure}

Figure~\ref{fig:CRVRs_tN} shows various kinematic distributions in the CRs associated with the $\tNmed$ SR. The backgrounds are scaled with normalisation factors obtained from a simultaneous likelihood fit of the CRs, described in Section~\ref{sec:results}. 

A set of VRs associated with the corresponding CRs is defined by modifying the requirements on the $\mt$, $\amtTwo$, and hadronic top-tagging variables. The semileptonic \ttbar\ validation region (T1LVR) and $W$+jets validation region (WVR) slide the \mt\ window from $30$--$90$\, $\GeV$ to $90$--$120$\,$\GeV$. The dileptonic \ttbar\ VR (T2LVR) inverts the requirement of the hadronic top-quark veto (so that a hadronic top-quark tag is required) and relaxes the requirement on $\amtTwo$. Since the \ttbar\ events are mostly dileptonic after the large \mt\ requirement, the purity of dileptonic \ttbar\ events remains high, despite the hadronic top-quark tag requirement. The relaxed \amtTwo\ requirement significantly reduces the potential signal contamination. There is no single-top $Wt$ VR (STVR) for these CRs. The \mt\ window for the STCR extends to 120\,$\GeV$ in order to increase the number of data events entering the CR.
 
In Figure~\ref{fig:CRVRs_tN}, various kinematic distributions in the VRs associated with \tNmed\ are compared to the observed data. The backgrounds are scaled with normalisation factors obtained from a simultaneous likelihood fit of the CRs, described in Section~\ref{sec:results}.

Tables~\ref{tab:CRs_tN} and \ref{tab:CRs_DM} detail the definitions of the CRs and VRs associated with the SRs $\tNmed$, $\tNhigh$, $\DMlow$, and $\DMhigh$.

The CRs and VRs associated with \DMlowloose\ are retained unchanged from the previous analysis, and are described in Table~\ref{tab:CRs_DMlowloose}. The \ttbar\ and $W$+jets backgrounds are estimated from a low \mt\ region, $\mt \in [30,90]$ \gev, with and without a $b$-tag requirement, respectively. The corresponding VRs are defined with $\mt \in [90,120]$ \gev. The single-top $Wt$, and $\ttbar Z$ backgrounds are estimated using the same strategy as the rest of the regions described in this section.

\begin{table}[t]
  \centering
  \caption{
  Overview of the selections for the $\tNmed$ and $\tNhigh$ signal regions as well as the associated control and validation regions. The control regions include the semileptonic \ttbar\ control region (T1LCR), the dileptonic \ttbar\ control region (T2LCR), the $W$+jets control region (WCR), and the single-top $Wt$ control region (STCR).  The validation regions include the semileptonic \ttbar\ validation region (T1LVR), the dileptonic \ttbar\ validation region (T2LVR), and the $W$+jets validation region (WVR). List values are provided in between parentheses and square brackets denote intervals. 
  The veto on the reclustered hadronic top-quark candidate is satisfied for events where no reclustered jet candidate is found, or where the mass of the hadronic top-quark candidate (\mTopRecluster) is below the specified tag threshold. For the WCR, $\DeltaR$($b_1$,$b_2$) $<1.2$ is not required when the event has only one $b$-tagged jet.
  The selection of the $\ttbar+V$ control region (TZCR) is detailed in Section~\ref{subsec:ttZ}.
  }
  \vspace{3mm}
{\footnotesize
{\renewcommand{\arraystretch}{1.1}
  \begin{tabular*}{\textwidth}{@{\extracolsep{\fill}}| l | ccccc |}
                \hline
    \hline
                                      & \tNmed        & T1LCR/VR             & T2LCR/VR         & WCR/VR                 & STCR \\ \hline
                Preselection                            & \multicolumn{5}{ c |}{high-\met\ preselection} \\ \hline
    Number of (jets,~$b$-tags)     & ($\ge4$,~$\ge1$)  & ($\ge4$,~$\ge1$)     & ($\ge4$,~$\ge1$) & ($\ge4$,~$\ge1$)      & ($\ge4$,~$\ge2$) \\
    Jet $\pt$ \,[\GeV]       & \multicolumn{5}{ c |}{$>(60,~50,~40,~40)$} \\
    $b$-tagged jet $\pt$  \,[\GeV]     & \multicolumn{4}{ c |}{$>25$}                                                       & $>(25,~25)$\\
    \hline
    \met \,[\GeV]                     & \multicolumn{5}{ c |}{$>250$} \\
    \perpmet \,[\GeV]                     & \multicolumn{5}{ c |}{$>230$} \\
    \mt \,[\GeV]                       & $>160$       & $[30,90]$ / $[90,120]$ & $>120$           & $[30,90]$ / $[90,120]$  & $[30,120]$ \\
    \HTmissSig                        & $>14$       & $>10$              & $>10$            & $>10$                 & $>10$ \\
    \mTopRecluster \,[\GeV]                 & $>150$       & $>150$               & top veto / $>150$  & top veto              & top veto \\
    \amtTwo \,[\GeV]                        & $>175$       & $<200$               & $<200$ / $<130$    & $>200$                & $>200$ \\
    $\DeltaR(b,\ell)$              & $<2.0$       & --                   & --               & --                    & -- \\
    $\DeltaR$($b_1$,$b_2$)             & --               & --                   & --               & $<1.2$                & $>1.2$ \\
    Lepton charge                     & --             & --              & --               & $+1$                  & -- \\
        $|\Delta\phi(j_{1,2},\Ptmiss)|$          & \multicolumn{5}{ c |}{$>0.4$}    \\ 
        $\mtTwoTau$ based $\tau$-veto\,[\GeV]   & \multicolumn{5}{ c |}{$>80$}     \\ 
    \hline \hline
                                       & \tNhigh        & T1LCR/VR       & T2LCR/VR         & WCR/VR                 & STCR \\ \hline
                Preselection                            & \multicolumn{5}{ c |}{high-\met\ preselection} \\ \hline
    Number of (jets,~$b$-tags)     & ($\ge4$,~$\ge1$)  & ($\ge4$,~$\ge1$)     & ($\ge4$,~$\ge1$) & ($\ge4$,~$\ge1$)      & ($\ge4$,~$\ge2$) \\
    Jet $\pt$ \,[\GeV]       & \multicolumn{5}{ c |}{$>(100,~80,~50,~30)$} \\
    $b$-tagged jet $\pt$  \,[\GeV]     & \multicolumn{4}{ c |}{$>25$}                                                       & $>(25,~25)$\\
    \hline
    \met \,[\GeV]                     & $>550$       & $>350$              & $>350$           & $>350$                & $>350$ \\
    \mt \,[\GeV]                       & $>160$       & $[30,90]$ / $[90,120]$ & $>120$           & $[30,90]$ / $[90,120]$  & $[30,120]$ \\
    \HTmissSig                        & $>27$       & $>10$              & $>10$            & $>10$                 & $>10$ \\
    \mTopRecluster \,[\GeV]                 & $>130$       & $>130$               & top veto / $>130$  & top veto              & top veto \\
    \amtTwo \,[\GeV]                        & $>175$       & $<200$               & $<200$ / $<130$    & $>200$                & $>200$ \\
    $\DeltaR(b,\ell)$              & $<2.0$       & --                   & --               & --                    & -- \\
    $\DeltaR$($b_1$,$b_2$)             & --               & --                   & --               & $<1.2$                & $>1.2$ \\
    Lepton charge                     & --             & --              & --               & $+1$                  & -- \\
        $|\Delta\phi(j_{1,2},\Ptmiss)|$          & \multicolumn{5}{ c |}{$>0.4$}    \\ 
        $\mtTwoTau$ based $\tau$-veto\,[\GeV]   & \multicolumn{5}{ c |}{$>80$}     \\ 
    \hline \hline
  \end{tabular*}
}
}
  \label{tab:CRs_tN}
\end{table}

\begin{table}[t]
  \centering
  \caption{
  Overview of the selections for the $\DMlow$ and $\DMhigh$ signal regions as well as the associated control and validation regions. The control regions include the semileptonic \ttbar\ control region (T1LCR), the dileptonic \ttbar\ control region (T2LCR), the $W$+jets control region (WCR), and the single-top $Wt$ control region (STCR).  The validation regions include the semileptonic \ttbar\ validation region (T1LVR), the dileptonic \ttbar\ validation region (T2LVR), and the $W$+jets validation region (WVR). List values are provided in between parentheses and square brackets denote intervals. 
  The veto on the reclustered hadronic top-quark candidate is satisfied for events where no reclustered jet candidate is found, or where the mass of the hadronic top-quark (\mTopRecluster) is below a certain threshold. For the WCR, $\DeltaR$($b_1$,$b_2$) $<1.2$ is not required when the event has only one $b$-tagged jet.
  The selection of the $\ttbar+V$ control region (TZCR) is detailed in Section~\ref{subsec:ttZ}.
  }
  \vspace{3mm}
{\footnotesize
{\renewcommand{\arraystretch}{1.1}
  \begin{tabular*}{\textwidth}{@{\extracolsep{\fill}}| l | ccccc |}
                \hline \hline
                                       & \DMlow        & T1LCR/VR             & T2LCR/VR         & WCR/VR                & STCR \\ \hline
                Preselection                            & \multicolumn{5}{ c |}{high-\met\ preselection} \\ \hline
    Number of (jets,~$b$-tags)     & ($\ge4$,~$\ge1$)  & ($\ge4$,~$\ge1$)     & ($\ge4$,~$\ge1$) & ($\ge4$,~$\ge1$)     & ($\ge4$,~$\ge2$) \\
    Jet $\pt$  \,[\GeV]       & \multicolumn{5}{ c |}{$>(120,~85,~65,~60)$} \\
    $b$-tagged jet $\pt$  \,[\GeV]     & \multicolumn{4}{ c |}{$>25$}                                                       & $>(25,~25)$\\
    \hline
    \met \,[\GeV]                     & $>320$       & $>250$              & $>230$           & $>250$               & $>250$ \\
    \mt \,[\GeV]                       & $>170$       & $[30,90]$ / $[90,120]$ & $>120$           & $[30,90]$ / $[90,120]$ & $[30,120]$ \\
    \HTmissSig                        & $>14$       & $>10$              & $>10$            & $>10$                & $>10$ \\
    \mTopRecluster \,[\GeV]                 & $>130$       & $>130$               & top veto / $>130$  & top veto             & top veto \\
    \amtTwo \,[\GeV]                        & $>160$       & $<200$               & $<160$           & $>160$               & $>200$ \\
    \dPMETlep                        & $>1.2$        & --                   & $>1.2$         & --                   & -- \\
    \minDeltaPhi                & $>1.0$        & --              & --               & --                   & -- \\
    $\DeltaR$($b_1$,$b_2$)             & --               & --                   & --               & $<1.2$               & $>1.2$ \\
    Lepton charge                     & --             & --              & --               & $+1$                 & -- \\
        $|\Delta\phi(j_{1,2},\Ptmiss)|$          & \multicolumn{5}{ c |}{$>0.4$}                  \\ 
        $\mtTwoTau$ based $\tau$-veto\,[\GeV]   & \multicolumn{5}{ c |}{$>80$}                   \\ 
    \hline \hline
                                       & \DMhigh        & T1LCR/VR       & T2LCR/VR         & WCR/VR                & STCR \\ \hline
                Preselection                            & \multicolumn{5}{ c |}{high-\met\ preselection} \\ \hline
    Number of (jets,~$b$-tags)     & ($\ge4$,~$\ge1$)  & ($\ge4$,~$\ge1$)     & ($\ge4$,~$\ge1$) & ($\ge4$,~$\ge1$)     & ($\ge4$,~$\ge2$) \\
    Jet $\pt$ \,[\GeV]       & \multicolumn{5}{ c |}{$>(125,~75,~65,~25)$} \\
    $b$-tagged jet $\pt$  \,[\GeV]     & \multicolumn{4}{ c |}{$>25$}                                                       & $>(25,~25)$\\
    \hline
    \hline
    \met \,[\GeV]                     & $>380$       & $>280$              & $>280$           & $>280$               & $>280$ \\
    \mt \,[\GeV]                       & $>225$       & $[30,90]$ / $[90.120]$ & $>120$           & $[30,90]$ / $[90,120]$ & $[30,120]$ \\
    \mTopRecluster \,[\GeV]                 & $>130$       & $>130$               & top veto / $>130$  & top veto             & top veto \\
    \amtTwo \,[\GeV]                        & $>190$       & $<200$               & $<200$ / $<190$    & $>190$               & $>200$ \\
    \dPMETlep                        & $>1.2$        & --                   & $>1.2$         & --                   & -- \\
    \minDeltaPhi                & $>1.0$        & $>1.0$              & --               & $>1.0$               & -- \\
    $\DeltaR$($b_1$,$b_2$)             & --               & --                   & --               & $<1.2$               & $>1.2$ \\
    Lepton charge                     & --             & --              & --               & $+1$ / --            & -- \\
        $|\Delta\phi(j_{1,2},\Ptmiss)|$         & \multicolumn{3}{ c }{$>0.4$}                                & $>0.4$ / --          & $>0.4$  \\ 
        $\mtTwoTau$ based $\tau$-veto\,[\GeV]   & \multicolumn{5}{ c |}{$>80$}                   \\ 
    \hline \hline
  \end{tabular*}
}
}
  \label{tab:CRs_DM}
\end{table}

\begin{table}[t]
  \centering
  \caption{
  Overview of the selections for the $\DMlowloose$ signal region as well as the associated control and validation regions. The control regions include the \ttbar\ control region (TCR), the $W$+jets control region (WCR), and the single-top $Wt$ control region (STCR). The validation regions include the \ttbar\ validation region (TVR) and the $W$+jets validation region (WVR). List values are provided in between parentheses and square brackets denote intervals. 
  The selection of the $\ttbar+V$ control region (TZCR) is detailed in Section~\ref{subsec:ttZ}.
  }
  \vspace{3mm}
{\footnotesize
{\renewcommand{\arraystretch}{1.1}
  \begin{tabular*}{\textwidth}{@{\extracolsep{\fill}}| l | cccc |}
                \hline \hline
                                       & \DMlowloose        & TCR/VR              & WCR/VR                & STCR \\ \hline
                Preselection                            & \multicolumn{4}{ c |}{high-\met\ preselection} \\ \hline
    Number of (jets,~$b$-tags)     & ($\ge4$,~$\ge1$)      & ($\ge4$,~$\ge1$) & ($\ge4$,~$= 0$)     & ($\ge4$,~$\ge2$) \\
    Jet $\pt$  \,[\GeV]       & \multicolumn{4}{ c |}{$>(60,~60,~40,~25)$} \\
    $b$-tagged jet $\pt$  \,[\GeV]     & \multicolumn{3}{ c |}{$>25$}                                                       & $>(25,~25)$\\
    \hline

    \met \,[\GeV]                     & $>300$     & {$>230$}   & {$>230$}   & {$>230$} \\
    \HTmissSig                     & $>14$          & {$>8$}      & {$>8$}        & {$>8$} \\   
    \mt \,[\GeV]                 & $>120$     & {[30,90] / [90,120]}        & {[30,90] / [90,120]}        & {[30,120]} \\
    \amtTwo  \,[\GeV]              & $>140$     & {$[100,200]$} & {$>100$}   & {$>200$} \\
    \minDeltaPhi              & $>1.4$      & $>1.4$  & $>1.4$       & $>1.4$ \\
    \dPMETlep         & $>0.8$     & $>0.8$   & $>0.8$        & -- \\
    $\DeltaR(b_{1}, b_{2})$     & --       & --        & --         & {$>1.8$} \\
        $\mtTwoTau$ based $\tau$-veto\,[\GeV]   & \multicolumn{4}{ c |}{$>80$}                   \\ 
    \hline \hline
  \end{tabular*}
}
}
  \label{tab:CRs_DMlowloose}
\end{table}

%% file: texfiles/bkg_tN_diag.tex
\subsection{BDT analyses}
\label{subsec:BDTCR}

For the signal regions $\tNdiaglow$, $\tNdiagmed$ and $\tNdiaghigh$, control regions use the signal selections but change the requirements on the BDT output scores. Due to its large fractional contribution, only the $\ttbar$ background is constrained using data, with all other backgrounds using predictions from samples of simulated events.

Although the main background is always the \ttbar\ process in all three SRs, the fraction of dileptonic \ttbar\ events varies. Therefore, a different strategy is employed for each SR.

For the signal regions \tNdiaglow\ and $\tNdiaghigh$, the $\ttbar$ background is treated as a single component, with a single normalisation factor being derived. One \ttbar\ control region (TCR) is used for $\tNdiaglow$, while three control-region bins (TCR1, TCR2, and TCR3) are used for $\tNdiaghigh$ in order to improve the stability of the simultaneous fit by reducing the correlation between the signal and \ttbar\ background.

\begin{figure}[!htb]
\centering
\includegraphics[width=.40\textwidth]{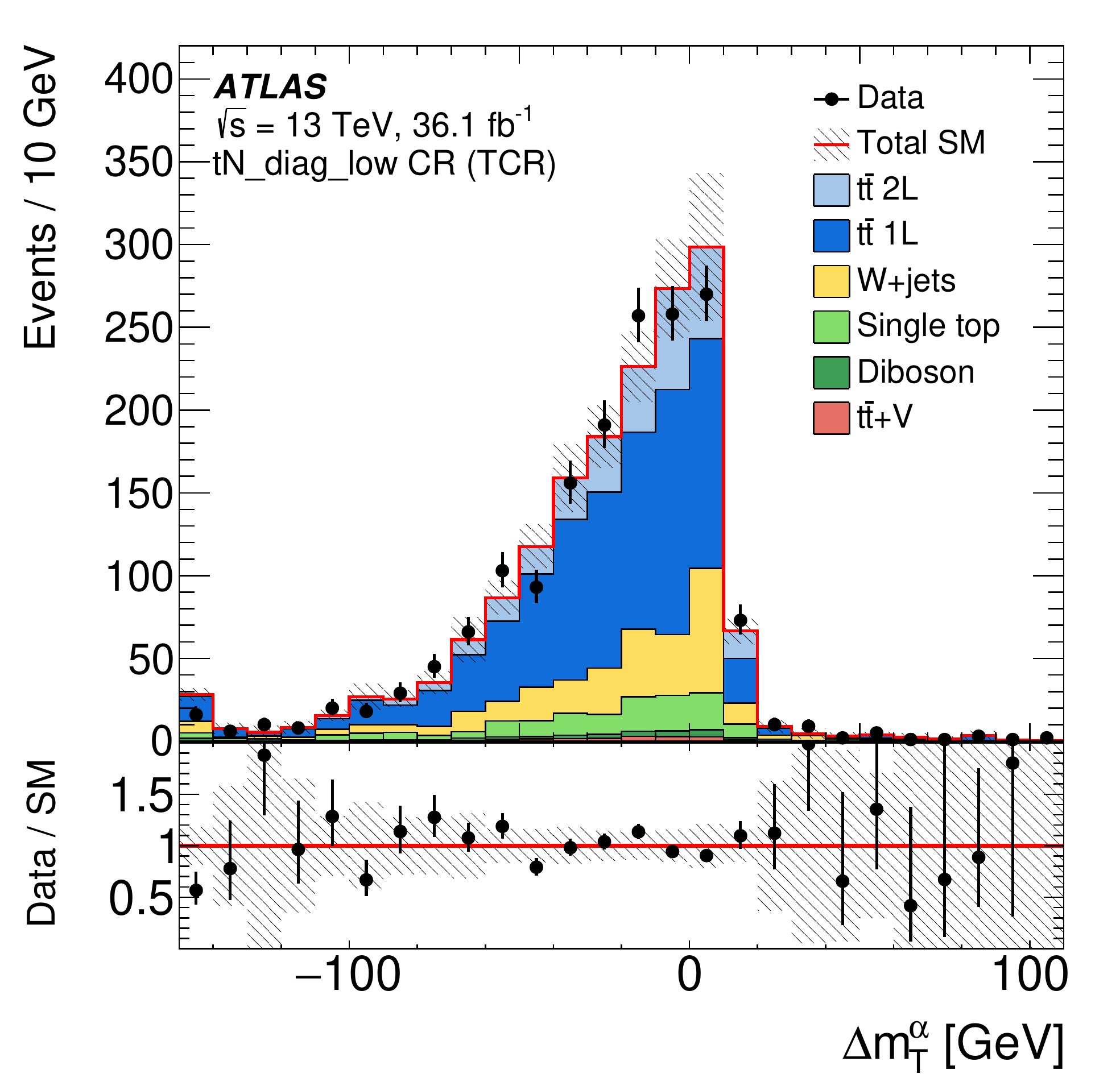} 
\includegraphics[width=.40\textwidth]{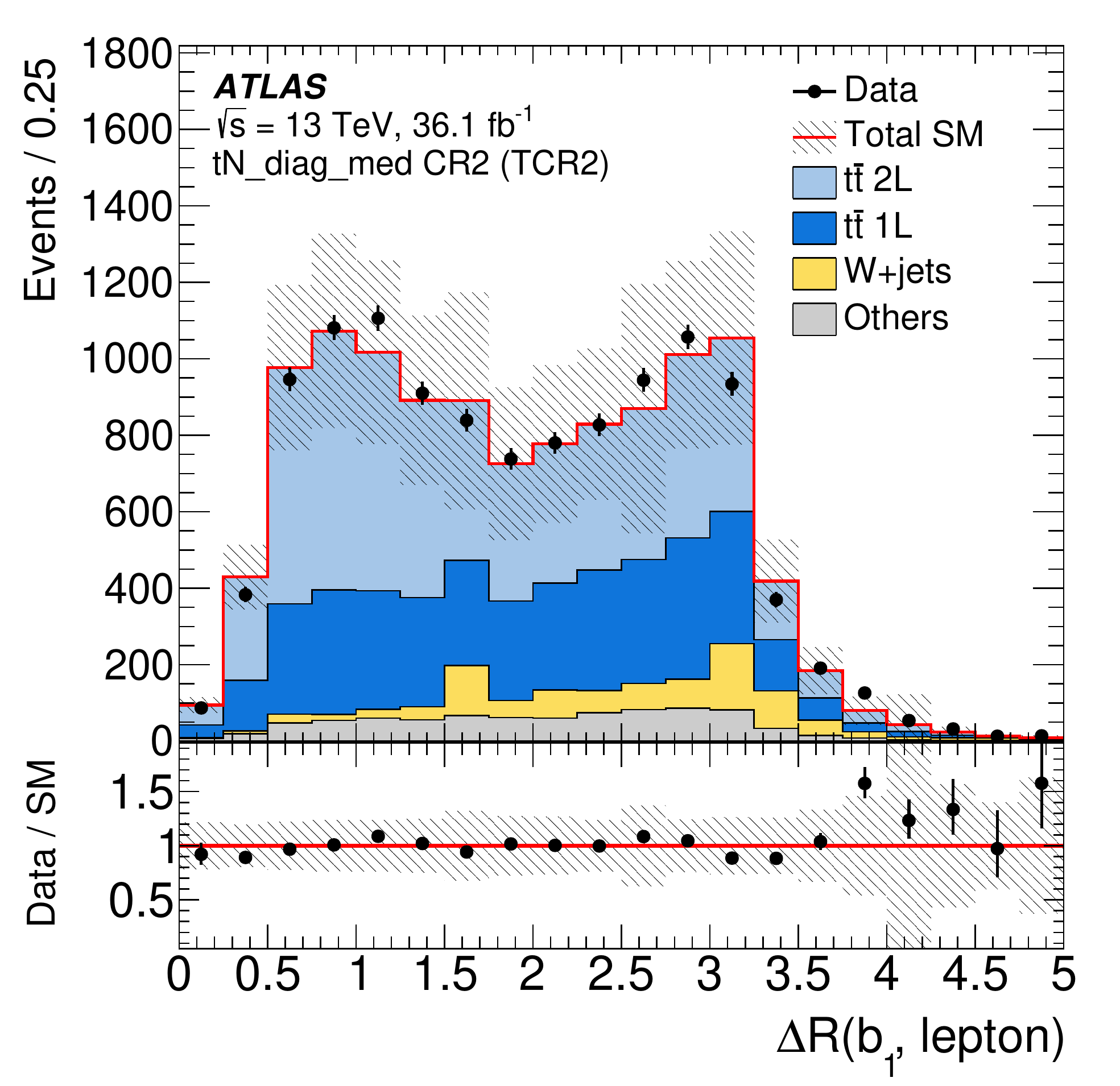}
\caption{Kinematic distributions in the \tNdiaglow\ and \tNdiagmed\ control regions: (left) $\Delta \mt^{\alpha}$ in the \tNdiaglow\ top control region (TCR) and (right) $\DeltaR(b_1,\ell)$ in the $\tNdiagmed$ top control region (TCR2). 
Values of $m_{\tone}=200$\,\GeV\ and $m_{\ninoone}=27$\,\GeV\ are used, resulting in $\alpha=0.135$.
The $\ttbar$ background is scaled by a normalisation factor obtained from the control region. The category labelled `Others' stands for minor SM backgrounds that contribute less than 5\% of the total SM background. The hatched area around the total SM prediction and the hatched band in the Data/SM ratio include statistical and experimental uncertainties. The last bin contains overflows.}
\label{fig:BDT_190_CR}
\end{figure}

For \tNdiagmed, the $\ttbar$ background is split into semileptonic and dileptonic \ttbar\ contributions. Two control-region bins (TCR1 and TCR2, enriched in dileptonic and semileptonic \ttbar\ events respectively) are defined to constrain the $\ttbar$ background and determine two separate normalisation factors for its two components in all fits to the data. Selected kinematic distributions in the \tNdiaglow\ and \tNdiagmed\ CRs are shown in Figure~\ref{fig:BDT_190_CR}.

An overview of the CR selections for the BDT analyses can be found in Table~\ref{tab:BDT_CR_definitions}.

\begin{table}[t]
        \centering
        \caption{Overview of signal region and control region definitions for the BDT analyses targeting the compressed bino LSP scenarios. The selections described in Table~\ref{tab:SRs_tN_diag_low_high} are applied, except for the BDT score. 
  Square brackets denote intervals.
  }
        \vspace{3mm}
{\footnotesize
{\renewcommand{\arraystretch}{1.1}
        \begin{tabular*}{\textwidth}{@{\extracolsep{\fill}}| l | c | c c | c c c|}
                \hline
                \hline
                Signal Region & \tNdiaglow   & \multicolumn{2}{c|}{\tNdiagmed} & \multicolumn{3}{c|}{\tNdiaghigh} \\ 
                BDT score &  $\geq 0.55$ &  \multicolumn{2}{c|}{$\geq 0.4$}  &  \multicolumn{3}{c|}{$\geq 0.6$} \\
                \hline
                Associated CRs  &  TCR & TCR1 & TCR2  &  TCR1 & TCR2 & TCR3 \\
                BDT score &  $[-1,\ 0.1]$ & $[-1,\ -0.4]$ & $[-0.4,\ 0.4]$  &  $[-1,\ -0.5]$ & $[-0.5,\ 0]$ & $[0,\ 0.4]$ \\
                \hline \hline
        \end{tabular*}
}
}
        \label{tab:BDT_CR_definitions}
\end{table}

%% file: texfiles/bkg_bWN.tex
\subsection{\threeBody\ analysis}
\label{subsec:bWNCR}

Almost all of the background in the \bWN\ SR consists of dileptonic \ttbar\ events (where one of the leptons is lost or a hadronically decaying $\tau$ lepton). Therefore, a single high-purity TCR is defined by relaxing the selection requirements on \met\ and \amtTwo. In addition, the requirement on $\dPMETlep$ is inverted to reduce the potential signal contamination. The TVR is defined by sliding the \amtTwo\ window to $110$--$130$\,$\GeV$ in order to validate the background normalisation obtained from the TCR.

\begin{figure}[htbp]
  \centering
  \includegraphics[width=.40\textwidth]{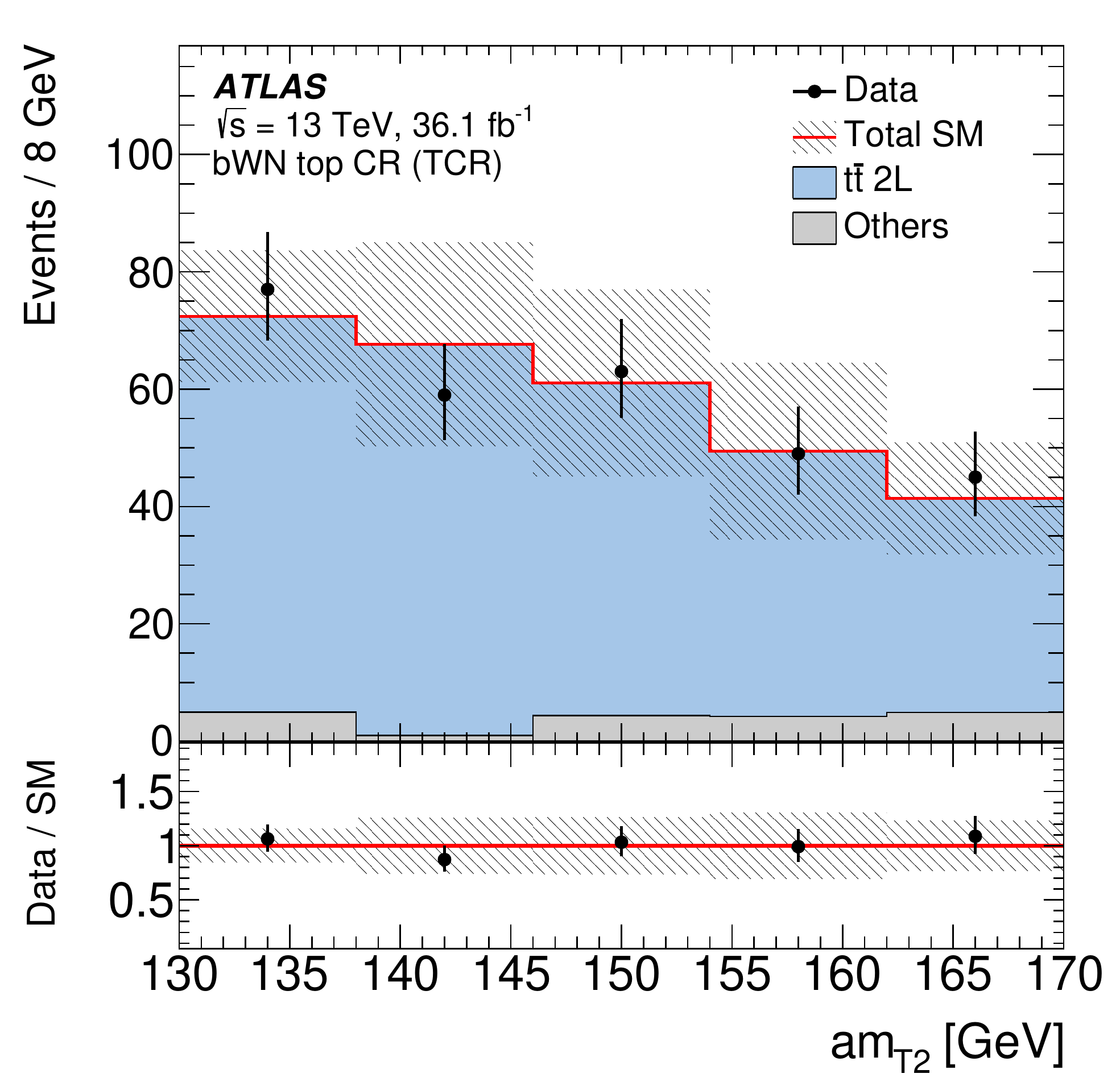}
  \includegraphics[width=.40\textwidth]{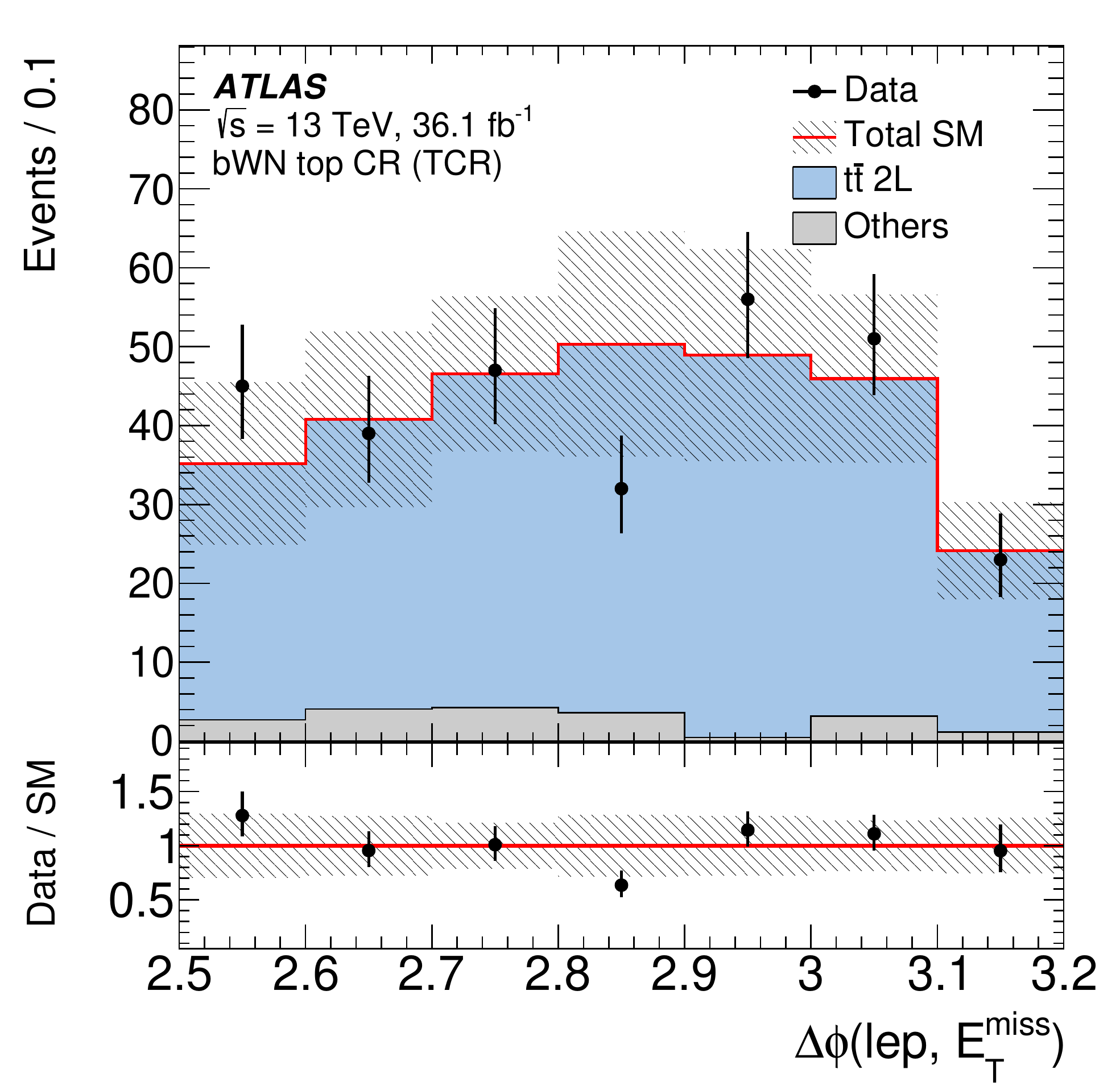}
  \caption{Kinematic distributions in the \bWN\ top control region (TCR): (left) $\amtTwo$ and (right) $\dPMETlep$. The \ttbar\ process is scaled by a normalisation factor obtained in the corresponding control region. The category labelled `Others' stands for minor SM backgrounds that contribute less than 5\% of the total SM background. The hatched area around the total SM prediction and the hatched band in the Data/SM ratio include statistical and experimental uncertainties. The last bin contains overflows.}
  \label{fig:CRs_bWN}
\end{figure}

Figure~\ref{fig:CRs_bWN} shows kinematic distributions in the CRs associated with the $\bWN$ SR. Table~\ref{tab:CRs_bWN} details the corresponding CR and VR selections together with the SR selection.

\begin{table}[t]
  \centering
  \caption{Overview of the selections for the \bWN\ signal region and associated control and validation regions. List values are provided in between parentheses and square brackets denote intervals. 
  }
  \vspace{3mm}
{\footnotesize
{\renewcommand{\arraystretch}{1.1}
  \begin{tabular*}{\textwidth}{@{\extracolsep{\fill}}| l | cc |}
                \hline
    \hline
                                & \bWN           & TCR/VR                    \\ \hline
                Preselection                            & \multicolumn{2}{ c |}{high-\met\ preselection}   \\ \hline
                Number of (jets,~$b$-tags)              & ($\ge4$,~$\ge1$)     & ($\ge4$,~$\ge1$)          \\
    Jet $\pt$  \,[\GeV]       & \multicolumn{2}{ c |}{$>(50,~25,~25,~25)$}       \\
    $b$-tagged jet $\pt$  \,[\GeV]     & \multicolumn{2}{ c |}{$>25$}                     \\
                \hline
                \met \,[\GeV]                           & $>300$               & $>230$                    \\
                \mt \,[\GeV]                            & $>130$               & $>130$                    \\
                \amtTwo  \,[\GeV]                       & $<110$               & $[130,170]$ / $[110,130]$ \\
                \dPMETlep                               & $<2.5$               & $>2.5$                    \\
        $|\Delta\phi(j_{1,2},\Ptmiss)|$         & \multicolumn{2}{ c |}{$>0.4$}                    \\ 
        $\mtTwoTau$ based $\tau$-veto\,[\GeV]   & \multicolumn{2}{ c |}{$>80$}                     \\ 
    \hline \hline
  \end{tabular*}
}
}
  \label{tab:CRs_bWN}
\end{table}

%% file: texfiles/bkg_bC2x.tex
\subsection{Hadronic $W$-tagging approach}
\label{subsec:WtagCR}

Control regions for the $\bCdiag$ and $\bCmed$ SRs exploit hadronic $W$-boson tagging ($W$-tagging) with the \mWRecluster\ variable, closely following the strategy described in Section~\ref{subsec:toptagCR}. The CRs invert two out of three requirements on \mt, \amtTwo, and the hadronic $W$-boson candidate mass. 

For the \bCbv\ SR, since the veto on $b$-tagged jets is required in the signal-region selection, a different CR strategy is used. The WCR and TCR remove the selection requirement on $\dPMETlep$ and select a \mt\ window of $30$--$90$\,\GeV\ 
to increase the number of events in the region while suppressing potential signal contamination. A $b$-tagged jet is further required in the TCR to improve the purity of \ttbar\ events.

\begin{figure}[htbp]
  \centering
  \includegraphics[width=.40\textwidth]{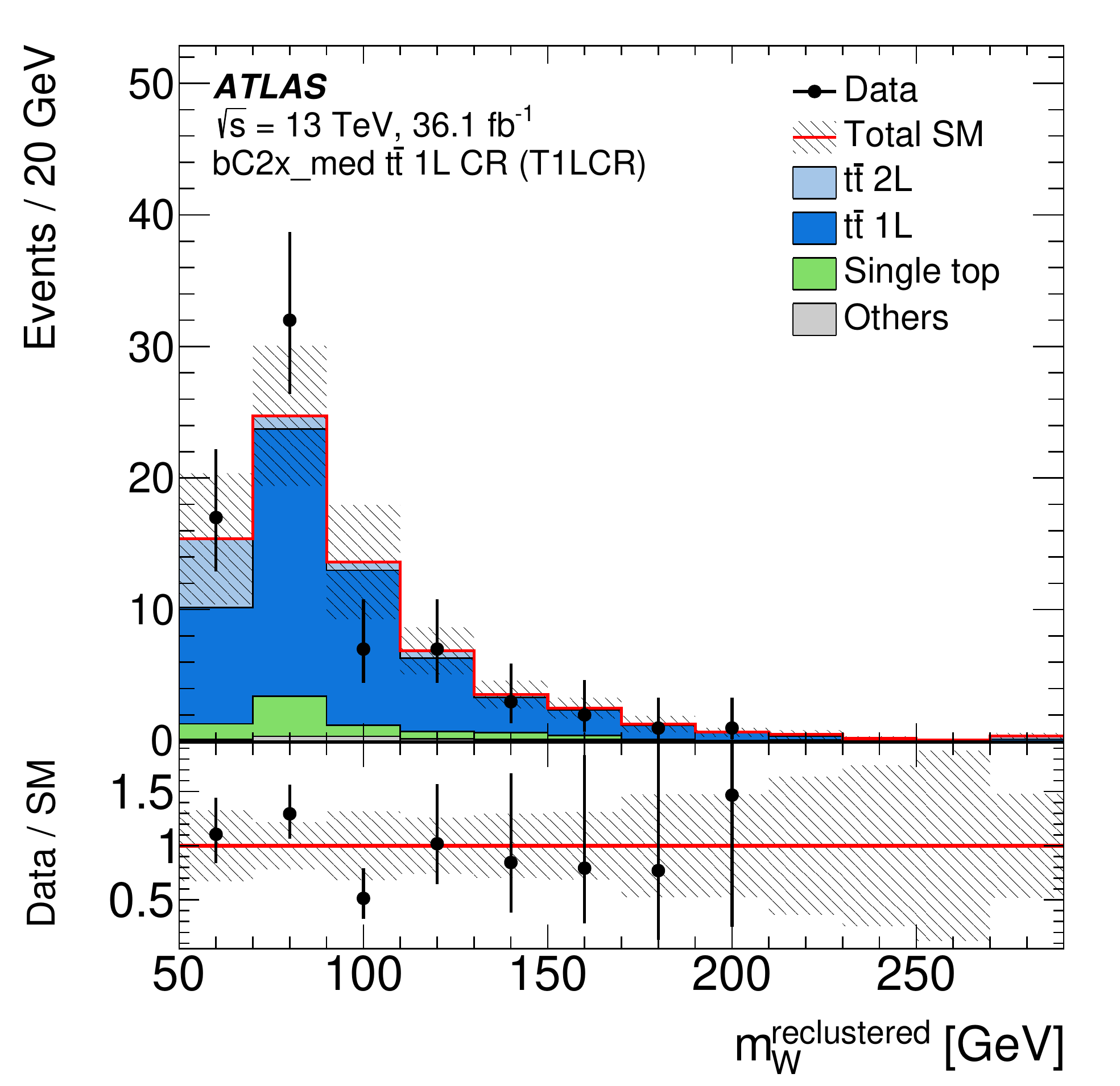}
  \includegraphics[width=.40\textwidth]{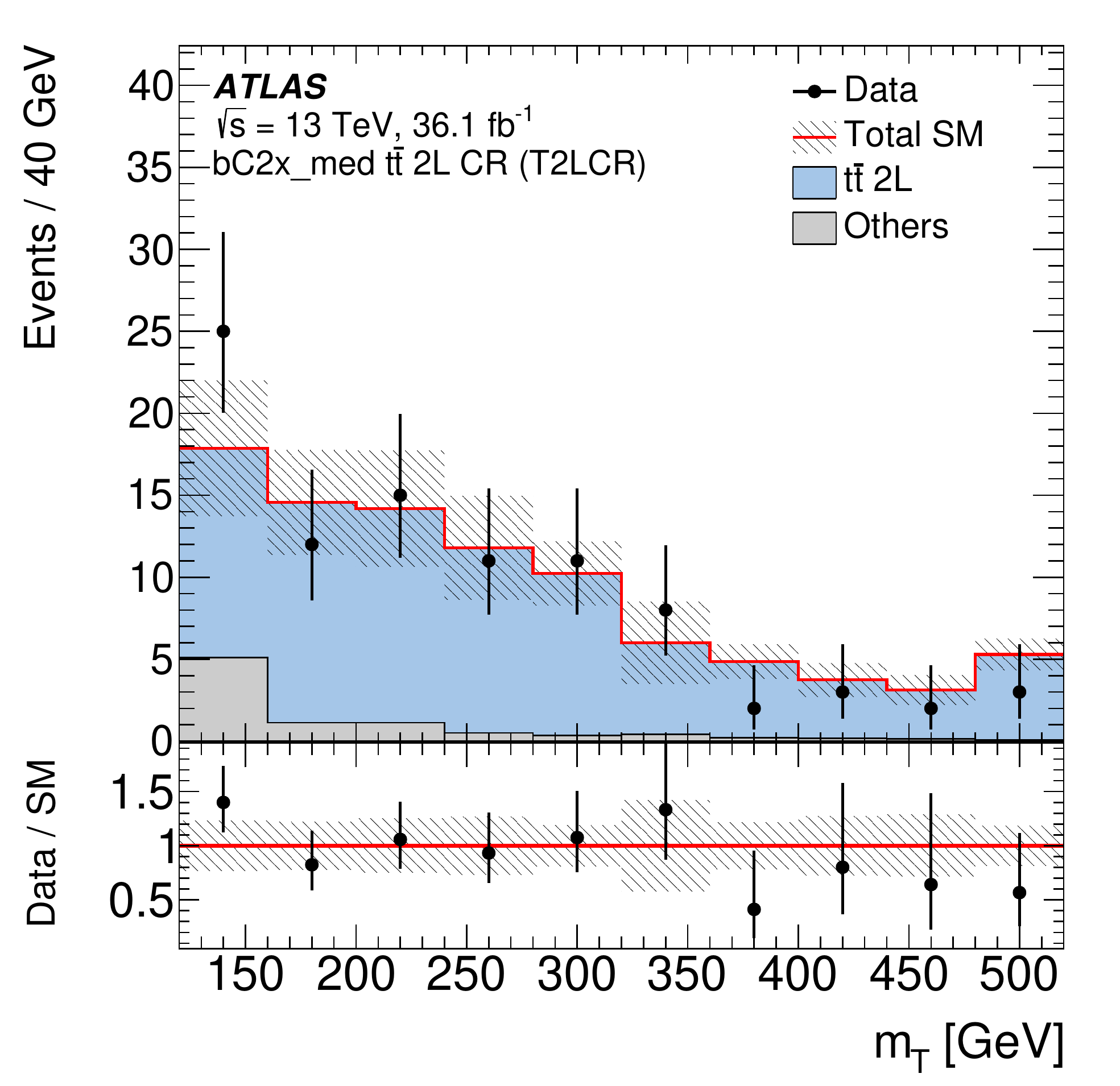}
  \includegraphics[width=.40\textwidth]{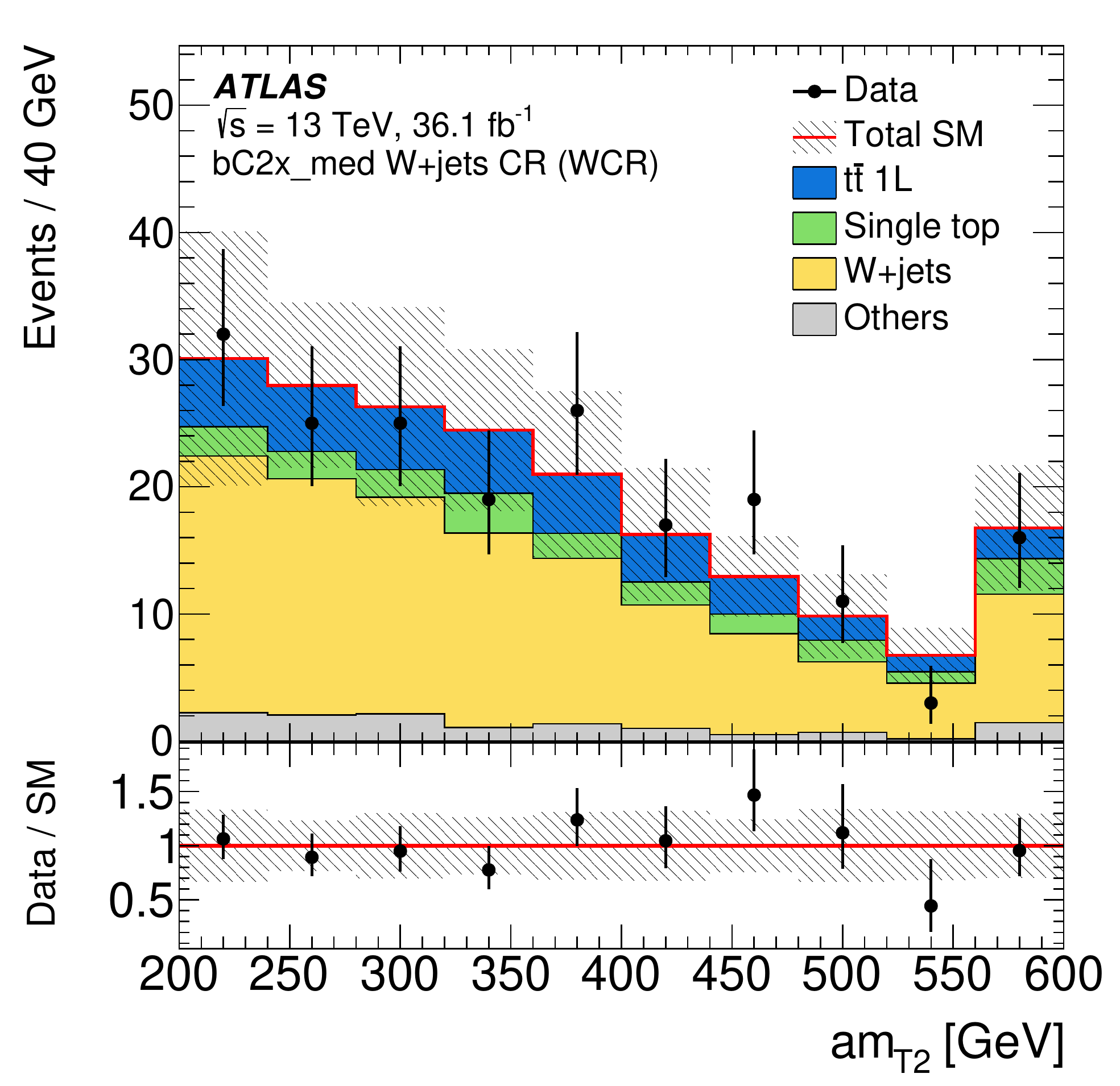}
  \includegraphics[width=.40\textwidth]{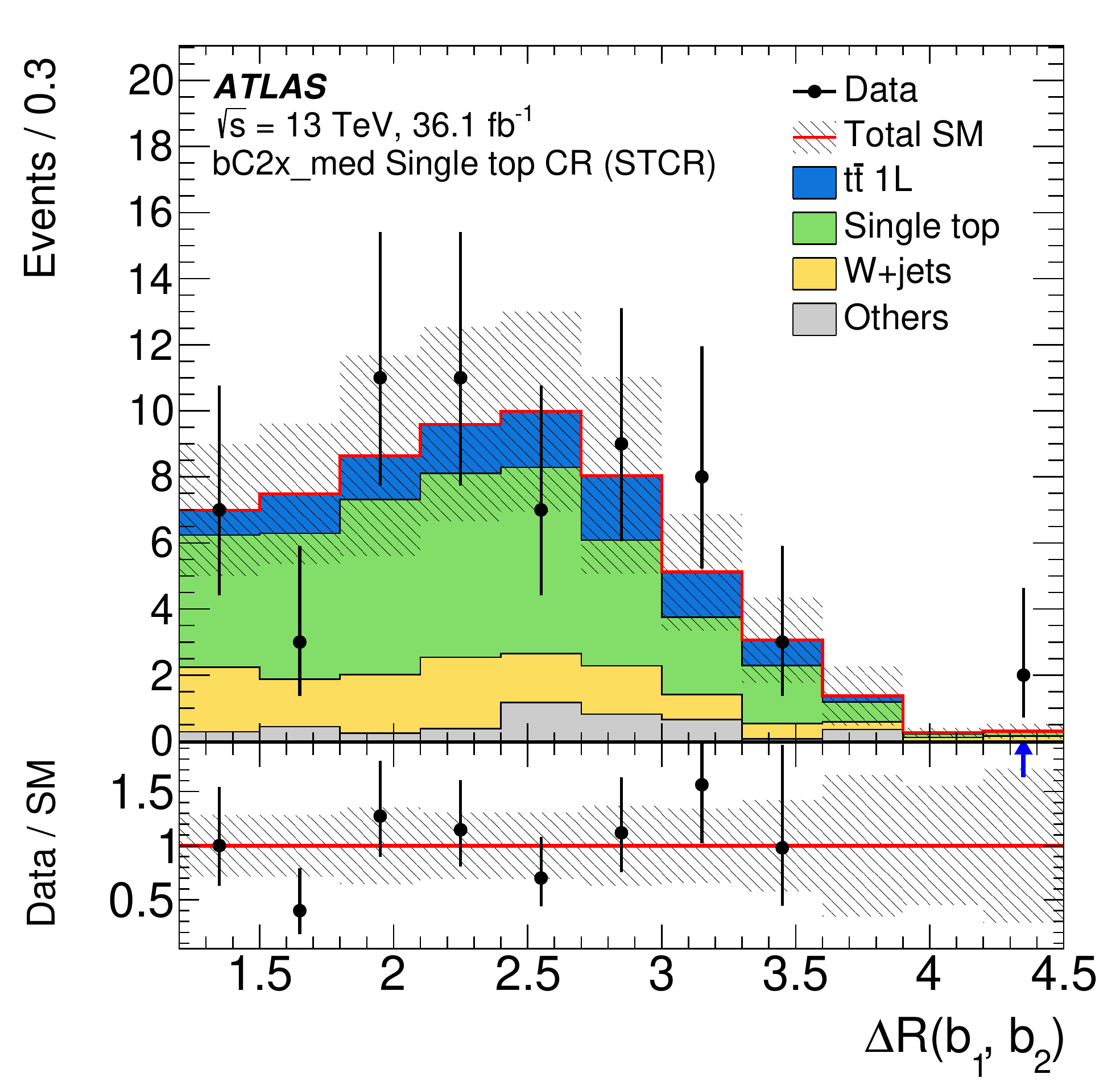}
  \caption{Kinematic distribution of the \bCmed\ control regions: (top left) reclustered jet mass (\mWRecluster) in the semileptonic \ttbar\ control region, (top right) \mt\ in the dileptonic \ttbar\ control region, (bottom left) \amtTwo\ in the $W$+jets control region, and (bottom right) $\DeltaR$($b_1$,$b_2$) in the single-top control region. Each of these backgrounds is scaled by normalisation factors obtained from the corresponding control region. The category labelled `Others' stands for minor SM backgrounds that contribute less than 5\% of the total SM background. The hatched area around the total SM prediction and the hatched band in the Data/SM ratio include statistical and experimental uncertainties. The last bin contains overflows. 
  }
  \label{fig:CRs_bC2x}
\end{figure}

Figure~\ref{fig:CRs_bC2x} shows selected kinematic distributions in associated CRs for $\bCmed$. 

A set of VRs associated with the CRs is defined following the approach taken for the top-tagging VRs in Section~\ref{subsec:toptagCR}, i.e.~by modifying the requirements on the $\mt$, $\amtTwo$, and hadronic $W$-tagging variables. Tables~\ref{tab:CRs_bC2x} and~\ref{tab:CRs_bCbv} detail the CR and VR selections for the corresponding SRs.

\begin{table}[t]
  \centering
  \caption{
  Overview of the selections for the $\bCdiag$ and $\bCmed$ signal regions as well as the associated control and validation regions. The control regions include the semileptonic \ttbar\ control region (T1LCR), the dileptonic \ttbar\ control region (T2LCR), the $W$+jets control region (WCR), and the single-top $Wt$ control region (STCR).  The validation regions include the semileptonic \ttbar\ validation region (T1LVR), the dileptonic \ttbar\ validation region (T2LVR), and the $W$+jets validation region (WVR). List values are provided in between parentheses and square brackets denote intervals. 
  The veto on the reclustered hadronic $W$-boson candidate is satisfied for events where no reclustered jet candidate is found, or where the mass of the hadronic top-quark candidate (\mTopRecluster) is below the specified tag threshold. For the WCR, $\DeltaR$($b_1$,$b_2$) $<1.2$ is not required when the event has only one $b$-tagged jet.
  The selection of the $\ttbar+V$ control region (TZCR) is detailed in Section~\ref{subsec:ttZ}.
  }
  \vspace{3mm}
{\footnotesize
{\renewcommand{\arraystretch}{1.1}
  \begin{tabular*}{\textwidth}{@{\extracolsep{\fill}}| l | ccccc |}
    \hline
                                                &    \bCdiag       &        T1LCR/VR      &     T2LCR/VR     &         WCR/VR       &         STCR      \\ \hline
                Preselection                                & \multicolumn{5}{ c |}{high-\met\ preselection} \\ \hline
    Number of (jets,~$b$-tags)         & ($\ge4$,~$\ge2$) & ($\ge4$,~$\ge2$)     & ($\ge4$,~$\ge2$) & ($\ge4$,~$\ge1$)     & ($\ge4$,~$\ge2$)  \\
    Jet $\pt $ \,[\GeV]           & \multicolumn{5}{ c |}{ $>(75,~75,~75,~30)$ } \\
    $b$-tagged jet $\pt $ \,[\GeV]         & $>(30,~30)$      & $>(30,~30)$         & $>(30,~30)$      & $>(30,~-)$           & $>(30,~30)$       \\
    \hline
    \met \,[\GeV]                         & \multicolumn{5}{ c |}{ $>230$} \\
    \HTmissSig                     & $>13$             & $>13$            & $>10$       & $>13$              & $>10$          \\ 
    \mt \,[\GeV]                 & $>180$          & $[30,90]$ / $[90,120]$ & $>120$        & $[30,90]$ / $[90,120]$ & $[30,120]$       \\
    \amtTwo  \,[\GeV]            & $>175$          & $<200$         & $<200$ / $<130$    & $>200$           & $>200$       \\
    $\minDeltaPhi(i=1,2)$                  & \multicolumn{5}{ c |}{ $>0.7$ } \\ 
    \mWRecluster \,[\GeV]          & $>50$           & $>50$            & $W$ veto / $>50$   & $W$ veto            & $W$ veto              \\
    $\DeltaR(b_1,b_2)$           & --               & --                   & --               & $<1.2$               & $>1.2$            \\
    Lepton charge                               & --               & --                   & --               & $=+1$                & --            \\
        $|\Delta\phi(j_{1,2},\Ptmiss)|$              & \multicolumn{5}{ c |}{$>0.4$}    \\ 
        $\mtTwoTau$ based $\tau$-veto\,[\GeV]       & \multicolumn{5}{ c |}{$>80$}     \\ 
    \hline \hline
                                                &    \bCmed        &        T1LCR/VR      &     T2LCR/VR     &         WCR/VR       &         STCR      \\ \hline
                Preselection                                & \multicolumn{5}{ c |}{high-\met\ preselection} \\ \hline
    Number of (jets,~$b$-tags)         & ($\ge4$,~$\ge2$) & ($\ge4$,~$\ge2$)     & ($\ge4$,~$\ge1$) & ($\ge4$,~$\ge1$)     & ($\ge4$,~$\ge2$)  \\
    Jet $\pt$ \,[\GeV]           & \multicolumn{5}{ c |}{ $>(200,~140,~25,~25)$  } \\ 
    $b$-tagged jet $\pt$ \,[\GeV]       & $>(140,~140)$     & $>(140,~140)$        & $>(140,~-)$    & $>(140,~-)$          & $>(140,~140)$      \\
    \hline
    \met \,[\GeV]                         & \multicolumn{5}{ c |}{ $>230$} \\
    \HTmissSig               & {$>10$}           & $>10$            & $>10$          & $>10$              & $>6$          \\
    \mt \,[\GeV]                 & $>120$           & $[30,90]$ / $[90,120]$ & $>120$       & $[30,90]$ / $[90,120]$  & $[30,120]$       \\
    \amtTwo  \,[\GeV]            & {$>300$}          & $<200$         & $<200$ / $<130$   & $>200$           & $>200$       \\
    $\minDeltaPhi(i=1,2)$                  & \multicolumn{5}{ c |}{ $>0.9$ } \\
    \mWRecluster \,[\GeV]          & $>50$         & $>50$            & $W$ veto / $>50$  & $W$ veto            & $W$ veto              \\
    $\DeltaR(b_1,b_2)$           & --               & --                   & --              & $<1.2$             & $>1.2$            \\
    Lepton charge                               & --               & --                   & --              & $=+1$                 & --                \\
        $|\Delta\phi(j_{1,2},\Ptmiss)|$              & \multicolumn{5}{ c |}{$>0.4$}    \\ 
        $\mtTwoTau$ based $\tau$-veto\,[\GeV]       & \multicolumn{5}{ c |}{$>80$}     \\ 
    \hline \hline
  \end{tabular*}
}
}
  \label{tab:CRs_bC2x}
\end{table}

\begin{table}[t]
  \centering
  \caption{
  Overview of the selections for the $\bCbv$ signal region, as well as the associated control regions for \ttbar\ (TCR) and $W$+jets (WCR), and the validation regions targeting \ttbar\ (TVR) and $W$+jets (WVR) backgrounds. List values are provided in between parentheses and square brackets denote intervals. 
  }
  \vspace{3mm}
{\footnotesize
{\renewcommand{\arraystretch}{1.1}
  \begin{tabular*}{\textwidth}{@{\extracolsep{\fill}}| l | ccc |}
    \hline
    \hline
                                              & \bCbv       &        TCR/VR         &      WCR/VR             \\ \hline
                Preselection                              & \multicolumn{3}{ c |}{high-\met\ preselection}                        \\ \hline
      Lepton \pt \, [\GeV]                      & \multicolumn{3}{ c |}{$>60$}                                          \\ 
      Number of (jets,~$b$-tags)             & ($\ge2$,~$=0$)      & ($\ge2$,~$\ge1$)      & ($\ge2$,~$=0$)          \\
      Jet $\pt$ \,[\GeV]                        & \multicolumn{3}{ c |}{ $>(120,~80)$}                                  \\
      $b$-tagged jet $\pt$ \,[\GeV]             & --                  & $>25$                 & --                      \\
    \hline
      \met \,[\GeV]                             & \multicolumn{3}{ c |}{$>360$}                                         \\ 
      \HTmissSig                                & \multicolumn{3}{ c |}{$>16$}                                          \\
      \mt \,[\GeV]                               & $>200$             & $[30,90]$ / $[90,120]$ & $[30,90]$ / $[90,120]$  \\
      $\minDeltaPhi(i=1)$                       & \multicolumn{3}{ c |}{$>2.0$}                                         \\ 
      $\minDeltaPhi(i=2)$                       & \multicolumn{3}{ c |}{$>0.8$}                                         \\ 
      $\dPMETlep$                          & $>1.2$             & --                     & --                      \\
      $\mWRecluster$ \,[\GeV]                   & \multicolumn{3}{ c |}{$[70,100]$}                                     \\ 
        $|\Delta\phi(j_{1,2},\Ptmiss)|$           & \multicolumn{3}{ c |}{$>0.4$}                                         \\ 
    \hline \hline
  \end{tabular*}
}
}
  \label{tab:CRs_bCbv}
\end{table}

%% file: texfiles/bkg_bCsoft.tex
\subsection{Soft-lepton analyses}
\label{subsec:softleptonCR}

For the soft-lepton SRs ($\bCsoftdiag$, $\bCsoftmed$, $\bCsofthigh$, and \bffN), a single TCR, dominated by semileptonic \ttbar\ events, is defined for the \ttbar\ background since the fraction of dileptonic \ttbar\ background is small compared to the other SRs because there is no \mt\ requirement. 

For $\bCsoftmed$ and $\bCsofthigh$ SRs, three CRs (TCR, WCR, and STCR) are defined by inverting the requirements on $\amtTwo$, $\pt^{\ell}$/$\met$, and the number of $b$-tagged jets, while requiring the same $\pt^W$ threshold as the corresponding SR to ensure similar kinematics in the SR and CRs for the $\pt$ of the top quark and the $W$ boson, which might be poorly modelled by the simulation. The TCR is designed by inverting the selection requirement on \amtTwo\ and relaxing the $\pt^{\ell}$/$\met$ requirement to minimise potential signal contamination while improving the purity. Similarly, the WCR and STCR are defined by relaxing $\pt^{\ell}$/$\met$, and requiring exactly one or at least two $b$-tagged jets respectively.

For the $\bCsoftdiag$ SR, the CR strategy using top-tagging is employed, based on the \mTopRecluster\ variable as described in Section~\ref{subsec:toptagCR}. The TCR is defined by requiring a tagged hadronic top-quark candidate and relaxing the requirement on \mt\ to increase the number of \ttbar\ events, while the WCR is defined by requiring a hadronic top-quark veto. For the WCR, an additional requirement is imposed on $\dphiBPtmissMin$ to increase the purity of $W$+jets events. A STCR is not defined for this SR, as the $Wt$ contribution is small compared to other backgrounds. The CRs for the \bffN\ SR are identical to those for \bCsoftdiag\ because of the similarity in the SR selections.

\begin{figure}[htbp]
  \centering
  \includegraphics[width=.40\textwidth]{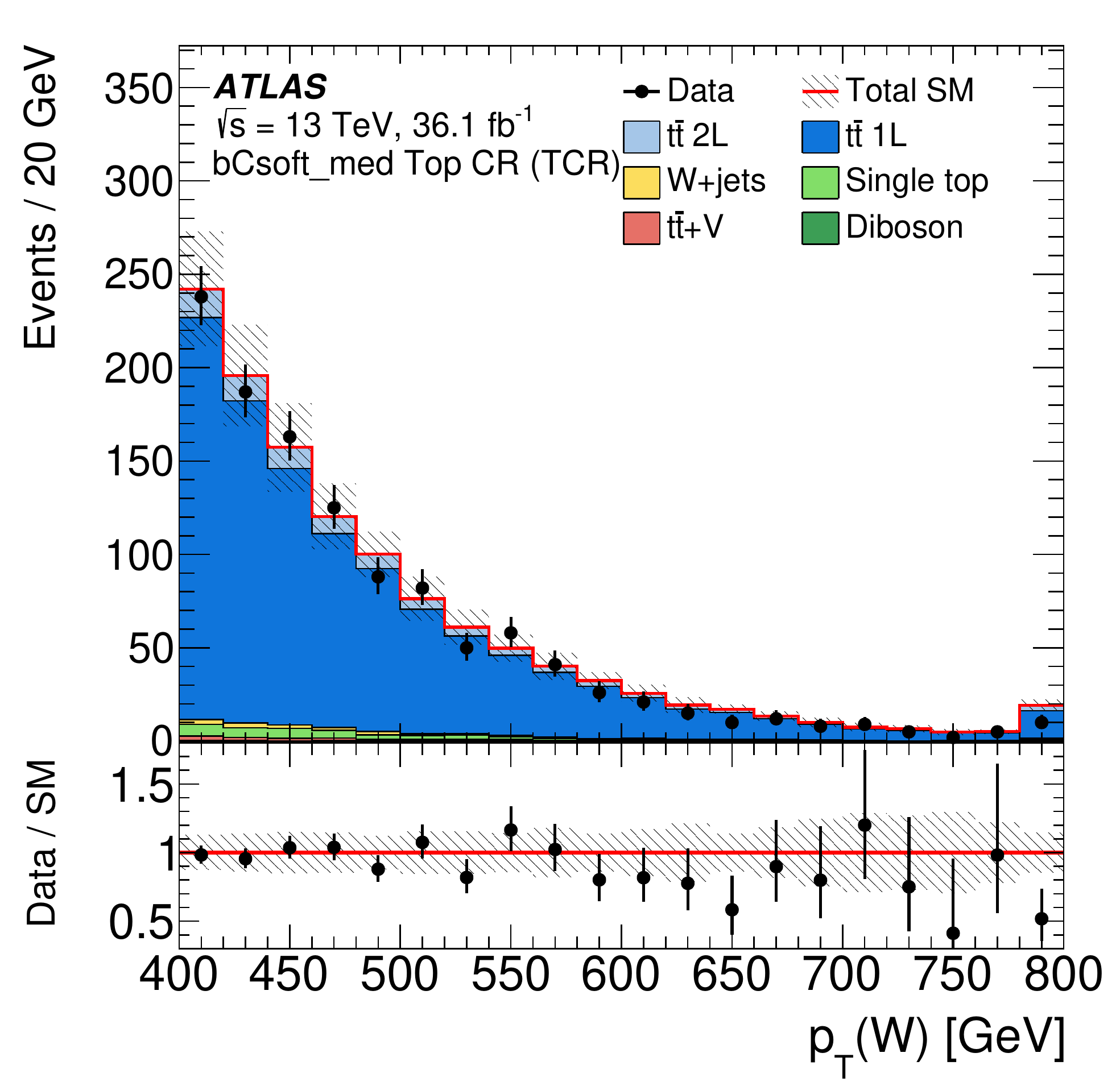}
  \includegraphics[width=.40\textwidth]{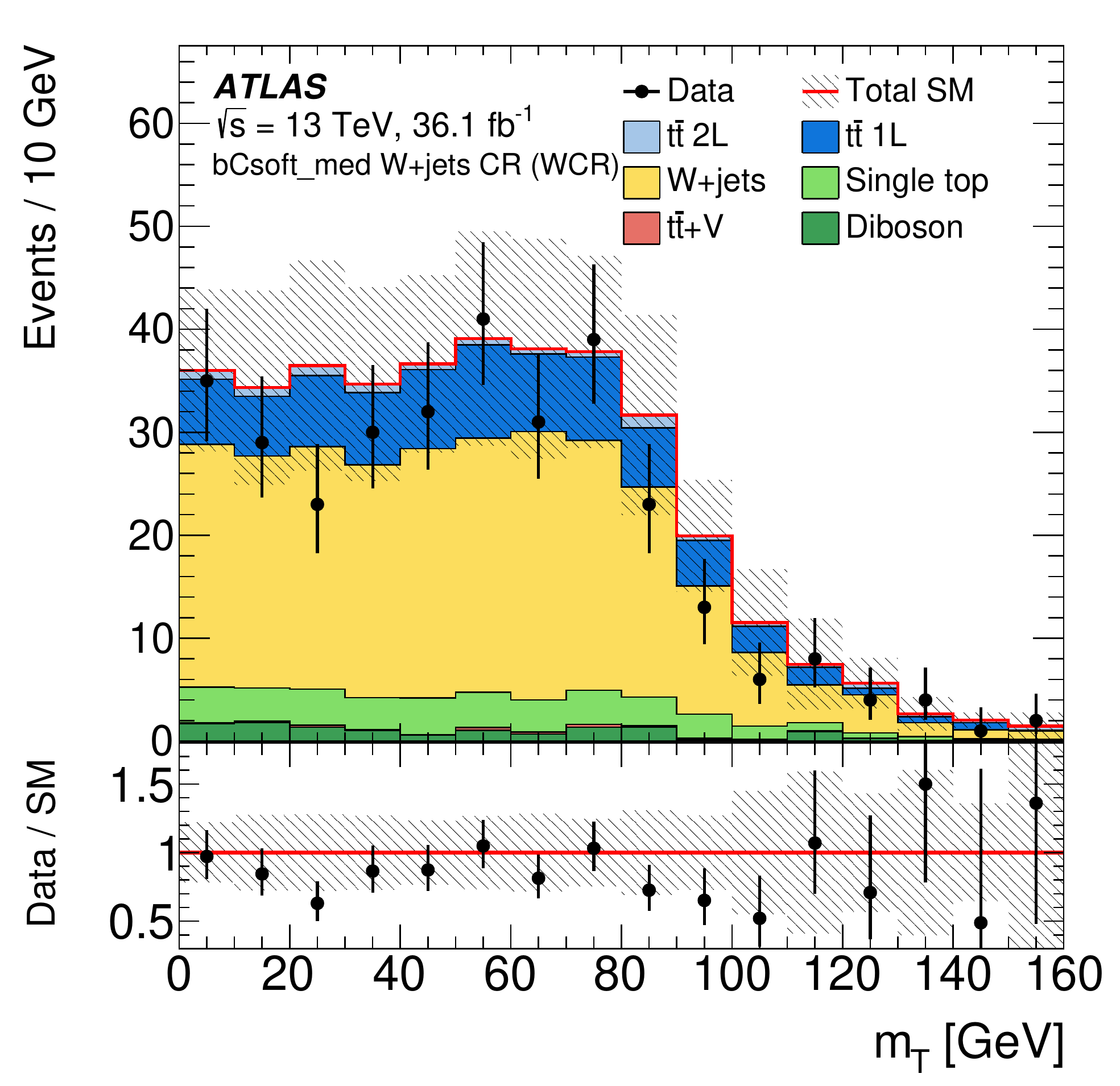}
  \includegraphics[width=.40\textwidth]{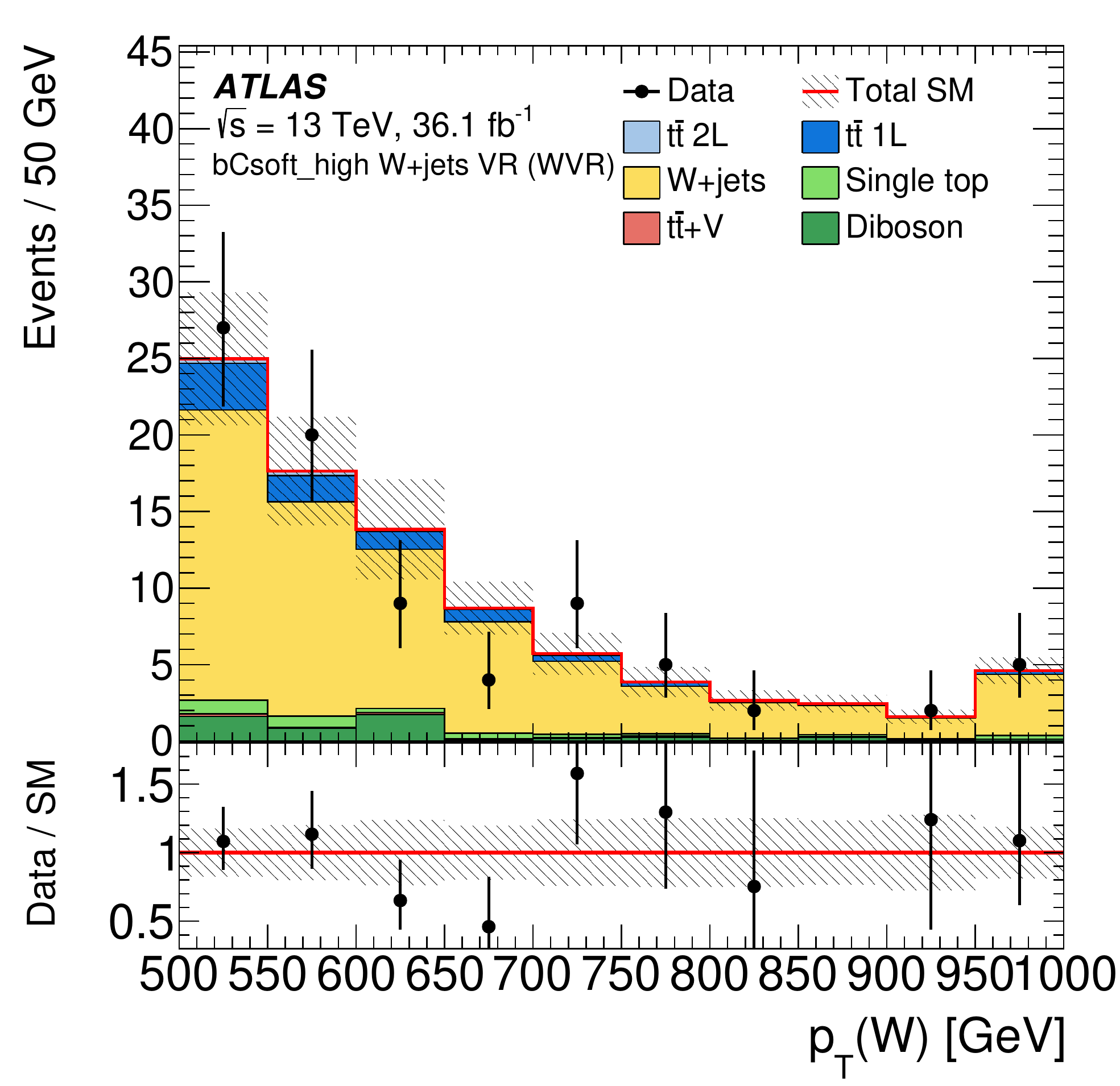}
  \includegraphics[width=.40\textwidth]{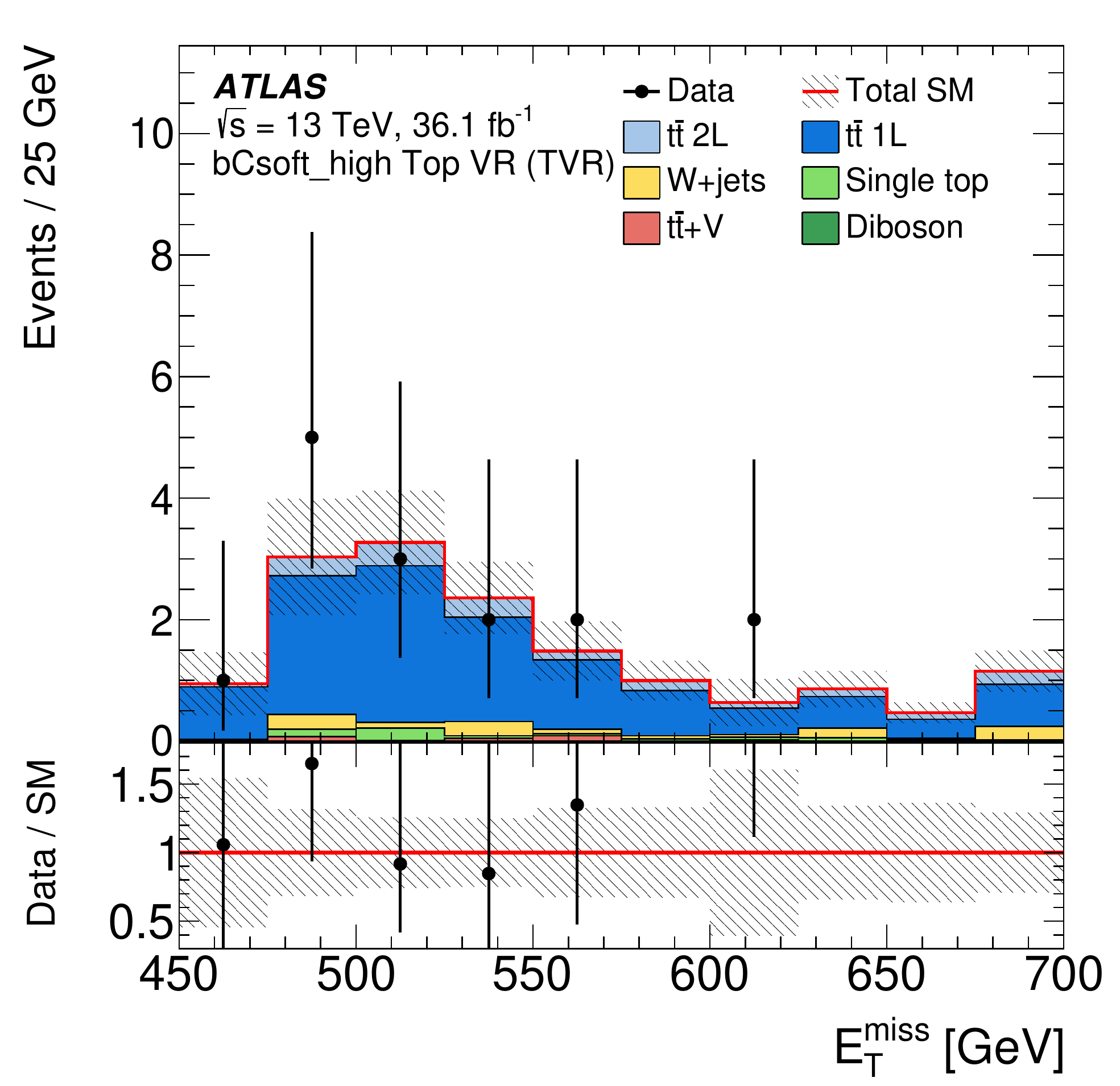}
  \caption{Kinematic distributions in the control regions associated with $\bCsoftmed$ and the validation regions associated with $\bCsofthigh$: (top left) $\pt^{W}$ in the top control region, (top right) \mt\ in the $W$+jets control region, 
(bottom left) \pt\ of the leptonically decaying $W$ boson ($\pt^W$) in the $W$+jets validation region, and 
(bottom right) \met\ in the top validation region. Each of the backgrounds is scaled by a normalisation factor obtained from the corresponding control region. The hatched area around the total SM prediction and the hatched band in the Data/SM ratio include statistical and experimental uncertainties. The last bin contains overflows.}
  \label{fig:CRVRs_bCsoft}
\end{figure}

Figure~\ref{fig:CRVRs_bCsoft} shows selected kinematic distributions in the CRs associated with $\bCsoftmed$. The backgrounds are scaled with normalisation factors obtained from the simultaneous likelihood fit of the CRs as described in Section~\ref{sec:results}. 

A set of VRs associated with corresponding CRs is also defined by inverting the requirement on $\pt^{\ell}$/$\met$. For the soft-lepton SRs, an STVR is defined together with the TVR and WVR. In Figure~\ref{fig:CRVRs_bCsoft}, selected kinematic distributions in the VRs associated with \bCsofthigh\ are compared to the observed data. The backgrounds are scaled with normalisation factors. Tables~\ref{tab:CRs_bCsoft_diag} and \ref{tab:CRs_bCsoft} detail the soft-lepton CR and VR selections.

\begin{table}[t!]
  \centering
  \caption{Overview of the selections for the $\bCsoftdiag$ and $\bffN$ signal regions, as well as the associated control regions for \ttbar\ (TCR) and $W$+jets (WCR), and the validation regions targeting \ttbar\ (TVR) and $W$+jets (WVR) backgrounds. List values are provided in between parentheses and square brackets denote intervals. 
  The veto on the reclustered hadronic top-quark candidate is satisfied for events where no reclustered jet candidate is found, or where the mass of the hadronic top-quark candidate (\mTopRecluster) is below a certain threshold. The leading jet is required not to be $b$-tagged in all regions.
  }
  \vspace{3mm}
{\footnotesize
{\renewcommand{\arraystretch}{1.1}
  \begin{tabular*}{\textwidth}{@{\extracolsep{\fill}}| l | ccc |}
                \hline
    \hline
                                      & $\bCsoftdiag$/$\bffN$ & TCR/VR                & WCR/VR               \\ \hline
                Preselection                            & \multicolumn{3}{ c |}{soft-lepton preselection}                   \\ \hline
    Number of (jets,~$b$-tags)     & ($\ge2$,~$\ge1$)  & ($\ge2$,~$\ge1$)      & ($\ge2$,~$=1$)        \\
    Jet $\pt$ \,[\GeV]       & \multicolumn{3}{ c| }{$>(120,~25)$}                                \\
    $b$-tagged jet $\pt$ \,[\GeV]     & \multicolumn{3}{ c| }{$>25$}                                       \\
    \hline
    \met \,[\GeV]                     & \multicolumn{3}{ c| }{$>300$}                                     \\
    \mt \,[\GeV]                       & $<50$ / $<160$    & $<160$                & $<160$                \\
    $\pt^{\ell}$/$\met$              & $<0.02$       & $[0.03,0.10]$ / $<0.03$ & $[0.03,0.10]$ / $<0.03$ \\
    \mTopRecluster \,[\GeV]                 & top veto       & $>150$                & top veto              \\
    \dphiBPtmissMin                & $<1.5$        & $<1.5$               & $>1.5$                \\
        $|\Delta\phi(j_{1,2},\Ptmiss)|$          & \multicolumn{3}{ c |}{$>0.4$}    \\ 
    \hline \hline
  \end{tabular*}
}
}
  \label{tab:CRs_bCsoft_diag}
\end{table}

\begin{table}[t]
  \centering
  \caption{
  Overview of the selections for the $\bCsoftmed$ and $\bCsofthigh$ signal regions, as well as the associated control regions for \ttbar\ (TCR) and $W$+jets (WCR), and the validation regions targeting \ttbar\ (TVR) and $W$+jets (WVR) backgrounds. 
  List values are provided in between parentheses and square brackets denote intervals.
  }
  \vspace{3mm}
{\footnotesize
{\renewcommand{\arraystretch}{1.1}
  \begin{tabular*}{\textwidth}{@{\extracolsep{\fill}}| l | cccc |}
                \hline
    \hline
                                      & \bCsoftmed        & TCR/VR           & WCR/VR        & STCR/VR             \\ \hline
                Preselection                            & \multicolumn{4}{ c |}{soft-lepton preselection}                                  \\ \hline
    Number of (jets,~$b$-tags)     & ($\ge3$,~$\ge2$)  & ($\ge3$,~$\ge2$) & ($\ge3$,~$=1$)      & ($\ge3$,~$\ge2$)    \\
    Jet $\pt $ \,[\GeV]       & \multicolumn{4}{ c |}{$>(120,~60,~40,~25)$}                                      \\
    $b$-tagged jet $\pt$ \,[\GeV]           & $>(120,~60)$      & $>(120,~60)$     & $>120$              & $>(120,~60)$        \\
    \hline
    \met \,[\GeV]                     & \multicolumn{4}{ c |}{$>230$}                                                    \\
    \mt \,[\GeV]                      & \multicolumn{4}{ c |}{$<160$}                                                    \\
    $\pt^W$ \,[\GeV]                      & \multicolumn{4}{ c |}{$>400$}                                                    \\
    $\pt^{\ell}$/$\met$              & $<0.03$       & $>0.03$ / $<0.03$  & $>0.20$ / $[0.1,0.2]$ & $>0.20$ / $[0.1,0.2]$ \\
                \amtTwo  \,[\GeV]                       & $>200$            & $<200$           & $>200$              & $>200$              \\
    \dphiBPtmissMin                & $>0.8$        & --          & $[0.8,2.5]$         & $>0.8$              \\
                $\Delta R(b_1,b_2)$                     & --                & --               & --                  & $>1.2$              \\
        $|\Delta\phi(j_{1,2},\Ptmiss)|$          & \multicolumn{4}{ c |}{$>0.4$}    \\ 
    \hline \hline
                                      & \bCsofthigh       & TCR/VR           & WCR/VR        & STCR/VR             \\ \hline
                Preselection                            & \multicolumn{4}{ c |}{soft-lepton preselection}                                  \\ \hline
    Number of (jets,~$b$-tags)     & ($\ge2$,~$\ge2$)  & ($\ge2$,~$\ge2$) & ($\ge2$,~$=1$)      & ($\ge2$,~$\ge2$)    \\
    Jet $\pt$ \,[\GeV]       & \multicolumn{4}{ c |}{$>(100,~100)$}                                             \\
    $b$-tagged jet $\pt$ \,[\GeV]     & \multicolumn{4}{ c |}{$>(100,~100)$}                                             \\
    \hline
    \met \,[\GeV]                     & \multicolumn{4}{ c |}{$>230$}                                                    \\
    \mt \,[\GeV]                      & \multicolumn{4}{ c |}{$<160$}                                                    \\
    $\pt^W$ \,[\GeV]                      & \multicolumn{4}{ c |}{$>500$}                                                    \\
    $\pt^{\ell}$/$\met$              & $<0.03$       & $>0.10$ / $<0.10$  & $[0.1,0.4]$ / $<0.10$ & $>0.30$ / $[0.1,0.3]$ \\
                \amtTwo  \,[\GeV]                       & $>300$            & $<300$           & $>300$              & $>300$              \\
    \dphiBPtmissMin                     & \multicolumn{4}{ c |}{$>0.4$}                                                    \\
                $\Delta R(b_1,b_2)$                     & $>0.8$            & $>0.8$           & --                  & $>0.8$              \\
                $\DeltaR(b,\ell)$                       & --                & --               & $>0.8$              & -- \\
        $|\Delta\phi(j_{1,2},\Ptmiss)|$          & \multicolumn{4}{ c |}{$>0.4$}    \\ 
    \hline \hline
  \end{tabular*}
}
}
  \label{tab:CRs_bCsoft}
\end{table}

%% file: texfiles/bkg_ttZ.tex
\subsection{Control regions for $\ttbar + V$}
\label{subsec:ttZ}

Top-quark pair production in association with a $Z$ boson that decays into neutrinos is an irreducible background to the \ttbar+\met\ signature.
In order to estimate the $\ttbar+Z$ contribution in the SRs, $Z$-boson decays into charged leptons are exploited to define high-purity CRs (TZCR).
The $\ttbar+V$ CRs require exactly three loose signal leptons, at least one of which must also satisfy the tight criteria.
Two leptons are required to have same flavour and opposite charge, and the mass of the dilepton system ($m_{\ell \ell}$) is required to be in the range $81~\GeV < m_{\ell \ell} < 101~\GeV$.
If more than one same-flavour and opposite-charge pairing is possible, the pair with a mass closest to $m_Z$ is chosen.
In addition, at least four jets, one of which is $b$-tagged, are required.
The minimum jet \pt\ of the four leading jets is required to match the thresholds used in the corresponding SR.
The diboson process ($WZ \rightarrow \ell\nu\ell\ell$) is a dominant background in the TZCR, and is normalised to data in a region identical to the TZCR, except for the requirement that no jet is $b$-tagged.
A constant diboson normalisation factor of 0.8, derived in this region, is applied to all TZCRs.

\begin{figure}[htb]
  \centering
  \includegraphics[width=.40\textwidth]{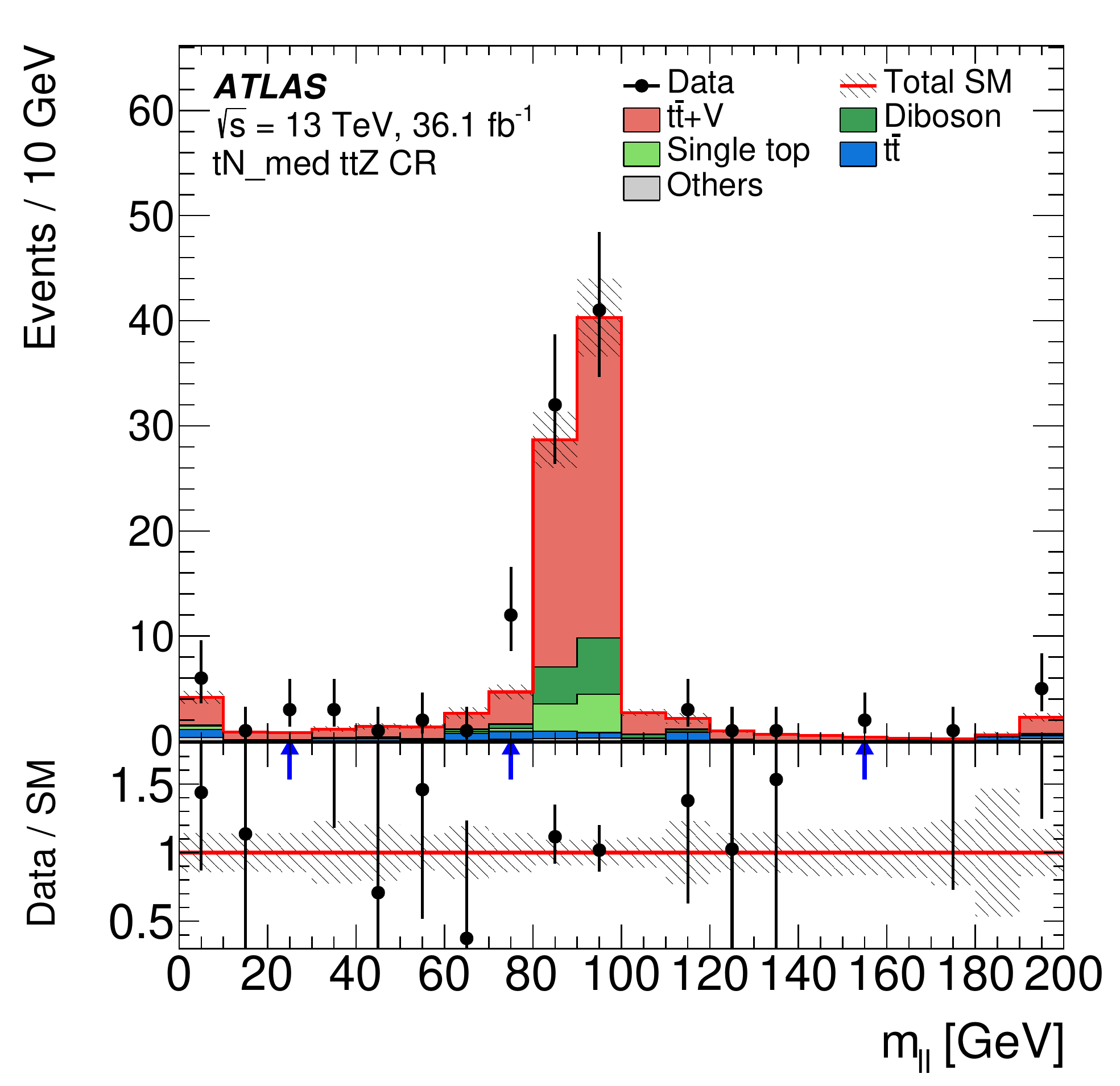}
  \includegraphics[width=.40\textwidth]{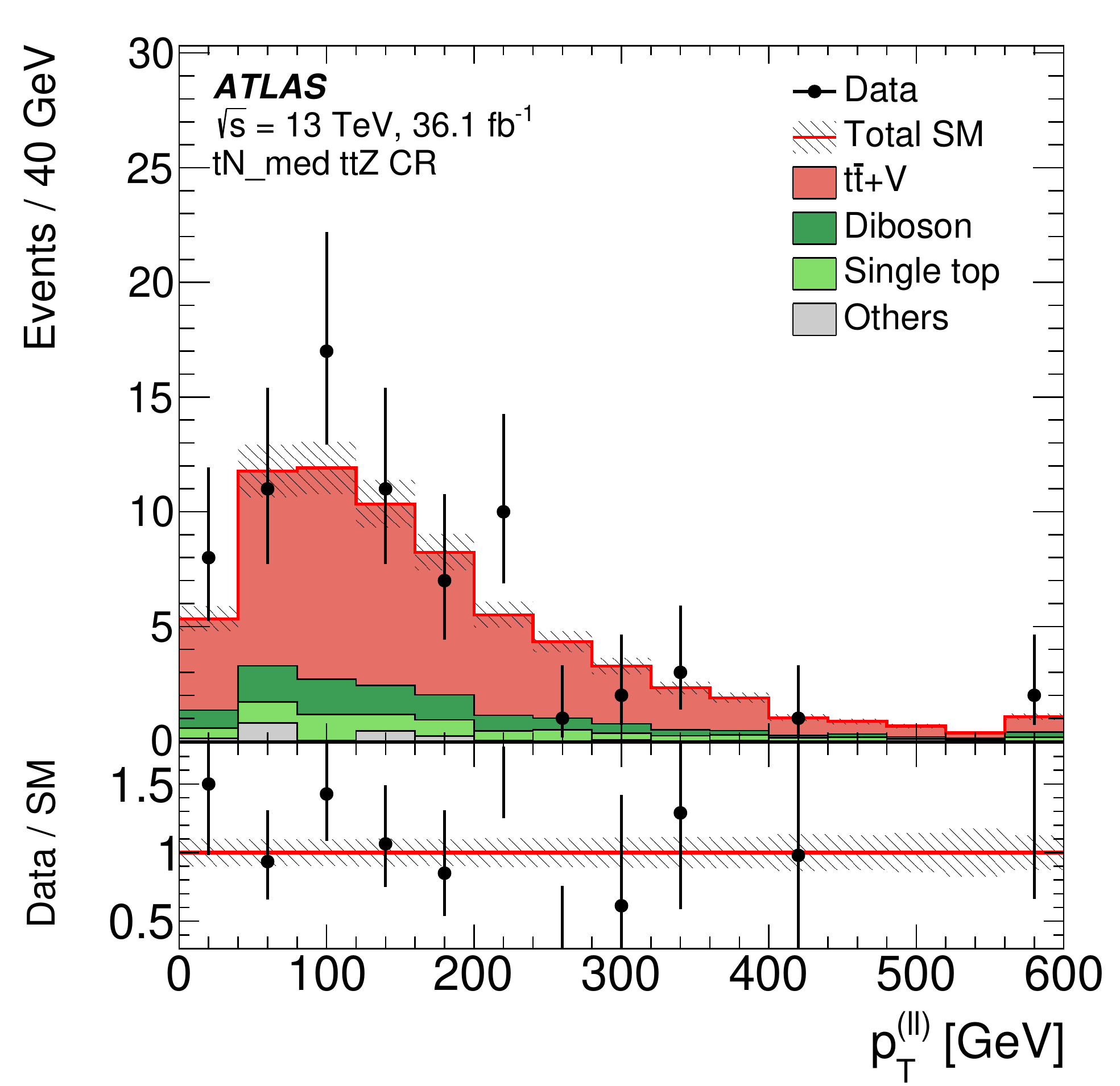}
  \caption{Distribution of (left) the dilepton mass and (right) $\pt^{\ell\ell}$ corresponding to the \pt\ of the reconstructed $Z$ boson in the $\ttbar + Z$ control region (TZCR) associated with the $\tNmed$ signal region. The $\ttbar + Z/W$ processes are normalised in the TZCR. The diboson background is normalised to data events with zero $b$-tagged jets. The hatched area around the total SM prediction and the hatched band in the Data/SM ratio include statistical and experimental uncertainties. The last bin contains overflows.
  }
  \label{fig:CR_ttZ}
\end{figure}

The $\ttbar +Z$ control region is defined for SRs where the $\ttbar+Z$ contribution is sizeable: $\tNmed$, $\tNhigh$, $\bCmed$, $\bCdiag$, $\DMlowloose$, $\DMlow$, and $\DMhigh$. The purity of the TZCR is $\approx 75\%$, with remaining events due to diboson and $tZ$ single-top production. Figure~\ref{fig:CR_ttZ} shows the $\pt^{\ell\ell}$ distribution in the TZCR associated with \tNmed, as well as $m_{\ell \ell}$ prior to requiring $81~\GeV < m_{\ell \ell} < 101~\GeV$. The $\pt^{\ell\ell}$ distribution serves as a proxy for the \met\ distribution in $\ttbar+Z(\nu\nu)$ events. The $\ttbar+Z(\ell\ell)$ method is cross-checked with an alternative method using the $\ttbar+\gamma$ process.  
The normalisation factors obtained from the $\ttbar+\gamma$ events are found to be consistent with those from the $\ttbar+Z(\ell\ell)$ method.

%% file: texfiles/systematics.tex
\section{Systematic uncertainties}
\label{sec:systematics}
The systematic uncertainties in the signal and background estimates arise both from experimental sources and from the uncertainty in the theoretical predictions and modelling. Since the yields from the dominant background sources, \ttbar, single-top $Wt$, $\ttbar+V$, and $W$+jets, are normalised to data in dedicated CRs, the uncertainties for these processes affect only the extrapolation from the CRs into the SRs (and amongst the various CRs), but not the overall normalisation. The systematic uncertainties are included as nuisance parameters with Gaussian constraints and profiled in the likelihood fits. The uncertainties are not reduced as a result of the profiling.

The dominant experimental uncertainties arise from imperfect knowledge of the jet energy scale (JES) and jet energy resolution (JER)~\cite{ATL-PHYS-PUB-2015-015, PERF-2016-04}, as well as the modelling of the $b$-tagging efficiencies and mis-tag rates~\cite{ATLAS-CONF-2014-004,ATLAS-CONF-2014-046}.
From these sources, the resulting uncertainties expressed as relative uncertainties in the total predicted background yield in the SRs are in the range 1.4--7\% for JES, 1.5--7\% for JER, and 1.6--13\% for $b$-tagging.
Other sources of experimental uncertainty include the modelling of the lepton energy scales, energy resolutions, reconstruction and identification efficiencies, trigger efficiencies, 
and the modelling of pile-up and the integrated luminosity. These uncertainties have a small impact on the final results.

The uncertainty in the modelling of the single-top and \ttbar\ backgrounds include effects related to the MC event generator, the hadronisation modelling, and the amount of initial- and final-state radiation~\cite{ATL-PHYS-PUB-2016-004}.
The MC generator uncertainty is estimated by comparing events produced with \powbox+\HERWIGpp~v2.7.1 with either \MGaMC~v2.2.3+\HERWIGpp~v2.7.1 (NLO) or {\textsc{Sherpa}}~v2.2.1. Events generated with \powbox\ are showered and subsequently hadronized with either $\PYTHIA$6 or \HERWIGpp\ to estimate the effect from the modelling of the hadronisation.
The impact of altering the amount of initial- and final-state radiation is estimated from comparisons of \powbox+$\PYTHIA$6 samples with different parton-shower radiation, NLO radiation, and modified factorisation and renormalisation scales.
An additional uncertainty stems from the modelling of the interference between the \ttbar\ and $Wt$ processes. The uncertainty is estimated using inclusive $WWbb$ events, generated using \MGaMC~v2.2.3 (LO), which are compared with the sum of the resonant \ttbar\ and $Wt$ processes~\cite{ATL-PHYS-PUB-2016-004}.
The resulting uncertainties from all the aforementioned sources in the extrapolation factors from the \ttbar\ and $Wt$ CRs to the SRs are 10--45\% for \ttbar, and 10--47\% for $Wt$ events, where the latter is dominated by the interference term.

The uncertainty in the modelling of the $\ttbar+Z$ background is estimated from independent variations of the renormalisation and factorisation scales, and PDF variations. A MC generator uncertainty is estimated by comparing events produced with \MGaMC~v2.2.3+$\PYTHIA$8 (NLO) and {\textsc{Sherpa}}~v2.2.1. The resulting modelling-induced uncertainties in the extrapolation factor are 10--37\%, dominated by the MC generator comparison.

The uncertainty in the $W$+jets background from the choice of MC generator is estimated by comparing {\textsc{Sherpa}} with \MGaMC~v2.2.3+$\PYTHIA$8 (NLO).
In addition, the effects of varying the scales for the matching scheme related to the merging of matrix elements and parton showers, renormalisation, factorisation, and resummation are estimated. The total uncertainty is found to be 4--44\%.

The sources of uncertainty considered for the diboson background are the effects of varying the renormalisation, factorisation, and resummation scales. 
Since the diboson
background is not normalised in a CR, the analysis is also sensitive to the uncertainty in the total cross-section.
The resulting theoretical uncertainty ranges from 13 to 32\%.

For the BDT analyses, a systematic-smoothing procedure in BDT score is applied to evaluate the uncertainties in the modelling of \ttbar\ and single-top $Wt$ processes. The procedure gives a reliable estimate of the uncertainties despite statistical fluctuations in the background samples, based on merging statistically insignificant bins and smoothing the result with a Gaussian kernel.

The SUSY signal cross-section uncertainty is taken from an envelope of cross-section predictions using different PDF sets and factorisation and renormalisation scales, as described in Ref.~\cite{Borschensky:2014cia}, and the resulting uncertainties range from 13\% to 23\%. Dedicated uncertainties in the signal acceptance due to the modelling of additional radiation are considered for SRs relying on ISR. These are estimated from the variation of factorisation and renormalisation scales, and range from 10\% to 20\%.
The uncertainty in the DM production cross-section is estimated from the effect of varying the renormalisation, factorisation, and matching scales, as well as the PDF choice. The uncertainty is found to be between 12\% and 20\%. Experimental uncertainties in the signal acceptance have negligible impact on the final results.

Table~\ref{tab:systematics} summarises the dominant systematic uncertainties in selected signal regions. The dominant sources of uncertainty are background modelling and JES/JER uncertainties in most of SRs. The uncertainty related to the description of the $b$-tagging mis-tag rates in the simulation becomes large in the $\bCsoftmed$. This is because the single-top $Wt$ or semileptonic \ttbar\ background events above the \amtTwo\ kinematic endpoint often have an associated charm-quark misidentified as a $b$-jet, and thus the background yield is sensitive to the mis-tag modelling.

\begin{table}[t]
  \centering
  \caption{Summary of the dominant systematic uncertainties in the total predicted background yields, obtained by the background-only fits as described in Section~\ref{subsec:obs_data}, in several representative signal regions: $\tNmed$, $\bWN$, $\bCmed$, and $\bCsoftmed$. Numbers are given as percentages of the total background estimate.
  }
  \vspace{3mm}
{\renewcommand{\arraystretch}{1.1}
  \begin{tabular*}{\textwidth}{@{\extracolsep{\fill}}| l | rrrr c |}
                \hline\hline
    Signal Region Uncertainty (\%)                      & \tNmed    & \bWN     & \bCmed   & \bCsoftmed & \\ \hline
    \hline
    $\ttbar+Z$ normalisation    & 11\phantom{.}\phantom{0}  &  --  & 6.8   & --  & \\
    $\ttbar$ (2L) normalisation  & 4.7    &  7.5 & 3.3    & 2.6  & \\
    $Wt$  normalisation          & 3.0    &  --  & 17\phantom{.}\phantom{0}  & 3.4  & \\
    $W$+jets  normalisation      & 2.5    &  --  & 2.1    & 8.1  & \\
    \hline
    $\ttbar+Z$ modelling       & 11\phantom{.}\phantom{0}    &  2.3     &  1.2     & $<1.0$     & \\
    \ttbar\ radiation       &  4.3    & 13\phantom{.}\phantom{0}     &  1.9     & 4.6        & \\
    \ttbar\ generator       &  3.6    &  7.8     &  1.7     & 4.6        & \\
    \ttbar\ hadronisation      &  2.5    & 12\phantom{.}\phantom{0}     &  5.8     & 3.9        & \\
    $Wt$--$\ttbar$ interference     &  $<1.0$    &  $<1.0$  & 13\phantom{.}\phantom{0}     & $<1.0$     & \\
    Single-top generator       &  $<1.0$    &  $<1.0$  &  4.9     & $<1.0$     & \\
    Single-top hadronisation    &  $<1.0$  &  $<1.0$  & 11\phantom{.}\phantom{0}     & $<1.0$     & \\
    \hline
    JER           &  2.8    &  1.5     &  6.8     &  2.4       & \\
    JES           &  2.8    &  6.6     &  1.4     &  2.1       & \\ 
    Mis-$b$-tag ($c$-quark)     &  2.3    &  1.6     &  4.9     & 13\phantom{.}\phantom{0}       & \\
    Mis-$b$-tag (light quark)     &  2.0    &  $<1.0$  &  2.0     &  4.6       & \\
    Pile-up         &  2.5    &  1.2     &  3.8     &  2.0       & \\
    \hline
    Total systematic uncertainty    & 18\phantom{.}\phantom{0}     & 22\phantom{.}\phantom{0}     & 28\phantom{.}\phantom{0}     & 15\phantom{.}\phantom{0}       & \\
    \hline \hline
  \end{tabular*}
}
  \label{tab:systematics}
\end{table}

%% file: texfiles/results.tex
\section{Results} \label{sec:results}

\subsection{Observed data and predicted backgrounds}
\label{subsec:obs_data}

In order to determine the SM background yields in the SRs, a likelihood fit is performed for each SR. The fit is configured to use only the CRs to constrain the fit parameters corresponding to the normalisations of $\ttbar$, single-top, $W$+jets, and $\ttbar+V$ processes in the dedicated CRs. This fit configuration is referred to as the background-only fit.

\begin{figure}[htbp]
  \centering
  \includegraphics[width=.99\textwidth]{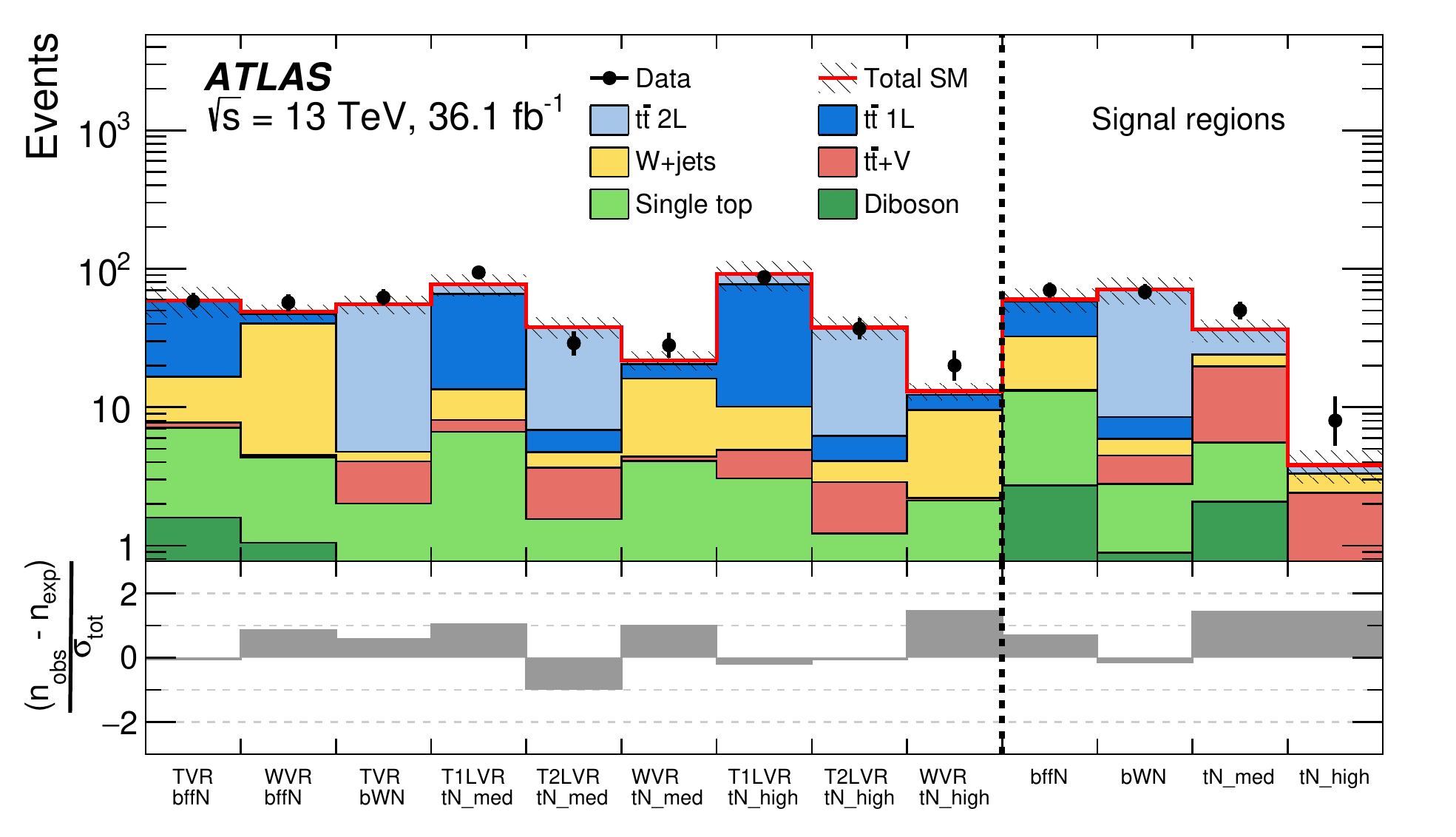} \\
  \includegraphics[width=.99\textwidth]{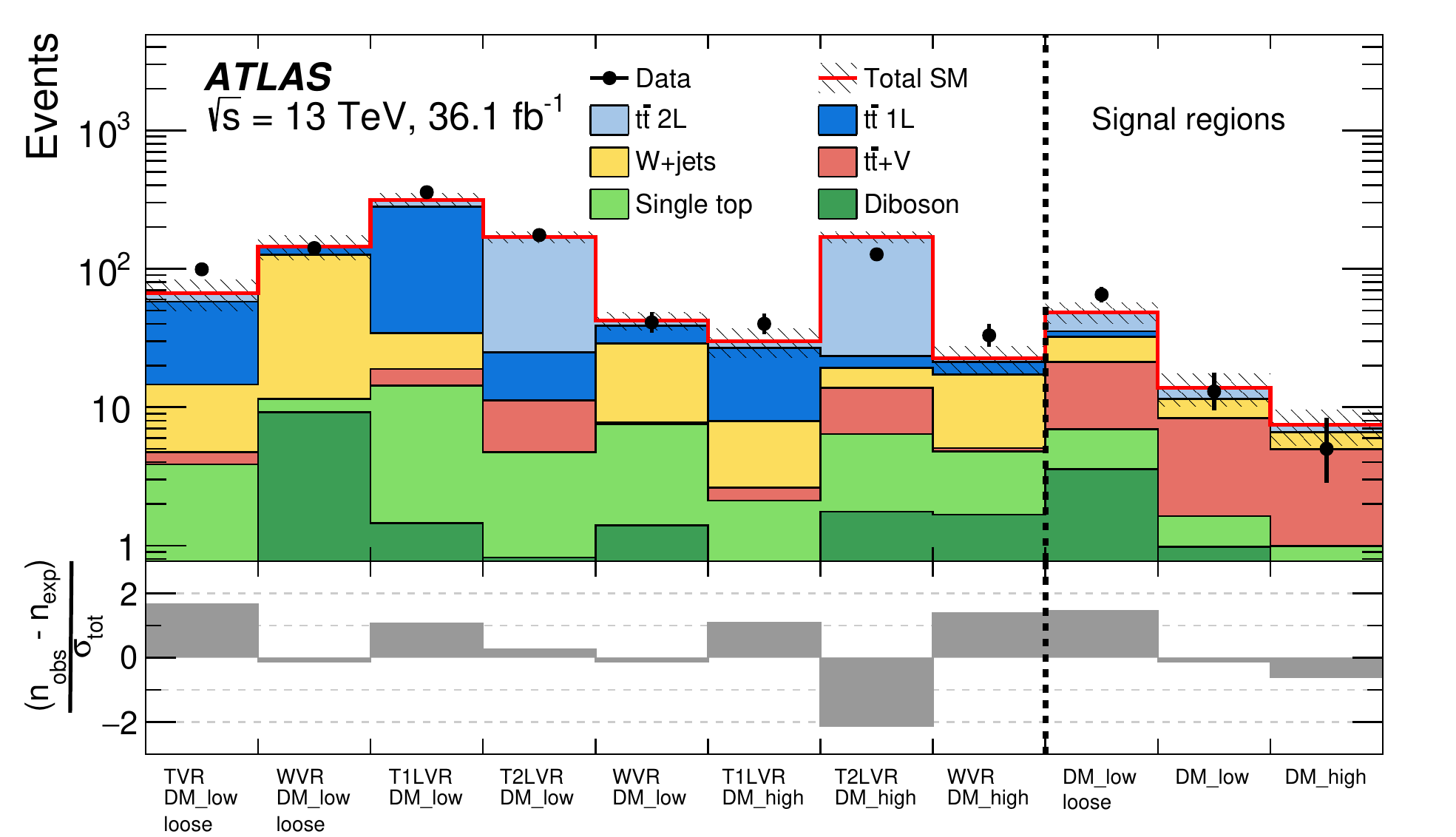}
  \caption{Comparison of the observed data ($n_\text{obs}$) with the predicted SM background ($n_\text{exp}$) in (top) the $\bffN$, $\bWN$, $\tNmed$ and $\tNhigh$ signal regions, and (bottom) the $\DMlowloose$, $\DMlow$, and $\DMhigh$ signal regions, and associated VRs. The background predictions are obtained using the background-only fit configuration, and the hatched area around the SM prediction includes all uncertainties. The bottom panels show the difference between data and the predicted SM background divided by the total uncertainty ($\sigma_\text{tot}$). 
  } 
  \label{fig:pulls-tN-DM}
\end{figure}

\begin{figure}[htbp]
  \centering
  \includegraphics[width=.99\textwidth]{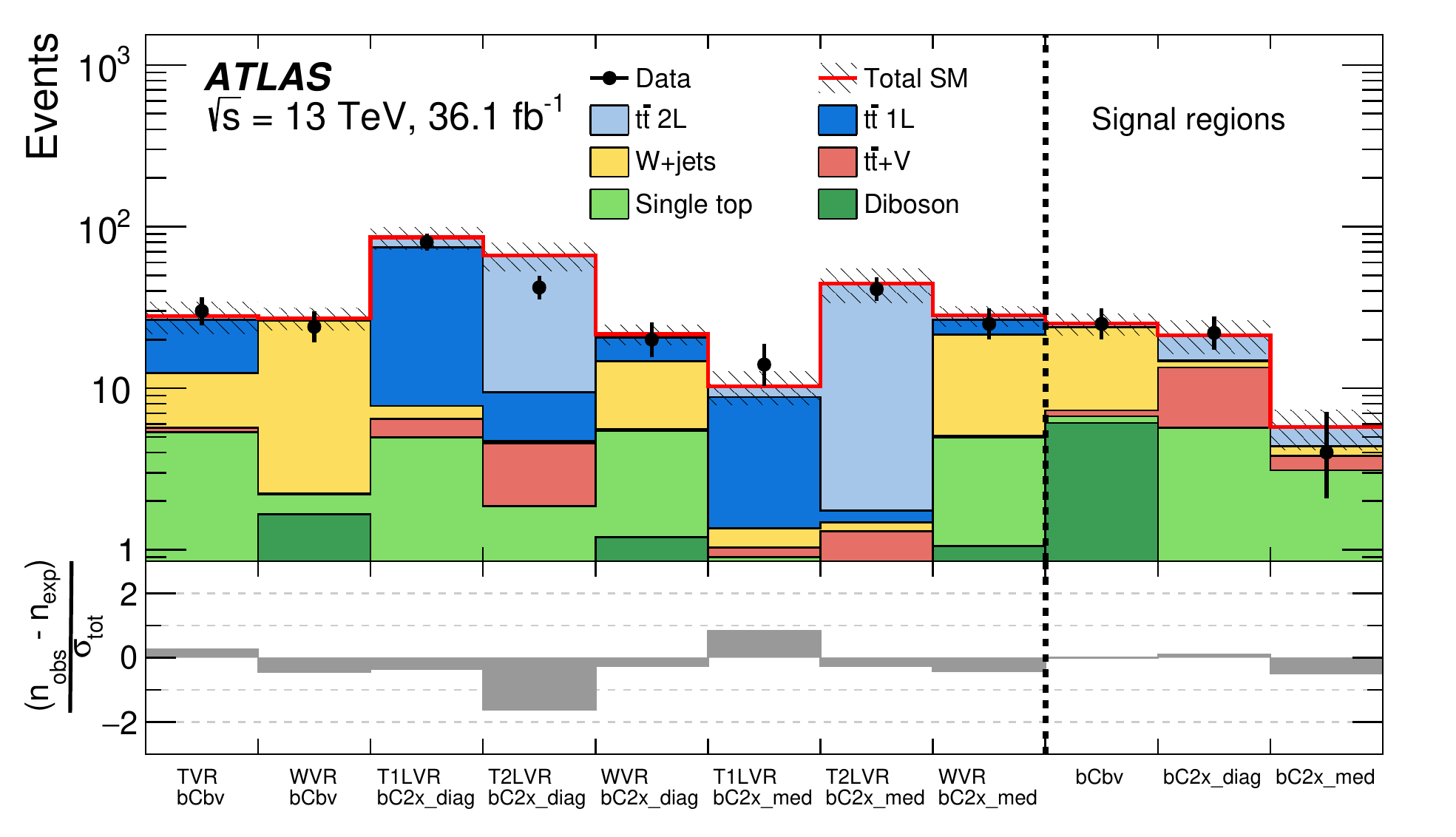} \\
  \includegraphics[width=.99\textwidth]{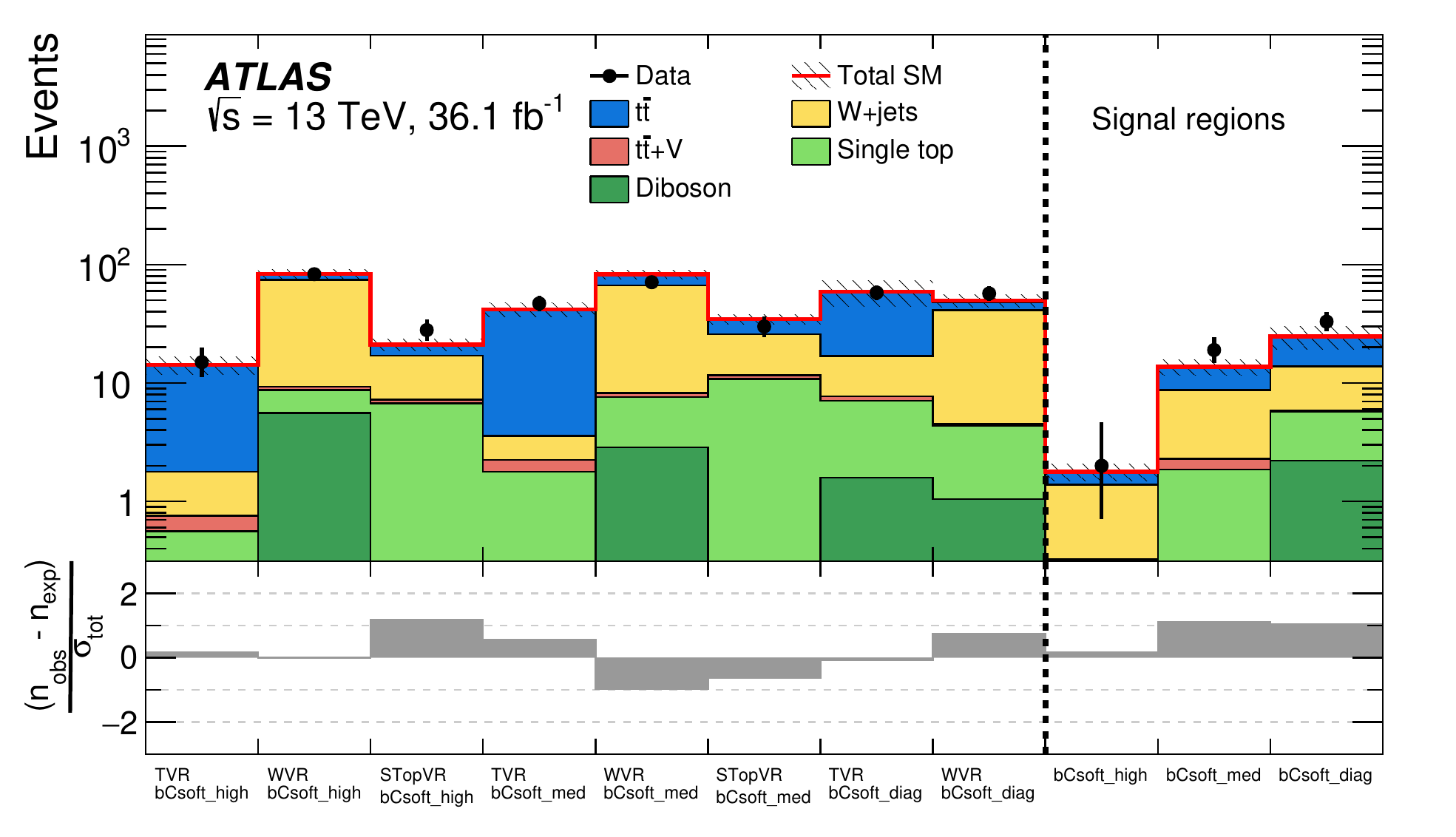}
  \caption{Comparison of the observed data ($n_\text{obs}$) with the predicted SM background ($n_\text{exp}$) in (top) the $\bCbv$, $\bCdiag$, and $\bCmed$ signal regions, (bottom) the $\bCsofthigh$, $\bCsoftmed$, and $\bCsoftdiag$ signal regions, together with associated VRs. The background predictions are obtained using the background-only fit configuration, and the hatched area around the SM prediction includes all uncertainties. The bottom panels show the difference between data and the predicted SM background divided by the total uncertainty ($\sigma_\text{tot}$). 
  } 
  \label{fig:pulls-bC}
\end{figure}

The number of observed events and the predicted number of SM background events from the background-only fits in all SRs and VRs are shown in Figures~\ref{fig:pulls-tN-DM} and \ref{fig:pulls-bC}. The SRs are not mutually exclusive and are therefore not statistically independent. In all SRs, the distributions indicate good agreement between the data and the SM background estimate. The largest excesses over the background-only hypothesis are 1.6\,$\sigma$ and 1.4\,$\sigma$, observed in \tNhigh\ and \tNmed, respectively. The previously observed excess in \DMlowloose\ is reduced with the inclusion of more data to the level of 1.5 $\sigma$.

The number of observed events together with the predicted number of SM background events in all 16 SRs are summarised in Tables~\ref{tab:discovery_SR_yields_tN} and \ref{tab:discovery_SR_yields_bC}, showing the breakdown of the various backgrounds that contribute to the SRs. The tables also list the results for the four fit parameters that control the normalisation of the four main backgrounds (normalisation factors, NFs), together with the associated fit uncertainties including the theoretical modelling uncertainties. In order to quantify the level of agreement of the SM background-only hypothesis with the observations in the SRs, a profile-likelihood-ratio test is performed. The resulting $p$-values ($p_0$) are also presented in the tables,
and are capped at 0.5. 
Model-independent upper limits on beyond-SM contributions are derived for each SR. A generic signal model is assumed that contributes only to  the SR and for which neither experimental nor theoretical systematic uncertainties except for the luminosity uncertainty are considered. All limits are calculated using the $\text{CL}_\text{s}$ prescription~\cite{Read:2002hq}. 
Table~\ref{tab:exclusion_SR_yields} details the number of observed events and the predicted number of SM background events for each bin of the shape-fit SRs. The NFs are compatible with unity in most cases, except for the single-top NFs in \bCsoftmed\ and $\bCsofthigh$. The single-top NFs are significantly below unity, possibly due to the effect of interference between the $Wt$ and \ttbar\ processes at NLO.

\begin{figure}[htbp]
  \centering
  \includegraphics[width=.40\textwidth]{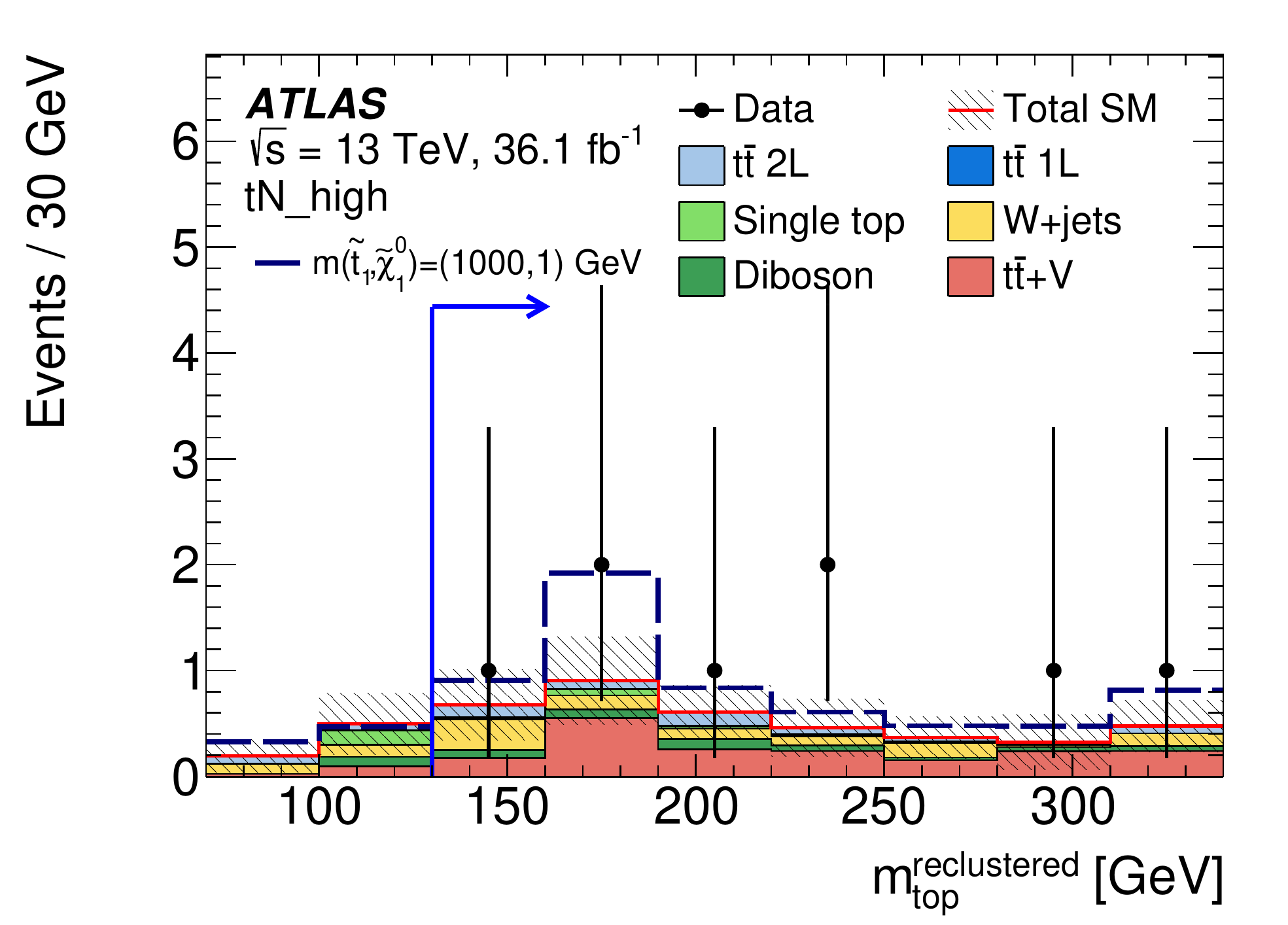}
  \includegraphics[width=.40\textwidth]{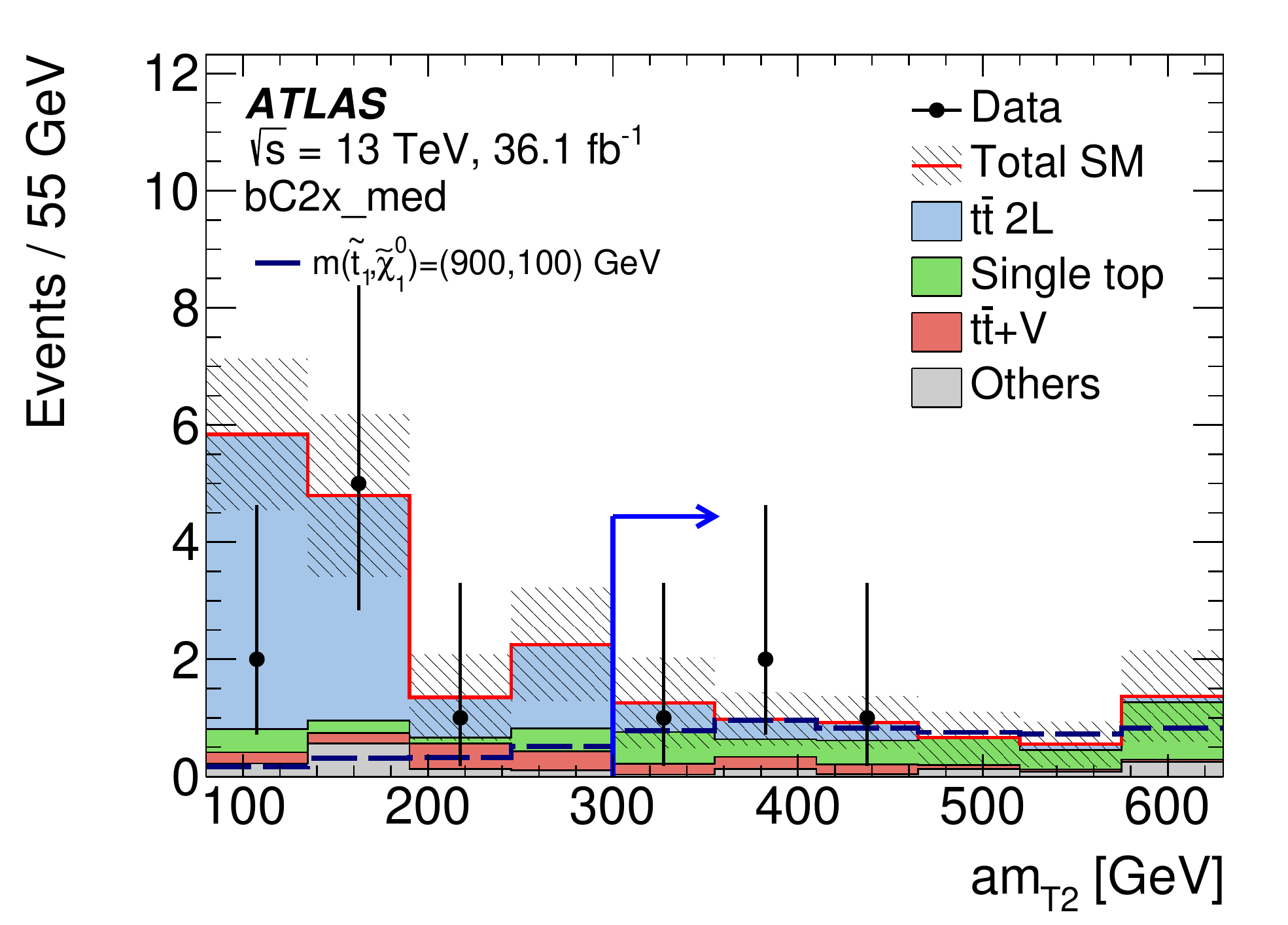}
  \includegraphics[width=.40\textwidth]{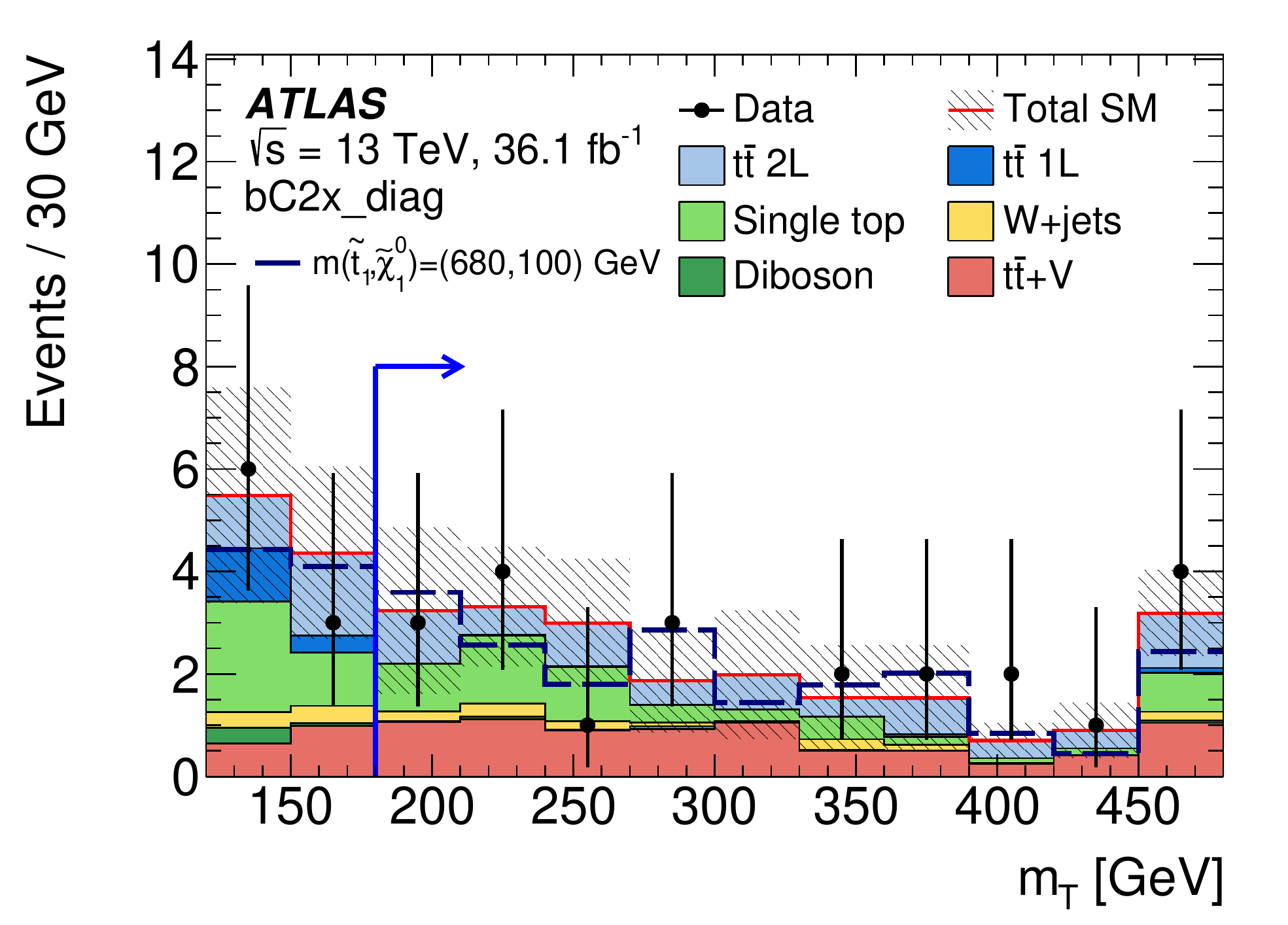}
  \includegraphics[width=.40\textwidth]{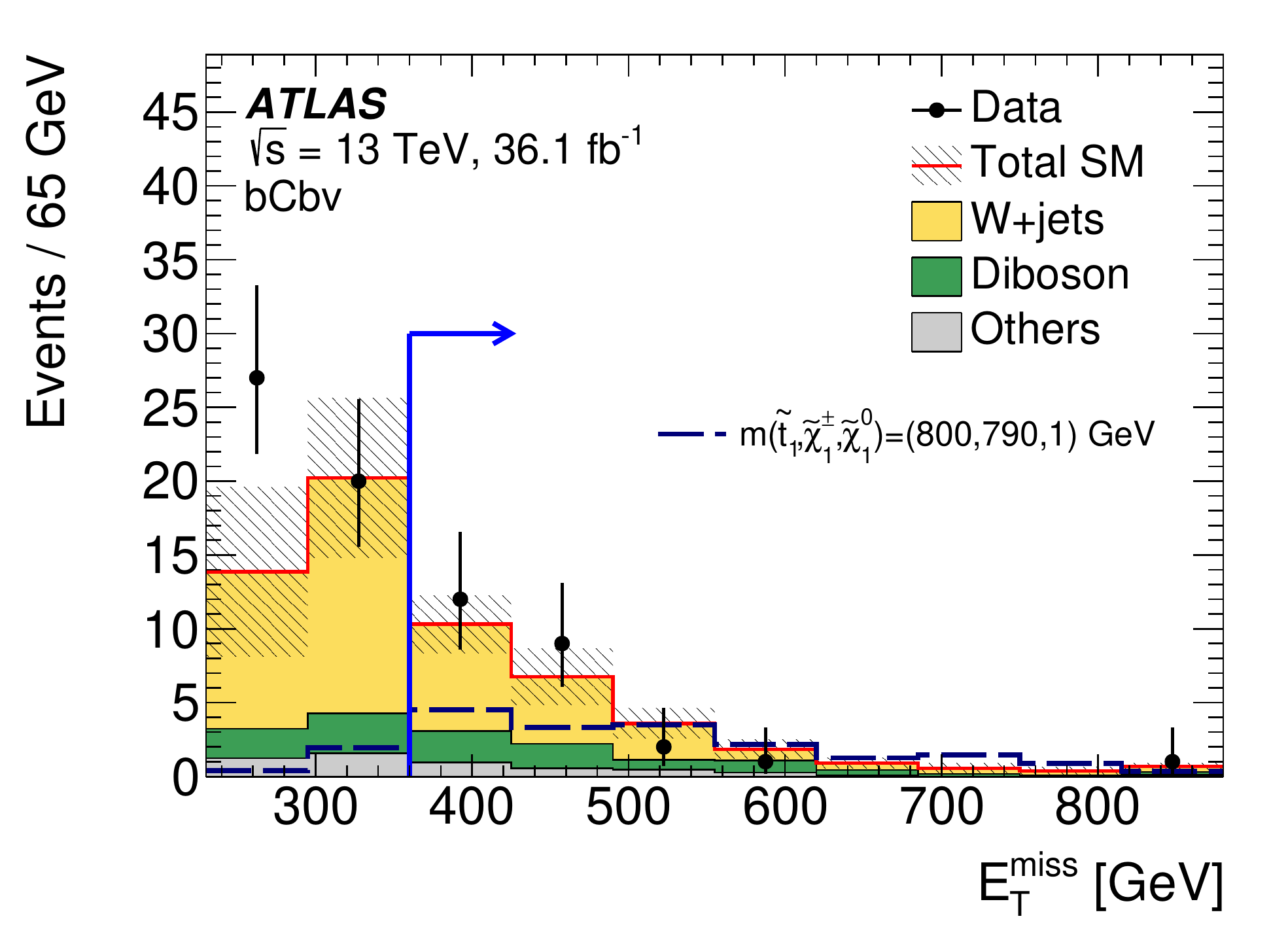}
  \includegraphics[width=.40\textwidth]{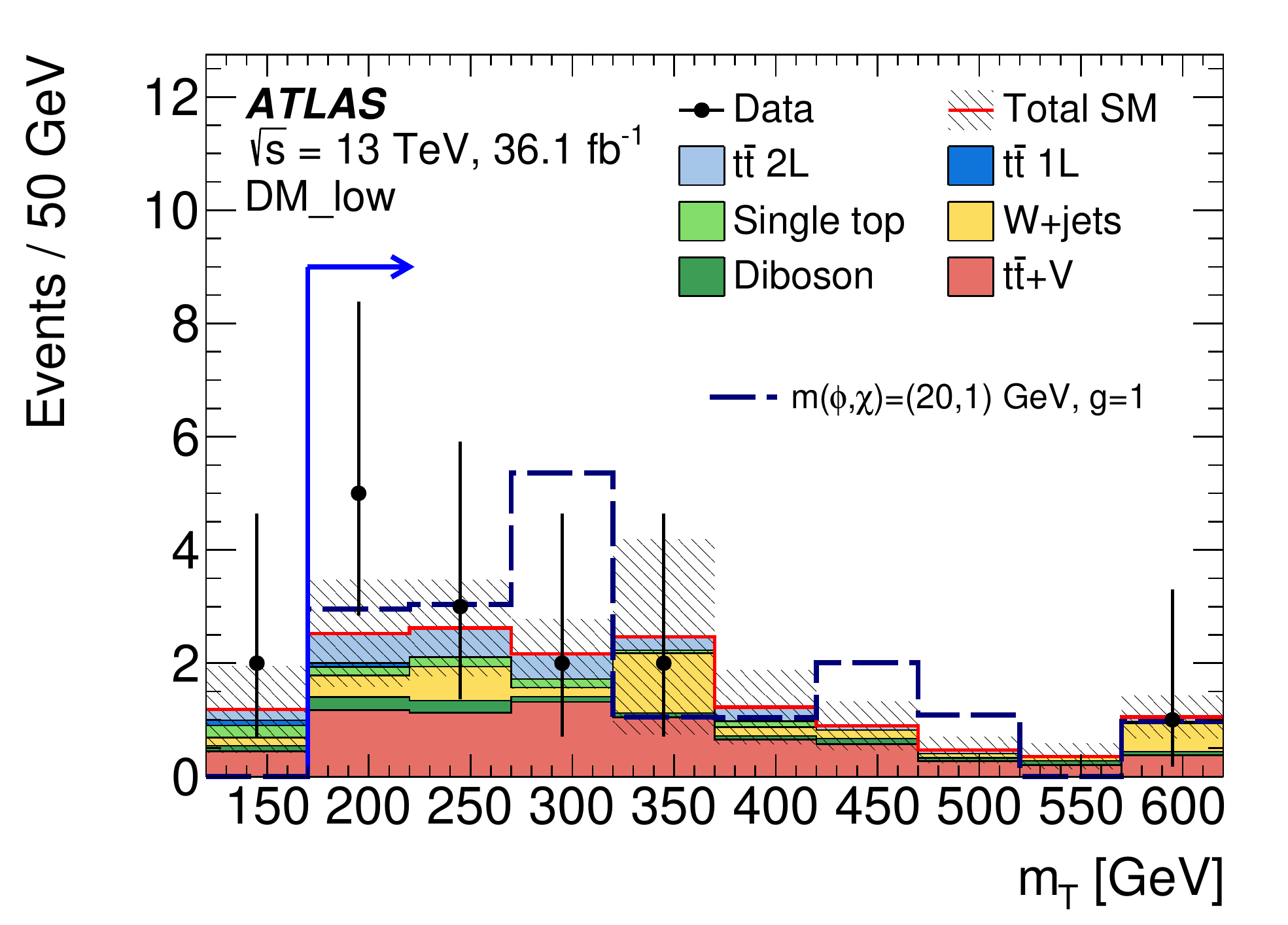}
  \includegraphics[width=.40\textwidth]{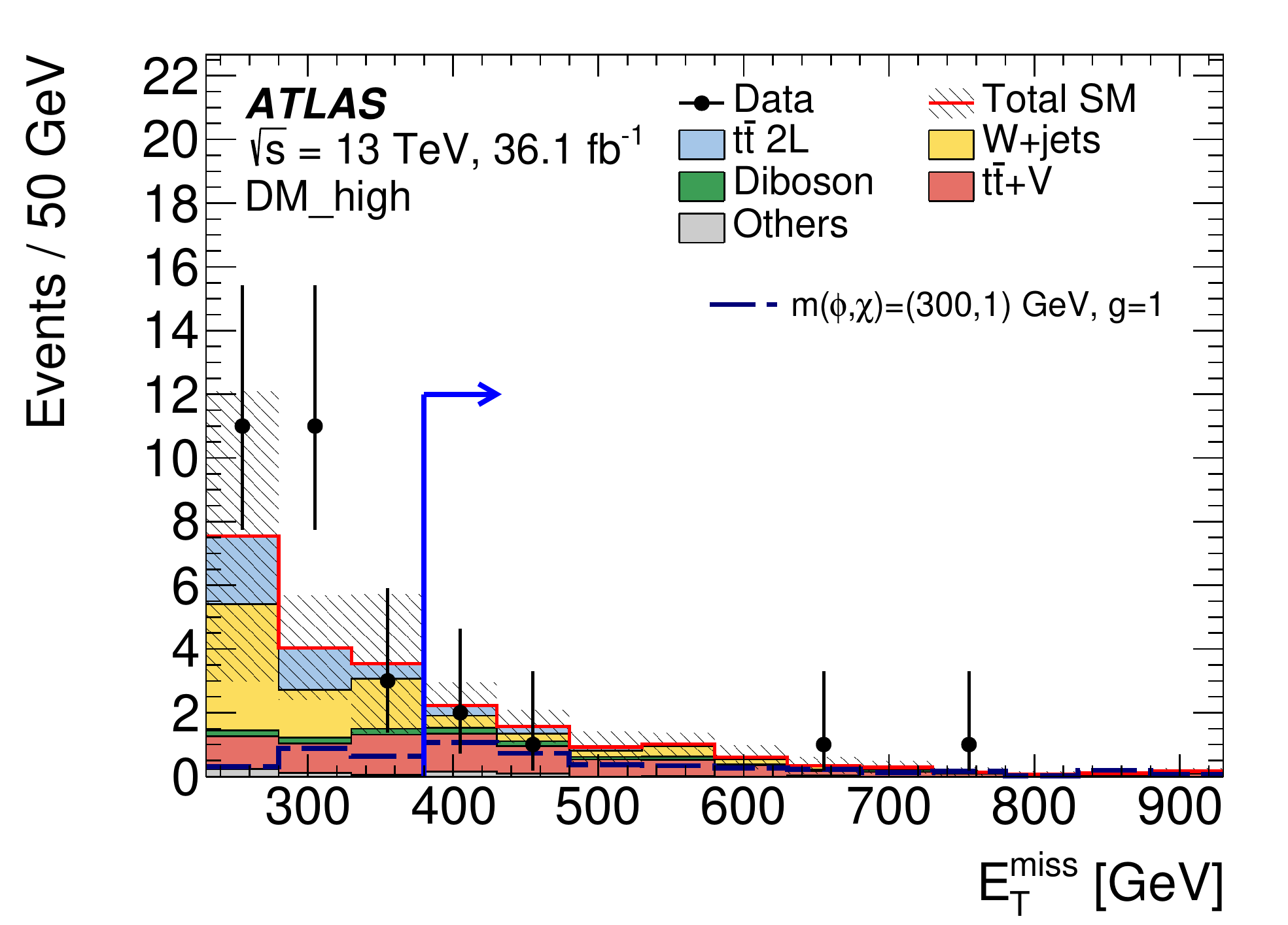}
  \caption{Kinematic distributions in the signal regions: (top left) \mTopRecluster\ in $\tNhigh$, (top right) \amtTwo\ in $\bCmed$, (middle left) \mt\ in $\bCdiag$, (middle right) \met\ in $\bCbv$, (bottom left) \mt\ in $\DMlow$, and (bottom right) \met\ in $\DMhigh$. The full event selection in the corresponding signal region is applied, except for the requirement (indicated by an arrow) that is imposed on the variable being plotted. The predicted SM backgrounds are scaled with the normalisation factors obtained from the corresponding control regions in Tables~\ref{tab:discovery_SR_yields_tN} and \ref{tab:discovery_SR_yields_bC}. 
In addition to the background prediction, a signal model is shown on each plot. In the DM+\ttbar\ signal model, a coupling of $g=1$ is assumed. 
The category labelled `Others' stands for minor SM backgrounds that contribute less than 5\% of the total SM background.
The hatched area around the total SM prediction includes statistical and experimental uncertainties. The last bin contains overflows. 
  }
  \label{fig:srs-cut-and-count}
\end{figure}

\begin{figure}[htbp]
  \centering
  \includegraphics[width=.40\textwidth]{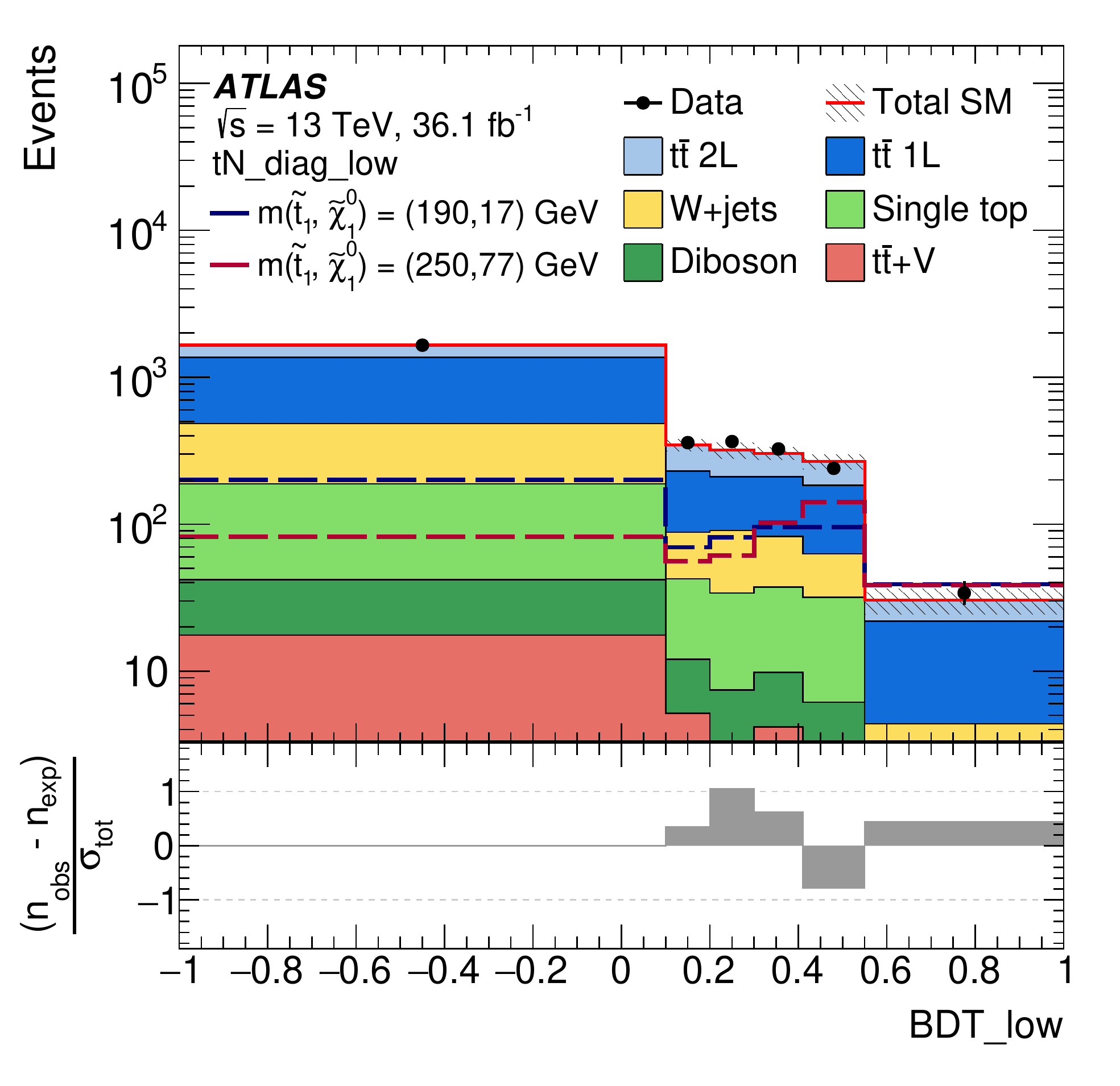}
  \includegraphics[width=.40\textwidth]{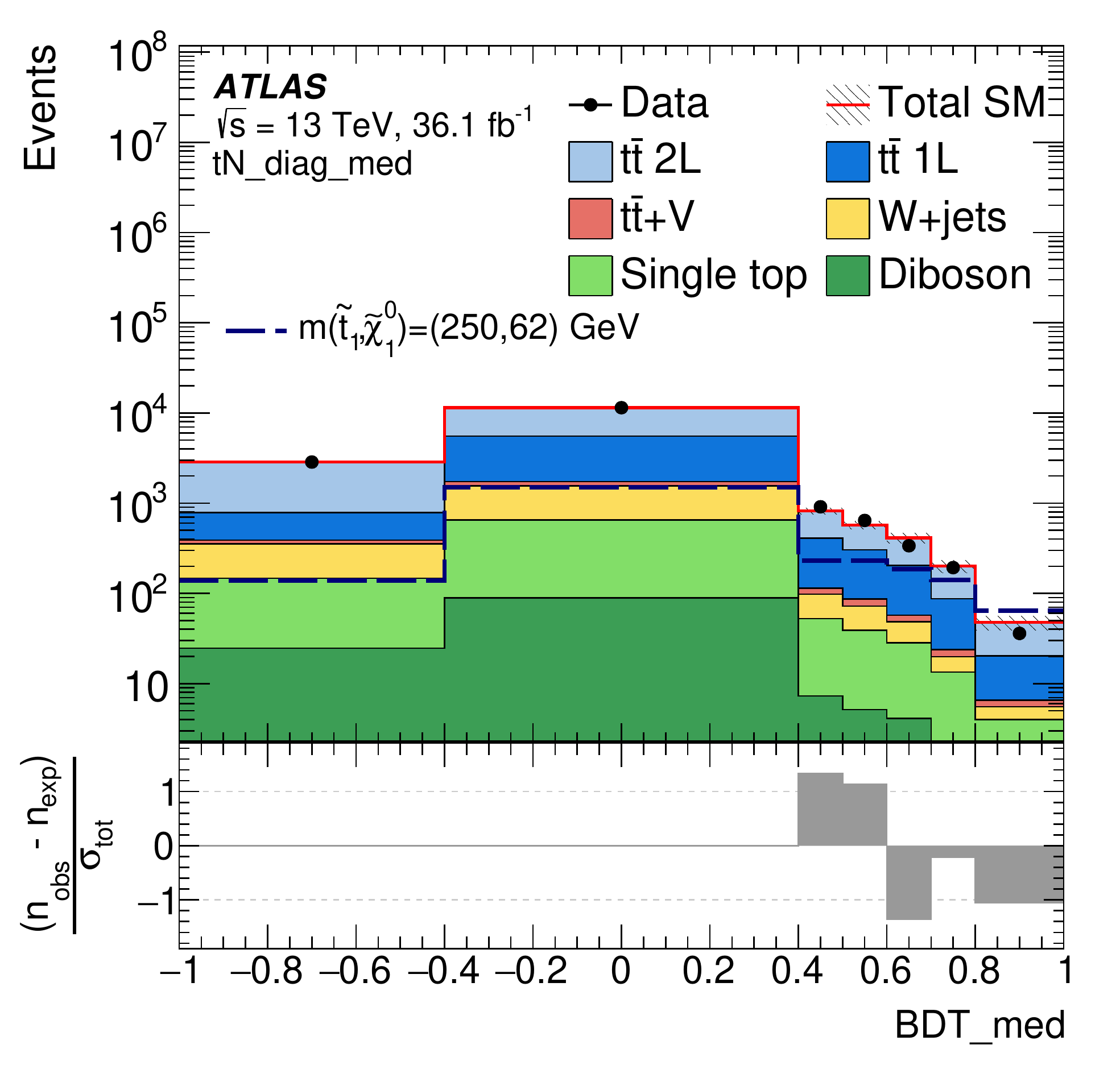}
  \includegraphics[width=.40\textwidth]{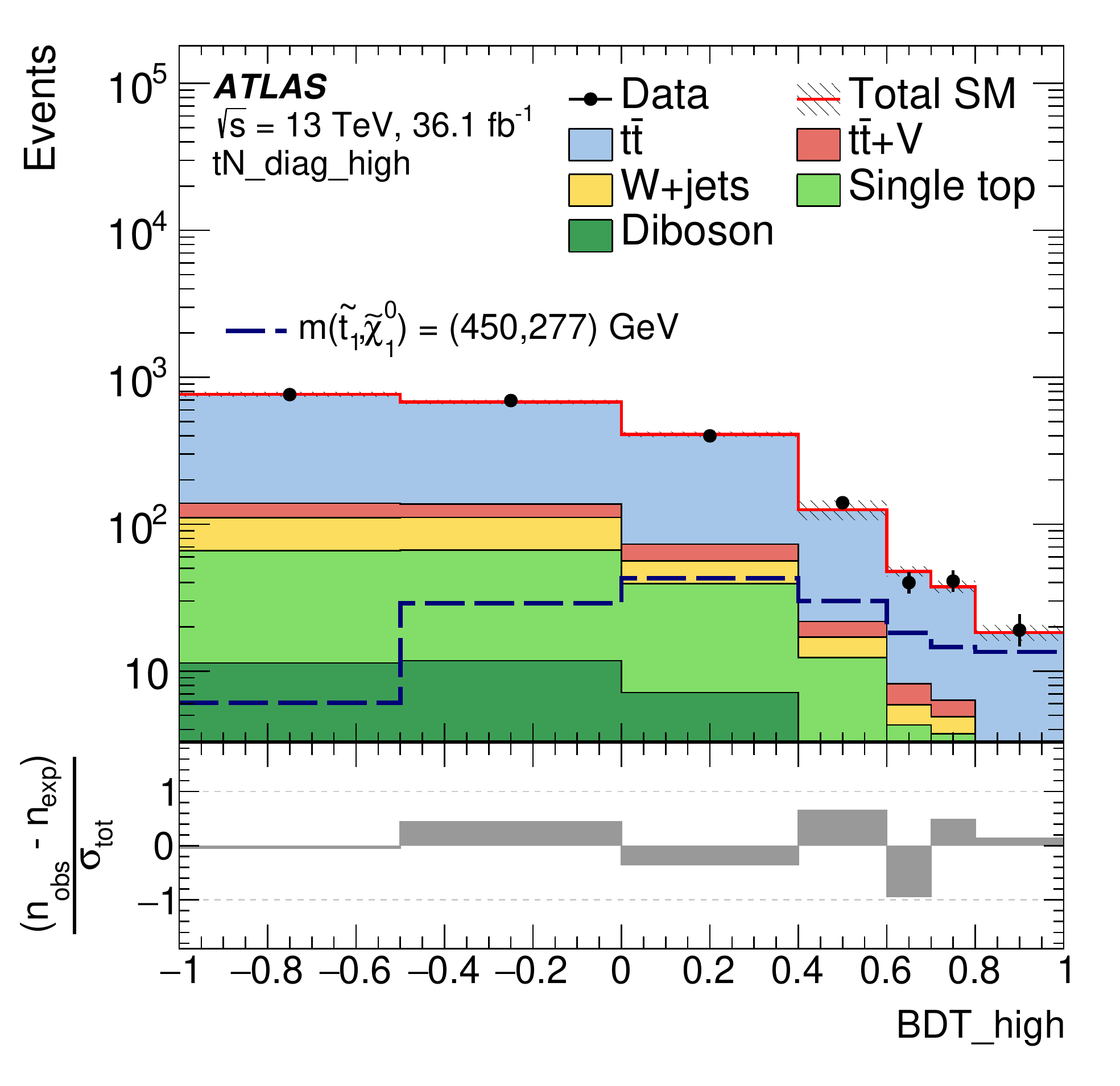}
  \caption{Distributions of BDT score for the \tNdiaglow\ (top left), $\tNdiagmed$ (top right), and \tNdiaghigh\ (bottom) regions. The SM background predictions are obtained using the background-only fit configuration, and the hatched area around the total SM background prediction includes all uncertainties. In addition to the background prediction, signal models are shown, denoted by $m(\tone,\ninoone)$. The bottom panels show the difference between data and the predicted SM background divided by the total uncertainty ($\sigma_\text{tot}$). 
  }
  \label{fig:srs-BDTs}
\end{figure}

Figures~\ref{fig:srs-cut-and-count}, \ref{fig:srs-BDTs}, and \ref{fig:srs-shapefits} show comparisons between the observed data and the SM background prediction with all SR selections applied except the requirement on the plotted variable. Good agreement is found between the observed data and the SM background prediction. The expected distributions from representative signal benchmark models are overlaid. 

\subsection{Exclusion limits}
\label{subsec:exclusion_limits}

No significant excess is observed, and exclusion limits are set based on profile-likelihood fits for the stop pair production models and the simplified model for top quarks produced in association with dark-matter particles.

The signal uncertainties and potential signal contributions to all regions are taken into account. All uncertainties except those in the theoretical signal cross-section are included in the fit. Exclusion limits at 95\% confidence level (CL) are obtained by selecting {\textit{a priori}} the signal region with the lowest expected $\text{CL}_\text{s}$ value for each signal model and the exclusion contours are derived by interpolating in the $\text{CL}_\text{s}$ value.

\begin{figure}[htbp]
  \centering
  \includegraphics[width=.99\textwidth]{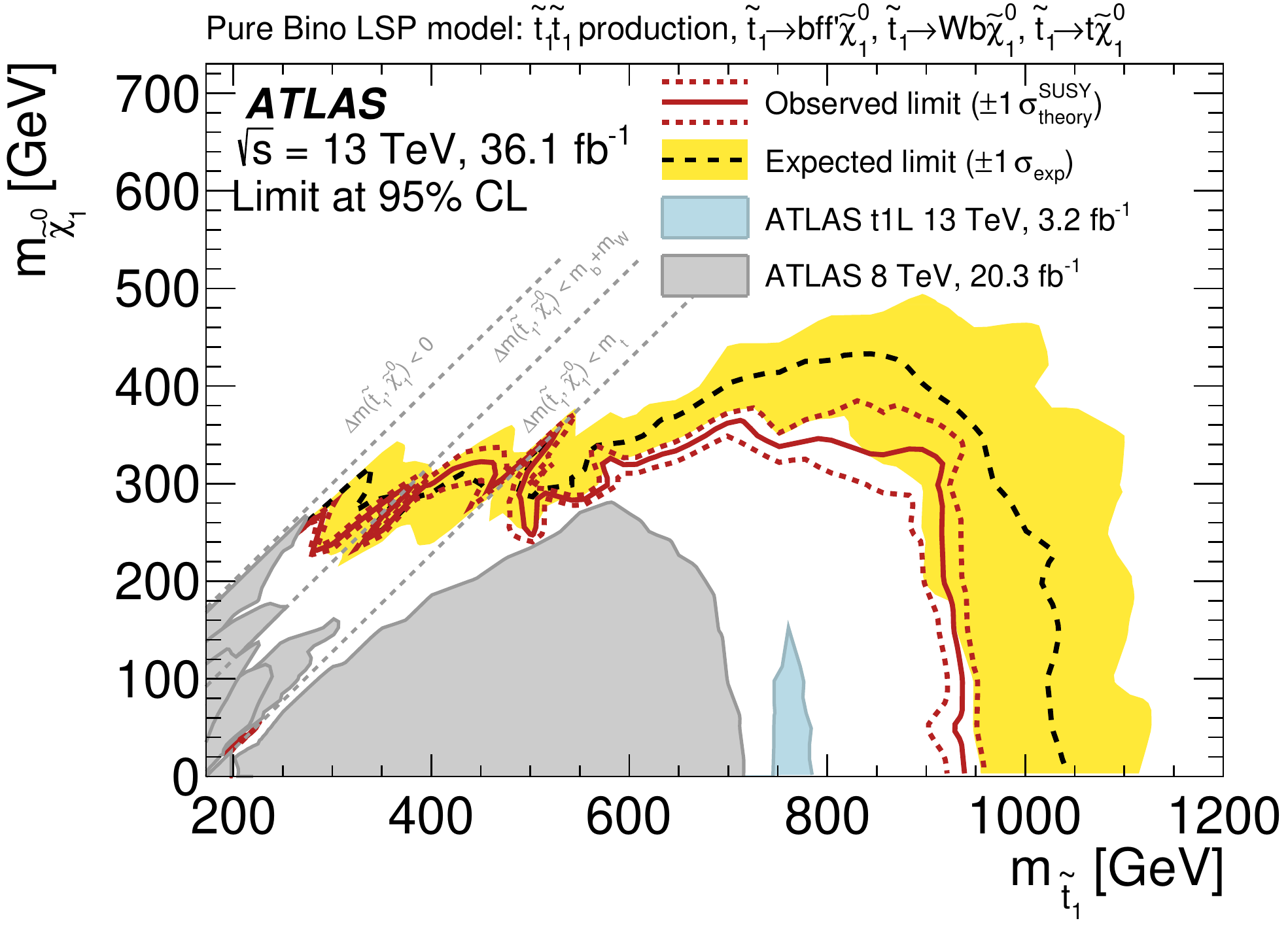}
  \caption{Expected (black dashed) and observed (red solid) 95\% excluded regions in the plane of $m_{\ninoone}$ versus $m_{\tone}$ for direct stop pair production assuming either $\topLSP$, $\threeBody$, or $\fourBody$ decay with a branching ratio of 100\%. The excluded regions from previous publications~\cite{SUSY-2013-15,SUSY-2015-02} are shown with the grey and blue shaded areas.
  } 
  \label{fig:contour-tN-comb}
\end{figure}

\begin{figure}[htbp]
  \centering
  \includegraphics[width=.99\textwidth]{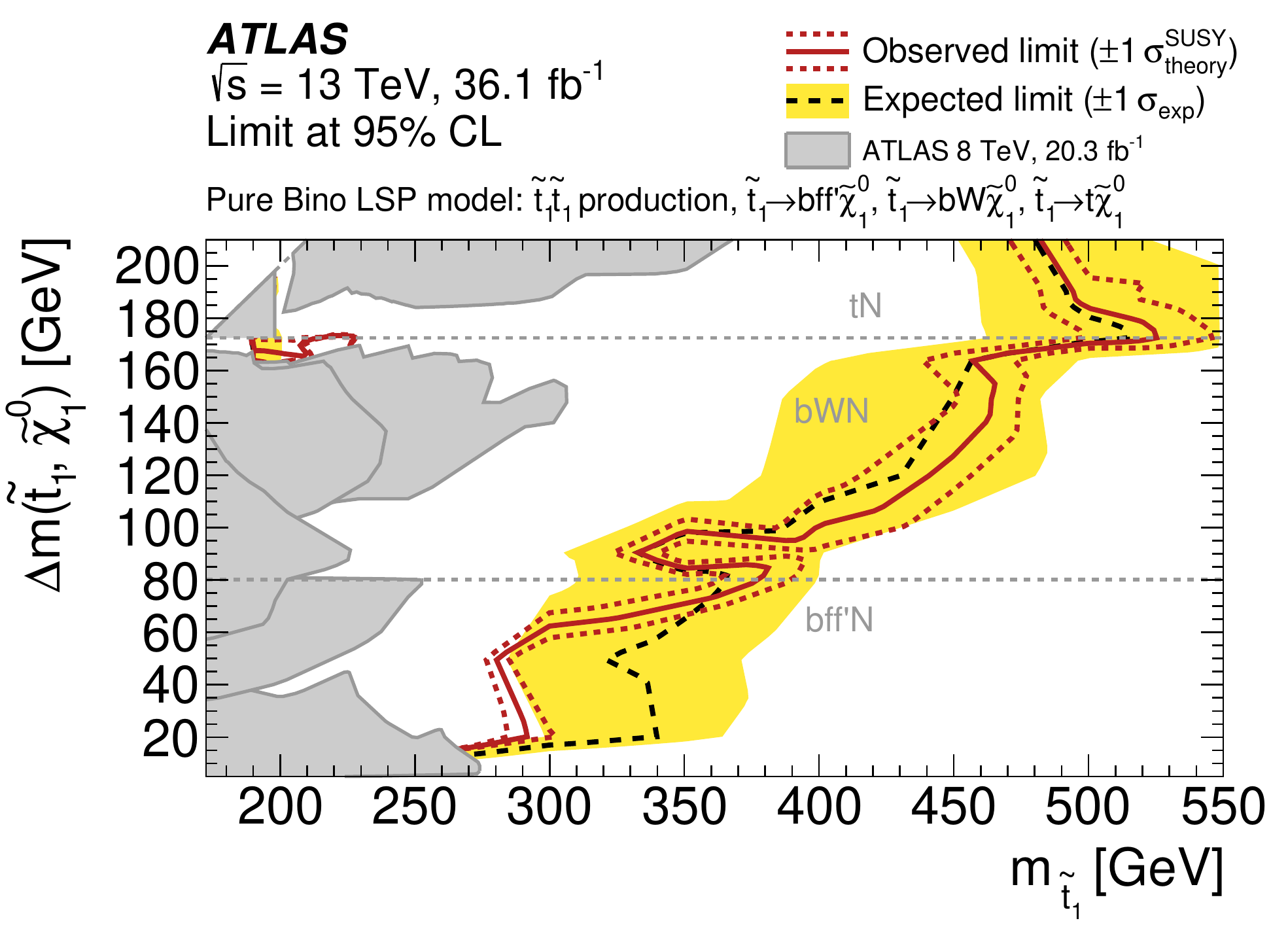}
  \caption{Expected (black dashed) and observed (red solid) 95\% excluded regions in the plane of $\Delta m(\tone,\ninoone)$ versus $m_{\tone}$ for direct stop pair production assuming either $\topLSP$, $\threeBody$, or $\fourBody$ decay with a branching ratio of 100\%. The excluded regions from previous publications~\cite{SUSY-2013-15,SUSY-2015-02} are shown with the grey shaded area.
  } 
  \label{fig:contour-tN-lowmass}
\end{figure}

Figure~\ref{fig:contour-tN-comb} shows the expected and observed exclusion contours as a function of stop and neutralino mass for the pure bino LSP scenario. The $\pm 1\,\sigma_{\textrm{exp}}$ uncertainty band indicates how much the expected limit is affected by the systematic and statistical uncertainties included in the fit. The $\pm 1\,\sigma_{\mathrm{theory}}^{\mathrm{SUSY}}$ uncertainty lines around the observed limit illustrate the change in the observed limit as the nominal signal cross-section is scaled up and down by the theoretical cross-section uncertainty. The exclusion limits are obtained under the hypothesis of mostly right-handed stops in the pure bino LSP scenario. 
Figure~\ref{fig:contour-tN-lowmass}  shows the expected and observed exclusion contours as a function of stop mass and the mass splitting $\Delta m(\tone,\ninoone)$, providing a greater level of detail for the transitions between the two-, three- and four-body decay regions. Stop masses above 195\,\gev\ are excluded for any value of the neutralino mass within the two-body decay region. The exclusion range extends to stop masses up to 480 \gev\ or higher depending on the neutralino mass.

The results improve upon previous exclusion limits by excluding the stop mass region up to 940\,\GeV\ for a massless lightest neutralino and assuming $\mathcal{B}(\topLSP) = 100\%$. In the three-body scenario, stop masses are excluded up to 500\,\GeV\ for a LSP mass of about 300\,\GeV. In the four-body scenario, stop masses are excluded up to 370\,\GeV\ for a mass-splitting between the stop and the LSP as low as 20\,\GeV. 

The non-excluded area between the three- and four-body decay regions is due to a reduction in search sensitivity as the kinematic properties of the signal change significantly when transitioning from a four-body to a three-body decay. In particular, approaching this boundary from the three-body side, the momenta of the two $b$-jets decrease to zero and hence the acceptance of the \pt\ requirement on the $b$-tagged jet in the \bWN\ signal region decreases rapidly. 

The kinematic properties change again at the kinematic boundary between the three-body and on-shell top-quark decay modes. When approaching this diagonal from the on-shell top-quark side, the search sensitivity usually worsens due to the difficulty in disentangling signal from the \ttbar\ background. However, the dedicated BDT analysis (here in particular \tNdiaghigh) recovers partly the sensitivity.

\begin{figure}[htbp]
  \centering
  \includegraphics[width=.99\textwidth]{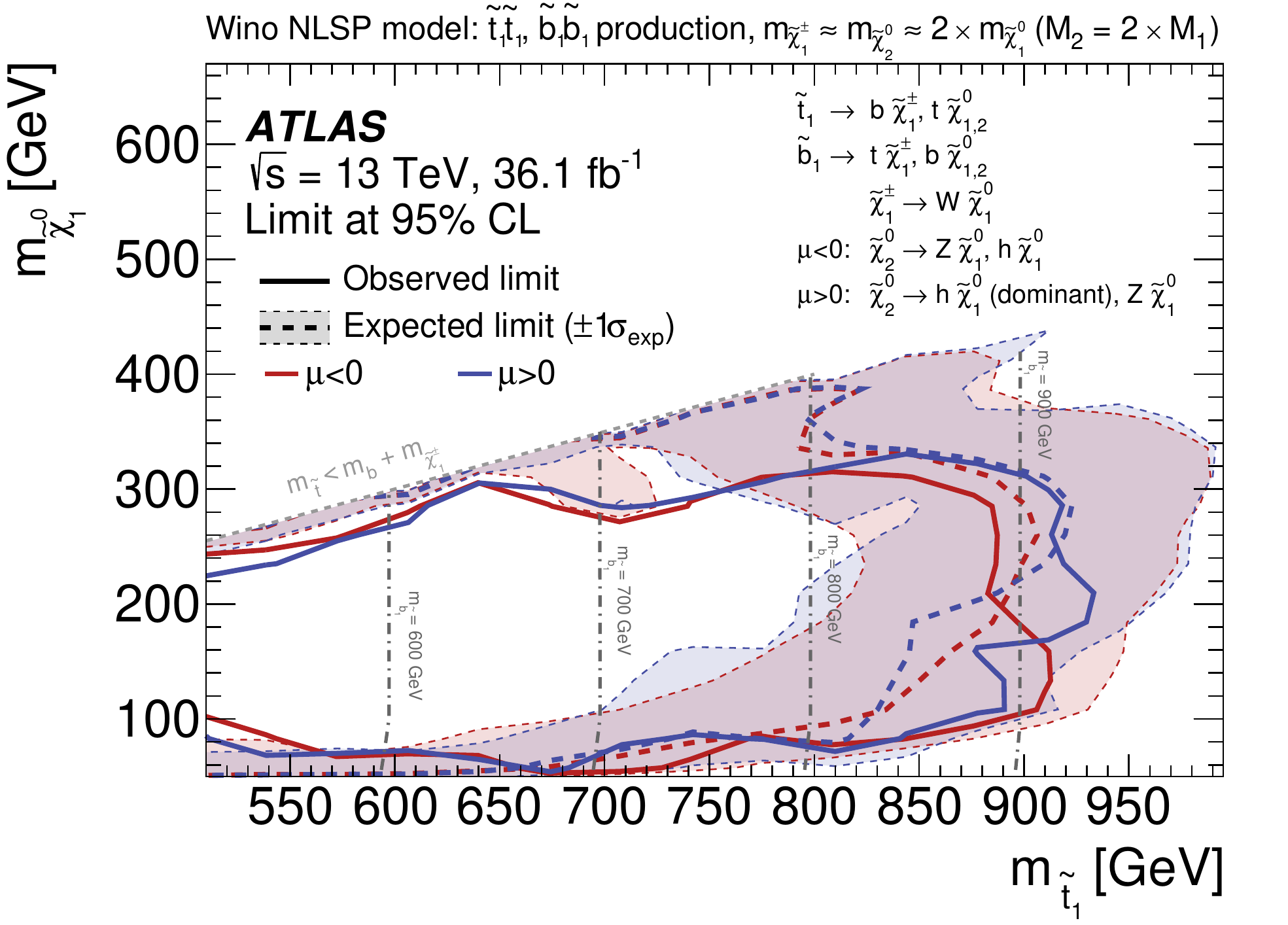}  
  \caption{Expected (dashed) and observed (solid) 95\% excluded regions in the plane of $m_{\ninoone}$ versus $m_{\tone}$ for direct stop/sbottom pair production in the wino NLSP model under the hypothesis of $m_{q3L}<m_{tR}$, where various decay modes ($\bChargino$, $\topLSP$, $\topNLSP$, $\tChargino$, $\bottomLSP$, and $\bottomNLSP$) are considered with different branching ratios for each signal point. The $\ninotwo$ decays into $\ninoone$ predominantly via either a $Z$ boson or a Higgs boson depending on the sign of the $\mu$ parameter. Contours for the $\mu>0$ and $\mu<0$ hypotheses are shown as blue and red lines, respectively. In this model, the $\chinoonepm$ and $\ninotwo$ masses are assumed to be nearly twice as large as the LSP ($\ninoone$) mass. The grey vertical dash-dotted lines show the corresponding sbottom mass. The dashed line $m_{\tone}=m_b+m_{\chinoonepm}$ is a physical boundary of the \bChargino\ decay.
}
  \label{fig:contour-winoNLSP}
\end{figure}

\begin{figure}[htbp]
  \centering
  \includegraphics[width=.99\textwidth]{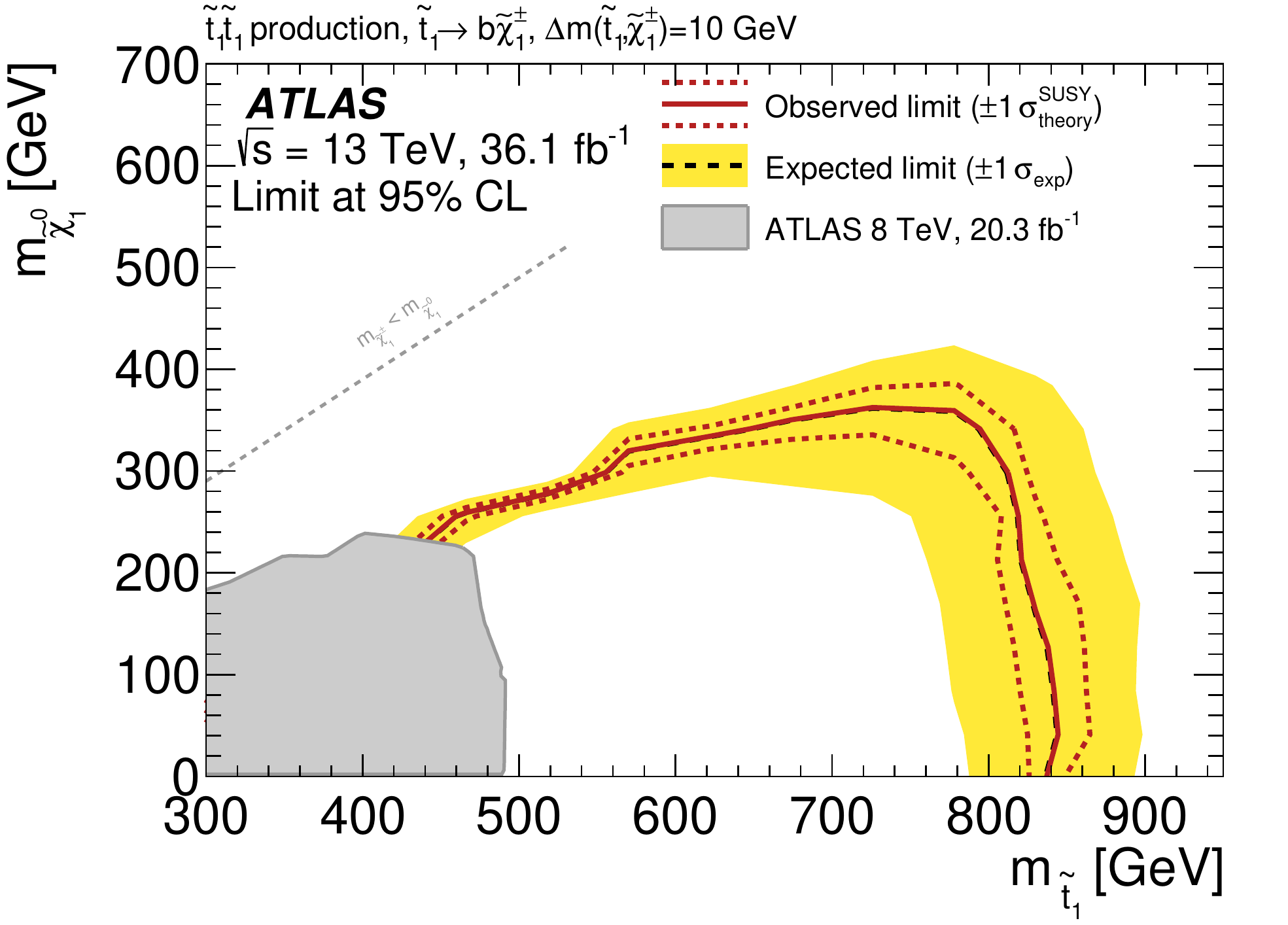}  
  \caption{Expected (black dashed) and observed (red solid) 95\% excluded regions in the plane of $m_{\ninoone}$ versus $m_{\tone}$ for direct stop pair production assuming the $b \chinoonepm$ decay with a branching ratio of 100\%. The chargino mass is assumed to be close to the stop mass, $m_{\chinoonepm} = m_{\tone} - 10$\,$\GeV$. 
}
  \label{fig:contour-bCbv}
\end{figure}

Limits are also set on the masses of the \tone\ and \ninoone\ in the wino NLSP scenario. Figure~\ref{fig:contour-winoNLSP} shows the exclusion contours based on the combination of all SRs targeting this scenario for positive and negative values of the $\mu$ parameter. 
The stop mass region up to 885\,\GeV\ (940\,\GeV) is excluded in scenarios with $\mu <0$ ($\mu > 0$) and a 200\,\GeV\ neutralino. 
Figure~\ref{fig:contour-bCbv} shows the exclusion limit for the simplified model \bChargino\ scenario with $m_{\tone}-m_{\chinoonepm}=10$\,$\GeV$. The stop mass region is excluded up to 840\,\GeV\ for a massless neutralino.
 
\begin{figure}[htbp]
  \centering
  \includegraphics[width=.99\textwidth]{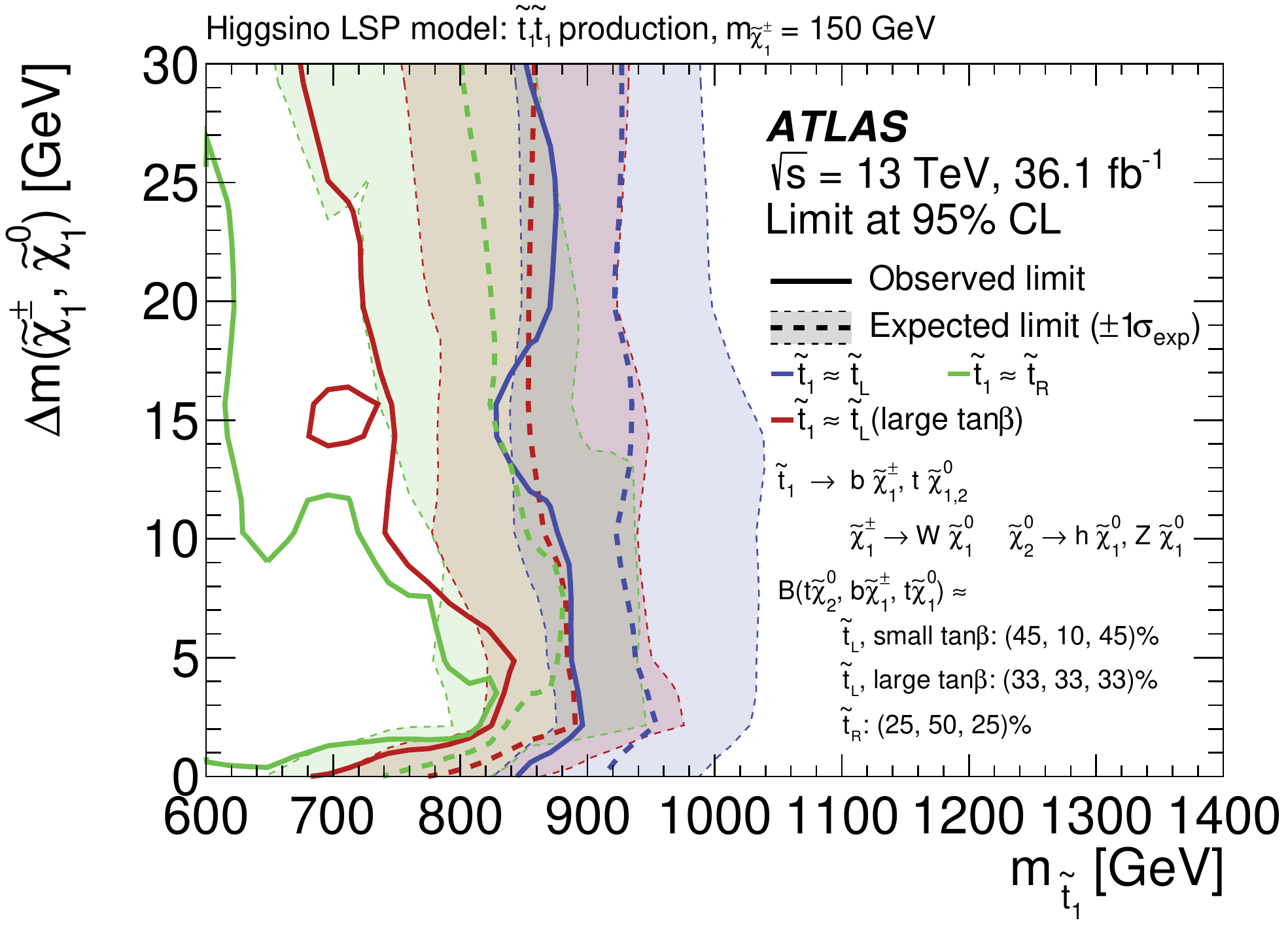}  
  \caption{Expected (dashed) and observed (solid) 95\% excluded regions in the plane of $\Delta m$ ($\chinoonepm$, $\ninoone$) versus  $m_{\tone}$  for direct stop pair production in the fixed $m_{\chinoonepm}=150$ \GeV\ higgsino LSP model where various decay modes ($\bChargino$, $\topLSP$, $\topNLSP$) are considered with different branching ratios, depending on the hypothesis being considered, and overlaid.
In this model, the mass relation of $\Delta m(\ninotwo,\ninoone)=2\times \Delta m(\chinoonepm,\ninoone)$ is assumed, varying $\Delta m(\chinoonepm,\ninoone)$ from 0\,\GeV\ to 30\,\GeV. For the region $\Delta m(\chinoonepm,\ninoone)<2$\,$\GeV$, only the $\topLSP$ decay is considered while the branching ratio is set to account for both the $\topLSP$ and $\topNLSP$ decays.
}
  \label{fig:contour-higgsinoLSP_dM}
\end{figure}

\begin{figure}[htbp]
  \centering
  \includegraphics[width=.99\textwidth]{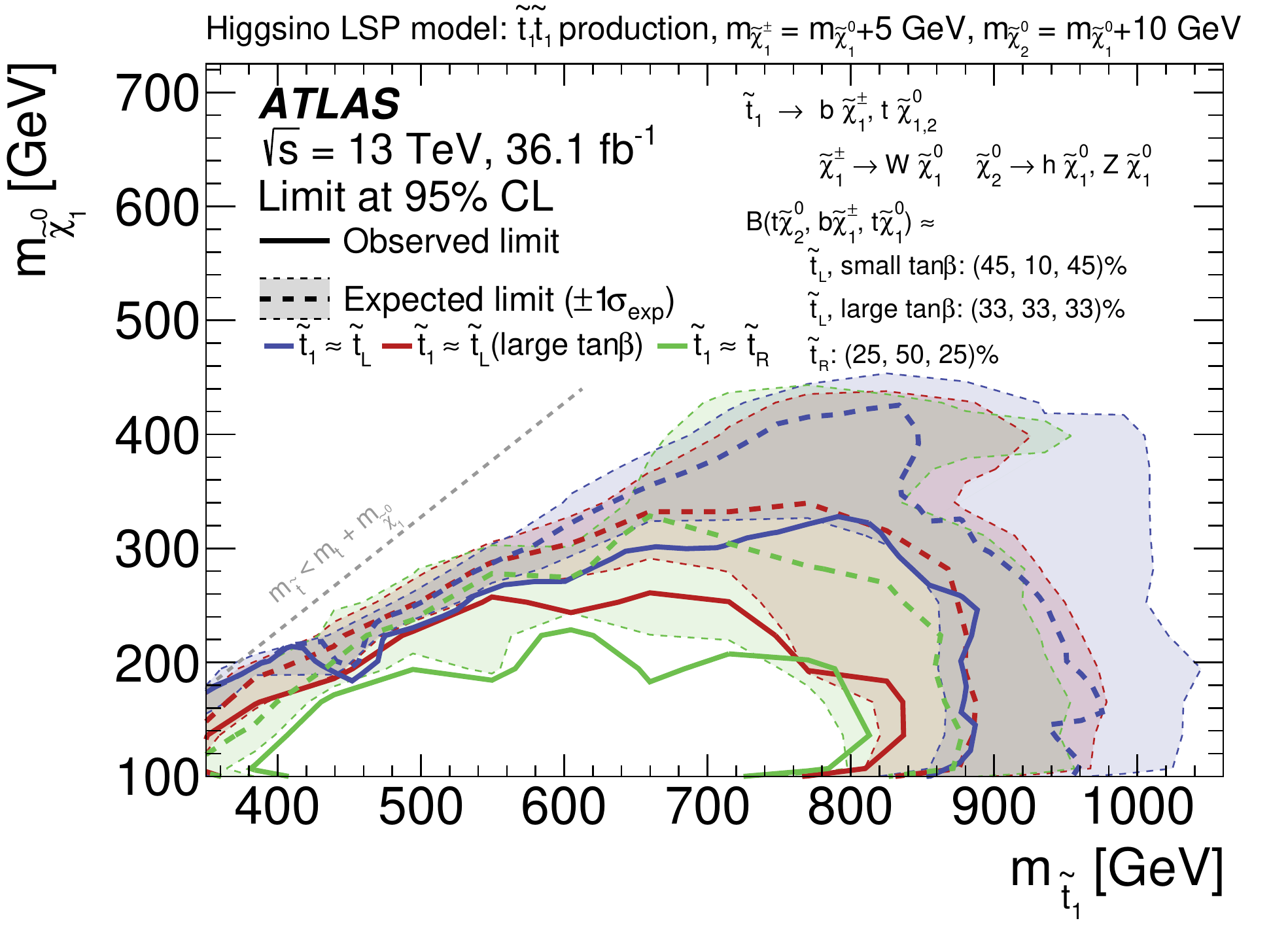}  
  \caption{Expected (dashed) and observed (solid) 95\% excluded regions in the plane of $m_{\ninoone}$ versus $m_{\tone}$ for direct stop pair production in the higgsino LSP model where various decay modes ($\bChargino$, $\topLSP$, $\topNLSP$) are considered with different branching ratios depending on the hypothesis being considered.
In this model, $\Delta m(\chinoonepm,\ninoone) =5$\,\GeV\ and $\Delta m(\ninotwo,\ninoone)=10$\,\GeV\ are assumed. 
  }
  \label{fig:contour-higgsinoLSP}
\end{figure}

Assuming the higgsino LSP scenario, limits are also set on the masses of the \tone\ and \ninoone.  Figures~\ref{fig:contour-higgsinoLSP_dM} and~\ref{fig:contour-higgsinoLSP} show the exclusion contours   for the three signal scenarios, $m_{tR}<m_{q3L}$, $m_{q3L}<m_{tR}$, and $m_{q3L}<m_{tR}$ with large $\tan\beta$, as described in Section~\ref{sec:simulation}. The results are based on the combination of  two orthogonal hard- and soft-lepton SRs. The stop decay branching ratios to $t \ninoone$, $b \chinoonepm$ and $t \ninotwo$ vary in these three scenarios. In the scenario with $m_{tR}<m_{q3L}$, the sensitivity is mostly driven by the $\bCsoftmed$ and $\bCsofthigh$ SRs, as the branching ratio of the $\bChargino$ decay (with soft leptons) is large, whereas the sensitivity is driven by the $\tNmed$ SR for the scenario with $m_{q3L}<m_{tR}$, as the branching ratios of the $\topLSP$ and $\topNLSP$ decays (with high-$\pt$ leptons from the leptonically decaying top quark) are dominant. The third scenario,  $m_{q3L}<m_{tR}$ with large $\tan\beta$, benefits from both the soft- and hard-lepton SRs, with equal branching ratios to all three decay modes.

\begin{figure}[htbp]
  \centering
  \includegraphics[width=.99\textwidth]{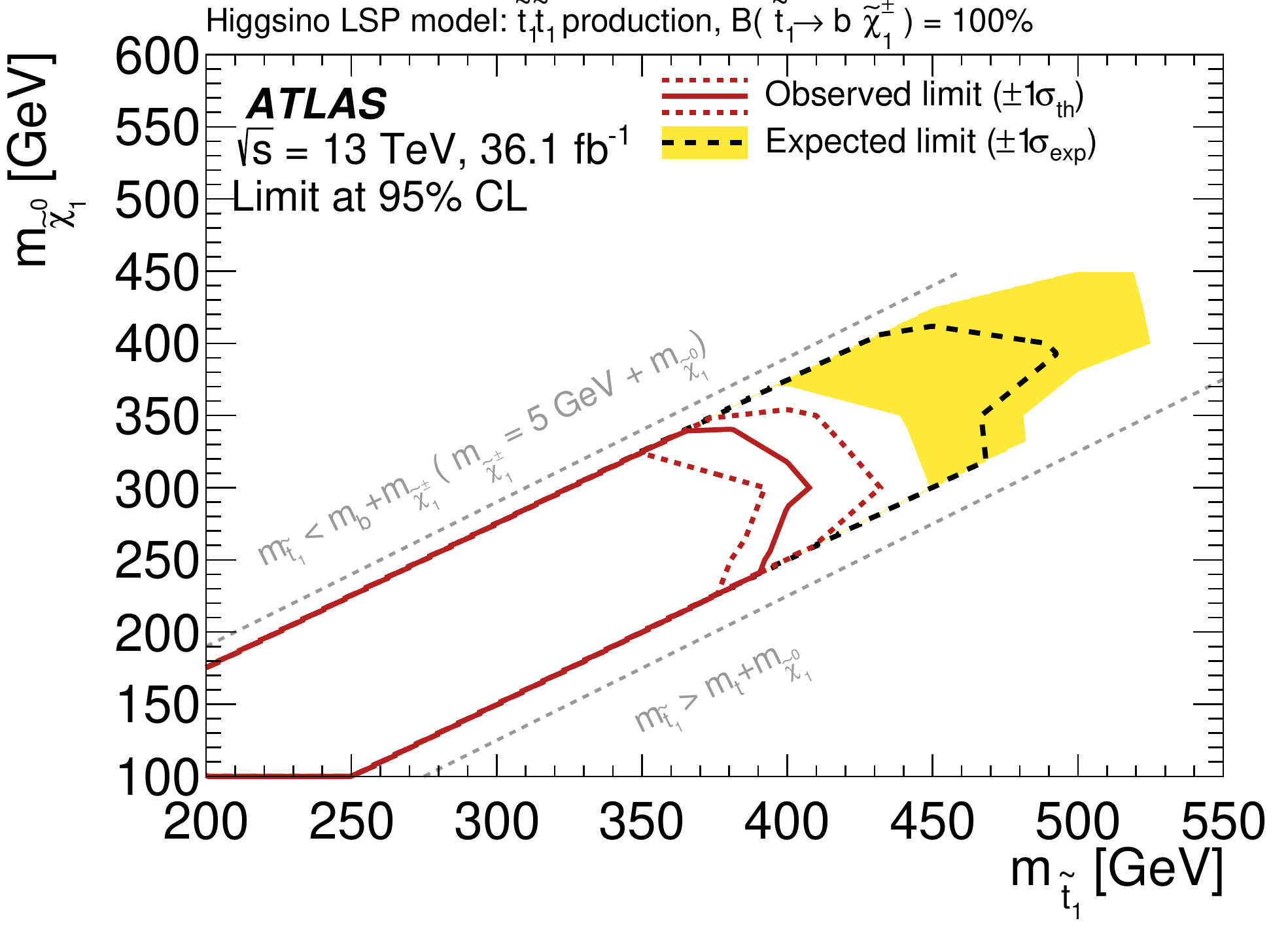}  
  \caption{Expected (black dashed) and observed (red solid) 95\% excluded regions in the plane of $m_{\ninoone}$ versus $m_{\tone}$ for direct stop pair production in the higgsino LSP model where only the $\bChargino$ decay mode is kinematically allowed due to the phase space constraint. In this model, $\Delta m(\chinoonepm,\ninoone)=5$\,\GeV\ is assumed.
  }
  \label{fig:contour-higgsinoLSP_diag}
\end{figure}

Figure~\ref{fig:contour-higgsinoLSP_diag} shows the region $m_b+m_{\chinoonepm} < m_{\tone} < m_\mathrm{top}+m_{\ninoone}$. Since the mass-splitting $\Delta m(\tone, \ninoone)$ is smaller than the top  mass a 100\% branching ratio to \bChargino\ is assumed, and the exclusion limit is set by a single soft-lepton SR, $\bCsoftdiag$. In the gaps between the exclusion contour and diagonal dashed lines indicating the kinematic boundaries ($m_{\tone}=m_b+m_{\chinoonepm}$ and $m_{\tone}=m_\mathrm{top}+m_{\ninoone}$), the assumption of a 100\% branching ratio may not be accurate due to phase-space effects, hence these gap regions are not considered in the interpretation.

In Figures~\ref{fig:contour-higgsinoLSP} and \ref{fig:contour-higgsinoLSP_diag}, $\Delta m(\chinoonepm,\ninoone)$ is fixed to 5\,\GeV\ and $\Delta m(\ninotwo,\ninoone)$ is fixed to 10\,\GeV. In Figure~\ref{fig:contour-higgsinoLSP_dM}, the mass relations $\Delta m(\ninotwo,\ninoone)=2\times \Delta m(\chinoonepm,\ninoone)$ and $m_{\chinoonepm}=150$\,\GeV\ are assumed, while $\Delta m(\chinoonepm,\ninoone)$ is varied in the range 0--30\,\GeV. For the region $\Delta m(\chinoonepm,\ninoone)<2$\,$\GeV$, only the $\topLSP$ process is simulated, with the branching ratio set to account for both the $\topLSP$ and $\topNLSP$ decays. In Figure~\ref{fig:contour-higgsinoLSP}, the stop mass region up to 890\,\GeV\ (800\,\GeV) is excluded in scenarios with $m_{q3L}<m_{R}$ ($m_{R}<m_{q3L}$).

\begin{figure}[htbp]
  \centering
  \includegraphics[width=.40\textwidth]{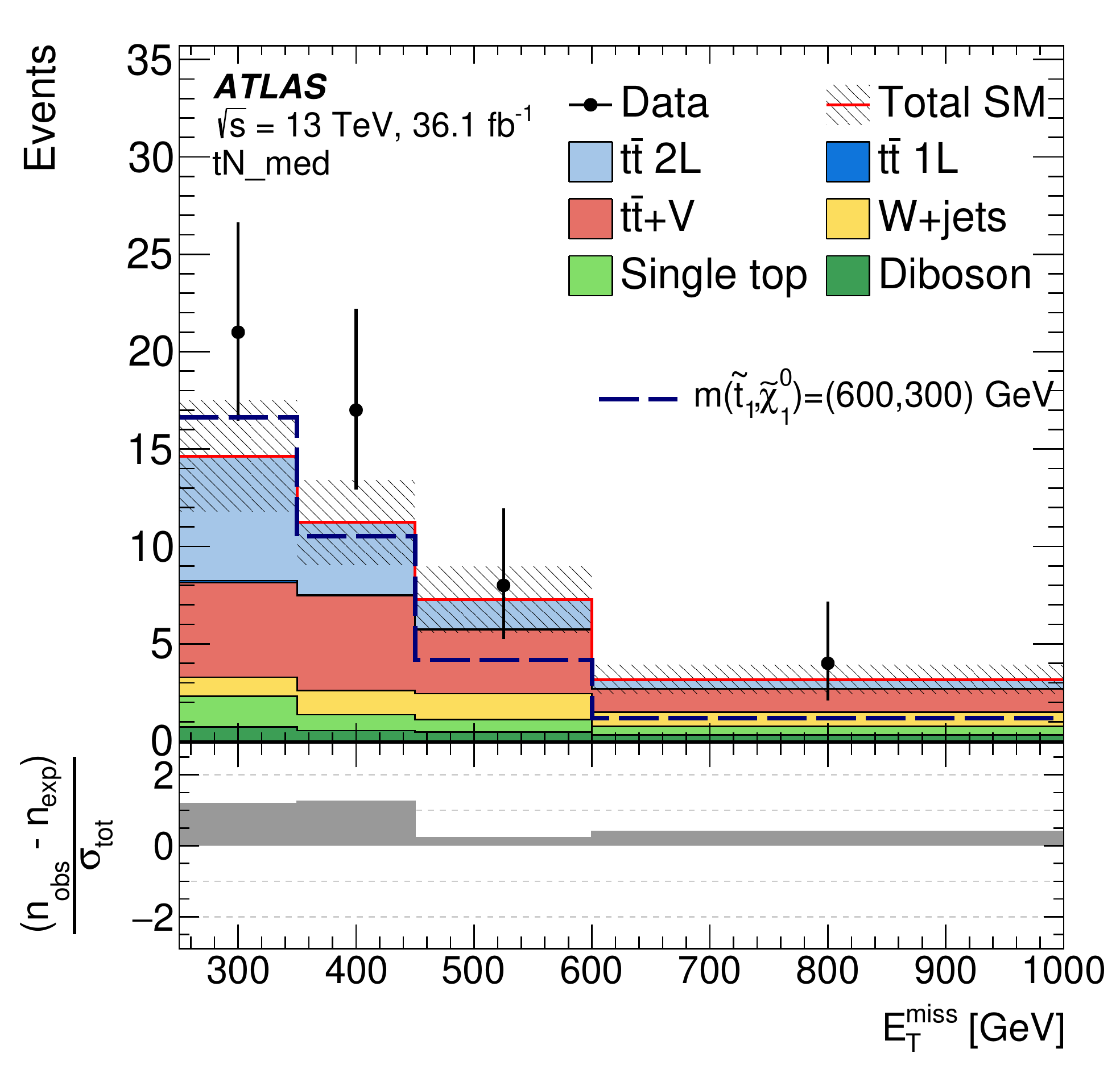}
  \includegraphics[width=.40\textwidth]{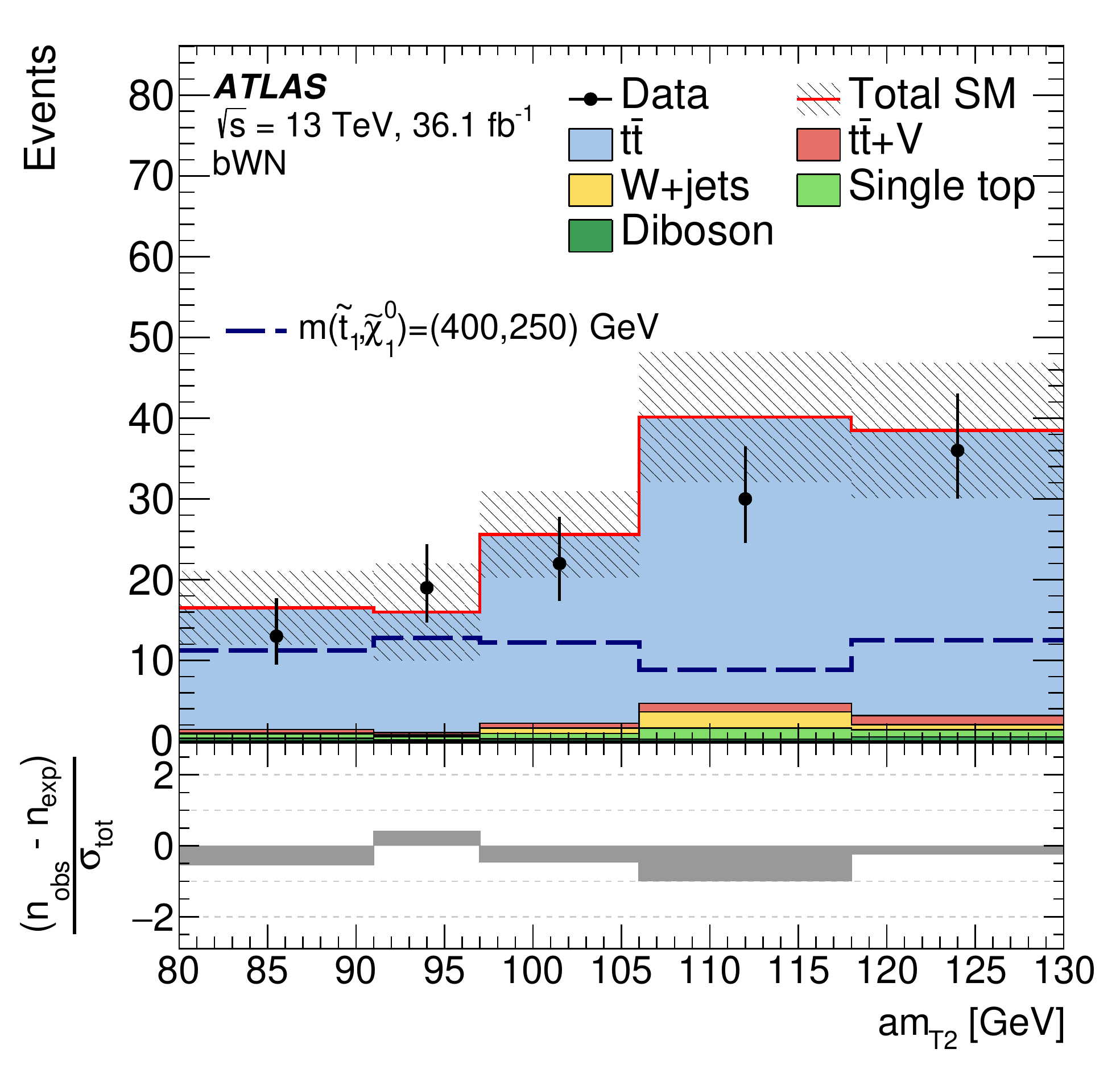}
  \includegraphics[width=.40\textwidth]{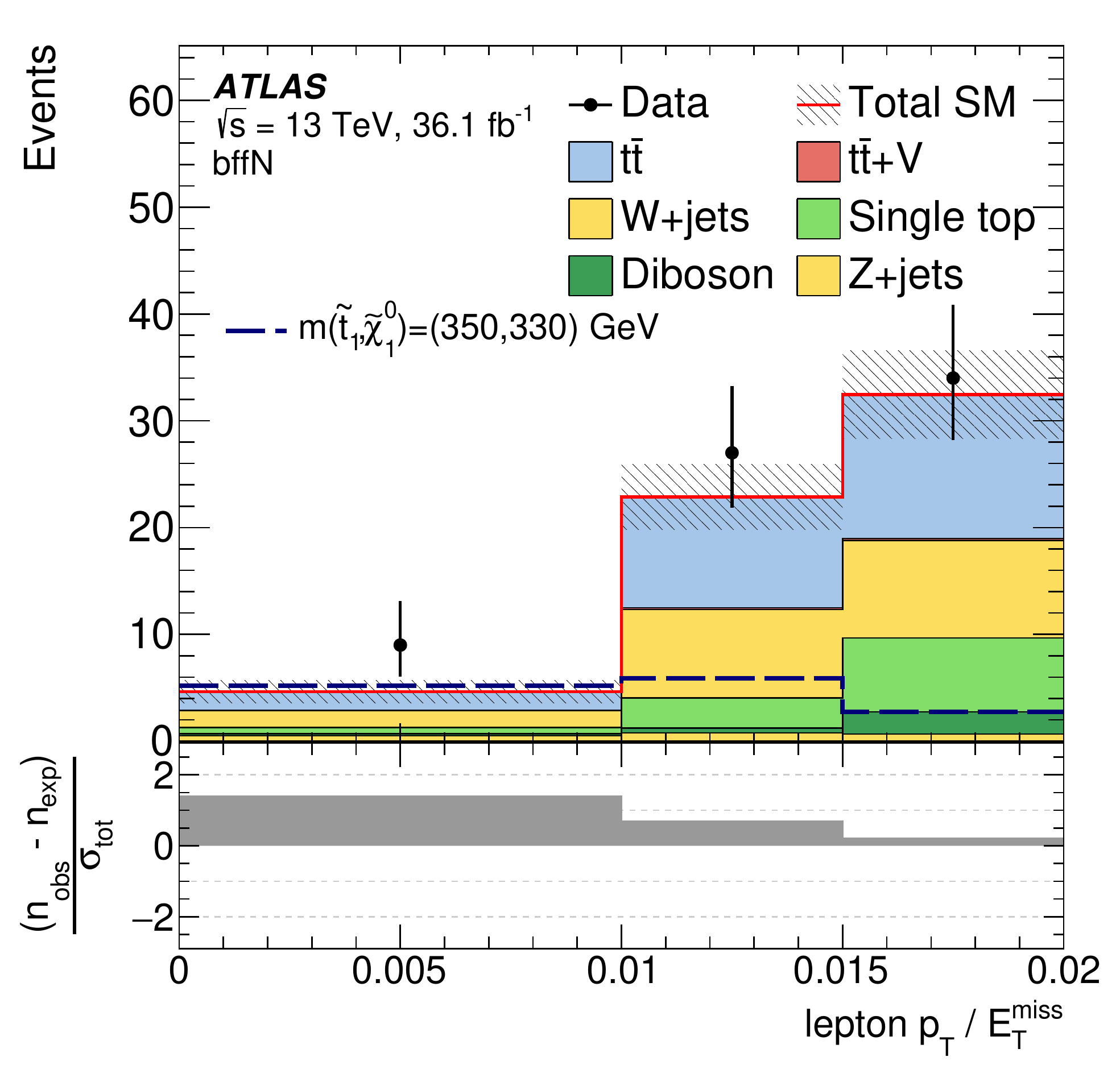}
  \includegraphics[width=.40\textwidth]{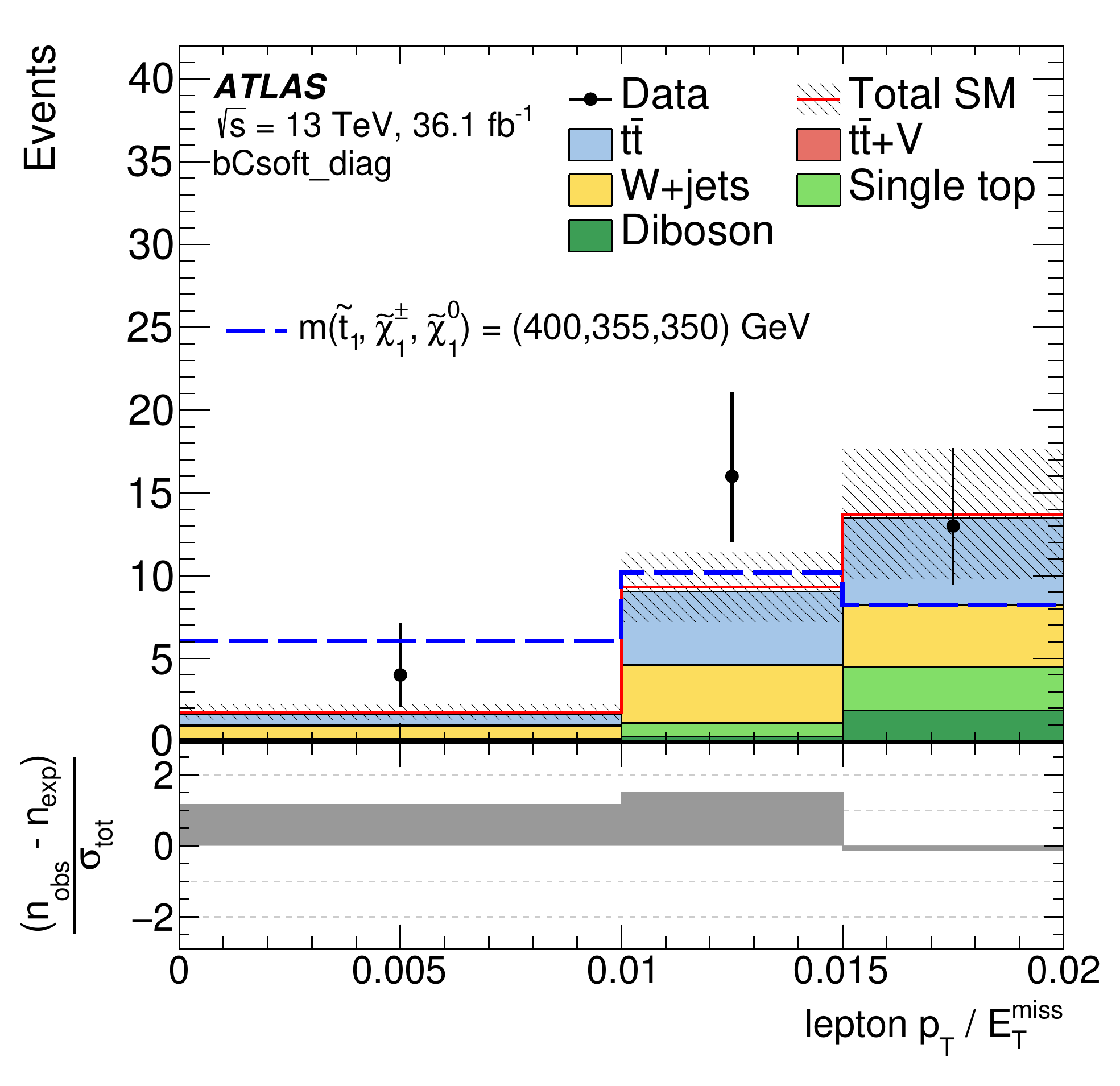}
  \includegraphics[width=.40\textwidth]{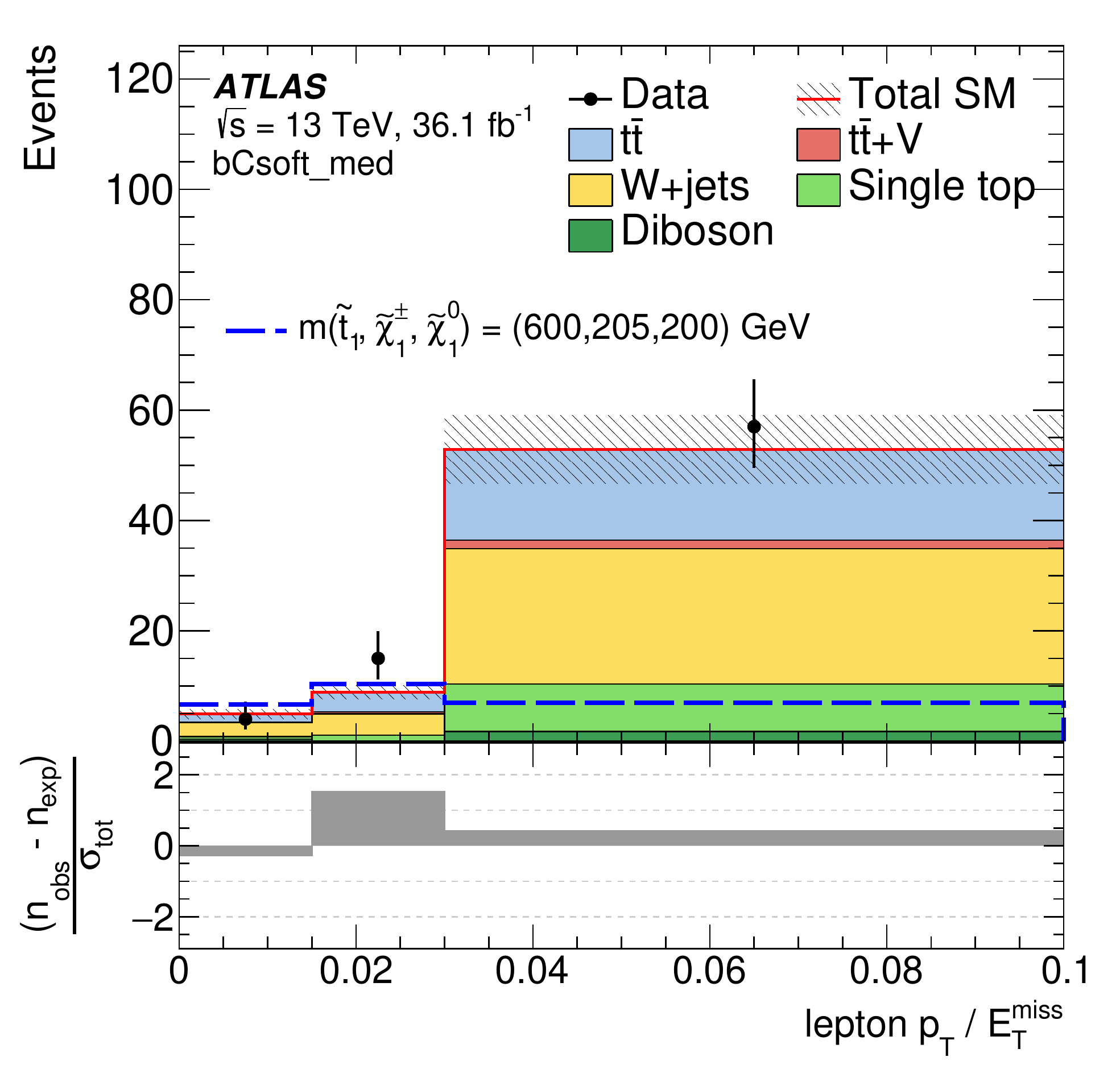}
  \includegraphics[width=.40\textwidth]{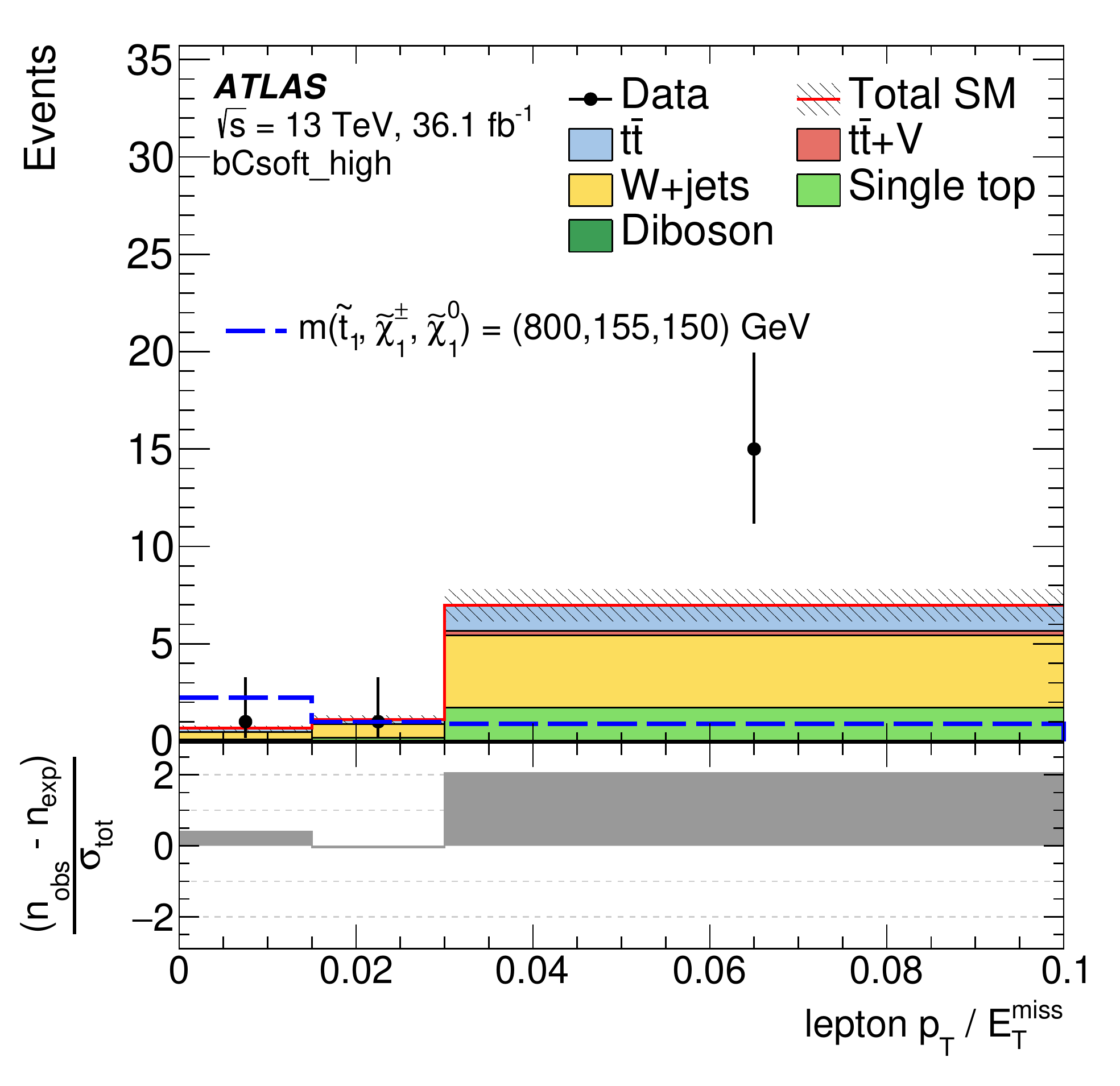}
  \caption{Kinematic distributions for the shape-fit analyses: (top left) \met\ in $\tNmed$, (top right) \amtTwo\ in $\bWN$, (middle left) $\pt^{\ell}$/$\met$ in $\bffN$, (middle right) $\pt^{\ell}$/$\met$ in $\bCsoftdiag$, (bottom left) $\pt^{\ell}$/$\met$ in $\bCsoftmed$, and (bottom right) $\pt^{\ell}$/$\met$ in $\bCsofthigh$. The full event selection in the corresponding signal region is applied, except for the requirement that is imposed on the variable being plotted.
The predicted SM backgrounds are scaled with the normalisation factors obtained from the corresponding control regions in Tables~\ref{tab:discovery_SR_yields_tN} and \ref{tab:discovery_SR_yields_bC}. The hatched area around the total SM prediction includes statistical and experimental uncertainties. The last bin contains overflows. Benchmark signal models are overlaid for comparison. The bottom panels show the difference between data and the predicted SM background divided by the total uncertainty ($\sigma_\text{tot}$). 
  }
  \label{fig:srs-shapefits}
\end{figure}

\begin{figure}[htbp]
  \centering
  \includegraphics[width=.99\textwidth]{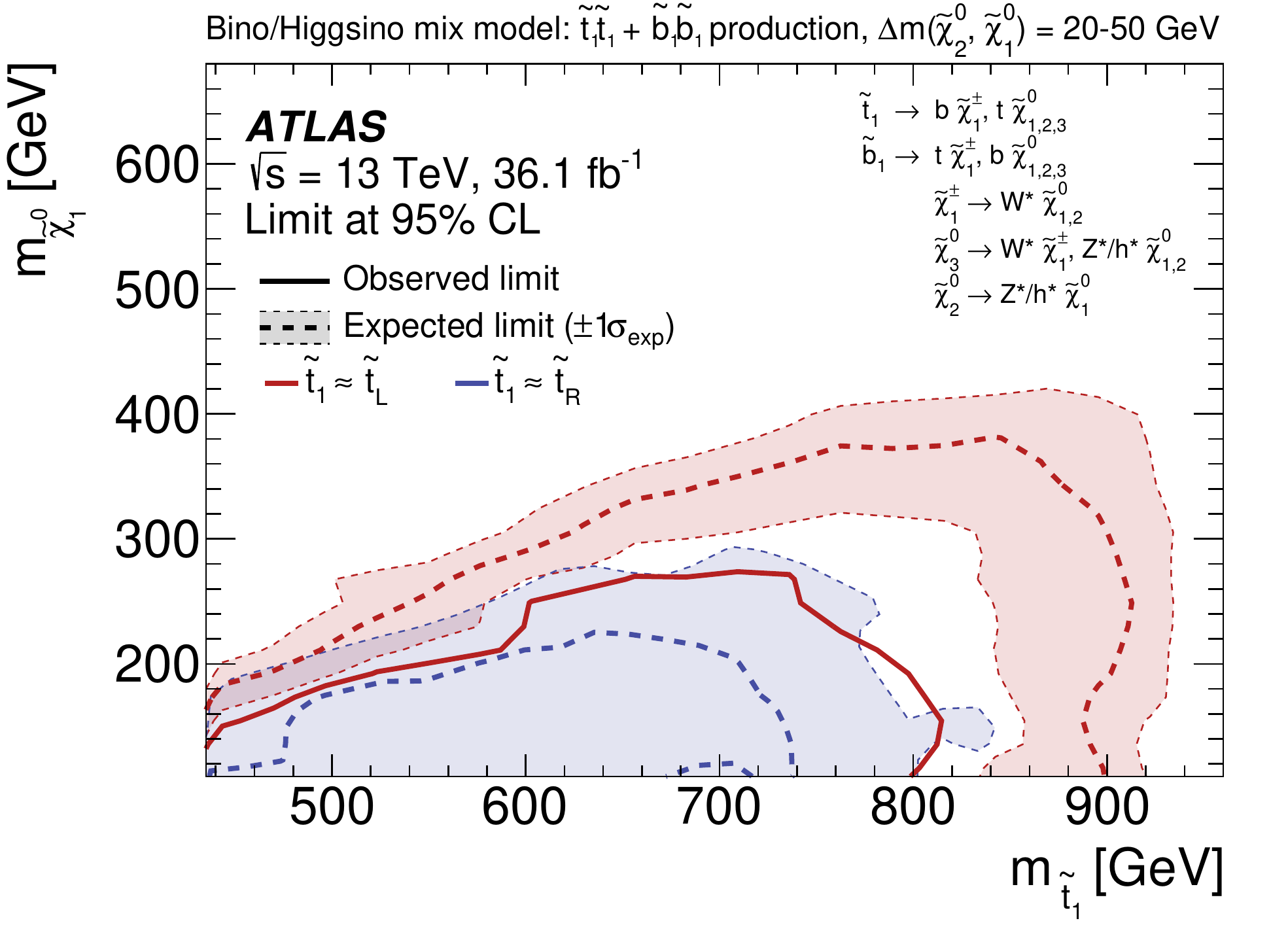}
  \caption{Expected (dashed) and observed (solid) 95\% excluded regions in the plane of $m_{\ninoone}$ versus $m_{\tone}$ for direct stop/sbottom pair production in the well-tempered neutralino model where various decay modes ($\bChargino$, $\topLSP$, $\topNLSP$, $\tChargino$, $\bottomLSP$, and $\bottomNLSP$) are considered with different branching ratios for each signal point. Contours for the $m_{q3L}<$$m_{tR}$ and $m_{q3L}>$$m_{tR}$ hypotheses are shown separately as red and blue lines, respectively. For the $m_{q3L}<$$m_{tR}$ hypothesis, both stop and sbottom pair production is considered while for the $m_{q3L}>$$m_{tR}$ hypothesis, only stop pair production is considered.
}
  \label{fig:contour-WellTempered}
\end{figure}

Limits are also set on the masses of the \tone\ and \ninoone\ in the well-tempered neutralino scenario as shown in Figure~\ref{fig:contour-WellTempered}. In the scenario with $m_{q3L}<m_{tR}$, the expected sensitivity is better than in the scenario with $m_{tR}<m_{q3L}$ as sbottom pair production can also contribute to the former, roughly doubling the signal acceptance. No observed limit is set in the $m_{tR}<m_{q3L}$ scenario, as a mild excess of data events is seen above the predicted SM background yield in the \bCsofthigh\ SR (shape-fit, as shown in Figure~\ref{fig:srs-shapefits}), which is the most sensitive SR in this scenario. On the other hand, the stop mass region up to 810\,\GeV\ is excluded in scenarios with $m_{q3L}<m_{tR}$. 

\begin{figure}[htbp]
  \centering
    \includegraphics[width=.49\textwidth]{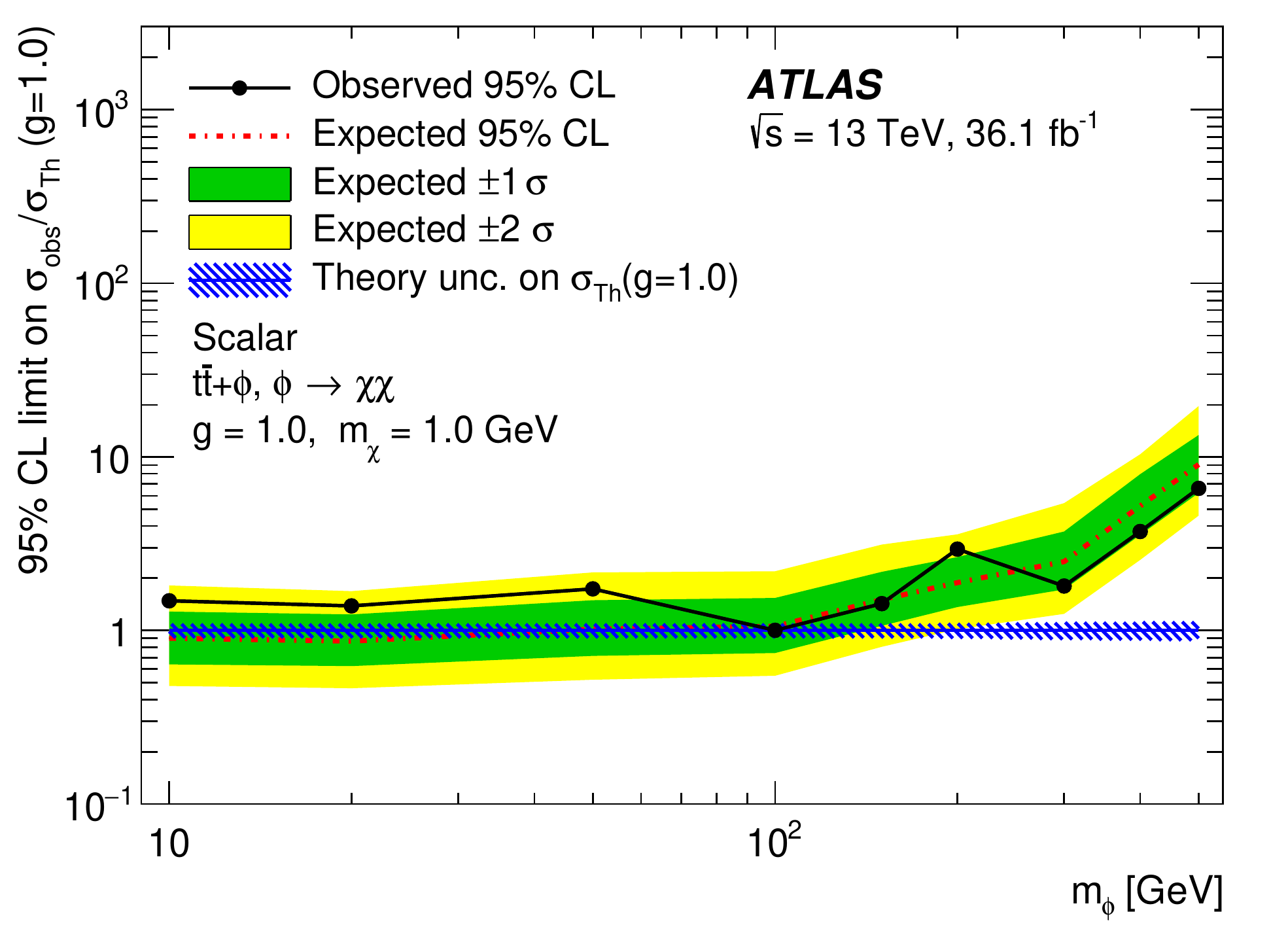}
    \includegraphics[width=.49\textwidth]{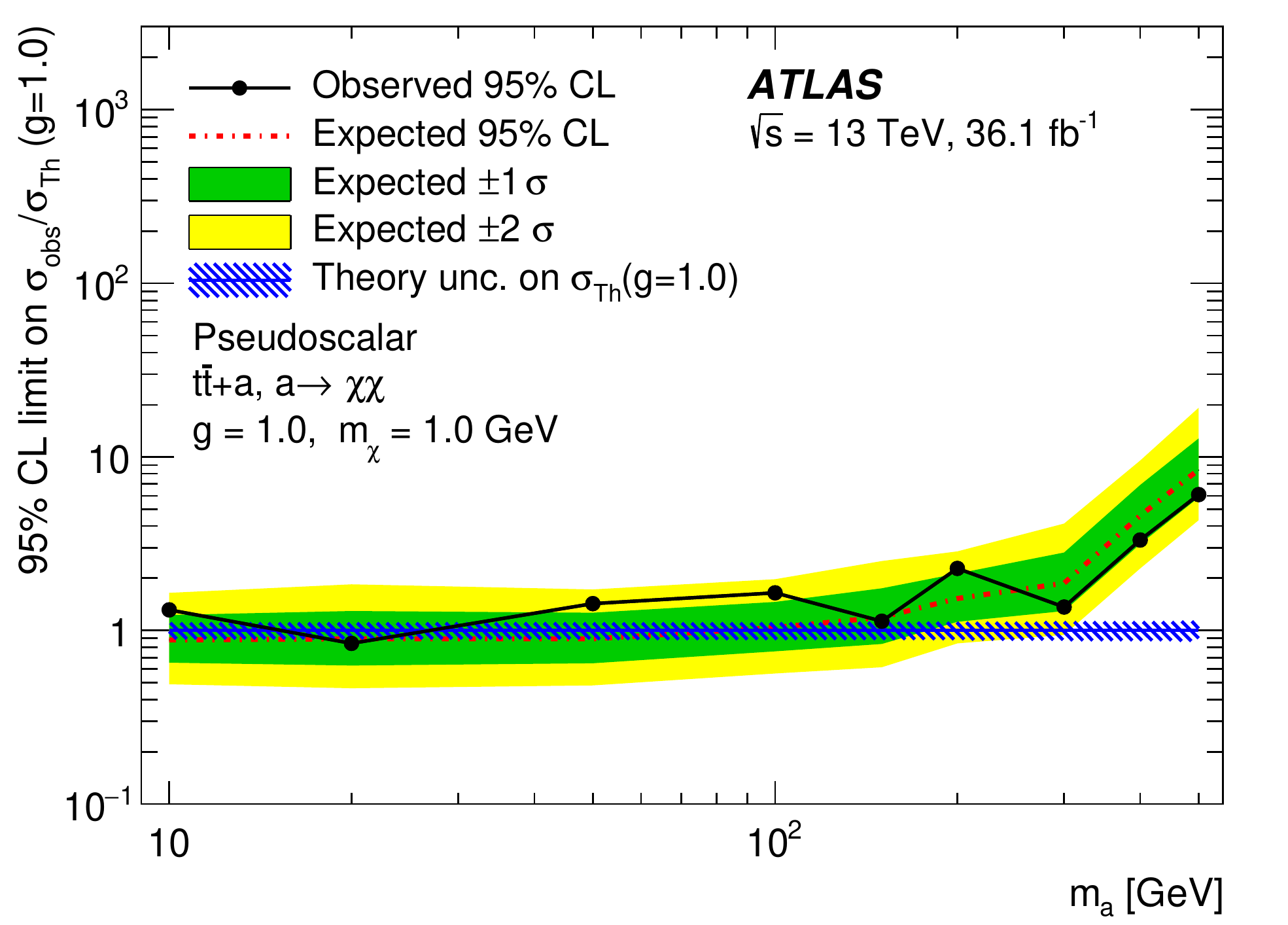}
    \includegraphics[width=.49\textwidth]{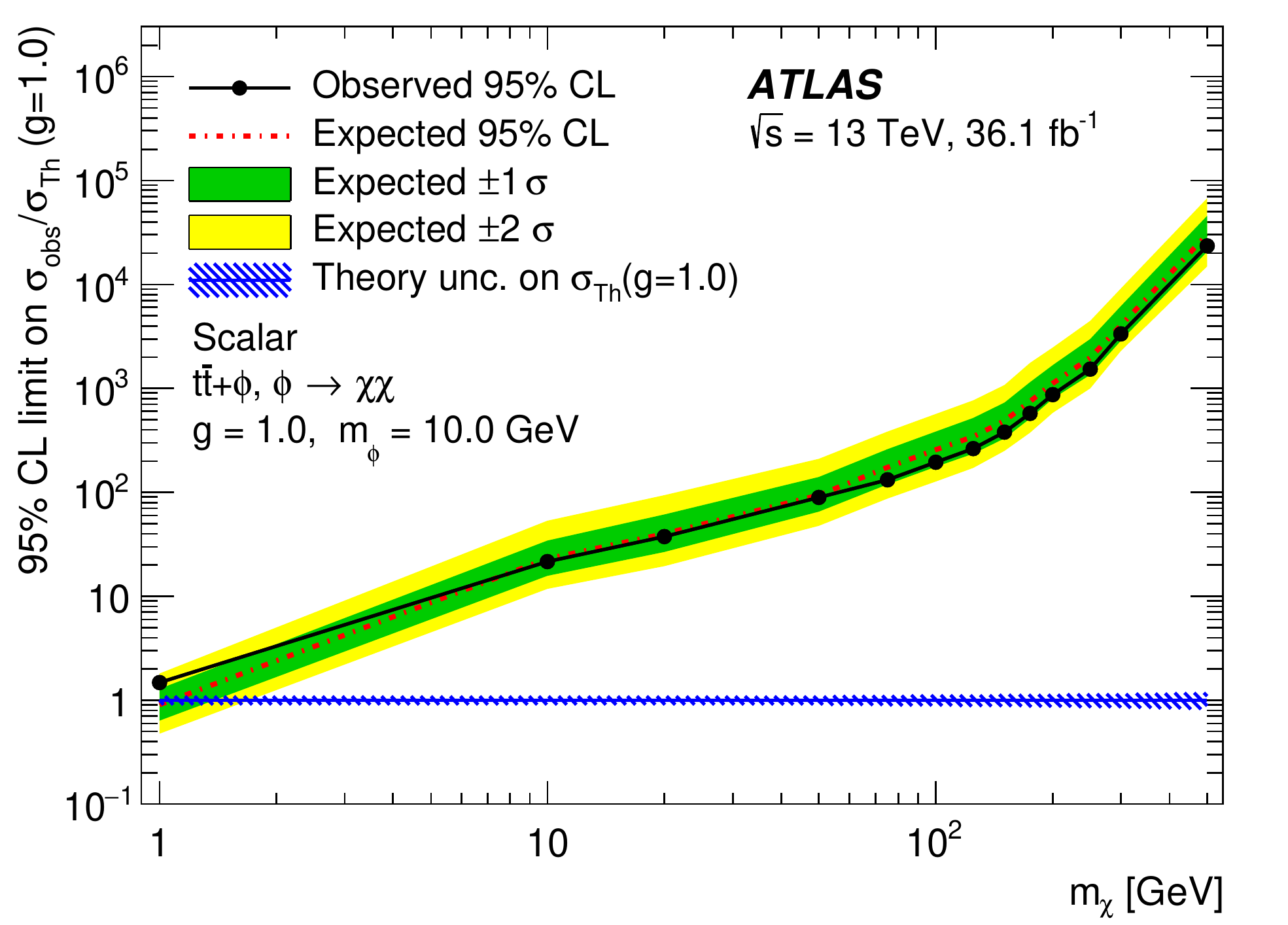}
    \includegraphics[width=.49\textwidth]{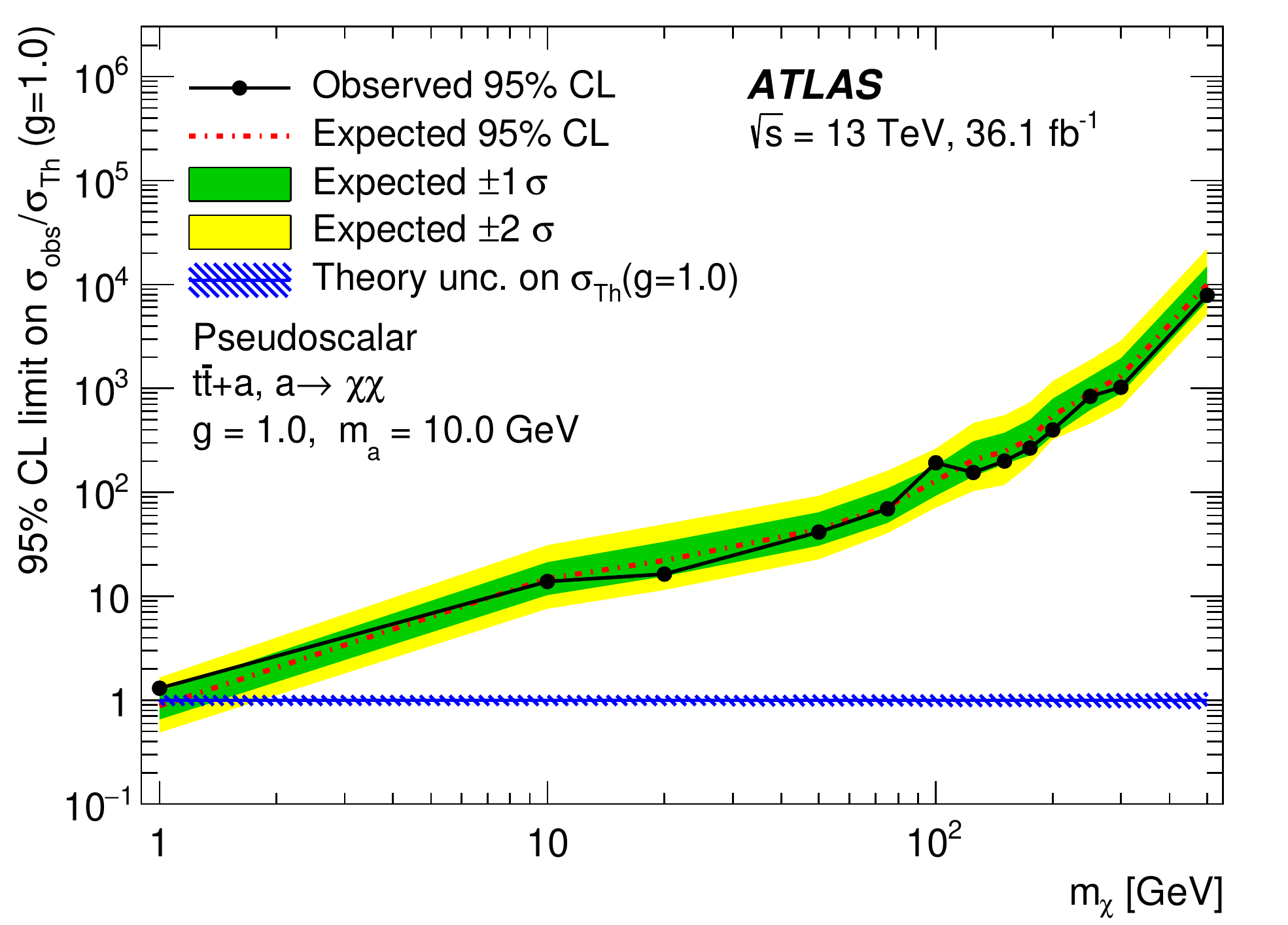}
  \caption{Upper limit on the ratio of the DM production cross-section to the simplified
  model expectation under the hypothesis of (left) a scalar or (right) a pseudoscalar mediator. 
  The limit is shown as a function of: (top) the mediator mass for a fixed mass of the DM candidate of 1\,$\GeV$, 
  or (bottom) the DM candidate mass for a fixed mediator mass of 10\,$\GeV$. 
  The coupling of the mediator to SM and DM particles is assumed to be $g=1$.
  }
  \label{fig:limit-DM}
\end{figure}

Figure~\ref{fig:limit-DM} shows the upper limit on the ratio of the DM+\ttbar\ production cross-section to the theoretical cross-section. Limits are shown under the hypothesis of a scalar or pseudoscalar mediator, and for a fixed DM candidate mass or for a fixed mediator mass. A scalar (pseudoscalar) mediator mass of around 100\,\GeV\ (20\,\GeV) is excluded at 95\% CL, assuming a 1 \GeV\ dark-matter particle mass and a common coupling of $g=1$ to SM and dark-matter particles.

\begin{samepage} 
\input{texfiles/res_discovery_tN.tex}

\end{samepage}
\begin{samepage}
\input{texfiles/res_discovery_bC.tex}

\end{samepage}
\begin{samepage}
\input{texfiles/res_exclusion.tex}

\end{samepage}

%% file: texfiles/res_discovery_tN.tex
\begin{landscape}
\begin{table}[hptb]
\begin{center}
\caption{The numbers of observed events in the pure bino LSP SRs together with the expected numbers of background events and their uncertainties as predicted by the background-only fits, the normalisation factors (NF) for the background predictions obtained in the fit, the probabilities (represented by $p_0$ values, and capped at 0.5) that the observed numbers of  events are compatible with the background-only hypothesis, and the expected ($N^{\textrm{limit}}_{\textrm{non-SM}}$ exp.) and observed ($N^{\textrm{limit}}_{\textrm{non-SM}}$ obs.) 95\% CL upper limits on the number of beyond-SM events. Some of the SRs where \ttbar\ background events are predominantly semileptonic or dileptonic have only one NF which is also applied to the subdominant \ttbar\ contribution.
Backgrounds with no associated NF are normalised with the theoretical cross-sections.
\label{tab:discovery_SR_yields_tN} 
}
\vspace{3mm}
{\small
\setlength{\tabcolsep}{0.2pc}
{\renewcommand{\arraystretch}{1.3}
\begin{tabular}{|l|c|c|c|c|c|c|c|}
\hline
\hline
Signal region     & \tNhigh  & \tNmed  & \tNdiaghigh & \tNdiagmed & \tNdiaglow & \bWN & \bffN \\
\hline
Observed          &  $8$    &   $50$  & $19$ & $115$ & $34$ & $68$ & $70$ 
\\
\hline
Total background  & $3.8\pm1.0$ & $36.3\pm6.6\phantom{0}$ & $18.3\pm2.2\phantom{0}$ & 
$115\pm31\phantom{0}$ & $30.3\pm5.9\phantom{0}$ & $71\pm16$ & $60.5\pm6.1\phantom{0}$ \\ \hline

$t \bar t\ 2\ell$ & $0.51\pm0.18$ & $12.1\pm2.9\phantom{0}$ & $15.2\pm2.4\phantom{0}$ & 
$65.1\pm9.4\phantom{0}$ & $8.5\pm2.3$ & $65\pm16$ & - \\

$t \bar t\ 1\ell$ & $0.020\pm0.001$ & $0.19\pm0.05$ & - & $35.0\pm8.9\phantom{0}$ & $17.5\pm4.1\phantom{0}$  & - & $25.5\pm5.5\phantom{0}$ \\

$\ttbar + V$   & $1.86\pm0.90$ & $14.2\pm5.5\phantom{0}$ & $0.68\pm0.37$ & 
$2.5\pm1.6$ & $0.34\pm0.20$ & $1.7\pm1.7$ & $0.35\pm0.06$ \\

Single top     & $0.13\pm0.10$ & $3.5\pm1.2$ & $1.5\pm1.2$ 
& $8.1\pm1.1$ & $2.3\pm1.2$ & $1.9_{-1.9}^{+2.0}$ & $10.3\pm4.4\phantom{0}$ \\
 
$W$+jets       & $0.88\pm0.24$ & $4.3\pm1.1$  & $0.70\pm0.56$ & 
$3.8\pm1.9$ & $1.7_{-1.7}^{+2.0}$ & $1.41\pm0.88$ & $19.6\pm4.9\phantom{0}$ \\

Diboson        & $0.42\pm0.16$ & $2.08\pm0.70$ & $0.21\pm0.11$ & 
$0.69_{-0.69}^{+0.73}$ & $0.07_{-0.07}^{+0.24}$ & $0.89\pm0.28$ & $2.72 
\pm0.99$ \\

$Z$+jets       & - & - & - & - & - & - & $1.9\pm1.8$  \\ 
\hline
 $ t \bar t\ 2\ell$ NF
 & $1.01\pm0.15$
 & $0.96\pm0.13$
 & $1.05\pm0.06$
 & $1.16\pm0.16$
 & -
 & $1.04\pm0.07$
 & - 
\\
 
$ t \bar t\ 1\ell$ NF
 & $0.97\pm0.08$ 
 & $1.05\pm0.09$ 
 & -
 & $1.16\pm0.28$
 & $0.85\pm0.10$
 & -
 & $0.73\pm0.11$
\\
 
$\ttbar + V$ NF 
 & $1.11\pm0.35$
 & $1.13\pm0.32$
 & -
 & - 
 & -
 & -
 & -
\\
 
Single top NF  
 & $0.64\pm0.37$
 & $1.19\pm0.37$
 & -
 & -
 & -
 & - 
 & -
\\

$W$+jets NF
 & $0.82\pm0.17$
 & $0.85\pm0.18$
 & -
 & -
 & -
 & - 
 & $1.19\pm0.26$
\\

\hline
$p_0$ (\,$\sigma$)& 0.05 (1.6) & 0.07 (1.4) & 0.44 (0.14) & 0.5 (0) & 0.33 (0.46) & 0.5 (0) & 0.17 (0.95) \\
$N^{\textrm{limit}}_{\textrm{non-SM}}$ exp.& 5.8 & 19  & 11 & 58 & 19 & 33 & 21 \\
$N^{\textrm{limit}}_{\textrm{non-SM}}$ obs. & 10 & 31 & 11 & 58 & 17 & 31 & 28 \\
\hline
\hline
\end{tabular}
}
}
\end{center}
\end{table}
\end{landscape}

%% file: texfiles/res_discovery_bC.tex
\begin{landscape}
\begin{table}[hptb]
\begin{center}
\caption{The numbers of observed events in DM+\ttbar, wino NLSP, bCbv, and higgsino LSP SRs together with the expected numbers of background events and their uncertainties as predicted by the background-only fits, the normalisation factors (NF) for the background predictions obtained in the fit, the probabilities (represented by $p_0$ values, and capped at 0.5) that the observed numbers of events are compatible with the background-only hypothesis, and the expected ($N^{\textrm{limit}}_{\textrm{non-SM}}$ exp.) and observed ($N^{\textrm{limit}}_{\textrm{non-SM}}$ obs.) 95\% CL upper limits on the number of beyond-SM events. Some of the SRs where \ttbar\ background events are predominantly semileptonic or dileptonic have only one NF which is also applied to the subdominant \ttbar\ contribution.
Backgrounds with no associated NF are normalised with the theoretical cross-sections.
\label{tab:discovery_SR_yields_bC} 
}
\vspace{3mm}
{\small
\setlength{\tabcolsep}{0.2pc}
{\renewcommand{\arraystretch}{1.3}
\begin{tabular}{|l|c|c|c|c|c|c|c|c|c|}
\hline
\hline
Signal region     & \DMhigh & \DMlow & \DMlowloose & \bCdiag & \bCmed & \bCbv & \bCsoftdiag & \bCsoftmed & \bCsofthigh \\ \hline
Observed          & $5$ & $13$ & $65$ & $22$ & $4$ & $25$ & $33$ & $19$ & $2$ \\
\hline
Total background  & $7.4\pm2.1$ & $13.8\pm3.6\phantom{0}$ & $48.3\pm8.2\phantom{0}$ & $21.3\pm5.0\phantom{0}$ & $5.8\pm1.6$ & $25.1\pm3.8\phantom{0}$ & $24.7\pm3.1\phantom{0}$ & $13.7\pm2.1\phantom{0}$ & $1.8\pm0.3$ \\ \hline

$t \bar t\ 2\ell$ & $0.82\pm0.27$ & $2.21\pm0.58$ & $16.0\pm5.7\phantom{0}$ & $6.4\pm2.4$ & $1.36\pm0.49$ & $1.25\pm0.65$ & - & - & - \\

$t \bar t\ 1\ell$ & $0.0\pm0.0$ & $0.07\pm0.03$ & - & $0.28\pm0.18$ & $0.04_{-0.04}^{+0.13}$ & - &  $10.3\pm2.4\phantom{0}$ & $4.9\pm1.5$  & $0.36\pm0.15$ \\

$\ttbar + V$   & $4.0\pm2.0$ & $6.7\pm3.2$ & $14.3\pm5.9\phantom{0}$ & $7.8\pm3.3$ & $0.71\pm0.38$&  $0.58\pm0.16$ & $0.14\pm0.06$ & $0.44\pm0.10$ & $0.05\pm0.02$ \\

Single top     & $0.33\pm0.16$ & $0.65\pm0.57$ & $3.4\pm1.3$ & $5.5\pm2.4$ & $3.0\pm1.5$ &$0.60\pm0.54$ & $3.5\pm1.5$ & $1.6\pm0.5$ & $0.23\pm0.11$ \\
 
$W$+jets       & $1.64\pm0.53$ & $3.2\pm1.3$ & $11.0\pm2.8\phantom{0}$ & $1.22\pm0.35$ & $0.54\pm0.14$ & $16.5\pm3.1\phantom{0}$ & $8.0\pm2.0$ & $6.4\pm2.0$ & $1.06\pm0.24$ \\

Diboson        & $0.66\pm0.21$ & $0.98\pm0.33$ & $3.6\pm1.3$ & $0.23\pm0.08$ & $0.07\pm0.04$ & $6.1\pm2.0$ & $2.21\pm0.93$ & $0.31\pm0.16$ & $0.04\pm0.01$ \\
$Z$+jets       & - & - & - & - & - & - &  $0.60\pm0.55$ & $0.17\pm0.16$ & $0.04\pm0.04$ \\

\hline
 $ t \bar t\ 2\ell$ NF
 &  $1.19\pm0.13$
 &  $1.06\pm0.12$ 
 &  $1.13\pm0.21$ 
 &  $1.28\pm0.17$
 &  $1.58\pm0.22$
 &  $0.78\pm0.28$
 & -
 & -
 & -
\\

$ t \bar t\ 1\ell$ NF
 &  $1.08\pm0.14$
 &  $0.95\pm0.04$ 
 &  -
 &  $0.96\pm0.08$
 &  $0.75\pm0.15$
 & -
 & $0.73\pm0.11$
 & $0.92\pm0.07$
 & $0.93\pm0.16$
\\

$\ttbar + V$ NF 
 & $0.98\pm0.38$
 & $1.06\pm0.38$
 & $1.10\pm0.32$
 & $1.18\pm0.39$
 & $0.95\pm0.52$
 & -
 & -
 & -
 & -
 \\

Single top NF  
 & $0.94\pm0.37$
 & $1.05\pm0.35$
 & $1.22\pm0.27$
 & $1.59\pm0.45$
 & $1.17\pm0.37$
 & - 
 & -
 & $0.47\pm0.14$
 & $0.37\pm0.15$
 \\
 
$W$+jets NF
 & $1.08\pm0.21$
 & $1.04\pm0.18$
 & $0.93\pm0.10$
 & $0.80\pm0.24$ 
 & $1.11\pm0.25$
 & $1.07\pm0.09$
 & $1.19\pm0.26$
 & $1.35\pm0.24$
 & $1.11\pm0.19$
 \\

\hline
$p_0$ (\,$\sigma$) & 0.5 (-) & 0.5 (-) & 0.07 (1.5) & 0.45 (0.11) & 0.5 (-) & 0.5 (-) & 0.09 (1.34)& 0.12 (1.17) & 0.44 (0.16) \\
$N^{\textrm{limit}}_{\textrm{non-SM}}$ exp.& 7.2 & 11 & 23 & 14 & 6.4 & 13  & 13 & 9.6 & 4.1 \\
$N^{\textrm{limit}}_{\textrm{non-SM}}$ obs. & 5.7 & 10 & 37 & 14 & 5.2 & 13 & 20 & 14 & 4.3 \\
\hline
\hline
\end{tabular}
}
}
\end{center}
\end{table}
\end{landscape}

%% file: texfiles/res_exclusion.tex
\begin{landscape}
\begin{table}[hptb]
\begin{center}
\caption{The numbers of observed events in each bin of the shape-fit SRs together with the expected numbers of total background events and their uncertainties as predicted by the background-only fits. The bin $i$ ($i=1$--$5$) corresponds to the $i$-th bin (from left to right) of the variable used in the shape-fit. The bin boundaries of the shape-fits are detailed in Table~\ref{tab:SRs_tN}, \ref{tab:SRs_tN_diag_low_high}, \ref{tab:SRs_other}, and \ref{tab:SRs_bCsoft}.
\label{tab:exclusion_SR_yields}
}
\vspace{3mm}
{\small 
\setlength{\tabcolsep}{0.2pc}
{\renewcommand{\arraystretch}{1.3}
\begin{tabular}{|l l l|c|c|c|c|c|}
\hline
\hline
Signal region & Fitted variable  &   & bin1  & bin2  & bin3 & bin4 & bin5 \\
\hline

\tNmed & \met  & Observed & $21$ & $17$ & $8$ & $4$ & -- \\
& & Total background  & $14.6 \pm 2.8$ & $11.2 \pm 2.2$ & $7.3 \pm 1.7$ & $3.16 \pm 0.74$ & -- \\
\hline

\tNdiaghigh & BDT\_high & Observed & $40$ & $41$ & $19$ & -- & -- \\
& & Total background & $47.3 \pm 3.6$          & $37.5 \pm 3.5$ & $18.3 \pm 2.2$ & -- & -- \\
\hline

\tNdiagmed & BDT\_med & Observed & $970$ & $678$ & $366$ & $211$ & $40$  \\
& & Total background & $886 \pm 83$ & $618 \pm 86$ & $440 \pm 71$ & $210 \pm 30$ & $51 \pm 10$ \\
\hline

\bWN & \amtTwo & Observed & $13$ & $19$ & $22$ & $30$ & $36$ \\
& & Total background & $16.5 \pm 4.5$ & $16.0 \pm 6.0$ & $25.6 \pm 5.3$ & $40.1 \pm 8.1$ & $38.5 \pm 8.3$ \\
\hline

\bffN & \lepPtoverMET & Observed & $9$ & $27$ & $34$ & -- & -- \\
& & Total background & $4.6 \pm 1.1$ & $22.9 \pm 3.1$ & $32.5 \pm 4.1$ & -- & -- \\
\hline

\bCsoftdiag & \lepPtoverMET & Observed & $4$ & $16$ & $13$  & -- & -- \\
& & Total background & $1.69 \pm 0.47$  & $9.3 \pm 2.1$  & $13.6 \pm 2.8$ & -- & -- \\
\hline

\bCsoftmed & \lepPtoverMET & Observed  & $4$ & $15$ & $57$ & -- & --\\
& & Total background & $4.92 \pm 0.90$ & $8.9 \pm 1.3$ & $52.9 \pm 6.2$ & -- & -- \\
\hline

\bCsofthigh & \lepPtoverMET & Observed & $1$ & $1$ & $15$  & -- & --\\ 
& & Total background & $0.67 \pm 0.13$ & $1.11 \pm 0.22$  & $6.98 \pm 0.81$ & -- & -- \\ 
\hline
\hline

\end{tabular}
}
}
\end{center}
\end{table}
\end{landscape}

%% file: texfiles/conclusion.tex
\section{Summary and conclusions}
\label{sec:conclusions}

This paper presents searches for direct top-squark pair production covering various SUSY scenarios and for a spin-0 mediator decaying into pair-produced dark-matter particles produced in association with \ttbar\ using the final state with one isolated lepton, jets, and \met. Thirteen signal-region selections are optimised for the discovery of a top-squark signature. The analysis also defines three signal-region selections for spin-0 mediator models.

The search uses \ourLumi\ of $pp$ collision data collected by the ATLAS experiment at the LHC at a centre-of-mass energy of $\sqrt{s}=13$\,\TeV. No significant excess is observed over the estimated Standard Model backgrounds. Exclusion limits at 95\% confidence level are derived for the considered models.

These results improve upon previous exclusion limits by excluding the top-squark mass region up to 940\,\GeV\ for a massless lightest neutralino and assuming a 100\% branching ratio for $\topLSP$. Exclusion limits are also improved in pMSSM models targeting various sparticle mass spectra. For the wino NLSP model, the top-squark mass region up to 885\,\GeV\ (940\,\GeV) is excluded in scenarios with $\mu <0$ ($\mu > 0$) and a 200\,\GeV\ neutralino. For the higgsino LSP model, the top-squark mass region up to 860\,\GeV\ (800\,\GeV) is excluded in scenarios with $m_{q3L}<m_{tR}$ ($m_{tR}<m_{q3L}$). 
Furthermore, in a model with well-tempered neutralinos, the top-squark mass region up to 810\,\GeV\ is excluded in scenarios with $m_{q3L}<m_{tR}$ while no limit is set in scenarios with $m_{tR}<m_{q3L}$. 

For the spin-0 mediator models, a scalar (pseudoscalar) mediator mass of around 100\,\GeV\ (20\,\GeV) is excluded at 95\% confidence level, assuming a 1 \GeV\ dark-matter particle mass and a common coupling to SM and dark-matter particles of $g=1$.

%% file: acknowledgements/Acknowledgements.tex

We thank CERN for the very successful operation of the LHC, as well as the
support staff from our institutions without whom ATLAS could not be
operated efficiently.

We acknowledge the support of ANPCyT, Argentina; YerPhI, Armenia; ARC, Australia; BMWFW and FWF, Austria; ANAS, Azerbaijan; SSTC, Belarus; CNPq and FAPESP, Brazil; NSERC, NRC and CFI, Canada; CERN; CONICYT, Chile; CAS, MOST and NSFC, China; COLCIENCIAS, Colombia; MSMT CR, MPO CR and VSC CR, Czech Republic; DNRF and DNSRC, Denmark; IN2P3-CNRS, CEA-DRF/IRFU, France; SRNSF, Georgia; BMBF, HGF, and MPG, Germany; GSRT, Greece; RGC, Hong Kong SAR, China; ISF, I-CORE and Benoziyo Center, Israel; INFN, Italy; MEXT and JSPS, Japan; CNRST, Morocco; NWO, Netherlands; RCN, Norway; MNiSW and NCN, Poland; FCT, Portugal; MNE/IFA, Romania; MES of Russia and NRC KI, Russian Federation; JINR; MESTD, Serbia; MSSR, Slovakia; ARRS and MIZ\v{S}, Slovenia; DST/NRF, South Africa; MINECO, Spain; SRC and Wallenberg Foundation, Sweden; SERI, SNSF and Cantons of Bern and Geneva, Switzerland; MOST, Taiwan; TAEK, Turkey; STFC, United Kingdom; DOE and NSF, United States of America. In addition, individual groups and members have received support from BCKDF, the Canada Council, CANARIE, CRC, Compute Canada, FQRNT, and the Ontario Innovation Trust, Canada; EPLANET, ERC, ERDF, FP7, Horizon 2020 and Marie Sk{\l}odowska-Curie Actions, European Union; Investissements d'Avenir Labex and Idex, ANR, R{\'e}gion Auvergne and Fondation Partager le Savoir, France; DFG and AvH Foundation, Germany; Herakleitos, Thales and Aristeia programmes co-financed by EU-ESF and the Greek NSRF; BSF, GIF and Minerva, Israel; BRF, Norway; CERCA Programme Generalitat de Catalunya, Generalitat Valenciana, Spain; the Royal Society and Leverhulme Trust, United Kingdom.

The crucial computing support from all WLCG partners is acknowledged gratefully, in particular from CERN, the ATLAS Tier-1 facilities at TRIUMF (Canada), NDGF (Denmark, Norway, Sweden), CC-IN2P3 (France), KIT/GridKA (Germany), INFN-CNAF (Italy), NL-T1 (Netherlands), PIC (Spain), ASGC (Taiwan), RAL (UK) and BNL (USA), the Tier-2 facilities worldwide and large non-WLCG resource providers. Major contributors of computing resources are listed in Ref.~\cite{ATL-GEN-PUB-2016-002}.

%% file: atlas_authlist.tex
 
\begin{flushleft}
{\Large The ATLAS Collaboration}

\bigskip

M.~Aaboud$^\textrm{\scriptsize 34d}$,    
G.~Aad$^\textrm{\scriptsize 99}$,    
B.~Abbott$^\textrm{\scriptsize 124}$,    
O.~Abdinov$^\textrm{\scriptsize 13,*}$,    
B.~Abeloos$^\textrm{\scriptsize 128}$,    
S.H.~Abidi$^\textrm{\scriptsize 165}$,    
O.S.~AbouZeid$^\textrm{\scriptsize 143}$,    
N.L.~Abraham$^\textrm{\scriptsize 153}$,    
H.~Abramowicz$^\textrm{\scriptsize 159}$,    
H.~Abreu$^\textrm{\scriptsize 158}$
% ...
\bigskip
\\

% ...
$^{13}$Institute of Physics, Azerbaijan Academy of Sciences, Baku; Azerbaijan.\\
% ...
$^{34}$$^{(a)}$Facult\'e des Sciences Ain Chock, R\'eseau Universitaire de Physique des Hautes Energies - Universit\'e Hassan II, Casablanca;$^{(b)}$Centre National de l'Energie des Sciences Techniques Nucleaires (CNESTEN), Rabat;$^{(c)}$Facult\'e des Sciences Semlalia, Universit\'e Cadi Ayyad, LPHEA-Marrakech;$^{(d)}$Facult\'e des Sciences, Universit\'e Mohamed Premier and LPTPM, Oujda;$^{(e)}$Facult\'e des sciences, Universit\'e Mohammed V, Rabat; Morocco.\\
% ...
$^{99}$CPPM, Aix-Marseille Universit\'e, CNRS/IN2P3, Marseille; France.\\
% ...
$^{124}$Homer L. Dodge Department of Physics and Astronomy, University of Oklahoma, Norman OK; United States of America.\\
% ...
$^{128}$LAL, Universit\'e Paris-Sud, CNRS/IN2P3, Universit\'e Paris-Saclay, Orsay; France.\\
% ...
$^{143}$Santa Cruz Institute for Particle Physics, University of California Santa Cruz, Santa Cruz CA; United States of America.\\
% ...
$^{153}$Department of Physics and Astronomy, University of Sussex, Brighton; United Kingdom.\\
% ...
$^{158}$Department of Physics, Technion, Israel Institute of Technology, Haifa; Israel.\\
$^{159}$Raymond and Beverly Sackler School of Physics and Astronomy, Tel Aviv University, Tel Aviv; Israel.\\
% ...
$^{165}$Department of Physics, University of Toronto, Toronto ON; Canada.\\
% ...
$^{a}$ Also at  Department of Physics, University of Malaya, Kuala Lumpur; Malaysia.\\
$^{b}$ Also at Borough of Manhattan Community College, City University of New York, NY; United States of America.\\
$^{c}$ Also at Centre for High Performance Computing, CSIR Campus, Rosebank, Cape Town; South Africa.\\
$^{d}$ Also at CERN, Geneva; Switzerland.\\
$^{e}$ Also at CPPM, Aix-Marseille Universit\'e, CNRS/IN2P3, Marseille; France.\\
% ...
$^{*}$ Deceased

\end{flushleft}
